\documentclass[preprint, 1p, showpacs, superscriptaddress, merge, sort&compress, times]{elsarticle}
\usepackage{pifont, geometry,fleqn, graphicx, txfonts}
\usepackage{amsmath}
\usepackage{amssymb}
\usepackage{amsfonts}
\usepackage{fontenc}

\usepackage{hyperref}
\hypersetup{
        colorlinks=true,
        linkcolor=blue
}
\newcommand{\lcol}[1]{{\color{blue} #1}}

\numberwithin{equation}{section}

\usepackage[vcentermath]{youngtab}  
\usepackage{multicol}        
\usepackage{multirow}        

\DeclareMathAlphabet\mathbfcal{OMS}{cmsy}{b}{n}

\newcommand{\Tr} {\operatorname{Tr}}
\newcommand{\subl}{\mathbb{L}}

\newcommand{\be}{\begin{equation}}
\newcommand{\ee}{\end{equation}}

\DeclareMathOperator{\tr}{tr}
\newcommand{\dq}{q}

\newcommand{\bra}[1]{\mbox{$\langle #1 |$}}

\newcommand{\ket}[1]{\mbox{$| #1 \rangle$}}
\newcommand{\braket}[2]{\mbox{$\langle #1  | #2 \rangle$}}

\newcommand{\fL}{\mathcal{L}}
\newcommand{\fO}{\mathcal{O}}
\newcommand{\bp}{\begin{pmatrix}}
\newcommand{\ep}{\end{pmatrix}}
\newcommand{\bea}{\begin{eqnarray}}
\newcommand{\eea}{\end{eqnarray}}
\newcommand{\vk}{\vec{k}}
\newcommand{\ve}{\vec{e}}
\newcommand{\vn}{\vec{n}}
\newcommand{\vq}{\vec{q}}

\newcommand{\vx}{\vec{x}}
\newcommand{\sig}{\sigma}
\newcommand{\fD}{\mathcal{D}}
\newcommand{\vsig}{\vec{\sig}}
\newcommand{\vy}{\vec{y}}

\usepackage{xspace}

\usepackage{tikz}
\usetikzlibrary{calc}
\usetikzlibrary{backgrounds}
\tikzset{spin/.style={circle=2pt,draw=black!100,fill=orange!80,inner sep=3pt}}

\newcommand{\id}{I}

\def\L{{\cal L}}

\def\tr{ \mbox{tr}}

\def\Tr{\mbox{Tr}}

\begin{document}
\title{Generalization of the Haldane conjecture to $\mbox{SU}(3)$ chains}
\date{\today} 

\author[epfl]{Mikl\'os Lajk\'o\corref{correspondingauthor}}
\author[ubc]{Kyle Wamer}
\author[epfl]{Fr\'ed\'eric Mila}
\author[ubc]{Ian Affleck}
\address[epfl]{Institute of Physics, Ecole Polytechnique F\'ed\'erale de Lausanne (EPFL), CH-1015 Lausanne, Switzerland}
\address[ubc]{Department of Physics and Astronomy, and Stewart Blusson Quantum Matter Institute, The University of British Columbia, Vancouver, B.C., Canada, V6T 1Z1}
\cortext[correspondingauthor]{Corresponding author}

\begin{abstract}
We apply field theory methods to $\mbox{SU}(3)$ chains in the symmetric representation, with $p$ boxes in the Young tableau, 
mapping them into a flag manifold nonlinear $\sigma$-model with a topological angle $\theta =2\pi p/3$. Generalizing the Haldane conjecture, 
we argue that the models are gapped for $p=3m$ but gapless for $p=3m\pm 1$ (for integer $m$), corresponding to a massless 
phase of the $\sigma$-model at $\theta =\pm 2\pi /3$. We confirm this with Monte Carlo calculations on the $\sigma$-model. 
\end{abstract}

   
\maketitle

\section{Introduction }
\label{sec:intro}
Almost thirty-five years ago, the ``Haldane conjecture'' \cite{ HaldanePRL1983, HaldanePLA1983} was a revolutionary discovery in both  condensed matter  and high energy physics. It was already well-known that antiferromagnetic chains did not have 
N\'eel-ordered ground states, due to the Mermin-Wagner-Coleman \cite{MerminWagner1966, Coleman1973} theorem forbidding spontaneous breaking of continuous symmetries in (1+1) dimensions. But Haldane argued that 
the behaviour was qualitatively different for integer and half-integer spin ($s$).  For half-integer spin there is power-law decay of the alternating spin correlations and gapless excitations.  
For integer spin there is exponential decay of the correlation function and a gap to all excited states. Previously it had generally been expected that they were gapless for all $s$, largely 
based on Bethe ansatz results for $s=1/2$. Haldane's argument  hinged on the large-$s$ limit, in which he mapped 
the low energy degrees of freedom of spin chains into the relativistic O(3) \mbox{nonlinear} $\sigma$-model with topological angle $\theta =2\pi s$. 
This model was quite popular in the high energy theory community at that time, as a simplified lower dimensional version of Quantum Chromodynamics (QCD)\cite{Elitzur1983}. Like QCD, the model 
is asymptotically free, with a renormalized coupling constant that flows to strong coupling at low energies, resulting in the perturbatively massless particles  becoming massive. 
The models also share an integer-valued topological charge and an associated topological angle. 
The O(3) nonlinear $\sigma$-model was already 
well-understood for $\theta=0$, due to its integrability, having a simple spectrum consisting of a massive triplet \cite{Zamolodchikov1979}.  While no exact results existed at that time for $\theta =\pi$, 
numerical results seemed to indicate that it remained massive \cite{BhanotNuclPhys1984, BhanotPRL1984}.  Haldane argued that, in fact, the model was massless for $\theta =\pi$, an unexpected result, which 
might have implications for QCD at $\theta =\pi$. Ironically, the surprise to the condensed matter community was the massive behaviour for integer $s$ and the surprise to the high 
energy community was the massless behaviour for $\theta =\pi$! 
In the following years, Haldane's results have been confirmed experimentally, with the measurement of gaps for quasi-1D spin-1 chains \cite{RenardRegnault2003review}, and numerically for  chains 
of $s=1$ and higher \cite{BotetPRB1983,NightingalePRB1986, KennedyJPhys1990,WhitePRB1993,Schollwockspin2PRB1996,todoPRL2001}.  The field theory prediction has also been confirmed numerically \cite{ WieseMC1995,Azcoiti2003, AllesPapa2008, Azcoiti2012,WiesePRD2012,AllesPapa2014} by Monte Carlo calculations, although the topological term  presents severe challenges since the Boltzmann 
factor is not positive-definite for $\theta \neq 0$.  A massless integrable model corresponding to $\theta=\pi$ was eventually found \cite{Zamolodchikov1992}.

Extension of these results to the group $\mbox{SU}(n)$ is of interest for several reasons. Cold atom experiments can realize $\mbox{SU}(n)$ chains with various representations \cite{WuPRL2003,HonerkampHofstetter2004,Cazalilla2009,gorshkov2010,BieriSU32012,Scazza2014,takahashi2012, Pagano2014,ZhangScience2014, CazalillaReyreview2014, Capponi_SUNreview_AnnPhys2016}. Mappings to relativistic field theories are also possible, in the limit of large representations, raising the possibility that other nonlinear $\sigma$-models might have massless 
phases driven by topological terms. A possible application of such field theory results to condensed matter physics exists. It was argued that the replica limit, $n\to 0$, 
of the $\mbox{U}(2n)/[\mbox{U}(n)\times \mbox{U}(n)]$ nonlinear $\sigma$-model with $\theta =\pi$ describes the delocalization transition in the integer quantum Hall effect \cite{LevinePRL1983,Affleck1986, FerdinandMirlinRevModPhys2008}. While this transition has been studied numerically \cite{ChalkerCoddington1988, HuckesteinKramerPRL1990, LeeWangKivelsonPRL1993,ZimbauerAnnPhys1994,ZimbauerJMathPhys1997,SlevinOhtsukiPRB2009,AmadoPRL2011}, no exact solution has yet been found for the critical exponents, despite thirty-five years of efforts. 

The goal of this paper is to extend Haldane's results to  $\mbox{SU}(3)$ chains with a particular set of representations  having a Young tableau consisting of a single row of $p$ boxes. We will 
argue that a gap exists for $p=3m$, where $m$ is an integer, but that the models are gapless for other values of $p$. Following Haldane, our approach is based 
on mapping the models into a relativistic quantum field theory at large $p$: the $\mbox{SU}(3)/[\mbox{U}(1)\times \mbox{U}(1)]$ nonlinear $\sigma$-model, defined on a space 
known as a ``flag manifold''. Since $\pi_2\big[ $\mbox{SU}(3)$/[\mbox{U}(1)\times \mbox{U}(1)]\big] = \mathbb{Z}\times \mathbb{Z}$, there are two topological angles which can appear in the Hamiltonian. We find that for translationally invariant systems the 
corresponding topological angles have equal and opposite values, $\pm \theta$, with $\theta =2\pi p/3$. We present Monte Carlo (MC) results indicating that the models are massive for $\theta =0$ 
but massless for $\theta =\pm 2\pi /3$. This leads to an extension of the ``Haldane conjecture'': we expect the SU(3) chains to be massive for $p=3m$ but 
massless in other cases.  We note that a gap for $p=3m$ was conjectured by Greiter {\it et al.}  \cite{GreiterRachelSchuricht2007,*GreiterRachel2007}.

A novel feature of this flag manifold nonlinear $\sigma$-model is that its Lagrangian contains an additional term, linear in both space and time derivatives, which is 
{\it not} a total derivative and therefore not a topological term and which is generated from the  $\mbox{SU}(3)$ chain models. 
We study this term using the renormalization group (RG) \cite{Polyakov:1975rr}, finding that the corresponding coupling constant is relevant at low energies.

\begin{figure}[h]
\begin{center}
\includegraphics[width=0.95\textwidth]{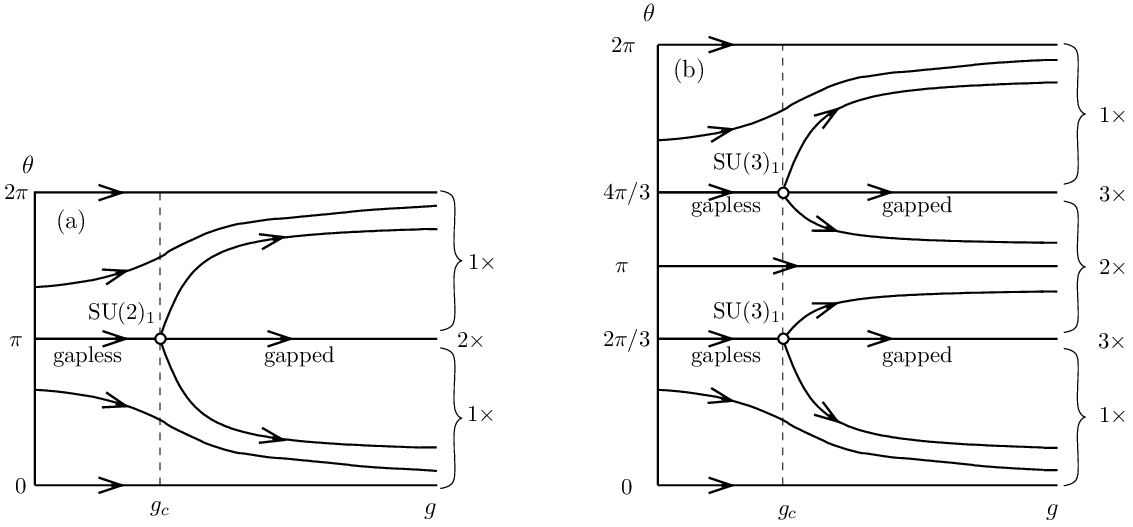}
\caption{(a) The renormalization group flow diagram of the $\mbox{O}(3)$ nonlinear $\sigma$-model, as proposed  in Ref.~\cite{ AffleckLesHouches1988}. At $\theta=\pi$ the system undergoes a phase transition from a gapless phase at $g<g_c$ into a gapped phase with a spontaneously broken $\mathbb{Z}_2$ symmetry at $g>g_c$. For $\theta\neq \pi$ the system is gapped with a unique ground state for all values of $g$.  (b) Proposed renormalization group flow diagram for the $\mbox{SU}(3)/[U(1)\times U(1)]$ nonlinear $\sigma$-model in the special case where the two topological angles are equal and opposite. 
At $\theta=2\pi/3$ and $4\pi/3$ the system undergoes a phase transition from a gapless phase at $g<g_c$ into a gapped phase with a spontaneously broken $\mathbb{Z}_3$ symmetry at $g>g_c$. For $2\pi/3<\theta <4\pi/3$ the system is gapped with a spontaneously broken $\mathbb{Z}_2$ symmetry, while for $\theta<2\pi/3$ and $\theta> 4\pi/3$ the system is gapped with a unique ground state for all values of $g$.} 
\label{fig:SU23RGflow}
\end{center}
\end{figure}

A  detailed picture of the RG flow has been obtained in the SU(2) case, with direct implications for spin chains \cite{AffleckLesHouches1988}. This is sketched in Fig.~\ref{fig:SU23RGflow}\lcol{a}. The massless critical 
point of the $\sigma$-model at $\theta =\pi$ was shown to correspond to a different nonlinear $\sigma$-model which is conformally invariant: the $\mbox{SU}(2)_1$ 
Wess-Zumino-Witten (WZW) model \cite{Affleck1986,ShankarReadNuclPhysB1990}. It is important to note that the $O(3)$ $\sigma$-model is not conformally invariant, but flows to this conformally 
invariant critical theory for $\theta =\pi$ and sufficiently small bare coupling, $g<g_c$. The appearance of this conformal field theory (CFT) is 
very natural, given the $\mbox{SU}(2)$ symmetry and the fact that the $\mbox{SU}(2)_k$ models for $k>1$ contain relevant operators allowed by symmetry \cite{AffleckHaldane1987}.
For $g>g_c$, the model goes into a gapped phase with a spontaneously 
broken $\mathbb{Z}_2$ symmetry.  In the spin chain, this symmetry corresponds to translation by one site, so the symmetry broken phase is dimerized \cite{MajumdarGhosh1969, Haldane_dimerization_PRB1982}.  The bare coupling constant, $g$, can be 
increased by adding frustrating antiferromagnetic next nearest neighbour exchange, $J_2$, which has been shown to produce this transition, with $g-g_c\propto J_2-J_{2c}$. \cite{Okamoto1992, EggertPRB1996}. Moving $\theta$ 
away from $\pi$ also produces a gapped phase. This can be achieved in the spin chain by adding alternating exchange interactions, breaking translation 
symmetry by hand. The scaling of the gap with $g-g_c$ and $\theta -\pi$ was predicted using the $\mbox{SU}(2)_1$ WZW model \cite{Affleck1986}. The transition along the $\theta =\pi$ line
is controlled by the marginal, symmetry preserving operator $\vec J_R\cdot \vec J_L$ where $\vec J_{R/L}$ are the right and left-moving current operators, 
with coupling constant $\propto g-g_c$.  One sign of the coupling is marginally irrelevant and the other marginally relevant, 
leading to the transition at $g=g_c$ and the gap is exponentially small in $g-g_c$. Moving $\theta$ away from $\pi$ corresponds to adding a term $(\theta -\pi )\hbox{tr} g$ to the effective Hamiltonian, 
where $g$ is the primary field of the WZW model, an $\mbox{SU}(2)$ matrix field of dimension $d=1/2$. Thus the gap is expected to scale as $|\theta -\pi |^{1/(2-d)}=|\theta -\pi |^{2/3}$, 
up to log corrections coming from the marginal operator. 
Our predicted phase diagram for the $\mbox{SU}(3)$ $\sigma$-model in the special case when the two topological angles are equal and opposite is sketched in Fig.~\ref{fig:SU23RGflow}\lcol{b}. 
We identify the critical theory at $\theta =\pm 2\pi /3$ with the $\mbox{SU}(3)_1$ WZW model.  We again expect a gapped phase for $g>g_c$ and for non-zero $\theta \mp 2\pi /3$ and  can again predict the gap scaling. A more general phase diagram in which the two topological angles can vary independently will be discussed in Sec.~\ref{sec:numerics}.  

There are three pieces of rigorous evidence for the $\mbox{SU}(2)$ phase diagram. One is the Bethe Ansatz solution for $s=1/2$ \cite{Hulthen1938}, giving the expected gapless ground state 
with no broken symmetries. Another is provided by the Lieb-Schultz-Mattis-Affleck (LSMA) theorem \cite{LSM1961,AffleckLieb1986} which proves that the model is either gapless 
or has a ground state degeneracy for half-integer (but not integer) spin.  The third is provided by the Affleck-Kennedy-Lieb-Tasaki (AKLT) models for integer spin \cite{AKLT1988}. 
The exact ground states were found for these models and seen to be gapped with no broken symmetries. We observe that these results carry over simply to  $\mbox{SU}(3)$. 
The $p=1$ case is the Sutherland model, solvable by Bethe ansatz \cite{Sutherland1975}, and known to have a gapless low energy theory corresponding to $\mbox{SU}(3)_1$ \cite{TsvelikWiegmann1983,AndreiFuruyaLowensteinRMP1983,AffleckSUn1988}.
The  LSMA theorem was proven for general $\mbox{SU}(n)$ and implies, for $\mbox{SU}(3)$,  either a gapless ground state or a ground state degeneracy for $p\neq 3m$ \cite{AffleckLieb1986}. The AKLT construction was also generalized to different   $\mbox{SU}(n)$ spin chains \cite{Katsura2008, QuellaPRB2013, LecheminantEPL2013,MorimotoFurusakiPRB2014}, in particular
 to the fully symmetric $\mbox{SU}(3)$ case  with   $p=3m$ by \citet{GreiterRachelSchuricht2007}, who constructed Hamiltonians whose exact ground states can be found and which appear to be gapped with no ground state degeneracy.

Several important apparent contradictions and open questions are raised by our results. According to the most recent numerical results \cite{NatafchainED2016, WeichselbaumPC}, there is no indication of a gap or finite correlation length for $p=3$. Besides, 
while numerical results are consistent with no gap for $p=2$, the corresponding critical exponents appear to be those of $\mbox{SU}(3)_2$, not $\mbox{SU}(3)_1$ \cite{NatafchainED2016, WeichselbaumPC}. (For a detailed discussion of numerical results, see Sec.~\ref{sec:conclusion}). Analogous to the $\mbox{SU}(2)$ case, 
we argue that the WZW models with $k>1$ are unstable, containing relevant operators allowed by symmetry, so $\mbox{SU}(3)_1$ appears to be the only viable candidate for the critical point \cite{AffleckSUn1988}.
We suspect that these two discrepancies may be a result of a long cross-over length scale beyond which the true low energy physics becomes observable. In this regard it is interesting to recall 
that the Haldane conjecture remained controversial for several years until reliable numerical results became available for sufficiently large systems, greatly 
aided by the development of the Density Matrix Renormalization Group (DMRG) technique \cite{WhiteDMRG1992, WhitePRB1993, Schollwockspin2PRB1996}. Finally, we have not been able to obtain the critical exponents of $\mbox{SU}(3)_1$ for the flag manifold $\sigma$-model because of the limitations of the MC approach in extracting the critical exponents.

In Sec.~\ref{sec:FW} we write the Hamiltonian for our $\mbox{SU}(3)$ lattice model and discuss flavour wave theory.  Although this erroneously predicts a classical ground state with spontaneously 
broken $\mbox{SU}(3)$ symmetry it still provides the starting point for the field theory. In Sec.~\ref{sec:rigresults} we review the Bethe ansatz integrable $p=1$ model, the LSMA theorem for the $\mbox{SU}(3)$ case 
and discuss the AKLT models of \citet{GreiterRachelSchuricht2007}.  Sec.~\ref{sec:FT} contains the derivation of the flag manifold $\sigma$-model at large $p$; we show that for translationally invariant spin models the two topological angles have opposite values $\pm \theta$ with 
$\theta =2\pi p/3$ and derive the unusual new term. In Sec.~\ref{sec:symms} we examine the symmetries of the $\sigma$-model, and their relation to the symmetries of the underlying spin model.  Sec.~\ref{sec:RG} contains our perturbative RG results.
Sec.~\ref{sec:numerics} discusses the general phase diagram spanned by the two topological angles based on calculations in the strong coupling limit and numerical Monte Carlo results. 
Sec.~\ref{sec:phasediag} discusses the nature of the gapless critical point, arguing that it should be the $k=1$ WZW model. 
Sec.~\ref{sec:conclusion}  contains conclusions and a discussion of open questions. Details and additional information can be found in a series of appendices.

\section{Flavour wave calculations}
\label{sec:FW}
\subsection{The model}

The goal of this paper is to investigate the properties of the SU(3) antiferromagnetic Heisenberg model with symmetric representations at each site, the generalization
of the SU(2) Heisenberg model with arbitrary spin. The model is defined by the Hamiltonian 
\begin{equation}
  {\cal H } = J \sum_i S^\alpha_\beta(i) S^\beta_\alpha(i+1) \;,
  \label{eq:H_nn}
\end{equation}
with $J>0$, and where the sum runs over the lattice sites $i$ and the repeated flavour indices $\alpha$
and $\beta$ that can take values 1, 2 or 3 corresponding to spins states $A$, $B$ or $C$. The operators $S_{\beta}^{\alpha}(l)$ are the generators of SU(3). 
They obey the SU(3) Lie algebra
\begin{equation}
  \left[S_{\beta}^{\alpha},S_{\beta'}^{\alpha'} \right] =
  S_{\beta}^{\alpha'}\delta_{\beta'}^{\alpha} - S_{\beta'}^{\alpha}\delta_{\beta}^{\alpha'}
  \label{eq:Scommrel}
\end{equation}
 where $\delta_{\beta}^{\alpha}$ is the Kronecker $\delta$ function.
 
The model is further specified by choosing the irreducible representation at each site. In this paper, we will concentrate on models 
with the same totally symmetric irreducible representation at each site. Fully symmetric representations correspond to Young-tableaux drawn with $p$ boxes arranged horizontally: \yng(1)\ ,\ \ \yng(2)\ ,\ \ \yng(3)\ ,\ \dots .
For $p=1$, this is the fundamental representation, the model is equivalent to quantum permutation of 3-flavour objects, and it is integrable by Bethe ansatz (see Sec.~\ref{sec:rigresults}).

For general $p$, the model can be reformulated using Schwinger bosons with three flavours. The generators at site $i$ can be written as

\begin{equation}
 S_{\beta}^{\alpha}(i) = b^{\dagger}_{\beta}(i) b^{\phantom{\dagger}}_{\alpha}(i)
 ,
 \label{eq:spindef1}
 \end{equation}
and the local Hilbert spaces are defined by putting $p$ bosons at each site, i.e.\ by the constraints 
\begin{equation}
b^{\dagger}_{\alpha}(i) b^{\phantom{\dagger}}_{\alpha}(i) =p.
\end{equation}
The resulting Hamiltonian is quartic in bosonic operators:
\begin{equation}
  {\cal H } = J \sum_i b^{\dagger}_{\beta}(i) b^{\phantom{\dagger}}_{\alpha}(i)b^{\dagger}_{\alpha}(i+1) b^{\phantom{\dagger}}_{\beta}(i+1).
\end{equation}

We note that the spin operators are usually defined to be traceless, i.e.\ $S^\alpha_\alpha$=0, which would correspond to the bosonic representation $\ S_{\beta}^{\alpha}(i) = b^{\dagger}_{\beta}(i) b^{\phantom{\dagger}}_{\alpha}(i) - p \delta_{\alpha,\beta}/n$. For convenience we use a slightly different definition in Eq.\ (\ref{eq:spindef1}), which leads to $S^\alpha_\alpha=p$. The difference between the two conventions  only gives constant terms in our calculations and doesn't change any of our conclusions.

\subsection{The classical limit}

In the following, we will investigate the properties of the model with the help of a field theory that describes fluctuations around a reference state that
should correspond to the ground state in the $p\rightarrow + \infty$ limit. As in SU(2), the candidate states are product wave functions in which, at
each site, all bosons are in the same state. This could be a pure flavour state, or a state corresponding to an SU(3) rotation in flavour space. If the coupling 
between two sites is antiferromagnetic, the energy is minimized if the flavour states are orthogonal, which will be achieved for instance if the states are
pure flavour states with different flavours. It the coupling is ferromagnetic, the energy is minimized if the flavour states are the same. Now, for the model of 
Eq.\ \eqref{eq:H_nn}, the classical ground state is not unique. The energy will be minimal as soon as neighbouring sites have orthogonal states, and this
can be achieved in an infinite number of ways ranging from the N\'eel state $ABABAB...$ to the three sublattice state $ABCABCABC...\,$. Further ground
states can be generated e.g.\ by changing locally $A$ or $B$ into $C$ in the N\'eel state, or by choosing after two consecutive orthogonal states, say $AB$, a state which is a rotation of the first one around the second one, i.e.\ a linear combination of $A$ and $C$. This is a major obstacle to the development of a
field theory to describe quantum fluctuations for two reasons. First of all, it is not clear {\it a priori} around which ground state the fluctuations should be introduced.
Secondly, and maybe more importantly, the semiclassical theory will have local zero modes, i.e.\ branches with vanishing velocity that make the theory
non relativistic.

There are several reasons however to believe that the three sublattice state $ABCABCABC...$ is the appropriate starting point. First or all, it is clear from the 
Bethe ansatz solution of the $p=1$ case that short-range order is of that type \cite{Sutherland1975}. Besides, in the large $p$ limit, it can be easily
shown that zero point fluctuations are minimal in that state because this is the only state where harmonic fluctuations are limited to pairs of neighbouring sites,
leading to a vanishing frequency for all modes, by contrast to any other state where at least some harmonic fluctuations will live on longer clusters, leading to
some non-vanishing frequencies \cite{CorbozPRX2012}. This suggests that the first effect of quantum fluctuations will be to select the three sublattice state by an `order-by-disorder' 
mechanism that generates effective additional couplings of order $1/p$ that lift the classical degeneracy. The simplest couplings that do the job are an antiferromagnetic coupling $J_2$ between next nearest neighbours or a ferromagnetic coupling $J_3$ between third neighbours. It will prove convenient to
keep both couplings for the discussion. So the model we will effectively study is defined by the Hamiltonian:
\begin{equation}
  {\cal H } = \sum_i ( J_1 {\cal H }_{i,i+1}+J_2 {\cal H }_{i,i+2}-J_3 {\cal H }_{i,i+3})
  \label{eq:H}
\end{equation}
with 
\begin{equation}
  {\cal H }_{i,j} = S^\alpha_\beta(i) S^\beta_\alpha(j)
\end{equation}
and with $J_1,J_2,J_3>0$. It should be kept in mind however that all the properties discussed in the rest of the paper are expected to apply
to the nearest neighbour Hamiltonian of Eq.\ \eqref{eq:H_nn}, and that the couplings $J_2,J_3=O(1/p)$ have been introduced as a first effect of 
quantum fluctuations in that model.

\subsection{Flavour wave theory}

For further reference, it will be useful to know the form of harmonic fluctuations around the $ABCABC...$ state, even if it is clear that these
fluctuations will destroy the order since we are in 1D. The calculation of these harmonic fluctuations is most easily done using 
linear flavour wave theory~\cite{N1984281,papa1988,JoshiZhang1999}, the extension of the usual SU(2) spin wave theory to $\mbox{SU}(N)$ models. 
At each site, the boson corresponding to the color of the classical ground state is condensed, and the generators are rewritten entirely in terms 
of the uncondensed bosons. Keeping only terms that are quadratic in the Hamiltonian, and after a Fourier transform on each sublattice, 
the Hamiltonian can be diagonalized by a Bogoliubov transformation, leading to
\begin{equation}
 \mathcal{H} = 
 p \sum_{k\in \textrm{RBZ}} \sum_{\alpha} \sum_{\beta\neq\alpha}
  \omega(k)
\tilde b^{(\alpha)\dagger}_{\beta,k} \tilde b^{(\alpha)}_{\beta,k}  + \text{const.}
   \;.
\end{equation}
In this expression, $k$ runs over the reduced Brillouin zone (RBZ) corresponding to the three site unit cell, and the operators $\tilde b^{(\alpha)\dagger}_{\beta,k},\tilde b^{(\alpha)}_{\beta,k} $
are Bogoliubov quasiparticles. The superscript $(\alpha)$ just keeps track of the sublattice. Details can be found in \ref{FWT}.
The dispersion of the flavour waves is given by
\begin{equation}
\begin{split}
  \omega(k)
= p\sqrt{\Big(J_1+J_2+2\big[1-\cos (3ka)\big]J_3\Big)^2- \Big(J_1^2+J_2^2+2J_1J_2\cos(3ka)\Big)}
\end{split}
\end{equation}
There are 6 degenerate branches in the reduced Brillouin zone, and 6 Goldstone modes. 
For small $k$, the dispersion is linear:
\begin{equation}
\begin{split}
  \omega(k)
\simeq 3 p \sqrt{J_1J_2+2J_1J_3+2J_2J_3} \ ka.
\end{split}
\end{equation}

\section{Rigorous results and $\mbox{SU}(3)_k$ critical points}
\label{sec:rigresults}

\subsection{Lieb-Schultz-Mattis-Affleck theorem}
\begin{figure}[htbp]
\begin{center}
\includegraphics[width=0.9\textwidth]{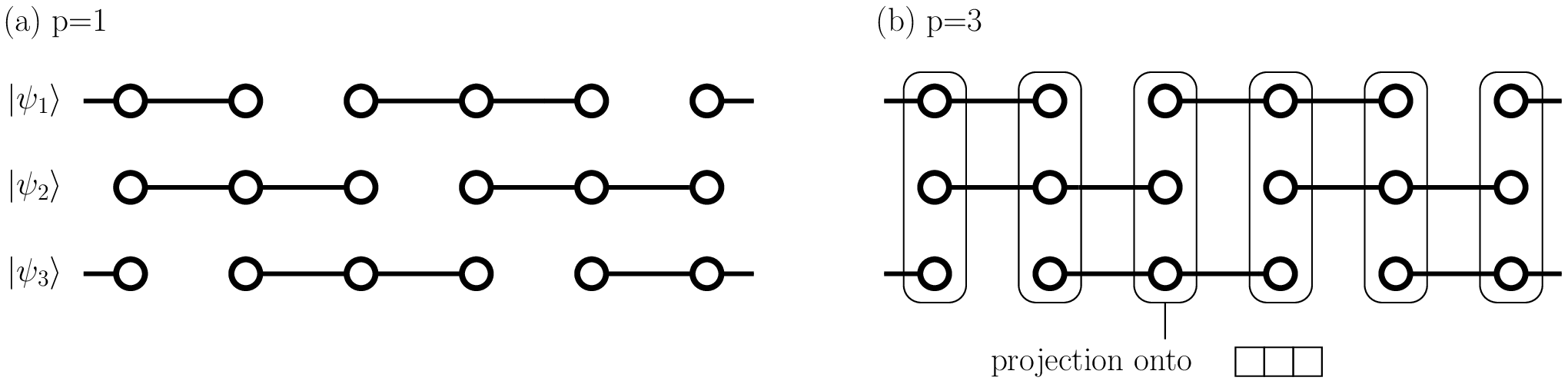}
\caption{Illustration of the exact ground states discussed by \citet{GreiterRachelSchuricht2007}. (a) Threefold degenerate trimerized ground states in the $p=1$ case, and (b) the uniqe ground state of an AKLT construction for the $p=3$ case.  See sections  III.A and VIII.B of Ref.\ \cite{GreiterRachelSchuricht2007} for the construction of the corresponding Hamiltonians.}
\label{fig:greiter_states}
\end{center}
\end{figure}

Let $\ket{\psi}$ be a ground state of the model defined in Eq.\ \eqref{eq:H_nn} on a system of length L (periodic boundary conditions assumed). Then we can obtain a low energy state by acting on $\ket{\psi}$ with the unitary operator \cite{AffleckLieb1986}:
\be U=\exp\left[i\frac{2\pi}{3L}\sum_{j=1}^LjQ_j\right],\ee
where $Q_j=S_1^1(j)+S_2^2(j)-2S_3^3(j)= p-3b_3^\dagger(j) b_3^{\phantom{\dagger}}(j)$ is a generator of $\mbox{SU}(3)$.  See  \ref{LSMA} for details. 
There we show that
\be \bra{\psi} T^{-1}UT \ket{\psi}=e^{i2\pi p/3} \bra{\psi} U \ket{\psi},\ee 
where $T$ is the operator which translates states by 1 site. 
Thus, translational invariance of $\ket{\psi}$ implies
\be  \bra{\psi} U \ket{\psi}=e^{i2\pi p/3} \bra{\psi} U \ket{\psi}.\ee 
This implies that $\bra{\psi}U \ket{\psi}=0$ for $p\neq 3m$, i.e.\ $U\ket{\psi}$ is a low energy state which is orthogonal to $\ket{\psi }$. This leaves two possibilities. If the ground state is unique, then there is a low energy excitation. 
Alternatively, there may be degenerate ground states in the thermodynamic limit, 
with the finite system containing an exponentially low energy excited state which is essentially a linear combinations of these 
ground states.  It can also be seen (\ref{LSMA}) that 
\be  \bra{\psi} U^2 \ket{\psi}=e^{i4\pi p/3} \bra{\psi} U \ket{\psi},\ee 
implying that $U^2 \ket{\psi}$ is another low energy state which is orthogonal to $\ket{\psi}$ and $U\ket{\psi}$, for $p\neq 3m$.  Furthermore, 
$\ket{\psi}$, $U\ket{\psi}$ and $U^2\ket{\psi}$ are all invariant under $T^3$, translation by 3 sites.
Thus, if there are no low energy excited states, we might expect a triplet of trimerized ground states, as illustrated in Fig.~\ref{fig:greiter_states}\lcol{a}. These 3 states map into each other under translations by 1 or 2 sites.  For a long finite system, we then expect linear combinations of these 3 ground states to give, to good approximation, the ground state and the two exponentially low-lying excited states as discussed above. 

We remark that, for $p=1$, a Hamiltonian was found by Greiter and Rachel \cite{GreiterRachelSchuricht2007} that has the simple trimer ground states.  Their Hamiltonian can be written as a sum of projection 
operators onto the total spin of each set of 4 nearest neighbour spins.  Noting that for the trimer ground states, 4 neighbouring spins can only be in the $3$ or $\bar 6$ 
representation, the Hamiltonian is chosen to give zero for those states and a positive energy for the other two possible representations that can occur from a product 
of 4 fundamental representations ($15$ and $15'$). These ground states have very short range correlations, vanishing for distances $>2$, and the model is expected to have a gap to all 
excitations.

\subsection{Bethe ansatz results and $\mbox{SU}(3)_k$ critical points}
The nearest neighbour $p=1$ model, was solved using the Bethe ansatz by Sutherland \cite{Sutherland1975} and its low energy degrees of freedom were shown to correspond to the $\mbox{SU}(3)_1$ WZW model \cite{TsvelikWiegmann1983,AndreiFuruyaLowensteinRMP1983,AffleckSUn1988}.
A simple way of understanding this result is to observe that, for $p=1$, we may represent the operators by 3 flavours of fermions\footnote{We remind the reader that for convenience we use a non-traceless convention for the spin operators}:
\be S^\alpha_\beta (j)=\psi_\beta^{\dagger}(j)\psi_\alpha^{\phantom{\dagger}} (j)
\ee
with the constraint of 1 particle on each site:
\be \psi_\alpha^{\dagger}(j)\psi_\alpha^{\phantom{\dagger}} (j)=1.\ee
The Hamiltonian becomes a simple exchange term:
\be \mathcal{H}=J\sum_j\psi_\beta^{\dagger}(j)\psi^{\phantom{\dagger}}_\alpha (j)\psi_\alpha^{ \dagger}(j+1)\psi^{\phantom{\dagger}}_\beta (j+1).\ee
We may obtain this model from an $\mbox{SU}(3)$ Hubbard model,
\be H=\sum_j\left\{-t[\psi_\alpha^{\dagger}(j)\psi_\alpha^{\phantom{\dagger}}(j+1)+h.c.]+U[\psi^{ \dagger}_\alpha(j)\psi_\alpha^{\phantom{\dagger}} (j)-1]^2\right\}\ee
in the limit $U/t\gg 1$. Starting at small $U$, we may take the continuum limit, giving 3 flavours of relativistic Dirac fermions.  We may then use non-abelian bosonization, 
which gives a charge boson plus the $\mbox{SU}(3)_1$ WZW model. The Hubbard interactions can be seen to gap the charge boson without effecting the low energy 
behavior in the spin sector, yielding the $\mbox{SU}(3)_1$ WZW model as the low energy effective theory.

Integrable $\mbox{SU}(3)$ spin chain models have been found for all $p$ \cite{AndreiJohannessonPLA1984, JohannessonPLA1986,JohannessonNuclPhysB1986}, with more complicated nearest neighbour interactions, and have been shown to correspond to the 
$\mbox{SU}(3)_p$ WZW models at low energies \cite{AffleckSUn1988,AlcarazMartins_JPhysA1989,FuhringerDMRG_AnnPhys2008}. This can again be understood from non-abelian bosonization. In this case we must introduce fermions with $p$ colours 
as well as $3$ flavours, and write a generalized Hubbard model. Non-abelian bosonization now gives the $\mbox{SU}(3)_p$ WZW model in the flavour sector, 
together with a charge boson and an $\mbox{SU}(p)_3$ WZW model for the colour degrees of freedom. However it is now seen that the Hubbard interactions will 
generally gap the flavour sector as well as the charge and colour sector, unless the interactions are fine-tuned.  This can be understood from the fact that 
the $\mbox{SU}(3)_p$ WZW models contain relevant operators allowed by symmetry for all $p>1$, which are expected to appear in the Hamiltonian and destabilize 
the critical theory. We understand the fine-tuned nature of the Bethe ansatz integrable models as, remarkably, corresponding to fine-tuning the field theory 
to eliminate all relevant operators. The only $\mbox{SU}(3)$ invariant relevant operator in the $\mbox{SU}(3)_1$ model is  $\tr(g)$ with dimension 2/3.  But the field theory 
representation implies that, under translation by 1 site, $g\to e^{i2\pi /3}g$, so this interaction is forbidden by translation symmetry, stablizing the $\mbox{SU}(3)_1$ critical point. 
For $p\neq 3m$, we may expect an RG flow from $\mbox{SU}(3)_p$ to $\mbox{SU}(3)_1$, consistent with the Zamolodchikov c-theorem \cite{Zamolodchikov1986}, which states that the central charge should decrease under RG flow. For $p=3m$ we expect an RG flow from $\mbox{SU}(3)_p$ to a gapped phase. 

\subsection{Affleck-Kennedy-Lieb-Tasaki states for $p=3m$}
\citet{GreiterRachelSchuricht2007} also found a Hamiltonian for $p=3m$ which has a unique, translationally invariant ground state, with spin correlations decaying with a finite 
correlation length. For the case $p=3$ this state is depicted in Fig.~\ref{fig:greiter_states}\lcol{b}. We regard each $p=3$ spin as consisting of three $p=1$ spins projected onto the symmetrized state. 
Then we combine three of these $p=1$ spins on three neighbouring sites to form a singlet. As can be seen from the figure, this state is actually translationally 
invariant since at each site, there is a trimer starting there and going to the right, a trimer starting there and going to the left, and a trimer centred there. 
The Hamiltonian is again written as a sum a projection operators, this time acting just on pairs of neighbouring sites. This time the energy is zero if two neighbouring sites 
are in the $\bar{10}$ or $27$ rep and is otherwise positive.
The model is expected to have an excitation gap. 
This construction can be straightforwardly generated to all cases where $p=3m$ by decomposing each spin into $3m$ $p=1$ spins and again drawing trimers, 
with $m$ going to the left, $m$ going to the right and $m$ centred at each site. 

Note that the above results are all consistent with the LSMA theorem. For $p=3m$ a gapped Hamiltonian can  be found which is translationally invariant but for $p=1$ a gapped 
Hamiltonian can only be found which is trimerized, breaking translational symmetry.

\section{$\mbox{SU}(3)/[\mbox{U}(1)\times\mbox{U}(1)]$ nonlinear $\sigma$-model}
 \label{sec:FT}
 
 Based on the discussion in Sec.~\ref{sec:FW} we consider an $\mbox{SU}(3)$ spin chain with the $p$-box symmetric representation at each site, and we investigate the low energy behaviour of a
Hamiltonian with antiferromagnetic nearest and next nearest and ferromagnetic third neighbour Heisenberg interactions defined as 
\begin{equation}
\begin{split}
\mathcal{H}= \sum_{i}  \left[J_1 S^\alpha_\beta(i) S^\beta_\alpha(i+1)+ J_2 S^\alpha_\beta(i) S^\beta_\alpha(i+2)- J_3 S^\alpha_\beta(i) S^\beta_\alpha(i+3)\right],
\end{split}
\label{eq:J1-J2model}
\end{equation}
keeping in mind that, in the large $p$ limit, the $J_2$ and $J_3$ terms can be considered as generated by quantum fluctuations. 
Here we only present the outline of our results, but we provide step by step calculations in \ref{sec:su3calc}. To serve as a comparison, we also provide similar calculations for the $\mbox{SU}(2)$ case in \ref{app:su2}. 

Using a spin coherent state path integral approach \cite{KlauderPRD1979,GnutzmannKus1998, ShibataTakagi1999, MathurSen2001}, one can write the imaginary time action of the model in Eq.\ \eqref{eq:J1-J2model}   as 
\begin{equation}
\begin{split}
S = \int\limits_0^\beta d\tau \sum_i \Bigg[ p^2\Big(&J_1  \left|\vec  \Phi^*(i)\cdot \vec \Phi(i+1,\tau) \right|^2 +J_2  \left|\vec  \Phi^*(i,\tau) \cdot \vec \Phi(i+2,\tau) \right|\\
&- J_3  \left|\vec  \Phi^*(i,\tau) \cdot \vec \Phi(i+3,\tau) \right|^2 \Big)
+p \Big(\vec{\Phi}^*(i,\tau) \cdot \partial_\tau \vec \Phi(i,\tau) \Big) \Bigg],
\end{split}
\label{eq:fullLUaction}
\end{equation}
where $\vec \Phi(i,\tau)$ is a three dimensional complex unit vector at site $i$ and imaginary time $\tau$, while $\beta$ is the inverse temperature.  For  antiferromagnetic $J_1,J_2$ and ferromagnetic $J_3$ the real part of the action is minimal for the classical three sublattice ground state manifold, which can be parametrized by a set of three orthogonal spin states corresponding to the rows of  $U$, a unitary matrix. Since the action is invariant under changing the overall phase of each of the three  spin states in $U$, we argue that ground state manifold is isomorphic to $\mbox{SU}(3)/[\mbox{U}(1) \times \mbox{U}(1)]$  \cite{UedaShannon2016}: 
 two phases can be changed independently, later referred to as gauge invariance  (see Sec~ \ref{subsec:gaugeinv}),  while the third phase is fixed by setting the determinant of $U$ to 1.  $\mbox{SU}(3)/[\mbox{U}(1) \times \mbox{U}(1)]$ has 6 generators, namely  the six off-diagonal Gell-Mann matrices corresponding to the six Goldstone modes discussed in the flavour-wave approximations in Sec.~\ref{sec:FW}.

 Considering low energy fluctuations around the classical ground state manifold, on the one hand the $U$ matrix can depend on the position,  corresponding to the slow joint rotation of the orthogonal states of the three site unit cell.  On the other hand the states inside a unit cell can also be non-orthogonal to each other. Accordingly, the low energy configurations can be described as \cite{SmeraldPRB2013, Smeraldthesis2013}
\begin{equation}
\label{eq:spinstates0}
\begin{split}
\left(\begin{array}{c}\vec \Phi_{1}^T(3j,\tau) \\ \vec\Phi_{2}^T(3j+1,\tau) \\ \vec \Phi_{3}^T(3j+2,\tau) \end{array}\right)
=L(j,\tau) U(j,\tau) \end{split}
\end{equation}
where the rows of $U(j,\tau)$ can be seen as three orthogonal states in unit cell $j$, 
\begin{equation}
U(j, \tau)= \left( \begin{array}{c} \vec  \phi_{1}^T(j,\tau) \\ \vec \phi_{2}^T(j,\tau) \\ \vec \phi_{3}^T(j,\tau) 
 \end{array}\right), 
 \label{eq:U0mx}
\end{equation} 
and $L(j,\tau)$ describes the transverse fluctuations, which make the  spin states non-orthogonal inside the unit cell: 
\begin{equation}
\arraycolsep=0.8pt
\begin{split}
L(j,\tau)=
\left( \begin{array}{ccc}
 \sqrt{1-\frac{a^2}{p^2}( |L_{12}|^2+|L_{13}|^2)}& \frac{a}{p} L_{12}&  \frac{a}{p} L_{13} \\
 \frac{a}{p}L_{12}^*& \sqrt{1-\frac{a^2}{p^2} (|L_{12}|^2+|L_{23}|^2)}& \frac{a}{p} L_{23} \\
 \frac{a}{p} L_{13}^*& \frac{a}{p} L_{23}^*&\sqrt{1-\frac{a^2}{p^2}( |L_{13}|^2+|L_{23}|^2)}\\
 \end{array}\right).
\end{split}
\end{equation}
For compactness we omitted the $(j, \tau)$ dependence of the the matrix elements of $L$. 
In this expression, $a$ stands for the lattice spacing and $p$ for the number of the boxes of the spin representation. The $a/p$ factors emphasize that these fluctuations are small, as large fluctuations are exponentially suppressed in the path integral. The ${L}$ matrix can be chosen to be hermitian as the skew-hermitian part would describe an infinitesimal joint rotation of the three spin states and could thus be merged into the unitary $U$ matrix.  The diagonal elements of $L$ have been chosen to keep the spin states normalized\footnote{Since the rows of  $U(j)$ form an orthonormal basis,  the spin states will be normalized if the rows of L are also normalized.}.
\begin{figure}
\begin{center}
\includegraphics[width=0.9\textwidth]{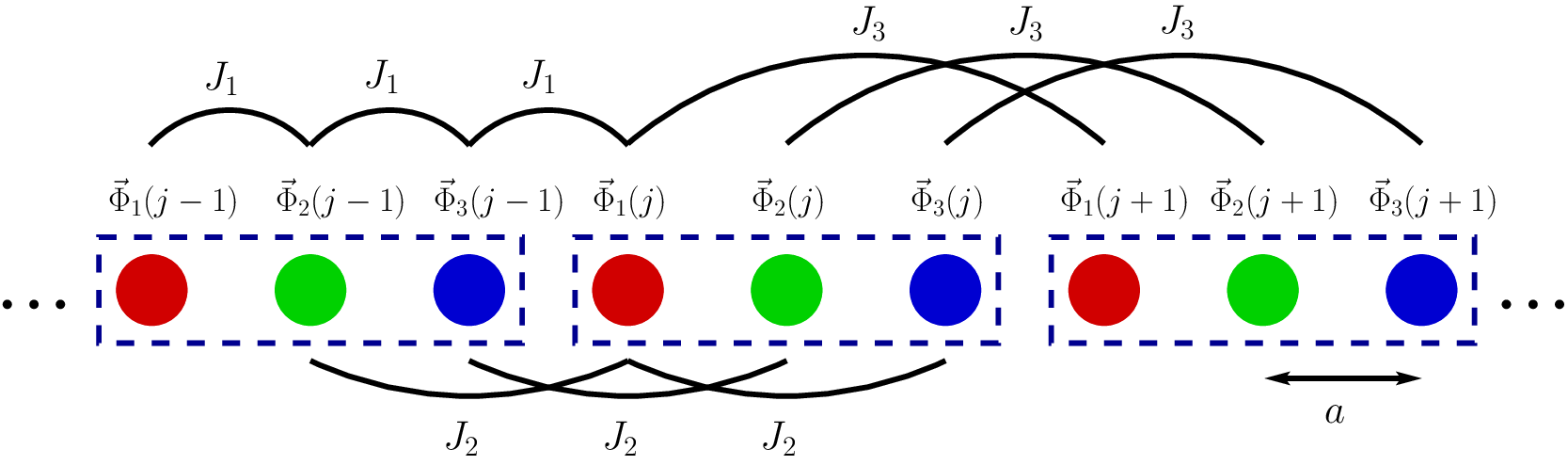}
\caption{Sketch of the low energy fluctuations around the classical three sublattice ground state. The spin states are given by Eq.\ \eqref{eq:spinstates0}. The lattice constant $a$ is the distance between neighbouring sites.
}
\label{fig:classicalGS}
\end{center}
\end{figure}

Substituting the above parametrization of the low-energy fluctuations into the action in  Eq.\ \eqref{eq:fullLUaction}, the functional integral over the $L$ variables can be carried out, leading to a form of the imaginary time action in terms of the $U(x, \tau)$ field only:
\begin{equation}
\begin{split}
S[U] = \int dx d\tau\Bigg( \sum_{n=1}^3  &\frac{1}{g} \Biggl[ v\,\tr   \left[  \Lambda_{n-1} U \partial_x U^\dagger \Lambda_{n} \partial_x U  U^\dagger \right]  +\frac{1}{v} \tr \left[  \Lambda_{n-1} U  \partial_\tau U^\dagger \Lambda_{n} \partial_\tau U  U^\dagger \right]  \Biggr]\\
& +i\sum_{n=1}^3\frac{\theta_n}{2\pi i}\varepsilon_{\mu\nu}\tr\left[\Lambda_n \partial_\mu U\partial_\nu U^\dagger\right] +i\frac{\lambda}{2\pi i} \varepsilon_{\mu\nu}\sum_{n=1}^3 \tr\left[ \Lambda_{n-1} U \partial_\mu U^\dagger \Lambda_{n} \partial_\nu U U^\dagger  \right]
 \Bigg),
\end{split}
\label{eq:fullaction}
\end{equation}
where the  $\Lambda_n$ matrices are defined by
\begin{equation}
\begin{split}
\Lambda_1=\left(\begin{array}{ccc} 1&0&0\\ 0&0&0\\ 0&0&0 \end{array}\right),\quad  \Lambda_2=\left(\begin{array}{ccc} 0&0&0\\ 0&1&0\\ 0&0&0 \end{array}\right), \quad
\Lambda_3=\left(\begin{array}{ccc} 0&0&0\\ 0&0&0\\ 0&0&1 \end{array}\right),
\end{split}
\label{eq:Lambdamx1}
\end{equation}
while  $\varepsilon_{\mu \nu}$ is the two dimensional Levi-Civita tensor ($\varepsilon_{x\tau} = -\varepsilon_{\tau x}=1$).   The coupling constant $1/g = p\sqrt{(J_1J_2+ 2J_3 J_1+2 J_3J_2)}/(J_1+J_2)$ and the velocity $v=3 a p \sqrt{J_1J_2 + 2J_3 J_1+2J_3J_2}$, in agreement with the flavour wave calculations in Sec.~\ref{sec:FW}. 

The imaginary term containing the $\theta_n$ parameters is topological, with the integer valued topological charges \cite{UedaShannon2016}
\begin{equation}
\begin{split}
Q_n =\frac{1}{2\pi i} \varepsilon_{\mu \nu} \int dxd\tau \tr\left[\Lambda_n\partial_\mu U\partial_\nu U^\dagger\right].
\end{split}
\label{eq:Qdef}
\end{equation}

The $\lambda$-term is also imaginary, but non-topological. In fact, the value of $\lambda$ is  non-universal:
\begin{equation}
\begin{split}
\lambda =p \frac{2\pi}{3}\frac{2J_2-J_1}{J_1+J_2} .
\end{split}
\end{equation}
For simplicity, let us introduce the notation 
\begin{equation}
\begin{split}
q_{mn}=\frac{1}{2\pi i}\int dx d\tau\, \varepsilon_{\mu\nu} \tr&\left[ \Lambda_m U \partial_\mu U^\dagger \Lambda_{n} \partial_\nu U U^\dagger  \right],
\end{split}
\label{eq:qterm}
\end{equation}
where $q_{mn}=-q_{nm}$. In terms of these quantities, the $\lambda$-term of the action can be written: $i\lambda\left( q_{12}+q_{23}+q_{31}\right)$. 
The topological charges can also be expressed using the  $q_{mn}$'s as
\begin{equation}
\begin{split}
Q_1=q_{12}+q_{13}\,, \quad
Q_2=q_{21}+q_{23}\,, \quad
Q_3=q_{31}+q_{32}\, . 
\end{split}
\label{eq:qQrel}
\end{equation}
The antisymmetry of the $q_{mn}$ implies that $Q_1+Q_2+Q_3=0$. So the action is invariant under a global shift of the topological angles, and one can set one
of them to 0. Unless specified otherwise, we will work with the convention $\theta_2=0$.

For the translationally invariant model of Eq.\ (\ref{eq:J1-J2model}), and with this convention, the topological angles are given by  $\theta_1 = -\theta_3 = p 2\pi/3$, and the 
action actually takes the form
\begin{equation}
\begin{split}
S[U] = \int dx d\tau\Bigg(& \sum_{n=1}^3  \frac{1}{g} \Biggl[ v\,\tr   \left[  \Lambda_{n-1} U \partial_x U^\dagger \Lambda_{n} \partial_x U  U^\dagger \right]  +\frac{1}{v} \tr \left[  \Lambda_{n-1} U  \partial_\tau U^\dagger \Lambda_{n} \partial_\tau U  U^\dagger \right]  \Biggr]\\
& +i\frac{\theta}{2\pi i}\varepsilon_{\mu\nu}\tr\left[(\Lambda_1-\Lambda_3) \partial_\mu U\partial_\nu U^\dagger\right] +i\frac{\lambda}{2\pi i} \varepsilon_{\mu\nu}\sum_{n=1}^3 \tr\left[ \Lambda_{n-1} U \partial_\mu U^\dagger \Lambda_{n} \partial_\nu U U^\dagger  \right]
 \Bigg),
\end{split}
\label{eq:action_theta}
\end{equation}
with a topological angle given by
\begin{equation}
\theta=\frac{2\pi}{3}p.
\end{equation}
It is the phase diagram of this action (with $\lambda=0$) that is sketched in Fig.~\ref{fig:SU23RGflow}\lcol{b}. 
To discuss the properties of that model, it will be useful to consider the general case of Eq.\ (\ref{eq:fullaction}) where the topological angles are free to vary.

The action in Eq.\ \eqref{eq:fullaction} can be written using the three orthogonal fields $\vec{\phi}_1,\vec{\phi}_2, \vec{\phi}_3$  forming the $U$ matrix.   Setting the velocity to $v=1$, the action can be rewritten as

\begin{equation}
\begin{split}
S = \int dx d\tau\Bigg(& \sum_{n=1}^3  \frac{1}{2g} \,   \left( \left| \partial_\mu \vec{\phi}_n^{\phantom{*}}\right|^2 -  \left| \vec{\phi}_n^*\cdot \partial_\mu \vec{\phi}_n^{\phantom{*}} \right|^2 \right) +i \sum_{n=1}^3\frac{\theta_n}{2\pi i}\varepsilon_{\mu\nu}\left(\partial_\mu \vec{\phi}_n^{\phantom{*}}\cdot \partial_\nu \vec{\phi}^*_n\right)\\
& + i\frac{\lambda}{2\pi i} \varepsilon_{\mu\nu}\sum_{n=1}^3 \left(  \vec{\phi}_{n+1}^* \cdot \partial_\mu \vec{\phi}_{n}^{\phantom{*}}  \right)\left( \vec{\phi}_{n+1}^{\phantom{*}} \cdot \partial_\nu \vec{\phi}_{n}^*\right)
 \Bigg)\\
\end{split}
\label{eq:fullactionphi}
\end{equation}
In this form, it is apparent that the action consists of three copies of a $\mbox{CP}^2$ field theory \cite{UedaShannon2016,PimenovPunk2017}, each with a  topological term, while the $\lambda$-term couples the three theories. Of course, they are also coupled due to the orthogonality constraint, which leads to $Q_1+Q_2+Q_3=0$, and which allows one in  general to set one of the topological angles to 0. 

If the $\lambda$-term is neglected, it is also possible to rewrite the action in terms of three gauge fields, $A_\mu^n$ \cite{DaddaLuscher1978, UedaShannon2016}:
\begin{equation}
  S=\int dx d \tau \Bigg(\sum_{n=1}^3\left[  \frac{1}{2g} \left|\left( \partial_\mu +iA_\mu^n\right) \vec{\phi}_n^{\phantom{*}}\right|^2+{i\theta_n\over 2\pi}\epsilon_{\mu \nu}\partial_\mu A_\nu^n\right] \Bigg).
\label{eq:cp2gaugeform}
\end{equation}
We may actually impose the constraint:
 \begin{equation} A_\mu^1(x)+A_\mu^2(x)+A_\mu^3(x)=0\ \  (\forall x,\mu ),
 \end{equation}
which follows from the orthogonality of the three $\vec \phi$ fields.  
The equivalence to Eq.\ \eqref{eq:fullactionphi}  follows by carrying out the functional integral over the $A_\mu^n$ fields (see \ref{sec:su3calc} for details). 
All three forms of the action in Eqs.\  \eqref{eq:fullaction},  \eqref{eq:fullactionphi} and  \eqref{eq:cp2gaugeform} are useful in different contexts of our study.

\section{General properties of the field theory}
\label{sec:symms}

In this section we briefly review the symmetries and other general properties of the field theory of Eq.\ \eqref{eq:fullaction} or Eq.\ \eqref{eq:fullactionphi}.  

\subsection{SU(3) symmetry}

Throughout this paper we only consider spin models with global $\mbox{SU}(3)$ symmetry,  hence the resulting  $\sigma$-models  are also invariant under SU(3) rotations. These are of the form $U'(x,\tau)= U(x,\tau)V$, or equivalently $\vec \phi_n'(x,\tau)= V^T\vec \phi(x,\tau)$ where the unitary V matrix clearly cancels out in every term of the action in Eq.\ \eqref{eq:fullaction} or Eq.\ \eqref{eq:fullactionphi}.

\subsection{Gauge invariance}
\label{subsec:gaugeinv}
The overall phases of the spin coherent states shouldn't change the form of the action. This manifests in the gauge invariance of the action in Eq.\ \eqref{eq:fullaction} under the  transformation $U'(x,\tau)= D(x,\tau) U(x,\tau)$, where 
\begin{equation} \label{eq:gauge}
\begin{split}
D (x,\tau)=\left( \begin{array}{ccc} e^{i\vartheta_1(x,\tau)}&0&0\\0&e^{i\vartheta_2(x,\tau)}&0\\0&0&e^{i\vartheta_3(x,\tau)}\end{array} \right)
\end{split}
\end{equation}
with $\vartheta_3(x,\tau)=-(\vartheta_1(x,\tau)+\vartheta_2(x,\tau))$.  In terms of the fields, this transformation corresponds to $\vec \phi'_n = e^{i \vartheta_n} \vec \phi_n$.
A proof of gauge invariance in this language can be found in \ref{gauge_invariance}. Gauge invariance is also evident in the formulation of 
Eq.\ \eqref{eq:cp2gaugeform}, where gauge fields are explicitly introduced. 

\subsection{Time reversal symmetry}
Another fundamental symmetry is  time reversal symmetry. The effect of  time reversal (in real time) 
is simply $TU(x,t)T=  U(x,-t)$, or equivalently $T \vec{\phi}_n(x,t)T=\vec{\phi}_n(x,-t)$, as well as complex conjugation of c-numbers: $i\to -i$. The first term in the action in Eq.\ \eqref{eq:fullactionphi} is clearly invariant under T-reversal. 
The  topological $\theta$-term  and the  $\lambda$-terms pick up a factor of i when going to real time, which makes these terms real (Hermitian). The $i$ to $-i$ transformation then compensates for 
$\partial_t\to -\partial_t$, also leaving these terms time reversal invariant. 

\subsection{$\mathbb{Z}_3$ symmetry}
 \label{subsec:Z3symm}
The field theory has an additional global $\mathbb{Z}_3$ symmetry:  $U'(x,\tau) = R_{\mathbb{Z}_3} \, U(x,\tau)$, with 
\begin{equation}
\begin{split}
R_{\mathbb{Z}_3}=\left(\begin{array}{ccc} 0&1&0\\0&0&1\\1&0&0\end{array}\right),
\end{split}
\label{eq:cyclicpermute}
\end{equation}
which  cyclically permutes the three $\vec \phi_n$ fields. This symmetry is a consequence of the invariance of the spin model under translation by one site. In the field theory derivation we  assumed a three sublattice ordered ground state, which is only suitable for spin models which are invariant under three site translation, hence $R_{\mathbb{Z}_3}^3 = I$  is a symmetry independently of the parameters of the field theory. 

It is clear in any formulation that the real part of the action in Eq.\ \eqref{eq:fullactionphi} is invariant under $\mathbb{Z}_3$ as long as the  coupling $g$ is the same for all  three $\mbox{CP}^2$ theories.
When $\theta_1=-\theta_3= p2 \pi/3$, the topological term is also invariant. Indeed it transforms as
\begin{equation}\begin{split} 
i p{2 \pi \over 3}(Q_1-Q_3) \to  ip{2 \pi \over 3}\big(Q_2-Q_1\big) =ip{ 2 \pi \over 3}\big(Q_1-Q_3\big)+ ip{2 \pi\over3} \big(Q_1+Q_2+Q_3\big)-i 3 p{2 \pi \over3}  Q_1.
\end{split}\end{equation}
Since $Q_1+Q_2+Q_3=0$ and $Q_1$ is integer-valued,  the second term of the right hand side is 0, and the third term gives an integer multiple of $2\pi$, leading to:
\begin{equation}
 e^{ i(2 p \pi /3)(Q_1-Q_3)} \equiv e^{i (2 p\pi /3)(Q_2-Q_1)}.
\end{equation}
Finally, the  $\lambda$-term is  clearly invariant under $\mathbb{Z}_3$ symmetry  as $q_{12}+q_{23}+q_{31}$ transforms into itself. 

\subsection{Parity symmetry}
\label{ssec:Psymm}
The action in Eq.\ \eqref{eq:fullactionphi} is invariant under parity symmetries as well, which correspond to  mirror symmetries between two  neighbouring sites in the spin model.  Since a three site translation symmetry is always conserved, there are three non-equivalent mirror symmetries in the spin model, resulting in three non-equivalent parity symmetries in the $\sigma$-model. Take for example a mirror plane between two sites in sublattices 1 and 3 (see Fig.~\ref{fig:Tbreak}). The corresponding parity symmetry of the field theory transforms $U$ matrices  as   $U'(x,\tau) = R_{13}\, U(-x,\tau)$, with 
\begin{equation}
\begin{split}
R_{13}=\left(\begin{array}{ccc} 0&0&1\\0&1&0\\1&0&0\end{array}\right).
\end{split}
\end{equation}
In terms of the fields it corresponds to $\vec \phi_{1(3)} (x, \tau) \to \vec \phi_{3(1)} (-x,\tau)$ and $\vec \phi_2 (x, \tau) \to \vec \phi_2(-x,\tau)$, i.e.\ it exchanges fields 1 and 3. This means that in the real part of the action the terms for the fields 1 and 3 will be exchanged, while the terms for field 2 will be unchanged. This is clearly a symmetry of the action of Eq.\ \eqref{eq:fullactionphi}.
 
The topological term transforms as $\theta (Q_1-Q_3) \to -\theta (Q_3-Q_1)$, which is the same as the original one. The extra $-$ sign appears as there is always exactly one spatial derivative in the expression of the topological charges (see Eq.\ \eqref{eq:Qdef}), and  the parity transformation inverts the $x$ coordinate. Similarly,  the  $\lambda$-term becomes  $- \lambda ( q_{21}+q_{32}+q_{13})$,  which is equal to the original term as $q_{mn}=-q_{nm}$.

The invariance of the action in Eq.\ \eqref{eq:fullactionphi} under the other two parity symmetries $R_{12}$  and $R_{13}$ can be shown 
in a similar way as for $R_{13}$. 
   
 \subsection{Breaking lattice symmetries and general form of the action}
 \label{subsec:symmbreak}

If the spin model is not invariant under the translation and the mirror symmetries, the $\mathbb{Z}_3$ and parity symmetries of the $\sigma$-model will be broken, and the action will take the form
\begin{equation}
\begin{split}
S =\int dx d\tau\Bigg[& \sum_{n=1}^3\Bigg(  \frac{v_n}{2g_{n}} \,   \left( \left| \partial_x^{\phantom{*}} \vec{\phi}_n^{\phantom{*}}\right|^2 -  \left| \vec{\phi}_n^* \cdot\partial_x^{\phantom{*}} \vec{\phi}_n^{\phantom{*}} \right|^2 \right)+\frac{1}{2 v_n g_{n}} \,   \left( \left| \partial_\tau^{\phantom{*}} \vec{\phi}_n^{\phantom{*}}\right|^2 -  \left| \vec{\phi}_n^* \cdot\partial_\tau^{\phantom{*}} \vec{\phi}_n^{\phantom{*}} \right|^2 \right)  \Bigg)\\
&+i\sum_{n=1}^3 \theta_n Q_n  + i \lambda (q_{12}+q_{23}+q_{31})\Bigg],
\label{eq:fullactionphigenform}
\end{split}
\end{equation}
In general, the three copies of the $\mbox{CP}^2$ model do not have the same coupling constants and velocities any more. The topological angles can also take arbitrary values, but since the topological charges still satisfy $Q_1+Q_2+Q_3=0$ because of the orthogonality of the fields, one can still set one of them to 0, and one is left with two independent topological charges, for instance $\theta_1$ and $\theta_3$. 
All these statements are illlustrated in the case where the nearest neighbor interaction takes three different values between each pair of sublattices discussed 
in \ref{app:genform}.  We also show that assuming $\mbox{SU}(3)$, gauge and time reversal invariance, Eq.\ \eqref{eq:fullactionphigenform} is indeed the most general form of the $\sigma$-model. 

\begin{figure}[h]
\begin{center}
\includegraphics[width=0.9\textwidth]{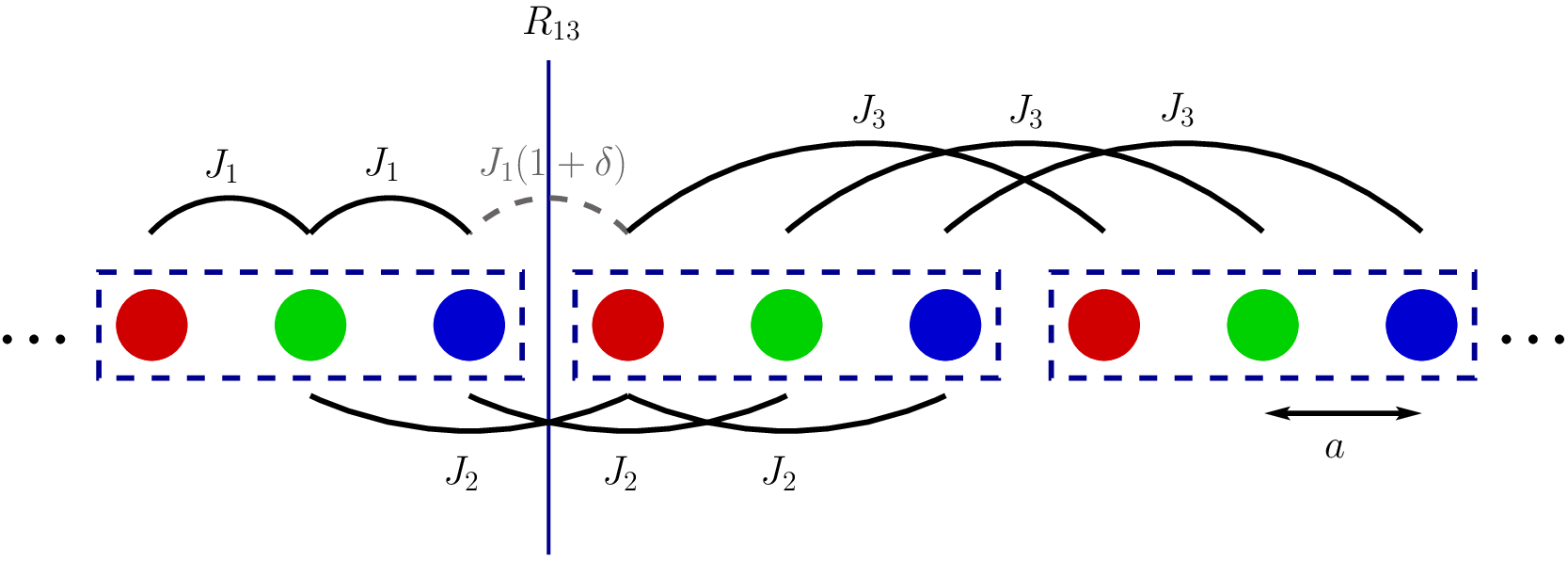}
\caption{Breaking translational symmetry of the system by weakening the nearest neighbour bonds between sublattices 1 and 3. The $R_{13}$ mirror symmetry and a three site translational symmetry still remains. } 
\label{fig:Tbreak}
\end{center}
\end{figure}

Here we present an  intermediate situation where the translation symmetry is broken but one of the mirror symmetries is still present, say $R_{13}$. This is achieved by weakening/strengthening the nearest neighbour bond 
between sublattices 1 and 3 as shown in Fig.~\ref{fig:Tbreak}. In this case the $\mbox{CP}^2$ theories of the fields $\vec{\phi}_1$
and $\vec{\phi}_3$ will have the same parameters, but not that describing $\vec{\phi}_2$. The topological angles are not fixed to  $\pm2p\pi/3$, but they still  satisfy the additional constraint $\theta_1=-\theta_3$ due to the $R_{13}$ symmetry, 
\begin{equation}
\begin{split}
\theta_1=-\theta_3= p \frac{2\pi}{3} \left(1 +\frac{J_2}{J_1+J_2} -\frac{J_2}{J_1 (1+\delta)+J_2} \right)
\end{split}
\end{equation}
If $\delta<0$, the bond is weakened and $\theta_1 =-\theta_3 < p2\pi/3$, while if $\delta>0$, the bond is strengthened and $\theta_1 =-\theta_3 > p2\pi/3$.   Since $\theta \neq p 2\pi/3$,  the topological term is no longer invariant under $\mathbb{Z}_3$, or under $R_{12}$ and $R_{23}$.

 \subsection{Additional remarks about the field theory}
 \label{sec:FTremarks}
 Due to the gauge invariance discussed above, this nonlinear $\sigma$-model is defined on the manifold $\mbox{SU}(3)/[\mbox{U}(1)\times \mbox{U}(1)]$, which is called a flag manifold. It was discussed 
 earlier by \citet{Bykov2013} in the context of $\mbox{SU}(n)$ chains with alternating representations on adjacent sites
  and by \citet{UedaShannon2016} in the context of two dimensional $\mbox{SU}(3)$ magnets at finite temperature.  
  
  We believe it is possible to factorize any $\mbox{SU}(3)$ matrix in the form
  \be U=\exp \big[i\sum_a\theta_aT_a\big ]\exp \big[i \sum_\alpha \theta_\alpha T_\alpha \big].
 \label{eq:diagoffdiagfactorizaiton} 
 \ee
  Here the $T_a$ are the 6 off-diagonal generators of $\mbox{SU}(3)$ and the $T_\alpha$ are the 2 diagonal generators. It would then follow, due to the gauge invariance discussed above, 
  that the diagonal factor could be dropped, with the off-diagonal generators giving the $\mbox{SU}(3)/[\mbox{U}(1)\times \mbox{U}(1)]$ manifold. Related factorizations have  been proven earlier \cite{murnaghan1962,Wilcox1967,Puri2001}.
   We give a proof of this factorization to third order in the $\theta_a$'s and $\theta_\alpha$'s in \ref{sec:factorization}. We have also checked numerically that for random unitary matrices a factorization according to Eq.\ \eqref{eq:diagoffdiagfactorizaiton} could always be  found.

  Since $\pi_2 \big[\mbox{SU}(3)/[\mbox{U}(1)\times \mbox{U}(1)]\big]=\mathbb{Z}\times \mathbb{Z}$, there 
  are 2 topological invariants, as discussed above. This field theory may be regarded as a natural generalization of the $\mbox{O}(3)$ $\sigma$-model, which is  defined on the manifold $\mbox{SU}(2)/\mbox{U}(1)$ and has 
  one topological invariant. An interesting difference is the presence of the ``$\lambda$-term'' discussed above, which respects all symmetries including Lorentz invariance. It has  
  the very unusual property of being antisymmetric in space and time derivatives, and consequently imaginary for imaginary time, but not being a total derivative (topological invariant). This means 
  that the coupling constant, $\lambda$, can renormalize perturbatively and could potentially have important effects on the behaviour of the model and therefore of the spin chains.  We analyze
  it using the perturbative renormalization group in the next section.

\section{Renormalization group analysis}
\label{sec:RG}

We rewrite the matrices $U$ in terms of the Gell-Mann matrices (GM) $T_A$, as defined in Eq.\ \eqref{gm}. Due to the factorization property of $\mbox{SU}(3)$ matrices (discussed in Sec.~\ref{sec:FTremarks} and \ref{sec:factorization}), and the gauge symmetry (see Eq.\ \eqref{eq:gauge}), we can replace $U$ with $e^{i\theta_aT_a}$ in Eq.\ \eqref{eq:fullaction}. Here and throughout, lowercase Roman letters index the off-diagonal GM, lowercase Greek letters index the diagonal GM, uppercase Roman letters index all eight GM, and repeated indices are summed over. After rescaling $\theta \to \sqrt{\frac{g}{2}}\theta_a$, we find 
\be \label{step1}
	\mathcal{L} = \frac{1}{2} \partial_\mu\theta_a\partial_\mu\theta_a + i\lambda g^{\frac{3}{2}} \epsilon_{\mu\nu} R_{abc}\theta_a\partial_\mu \theta_b\partial_\nu \theta_c
	+ gP_{abcd}\theta_a\theta_b\partial_\mu\theta_c\partial_\mu\theta_d
	+ \mathcal{O}(\lambda g^2) + \mathcal{O}(g^2)
\ee
where $R_{abc}$ and $P_{abcd}$ are tensors of real coefficients that can be expressed in terms of the SU(3) structure factors.  At this order in perturbation theory, the imaginary $\lambda$-term  term has coupling constant $\tilde \lambda = \lambda g^{3/2}$. We will calculate the beta function of $\tilde\lambda$, showing that it renormalizes to large values at large length scales. We'll also calculate the beta function of $g$, showing that it renormalizes to large values, so that our flag manifold $\sigma$-model is asymptotically free.

To perform the RG calculations, we rewrite the fields $U$ in terms of `slow' ($U_s)$ and `fast' ($U_f$) fields, as 
\be \label{decomp}
	U = U_f U_s
\ee
These `fast' fields have momentum modes restricted to a Wilson shell $[b\Lambda, \Lambda)$, where $\Lambda$ is a reduced cutoff, and $b\lesssim1$. The RG step of integrating over this shell is then equivalent to integrating out the fields $U_f$ from the theory. This factorization of $U$ is motivated by Polyakov's work on the nonlinear-$\sigma$-model, which found the model's RG equations  by a quadratic expansion of the `fast fields' \cite{Polyakov:1975rr}. In \ref{polyakov}, we show the equivalence between Polyakov's decomposition and Eq.\ \eqref{decomp} for the SU(2) case.

Now, we insert Eq.\ \eqref{decomp} into an equivalent form of the Lagrangian, derived in \ref{equivalent}:

\be \label{step2}
	\fL = \frac{1}{8g} \tr \partial_\mu [U^\dag T_\gamma U] \partial_\mu [U^\dag T_\gamma U] +\lambda \frac{\sqrt{3}}{2}\epsilon_{\mu\nu} \tr [\partial_\mu U U^\dag T_8 \partial_\nu U U^\dag T_3]
\ee
We've suppressed the topological term, as it does not contribute to the perturbative RG equations. The utility of this expression becomes apparent when we expand $U_f = e^{i\theta_a T_a}$, since we know how to express products of $T_a$ and $T_\gamma$ in terms of the SU(3) structure constants $f_{ABC}$ (see Appendix \ref{commute}). Such an expansion again follows from the gauge symmetry (Eq.\ \eqref{eq:gauge}) and the factorization of SU(3) matrices. Defining $M_A = U_s^\dag T_A U_s$ and $N_\mu = \partial_\mu U_s U_s^\dag$,  the first term in Eq.\ \eqref{step2} gives
\begin{equation}
\begin{split}
\frac{1}{8g}\Bigg[&\tr (\partial_\mu M_\gamma \partial_\mu M_\gamma) + 8 \partial_\mu\theta_a\partial_\mu\theta_a +4f_{a\gamma c}f_{b\gamma d}\theta_a\theta_b \tr (\partial_\mu M_c \partial_\mu M_d )
	+ 8if_{abe}\partial_\mu\theta_a \theta_b \tr  (N_\mu T_e) \\
	&+ 16if_{ab\gamma} \partial_\mu\theta_a\theta_b \tr ( N_\mu T_\gamma )
	- 2\theta_a\theta_b[f_{ab\gamma D}f_{\gamma bc} + f_{bcD}f_{\gamma ac}]\tr (\partial_\mu M_\gamma \partial_\mu M_D)\\
	&+\big(\text{linear terms in $\theta$}\big) + \fO(\theta^3)\Bigg]\,.
\end{split}
 \label{big1}
\end{equation}
In \ref{linear}, we argue that the linear terms can be dropped in the $\Lambda \to 0$ limit, 
 which follows from the fact that such terms produce four-derivative operators after the integration over $\theta_a$. The second term in Eq.\ \eqref{step2}, after dropping linear terms, gives 
\begin{equation}
 \label{big2}
 \begin{split}
	\epsilon_{\mu\nu} \lambda \frac{\sqrt{3}}{2}\Bigg[ &\tr \big( N_\mu T_8 N_\nu T_3\big) +\partial_\mu \theta_a \theta_b \tr \big(N_\nu T_b \left[  T_3 T_a T_8 -  T_8 T_a T_3\right]\big)
	\\&+ \frac{1}{2}\partial_\mu(\theta_a\theta_b) \tr \big(N_\nu [T_8T_aT_bT_3 - T_3T_aT_bT_8]\big) \\
	&+ \theta_a\theta_b \Big(4f_{b3d}f_{a8c}\tr \big(N_\mu T_c N_\nu T_d \big)+ \tr \big( N_\nu T_D N_\mu\left[ h_{ab8D} T_3 - h_{ab3D}T_8\right]\big)\Big)\\
	&+ 2i \partial_\mu\theta_a\theta_b \tr \big(T_a T_cN_\nu \left[ f_{b8c} T_3   - f_{b3c}T_8\right]\big)
	\Bigg]\, .
\end{split}
\end{equation}
We now Fourier transform the sum of Eq.\ \eqref{big1} and Eq.\ \eqref{big2}, and perform the Gaussian integral over $\theta_a$. Details of this step, which includes the application of numerous nontrivial identities involving the $f_{ABC}$, are included in \ref{beta}. The result is
\be \label{result}
	\fL_{\text{eff}} = \frac{1}{8g}\left(1 + \frac{5g \log b}{4\pi}\right) \tr \partial_\mu M_\gamma \partial_\mu M_\gamma 
	+ \lambda\left(1 + \frac{9g}{4\pi} \log b\right) \epsilon_{\mu\nu}\tr N_\mu T_8 N_\nu T_3 
	+ \fO(g) 
	+\fO(\lambda^2) + \fO(\lambda g)
\ee
where the $\log b$ dependence comes from the integral
\be
	\int \limits_{|k| \in [b\Lambda, \Lambda)} \frac{d^2k}{(2\pi)^2} \frac{1}{k^2} = -\frac{\log b}{2\pi}
\ee
Recalling the definitions of $M_A$ and $N_\mu$, we recognize that to our order of accuracy, $\fL_{\text{eff}}$ is of the same form as $\fL$, with $U$ replaced by $U_s$. This allows for the identification of the first two operators' factors with $\frac{1}{8g_{\text{eff}}}$ and $\lambda_{\text{eff}}$, respectively, leading to the following $\beta$ functions:
\be
	\beta_\lambda( \lambda,g )=   \frac{3g\lambda }{2\pi} 
	\hspace{10mm}
	\beta_g(\lambda,g) = -\frac{5g^2 }{4\pi}
	\hspace{15mm} \left( \beta_x:= \frac{dx}{d\log b}\right)
\ee
Therefore, $g$ flows to large values at large length scales. Moreover, since
\be
	\beta_{\tilde\lambda}(\lambda,g) = g^{\frac{3}{2}} \beta_\lambda + \frac{3}{2}g^{\frac{1}{2}} \lambda \beta_g
	= -\frac{3g}{8\pi} \tilde \lambda <0
\ee
the imaginary term in Eq.\ \eqref{step1} is also relevant.

\section{General phase diagram} 
\label{sec:numerics}
In the following, we analyze the general phase diagram of the flag manifold sigma model with $\lambda=0$. First we give an overview of the phase diagram based on considerations in the large $g$ limit and on the correspondence with $\mbox{SU}(3)$ spin models, then we present results of Monte Carlo simulations for finite couplings. 

\subsection{$g\to\infty$ limit}
\label{subsec:ginfty}
In this limit, with $\lambda$ set to zero, the action consists only of a topological term. Following the argument of \citet{SeibergPRL1984} and \citet{PlefkaSamuelPRD1997} for $\mbox{CP}^{N-1}$ models we study the $\theta_1 - \theta_3$ phase diagram.  As those authors discuss, a lattice version of the gauge field formulation of the action, Eq.\ (\ref{eq:cp2gaugeform}), 
 is most suitable for studying the strong coupling limit.  We work on a square lattice and introduce a unitary matrix, $U$ or equivalently 3 orthogonal vectors $\vec \phi_n$ 
 on every lattice site.  Gauge fields are introduced on the links of the lattice, giving the real part of the action \cite{Stone1981,DiVecchia1981,RabinoviciSamuel1981}:
 \be S_R=-{1\over 2g}\sum_{\vec r,n,\mu}\left[ \vec \phi_n^{\phantom{*}}(\vec  r)V^n(\vec r,\vec r+\vec \delta_\mu )\cdot \vec \phi_n^*(\vec r+\vec \delta_\mu )+c.c.\right] ,
 \label{eq:lattrealterm}\ee
 where   $\vec \delta_x, \vec\delta_\tau$ are unit vectors of the square lattice.
 $V^n(\vec r,\vec r+ \vec \delta_\mu )$ is a complex number of unit modulus, \be V^n(\vec r,\vec r+\vec \delta_\mu )=e^{iA^n(\vec r,\vec r+\vec \delta_\mu)}\ee
 and $A^n(\vec r,\vec r+\vec \delta_\mu )$ becomes the gauge field, $A^n_\mu (\vec r)$ in the continuum limit. 
 The topological term is written in terms of the product of the $V$ variables around a plaquette, namely
  \be S_I=i\sum_{\vec r}\sum_{n=1}^3{\theta_n\over 2\pi i}\ln V_P^n(\vec r),
   \label{eq:latttopterm}
 \ee
where
 \be V_P^n(\vec r)\equiv V^n(\vec r,\vec r+\vec \delta_x)V^n(\vec r+\vec \delta_x ,\vec r+\vec \delta_x+\vec \delta_\tau)V^n(\vec r+\vec \delta_x+\vec \delta_\tau,\vec r+\vec \delta_\tau)V^n(\vec r+\vec \delta_\tau ,\vec r).\ee
  The $\ln V_P^n$ contribution of each plaquette is restricted to lie in the range:
 \be -\pi < \frac{1}{i}\ln V_P^n\leq \pi \ee	
 and, as discussed in Sec.~\ref{sec:FT}, the constraint:
 \be \sum_{n=1}^3A^n(\vec r,\vec r+ \vec \delta_\mu )=0\ee
 is imposed on every link. 
 Defining the topological charge density
 \be q_n(\vec r)\equiv {1\over 2\pi i}\ln V_P^n(\vec r),\ee
 this translates to 
 \be \sum_{n=1}^3q_n(\vec r)=0\ \  (\forall \vec r). \ee
 This implies that the partition function is invariant under shifting all three topological angles by a common constant so, without loss of generality, we set $\theta_2=0$. 
 As discussed in \cite{SeibergPRL1984,PlefkaSamuelPRD1997}, periodic boundary conditions implies that the total topological charges are integers:
 \be Q_n\equiv \sum_{\vec r} q_n(\vec r)\in \mathbb{Z}.\ee
In the $g\to \infty$ limit, the partition function becomes:
\begin{equation}
\begin{split}
Z(\theta_1,\theta_3,g\to \infty ) ={}&{}\left\{\prod_{\vec r}\int\limits_{-1/2}^{1/2}   d\dq_1(\vec r) d\dq_2(\vec r) d\dq_3(\vec r)\,\delta [ \dq_1(\vec r)+\dq_2(\vec r)+\dq_3(\vec r)]\right\}
\\
&\exp\Big[  i\sum_{\vec r}\big( \theta_1q_1(\vec r)+\theta_3q_3(\vec r)\big)\Big] 
\sum_{Q_1}\delta \left( \sum_{\vec r}q_1(\vec r)-Q_1\right )\sum_{Q_3}\delta \left( \sum_{\vec r}q_3(\vec r)-Q_3\right ).
\end{split}
\end{equation}
Using the Fourier transform of the Dirac Comb,
\be \sum_Q\delta (x-Q)=\sum_me^{2\pi i mx},\ee
\be Z(\theta_1,\theta_3,g\to \infty )=\sum_{m_1,m_3\in \mathbb{Z}} z(\theta_1 +2\pi m_1,\theta_3 +2\pi m_3)^V
\label{eq:ZPBC}
\ee
where
\begin{equation}
\begin{split}
z(\theta_1,\theta_3) &= \, \int\limits_{-1/2}^{1/2}   d\dq_1 d\dq_2 d\dq_3\,\delta ( \dq_1+\dq_2+\dq_3)\exp\Big[  i(\theta_1q_1+\theta_3q_3) \Big] \\
&= \frac{2 \bigg((\theta_1-\theta_3) \cos \left(\frac{\theta_1-\theta_3}{2}\right)-\theta_1\cos \left(\frac{\theta_1}{2}\right)+\theta_3 \cos \left(\frac{\theta_3}{2}\right)\bigg)}{\theta_1 \theta_3(\theta_1-\theta_3)}
\end{split}
\end{equation}
and $V$ is the area (more precisely the number of plaquettes) of the system. In the thermodynamic limit ($V\to \infty$), and for any  value of the angles, the largest term in the sum will dominate it. Thus  the free energy density is
\begin{equation}
\begin{split}
f (\theta_1, \theta_3, g\to\infty)=-\frac{1}{V} \log \mathcal{Z}(\theta_1, \theta_3, g\to\infty) = -\log \Big( \max_{m,n} z(\theta_1+2\pi m, \theta_3 +2\pi n) \Big).
\end{split}
\label{eq:FPBC}
\end{equation}
\begin{figure}[h!]
\begin{center}
\includegraphics[width=0.9\textwidth]{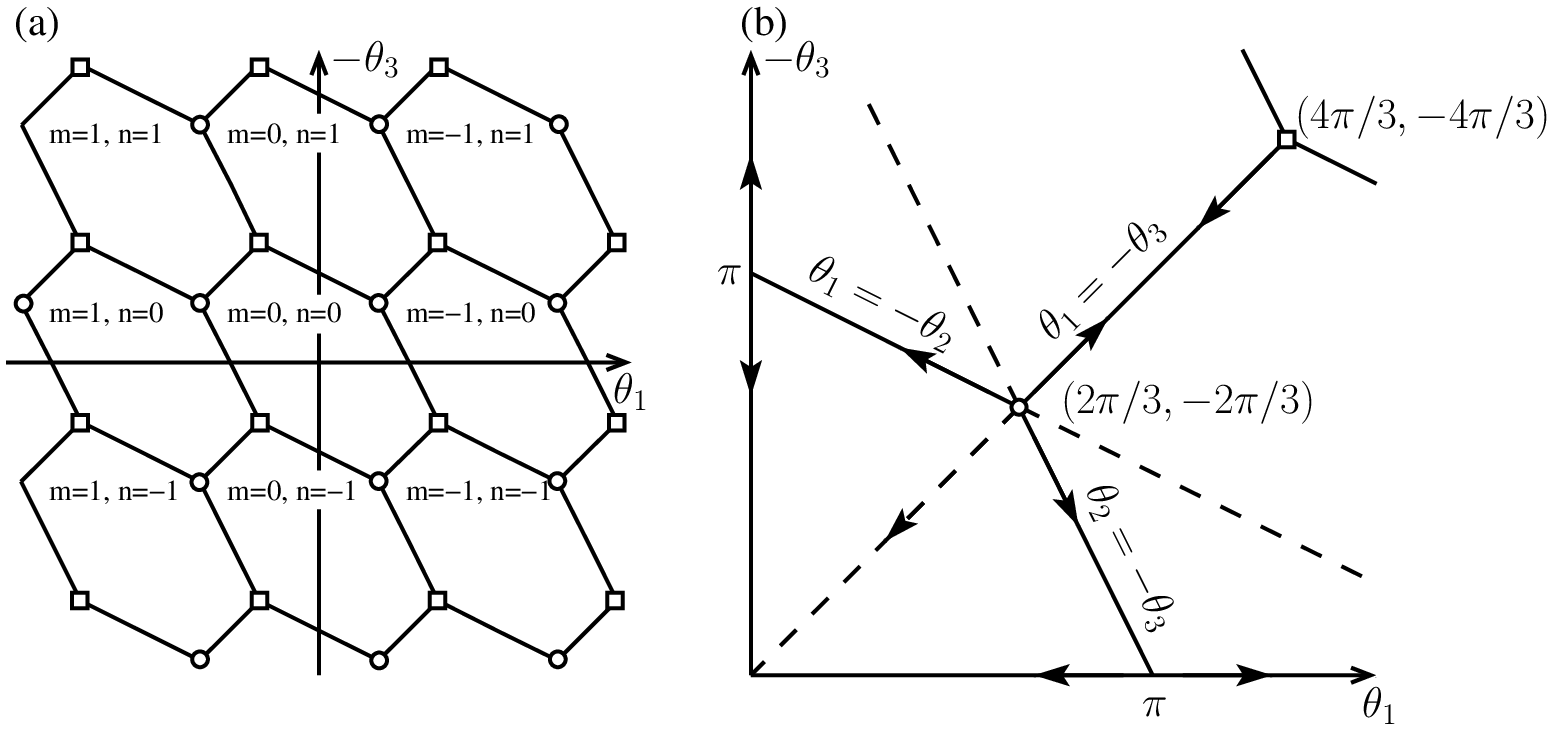}
\caption{(a)The different $\mathcal{R}_{m,n}$ sector on the $\theta_1 -\theta_3$ plane derived from the $g\to\infty$ calculations, and (b) a zoomed in version with the expected RG flow.  The phase diagram is similar for all g values, the flow in the $\theta_1 -\theta_3$ plane is everywhere complemented with a flow towards  $g\to\infty$.  The transition lines correspond to a conserved parity symmetry, which is spontaneously broken on the solid, but not dotted, lines. At the intersection of the lines, at $(2\pi/3 , -2\pi/3)$ a $\mathbb{Z}_3$ symmetry is also present.}
\label{fig:ththPD}
\end{center}
\end{figure}

In Fig.~\ref{fig:ththPD}\lcol{a}, we show the sectors $\mathcal{R}_{m,n}$ where the $z(\theta_1+2\pi m, \theta_3 +2\pi n)$ term is the largest. Moving from one $\mathcal{R}_{m,n}$ sector to an adjacent one, the free energy has  a cusp, i.e.\ a phase transition takes place. The  partition function and the free energy of Eqs.~\eqref{eq:ZPBC} and \eqref{eq:FPBC} are $2\pi$ periodic in both $\theta_1$ and $\theta_3$, as they should be since the topological charges are integer valued so that a $2\pi$ shift in the topological angle leaves the path integral (or more specifically $\exp(-S)$) invariant. Therefore we only discuss the phase diagram for the $0\leq \theta_1, -\theta_3 <2\pi$ region, referred to as {\it reduced phase diagram} in the following.

\begin{figure}[h!]
\begin{center}
\includegraphics[width=0.9\textwidth]{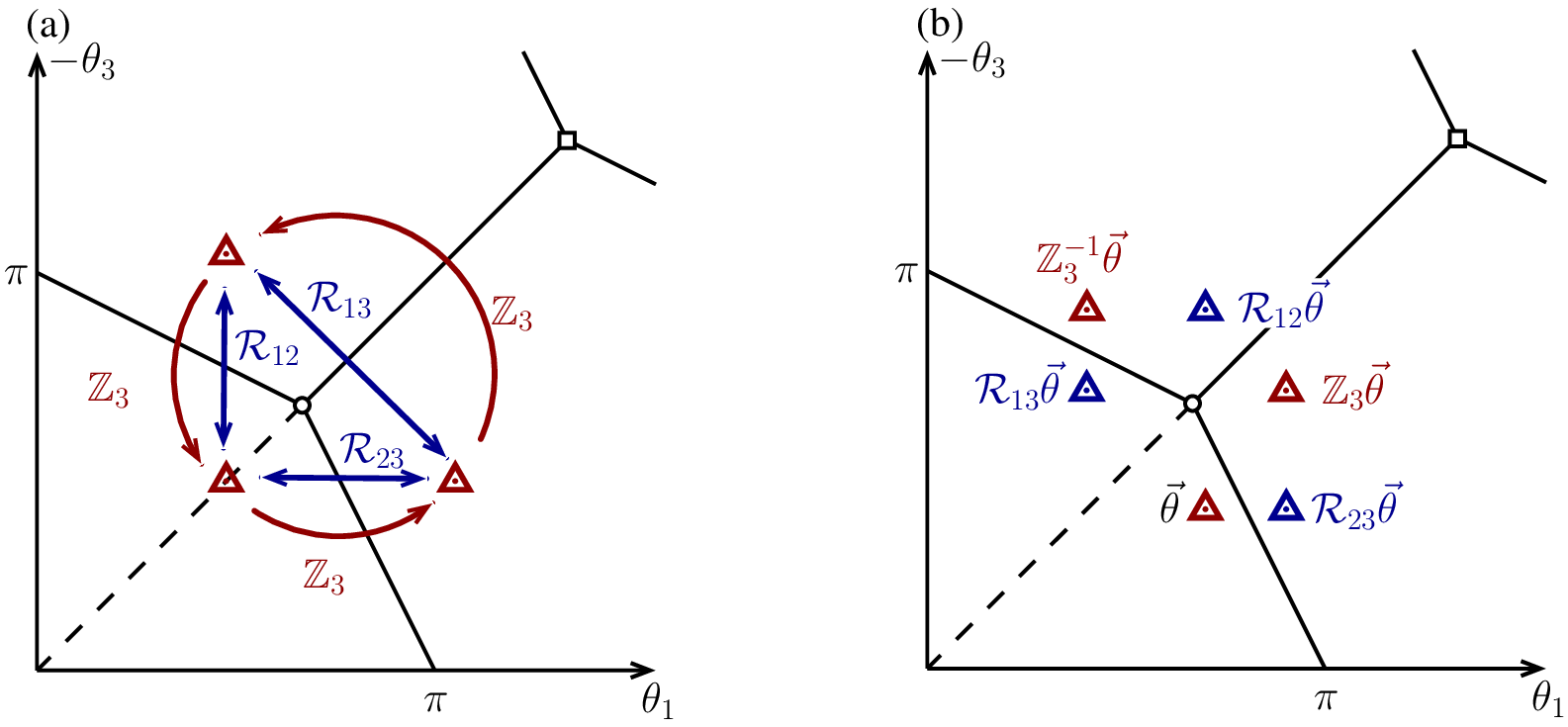}
\caption{The transformation of different points in the $\theta_1-\theta_3$ phase diagram under the $\mathbb{Z}_3$ and parity transformations. (a) A point on the $\theta_1=-\theta_3$ line is transformed into points on the $\theta_1=-\theta_2$ and $\theta_2=-\theta_3$ lines. Each point is conserved by one  of the three parity transformations. (b) Transformation of a generic point resulting in 6 equivalent points
 in the phase diagram (mod $2\pi$).}
 \label{fig:ththPDtrans} 
\end{center}
\end{figure}

There are three  high symmetry points in the reduced phase diagram defined by $\theta_1= -\theta_3 =  2m\pi/3 (\text{mod }2\pi)$ with $m=0,1,2$. At these points the action has a $\mathbb{Z}_3$ symmetry, which can be understood as a cyclic permutation of the three fields (see Sec.~\ref{subsec:Z3symm}). Additionally,  three parity symmetries are also present, each corresponding to the exchange of two out of the three fields (complemented by an invertion of the space coordinates as explained in Sec.~\ref{ssec:Psymm}). At each high symmetry point three special lines meet, each corresponding to the conservation of one of the three parity symmetries. Consider for example the $\theta_1=-\theta_3 = 2\pi/3$ point, and the parity symmetric lines meeting there, illustrated on Figs.\ \ref{fig:ththPD}, \ref{fig:ththPDtrans}. The  $\theta_1=-\theta_3$ line is invariant under the $R_{13}$ parity transformation defined in Sec.~\ref{ssec:Psymm}, which exchanges the fields $\phi_1$ and $\phi_3$. On the other two lines ($-\theta_3 = \pi -\theta_1/2 $ and $-\theta_3 = 2\pi-2 \theta_1$) the parities exchanging fields $\vec \phi_1\leftrightarrow \vec \phi_2$ and $\vec \phi_2\leftrightarrow \vec \phi_3$ are conserved, respectively\footnote{ Expressing the topological term with $Q_1$ and $Q_2$ along $-\theta_3 = \pi -\theta_1/2 $ one finds that  $\theta_1=-\theta_2\, (\text{mod } 2\pi)$, hence  the $R_{12}$ parity is conserved, while using $Q_2$ and $Q_3$ along  $-\theta_3 = 2\pi-2 \theta_1$   one finds that $\theta_2= -\theta_3 \, (\text{mod } 2\pi)$, hence  the $R_{23}$ parity is conserved.}. In one direction the parity symmetric lines are also transition lines between different $\mathcal{R}_{m,n}$ sectors until they reach another high symmetry point equivalent to $\theta_1=-\theta_3 = 4\pi/3$ (mod $2\pi$) (solid lines in Fig.~\ref{fig:ththPD}\lcol{b}), while in the other direction they run inside a sector towards a high symmetric point equivalent to $\theta_1=-\theta_3 = 0$ (dashed lines in Fig.~\ref{fig:ththPD}\lcol{b}). 
A general point  in the phase diagram transforms under the $\mathbb{Z}_3$ symmetry as
\begin{equation}
\begin{split}
 (\theta_1 Q_1 + \theta_3 Q_3 )  \underset{\mathbb{Z}_3^{\phantom{-1}}} \to(\theta_1 Q_2 +\theta_3 Q_1) &= (\theta_3-\theta_1)Q_1 - \theta_1 Q_3\\
  (\theta_1 Q_1 + \theta_3 Q_3 )  \underset{\mathbb{Z}_3^{-1}} \to(\theta_1 Q_3 +\theta_3 Q_2) &= -\theta_3Q_1 + (\theta_1-\theta_3) Q_3
\end{split}
\end{equation}
Under  the $\mathbb{Z}_3$ transformation  the high symmetry points are mapped into themselves modulo $2\pi$ as was discussed in Sec.~\ref{subsec:Z3symm}. The  parity symmetric lines meeting at a $\mathbb{Z}_3$  symmetric point   transform into each other modulo $2\pi$. For example, along the  $\theta_1=-\theta_3$ line the topological term is given by $ i \theta( Q_1- Q_3)$, which transforms to $i\theta (Q_2- Q_1)= i(-2 \theta Q_1-\theta Q_3)$ under  $\mathbb{Z}_3$. The transformed term  is along the $\theta_3= \theta_1/2$ line which is equivalent to the  $ \theta_3= \theta_1/2- \pi$ line  going through the $\theta_1=-\theta_3 =2\pi/3$ point. 
Similarly, under $\mathbb{Z}_3^{-1}$  the $ i \theta( Q_1- Q_3)$ term transforms into  $i\theta (Q_3- Q_2)= i( \theta Q_1+2\theta Q_3)$ along the  $ \theta_3 = 2\theta_1$ line,  which is equivalent to the $\theta_3 =  2\theta_1-2\pi $ line going through $\theta_1=-\theta_3 =2\pi/3$. 

In the same spirit, one can follow  the action of a parity transformation for a general point inside the reduced phase diagram:
\begin{equation}
\begin{split}
 (\theta_1 Q_1 + \theta_3 Q_3 )  &\underset{R_{13}} \to (-\theta_1 Q_3  - \theta_3 Q_1) \\
  (\theta_1 Q_1 + \theta_3 Q_3 )  &\underset{R_{12}} \to(-\theta_1 Q_2-\theta_3 Q_3) = \theta_1  Q_1 + (\theta_1-\theta_3) Q_3\\
    (\theta_1 Q_1 + \theta_3 Q_3 ) & \underset{R_{23}} \to(-\theta_1 Q_1 -\theta_3 Q_2) =( \theta_3-\theta_1) Q_1 + \theta_3 Q_3
\end{split}
\end{equation}
The high symmetry points are once again  invariant, while each parity transformation conserves one parity symmetric line and maps the other two into each other. For example the $\theta_1=-\theta_3$ line is invariant under $R_{13}$, and  maps to  $\theta_3 = \theta_1/2 \equiv  \theta_1/2-\pi$ or to $\theta_3=2\theta_1\equiv2\theta_1-2\pi$ under $R_{23}$ or $R_{12}$, respectively. All these transformation properties are illustrated in Fig.~\ref{fig:ththPDtrans}.

In the $g\to\infty$ limit  the system is gapped for all values of $\theta_1$ and $\theta_3$. This can be seen from the gauge field formulation of the action in Eq.\ \eqref{eq:cp2gaugeform} (or Eqs.\ \eqref{eq:lattrealterm} and \eqref{eq:latttopterm} for the lattice model). In the the $g \to \infty$ limit the action only consists of the topological term, which only depends on the gauge fields, thus the $\vec \phi_n(x)$ fields are correlation free. This means that the mass gap, which is the  inverse of the correlation length, diverges in this limit for all values of the topological angles. 

  At  $\theta_1 =-\theta_3=  2\pi/3$ this is accompanied by a spontaneous breakdown of the $\mathbb{Z}_3$ and parity symmetries. By calculating the expectation value of the topological charge densities \cite{SeibergPRL1984}
\begin{equation}
\begin{split}
\langle q_1 \rangle &= -i \frac{\partial f(\theta_1,\theta_3, g\to\infty)}{\partial \theta_1}\\
\langle q_3\rangle&= -i  \frac{\partial f(\theta_1,\theta_3, g\to\infty)}{\partial \theta_3}\\
\langle q_2\rangle=  -\langle q_1+q_3 \rangle &= i  \frac{\partial f(\theta_1,\theta_3, g\to\infty)}{\partial \theta_1} +i  \frac{\partial f(\theta_1,\theta_3, g\to\infty)}{\partial \theta_3},
\end{split}
\end{equation}
depending on from which sector we approach the high symmetry $\theta_1=-\theta_3=2\pi/3$ point we get
    \begin{equation}
    \begin{array}{l|ccc}
   \theta_1\to 2\pi/3&\multirow{2}{*}{$ \langle \dq_1\rangle$}&\multirow{2}{*}{$\langle \dq_2\rangle$}&\multirow{2}{*}{$\langle \dq_3\rangle $}\\
    \theta_3\to -2\pi/3&&\\
    \hline
   \text{from }  \mathcal{R}_{0,0}  &-i\frac{3}{2 \pi }+i\frac{\sqrt{3}}{8}&0& i\frac{3}{2 \pi }-i\frac{\sqrt{3}}{8}\\
     \text{from } \mathcal{R}_{-1,0} &i\frac{3}{2 \pi } - i\frac{\sqrt{3}}{8}& -i\frac{3}{2 \pi }+i\frac{\sqrt{3}}{8}&0\\
      \text{from } \mathcal{R}_{0,-1} &0&i\frac{3}{2 \pi }-i\frac{\sqrt{3}}{8}&-i\frac{3}{2 \pi }+i\frac{\sqrt{3}}{8}.
        \end{array}
    \end{equation}
    These three cases are connected by the $\mathbb{Z}_3$ transformation. Also under each parity symmetry one scenario is invariant, while the other two transform into each other. The situation is similar at $\theta_1=-\theta_3=4\pi/3$.  Along the transition lines running from  $\theta_1=-\theta_3=2\pi/3$ to a point equivalent to $\theta_1=-\theta_3=4\pi/3$, the free energy has a cusp and the remaining parity symmetry is also spontaneously broken. The transition across these lines is first order since the expectation values of the topological charge densities have a jump.  By contrast, along the  parity symmetric lines running from to $\theta_1=-\theta_3=2\pi/3$ to a point equivalent to $\theta_1=-\theta_3=0$, as well as at $\theta_1= -\theta_3=0$  itself, all symmetries are preserved since the free energy is continuously differentiable inside the $\mathcal{R}_{mn}$ sectors.

\subsection{ Finite g and connection with spin models}
\label{sec:finiteg}
 We believe that the structure of the phase diagram is the same for finite values of the coupling $g$ as well, since the transitions coincide with high symmetry lines/points, although the nature of the transitions can change. According to the discussion of Sec.~\ref{sec:symms}, the phase diagram of the $\sigma$-model can be illustrated by specific spin models. As mentioned, the  $\mathbb{Z}_3$ transformation corresponds to a translation by one site in the spin chain, while the parity symmetries correspond to the mirror symmetries between two neighbouring spins.  The high symmetry points of the phase diagram of the $\sigma$-model correspond to translationally invariant spin models; the $\theta_1=-\theta_3=0$ point corresponds to  spin models with $p =3 m$ boxes, while the $\theta_1=-\theta_3=2\pi/3$ and $\theta_1=-\theta_3=4\pi/3$ points describe the low energy behavior of spin chains with $p = 3m + 1$ and $p = 3m + 2$, respectively.

First, we focus on the neighbourhood of the $\theta_1=-\theta_3=2\pi/3$ point, which can be illustrated by $p=1$ spin models. 
From  Bethe ansatz results we know that the $p=1$ nearest neighbour Heisenberg model is  gapless. However, \citet{CorbozTsunetsugutrimerization} showed that  a transition to a trimerized phase occurs in the $J_1-J_2$ model.   This transition from a gapless to the trimerized phase suggests that a phase transition takes place for the $\sigma$-model at $\theta_1=-\theta_3=\pm2\pi/3$ at some critical coupling  $g_c$. Now the expression for the coupling constant derived in Sec.~\ref{sec:FT} ($1/g = p\sqrt{J_1J_2+ 2J_3 J_1+2 J_3J_2)}/(J_1+J_2)$) shows that $g$ is an increasing function of $J_2$ (for all $J_2>0$ if $J_3\ge J_1/2$ or for $J_2>(J_1^2-2J_1J_3)/(J_1+2J_1 J_3)$ otherwise). So we expect that for $g<g_c$ the system is gapless, while for $g>g_c$ it is gapped with a spontaneous breakdown of the $\mathbb{Z}_3$ symmetry, in agreement with the results of the previous section in the $g\to\infty$ limit. 

If the translational symmetry is explicitly broken, but a mirror symmetry is preserved in the nearest neighbour Heisenberg system ($g <g_c$), the corresponding $\sigma$-model is along one of the parity symmetric lines. 
For example,  changing the strength of the  bonds between sublattices 1 and 3, the $\sigma$-model is tuned along the $\theta_1=-\theta_3$ line (see Sec.~\ref{subsec:symmbreak}). If the bond is weakened, $\theta_1=-\theta_3<2\pi/3$, and we move towards the $\theta_1=-\theta_3=0$ point. This corresponds to a system with ...WSSWSS...\  bonds, where W stands for weaker and S  for stronger.  In this case $\mbox{SU}(3)$ singlets form on the sites connected by stronger bonds, and the system is gapped with a unique ground state. By contrast, if every third bond is strengthened (...SWWSWW...), pairs of sites connected by strong bonds will tend to form $\bar{3}$ states and  the system behaves as a $3\bar{3}$ chain, which spontaneously dimerizes with a finite gap \cite{Afflecknnbar1990,SorensenPRB1990}. This corresponds to a $\sigma$-model with $\theta_1=-\theta_3>2\pi/3$, i.e.\ along a transition line between sectors running from $\theta_1=-\theta_3=2\pi/3$  to $\theta_1=-\theta_3=4\pi/3$. 
If a different mirror symmetry is conserved in the spin model, the underlying $\sigma$-model moves along a different parity symmetric line. If both the translational and mirror symmetries are explicitly broken, the spin system is gapped, and therefore  away from the symmetric lines the $\sigma$-model is also expected to be  gapped.

Considering  the $p=1$, $J_1-J_2$ spin model with spontaneous translational symmetry breaking \cite{CorbozTsunetsugutrimerization} (i.e.\ $g>g_c$), if  every third nearest neighbour bond is weakened (strengthened),  one (two) out of the three ground states will be selected, and in both cases the system remains gapped. This, once again, corresponds to moving along one of the parity symmetric lines around $\theta_1=-\theta_3=2\pi/3$ in the $\sigma$-model phase diagram.  As a consequence we can expect that along the transition lines between different $\mathcal{R}$ sectors the system is gapped and twofold degenerate, while inside the sectors the system is gapped for any of value of $g$.  

In terms of the $\sigma$-model the neighbourhood of the $\theta_1=-\theta_3= 4\pi/3$ point is connected to that of the  $\theta_1=-\theta_3= 2\pi/3$ case by a complex conjugation, therefore the above considerations translate straightforwardly to the $\theta_1=-\theta_3= 4\pi/3$ case as well. This would correspond to spin systems with $p=3m+2$, which thus would show similar general behavior as $p=3m+1$ systems. 
 In the case of $\theta_1=-\theta_3=0$, i.e.\ for spin systems with $p=3m$, explicitly breaking the translational or mirror symmetries in the spin chain will tune the $\sigma$-model away from the $\theta_1=-\theta_3=0$ point, but it will still stay in the same phase. Thus, we expect that the $\sigma$-model is gapped for any value of the coupling inside the $\mathcal{R}$ sectors away from transition lines, even for $\theta_1=-\theta_3=0$.

\subsection{Monte Carlo simulations}
\label{sec:MC}

\begin{figure}[htbp]
\begin{center}
\includegraphics[width=0.9\textwidth]{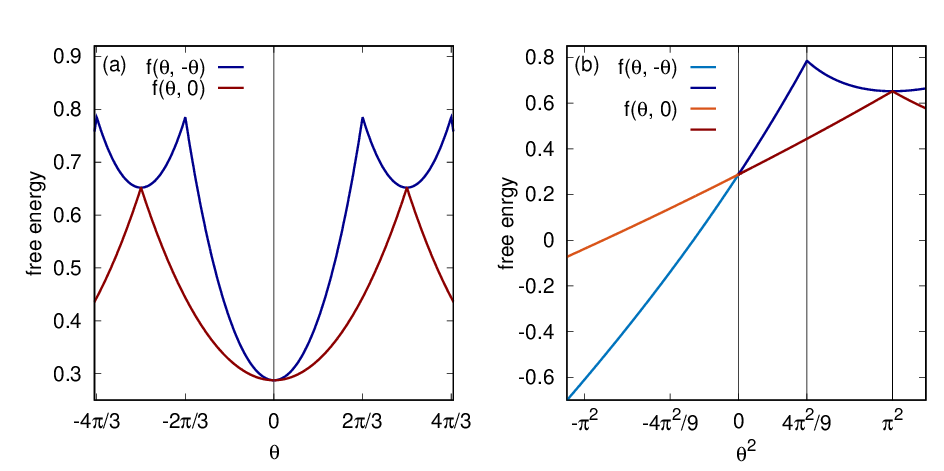}
\caption{(a) The free energy along cuts in the $\theta_1-\theta_3$ plane in the $g \to \infty$ limit. We show the $f(\theta,-\theta)$, and $f(\theta,0)$ cases for which we carried out MC extrapolation calculations at finite $g$ values. (b) Free energy as a function of $\theta^2$ for imaginary and real topological angles for the mentioned two cases at $g \to \infty$. It is clear that the extrapolation breaks down for real topological angles beyond a phase transition.}
\label{fig:Fextrapol}
\end{center}
\end{figure}

To confirm our predictions  for finite $g$ values, we  turn to classical Monte Carlo simulations. As mentioned above, we consider only the special case of $\lambda=0$. In the $\mbox{SU}(2)$ case, several methods have been developed to address the issue raised by the imaginary topological term \cite{WieseMC1995,ImachiYoneyamaPTP2006}. Here we choose the extrapolation scheme of \citet{AllesPapa2008}, which consists in carrying out simulations for imaginary topological angles $\theta = i\vartheta$, and in 
extrapolating those results to real angles. This extrapolation method can be illustrated in the $g\to\infty$ limit. 
Using the results of Sec.~\ref{subsec:ginfty},  the free energy for imaginary angles is given by
 \begin{equation}
\begin{split}
f (i \vartheta_1, i\vartheta_3, g\to\infty)=-\frac{1}{V} \log \mathcal{Z}(i\vartheta_1, i\vartheta_3, g\to\infty) = -\log \Big( \max_{m,n} z(i\vartheta_1+2\pi m, i\vartheta_3 +2\pi n) \Big),
\end{split}
\label{eq:FPBC2}
\end{equation}
In this case the $m=n=0$ term will always dominate. Therefore
\begin{equation}
\begin{split}
f (i \vartheta_1, i\vartheta_3, g\to\infty)
=-\log \left(\frac{2 \bigg(-(\vartheta_1-\vartheta_3) \cosh \left(\frac{\vartheta_1-\vartheta_3}{2}\right)+\vartheta_1\cosh \left(\frac{\vartheta_1}{2}\right)-\vartheta_3 \cosh \left(\frac{\vartheta_3}{2}\right)\bigg)}{\vartheta_1 \vartheta_3(\vartheta_1-\vartheta_3)} \right)
\end{split}
\end{equation}
In Fig.~\ref{fig:Fextrapol} we show the free energy  $f(a\theta, b\theta,g\to\infty)$  as a function of $\theta^2$ for different fixed  values of $a, b$. Since we know that the system undergoes a phase transition between different $\mathcal{R}_{m,n}$ sectors,  we can only hope to get information for points inside or at the boundary of $\mathcal{R}_{0,0}$. Fortunately due to the symmetries of the phase diagram, this is all that we need. For example, the high symmetry point $\theta_1=-\theta_3=2\pi/3$  can be reached by extrapolation  along the $\theta_1 =-\theta_3$  line ($a= -b =1$). The point $\theta_1=-\theta_3=\pi$ cannot be reached by extrapolating along this line since it is beyond a phase transition. However, this point is related to the point $\theta_1= \pi, \theta_3= 0$ by a  $\mathbb{Z}_3$ transformation, and this latter point is reachable by an extrapolation with $a=1, b=0$. In the  following  we present numerical Monte Carlo results along these two lines, but in general one can choose any values of $a,b$ to reach any point of the $\mathcal{R}_{0,0}$ sector via an extrapolation.

\begin{figure}[h]
\begin{center}
\includegraphics[width=0.6\textwidth]{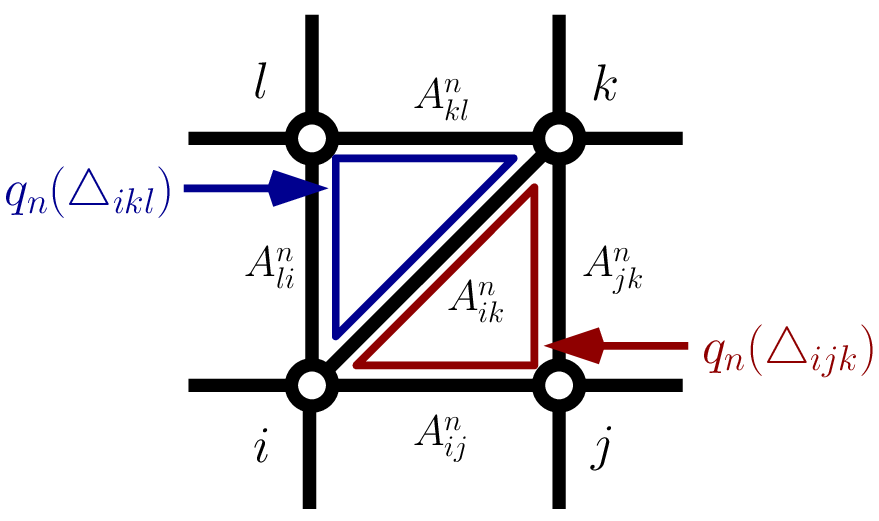}
\caption{Discretization of the 1+1 dimensional space-time as a square lattice. The   fields $\vec{\phi}_1(j), \vec{\phi}_2(j), \vec{\phi}_3(j)$ are defined on each lattice site $j$. They are mutually orthogonal,   giving the rows of the unitary matrix $U(j)$. The  topological term on the lattice is the sum of contributions from distinct triangles defined in Eq.\ \eqref{eq:topoterm_latt}. The index $n$ denotes the three fields.}
\label{fig:latticesplit}
\end{center}
\end{figure}

To implement the lattice Monte Carlo calculation, we discretize the action in Eq.\ \eqref{eq:fullactionphi} on a 1+1 dimensional square lattice for imaginary topological angles $\theta = i\vartheta$, extending the scheme of \citet{AllesPapa2008} to the SU(3) case. We discretize the real part of the action as \cite{DiVecchia1981, RabinoviciSamuel1981}
\begin{equation}
\begin{split}
\sum_{n=1}^{3} \sum_{\mu=x,\tau} \frac{1}{2g}\left(\left| \partial_\mu \vec{\phi}_n^{\phantom{*}}\right|^2 -  \left| \vec{\phi}_n^* \cdot \partial_\mu \vec{\phi}_n^{\phantom{*}} \right|^2 \right) \to -\frac{1}{2g} \sum_{n=1}^3 \sum_{\mu= x,\tau} \left| \vec\phi_n(\vec r_j)\cdot  \vec\phi_n( \vec r_j + \vec \delta_\mu)\right|^2 +\text{const.}\;,
\end{split}
\end{equation}
where  $r_j$ is a site on the discretized two dimensional space time and $ \vec \delta_x,  \vec \delta_\tau$ are the lattice unit vectors of the discretized square lattice.  One can show that this discretization gives back the continuum case in second order.  
The topological part of the action is discretized following the recipe of Berg and L\"uscher \cite{BergLuscher1981}. Every square plaquette is further split into two triangles, and three topological charge densities are defined on each triangle, as shown
in Fig.~\ref{fig:latticesplit}, by
 \begin{equation}
\begin{split}
\exp(i2\pi \, \dq_n(\triangle_{ijk}))  =  \frac{  \big(\vec\phi_n^*(\vec r_i)\cdot \vec\phi_n^{\phantom{*}}(\vec r_j) \big) \big(\vec\phi_n^*(\vec r_j)\cdot \vec\phi_n^{\phantom{*}}(\vec r_k) \big) \big(\vec\phi_n^*(\vec r_k) \cdot \vec\phi_n^{\phantom{*}}(\vec r_i) \big)}{ \big|\vec\phi_n^*(\vec r_i)\cdot \vec\phi_n^{\phantom{*}}(\vec r_j) \big| \big|\vec\phi_n^*(\vec r_j)\cdot \vec\phi_n^{\phantom{*}}(\vec r_k) \big|\big|\vec\phi_n^*(\vec r_k) \cdot \vec\phi_n^{\phantom{*}}(\vec r_i) \big|}.
\end{split}
\label{eq:topoterm_latt}
\end{equation}
 With this notation, the topological charge on the lattice system is calculated as $Q_n= \sum_{\triangle_{ijk}} \dq_n(\triangle_{ijk})$, where on each triangle the indices $i,j,k$ follow the same (e.g.\ counter-clockwise) direction. Each $\dq_n(\triangle)$ can take values between $\pm1/2$.  We note that this scheme is different from that of \citet{SeibergPRL1984} and \citet{PlefkaSamuelPRD1997} that
we used to discuss the $g\to \infty$ limit, but they give the same continuum limit (see \ref{sec:su3calc}).

We implemented a single site Metropolis type Monte Carlo algorithm \cite{Metropolis1953, AllesPapa2008,AllesPapa2014,BergLuscher1981} to measure the correlations of the discretized model.  In this method we sweep the lattice and on each site we replace the $U$ matrix with a uniformly generated \cite{BroznanPRD1988,MathurSen2001} one with a probability given by the Metropolis acceptance ratio. To reduce autocorrelation times and increase accuracy we also complement the single site update with a multigrid update method \cite{Hasenbuschmultigrid}, where we rotate  all matrices within $L_B \times L_B$ square blocks of increasing size. The rotation is done by selecting a  Gell-Mann generator $T^A$ and by updating the matrices in the block according to $U'(j)= U(j) \exp( i \varphi_{L_B} T^a)$, with an angle  $\varphi_{L_B}$  uniformly selected in $ [- \varphi_{max} 1/L_B^{1/2},\varphi_{max} 1/L_B^{1/2}]$ in order the keep the acceptance rate of the updates similar for all $L_B$. In this method one multigrid sweep consists of a single site sweep, then a sweep of non-overlapping blocks of size $L_B=2,4,8...$, respectively. For each $L_B$, the blocks are selected by a random shift. 

\begin{figure}[h]
\begin{center}
\includegraphics[width=0.8\textwidth]{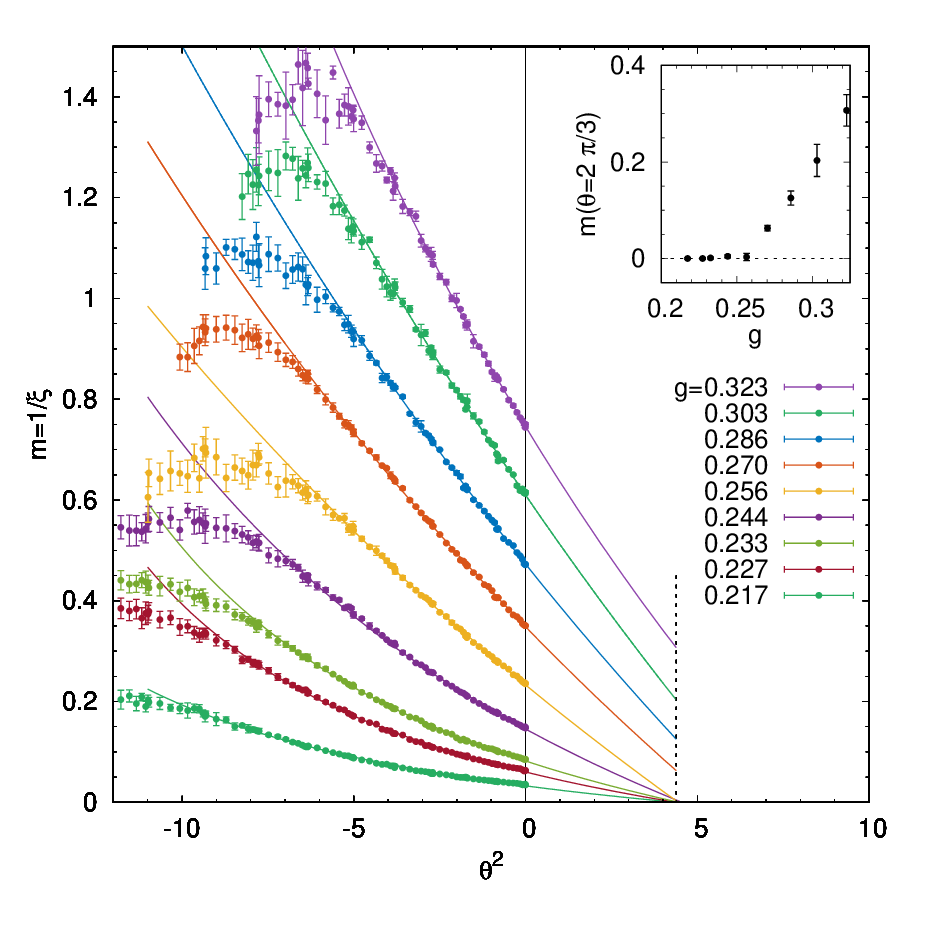}
\caption{Extrapolation of the inverse of the correlation length from Monte Carlo calculations of imaginary $\theta$ angles along the $\theta_1=-
\theta_3$ line, for a system of $192 \times 192$ sites. The inset shows the extrapolated value of the mass gap at $\theta=2\pi/3$.}
\label{fig:extrapolation}
\end{center}
\end{figure}

In the Monte Carlo algorithm we sample the time averaged correlation function \cite{SeibergPRL1984,PlefkaSamuelPRD1997}
\begin{equation}
\begin{split}
C(x)=\frac{1}{L}  \sum_{\tau} \Bigg\langle \left( \vec\phi_1^{\phantom{*}}(0,0) \cdot \vec\phi_1^* (x,\tau)  \right) \left(  \vec\phi_1^* (0,0) \cdot \vec\phi_1^{\phantom{*}}(x,\tau) \right) \Bigg\rangle 
\end{split}
\label{eq:corrfct}
\end{equation}
 and extract the correlation length by an exponential fitting. $C(x)$  is the generalization of the spin-spin correlation function to the $\mbox{SU}(n)$ case (see \ref{app:su2}  for details). 
   For each value of $\theta=i \vartheta$ and $g$ we sampled $2\times 10^5$ configurations with a sampling distance of 10 multigrid sweeps after $5\times10^4$ thermalizing multigrid sweeps. The numerical errors  were estimated by the binning method \cite{Troyerbinning2010}. 
In Fig.~\ref{fig:extrapolation} we show the inverse of the correlation length -- which is proportional to the mass gap -- as a function of $\theta^2$ for the extrapolation along the $\theta_1=-\theta_3$ line.  Following \citet{AllesPapa2008}, we extrapolate the mass gap to real $\theta$ values by fitting it with a function of the form $(c_1 +c_2 \theta^2)/(1+c_3\theta^2)$.

We find that upon increasing $\vartheta$ the mass gap has a change of behaviour. For small imaginary angles, the mass gap increases, i.e.\ the correlation length decreases. But, after a maximum in $1/\xi$, the mass gap decreases. In this region the system is characterized by a uniform saturated topological charge density. This is reminiscent of the results of \citet{ImachiYoneyamaPTP2006} for the $\mbox{CP}^2$ model, who argue that beyond some imaginary angle where the average topological charge of the system becomes comparable to its maximal value ($1/2$ per triangle), the partition function is no longer an analytic continuation of the real $\theta$ case. This means that in the extrapolation method one should  only consider imaginary angles smaller than this threshold.  This is why we chose as a limit for the fitting the inflection points in the mass gap results. 

The Monte Carlo results clearly show that the mass gap is finite for $\theta_1=-\theta_3=0$. This point corresponds to spin systems with  $p= 3 m$ boxes. This finding agrees with the proposal of  \citet{GreiterRachelSchuricht2007} for the possibility of gapped phases for such systems. 
However, for small $g$, the extrapolated mass gap vanishes around $\theta=2\pi/3$. So our Monte Carlo results predict that  spin representations with $p=3m+1$, or $ 3m+2$ are gapless. 
Upon increasing $g$, beyond $g_c\approx 2.55$, the extrapolated mass gap remains finite even at $\theta=2\pi/3$. We believe that this corresponds to a phase transition from a gapless into a gapped phase in accordance with the LSMA theorem.  This also agrees with our earlier conclusion that for $g\to\infty$ the system should be gapped.

\begin{figure}[h]
\begin{center}
\includegraphics[width=0.8\textwidth]{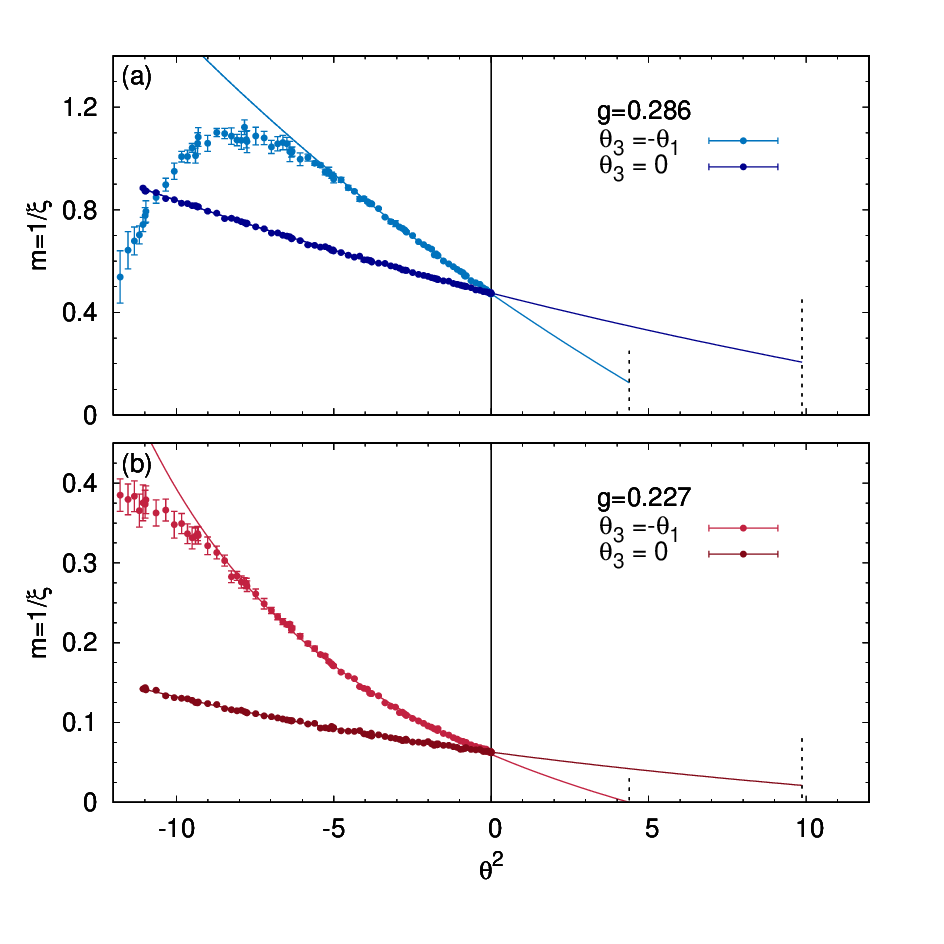}
\caption{Extrapolation of the inverse of the correlation length from Monte Carlo calculations of imaginary $\theta$ angles, for a system of $192 \times 192$ sites. For two couplings (we compare the extrapolations along the $\theta_3=0, \theta= \theta_1$ line with those along  the $\theta=\theta_1=-
\theta_3$ line.  (a) Presents a case, when $g>g_c$, while (b) shows a scenario where $g<g_c$.  It is clear, that in the  $\theta_3=0, \theta= \theta_1$ case the gap remains open at $\theta=\pi$ for both  $g>g_c$, and  $g<g_c$  . }
\label{fig:piextrapolation}
\end{center}
\end{figure}

We also followed the same extrapolation procedure along the $\theta_1 \neq 0, \theta_3=0$ line (see Fig.~\ref{fig:piextrapolation}). These results show that at $\theta_1=\pi, \theta_3=0$  (and thus for  $\theta_1=\pi, \theta_3=-\pi$)  the gap remains finite both below and above $g_c$, as predicted in Sec.~\ref{sec:finiteg}.   

\section{Phase diagram of the flag manifold $\sigma$-model } 
\label{sec:phasediag}

Our expected $\theta_1-\theta_3$ phase diagram for the flag manifold $\sigma$-model is summarized in Fig.~\ref{fig:ththPD}\lcol{b}, which must be complemented with an RG flow towards $g\to\infty$.  A  cut along the $\theta_1=-\theta_3=\theta$ line is shown in Fig.~\ref{fig:SU23RGflow}\lcol{b}  with the topological angle on the vertical axis and the coupling constant $g$ on the 
horizontal axis. We neglect the coupling constant $\lambda$ for the non-topological term linear in space and time derivatives. For generic values of $\theta_1$ and $\theta_3$, we expect the model to be gapped, with $g$ renormalizing to infinity. But for $\theta_1=-\theta_3 = \pm 2\pi /3$ we expect a critical point to occur 
at $g=g_c$, corresponding to the $\mbox{SU}(3)_1$ WZW model. It is important that the model has an extra symmetry at these values of $\theta_1,\theta_3$ which prevents 
it from renormalizing. This is the $\mathbb{Z}_3$ symmetry which permutes the three fields ($\vec \phi_1\to \vec \phi_2\to \vec \phi_3$),  discussed in Sec.~\ref{subsec:Z3symm}.

For $\theta_1=-\theta_3=\pm 2\pi /3$, the phase transition at $g=g_c$ is driven by the topological term $\lambda{'}J^A_RJ^A_L$.
In this term, $J_{R/L}$ are the right and left moving currents in the WZW model given by:
\be J_{R/L}^A\propto \hbox{tr} g^\dagger \partial_{\mp}gT^A\ee
where $\partial_\mp \equiv \partial_t\mp \partial_x$ and the $T^A$ are the generators of $\mbox{SU}(3)$. This interaction is marginally irrelevant 
for one sign of $\lambda '$ and marginally relevant for the other. Thus, we expect  $\lambda'$ to change sign at $g=g_c$ and for $g>g_c$ the gap to turn on exponentially slowly, as 
$\Delta \propto e^{-c/(g_c-g)}$. 

On the other hand, shifting $\theta_1, \theta_3$ slightly away from $\pm 2\pi /3$ corresponds to breaking the  $\mathbb{Z}_3$ symmetry.  We expect this symmetry to 
correspond to $g\to e^{i2\pi /3}g$ in the $\mbox{SU}(3)_1$ WZW model, the symmetry which forbids a $\tr g$ term in the effective Hamiltonian. When 
this symmetry is broken we expect a relevant perturbation $\propto \tr g$.  This operator has dimension $d=2/3$ so we 
expect the gap to scale as $|\theta -2\pi /3|^{1/(2-d)}=|\theta -2\pi /3|^{3/4}$, up to log corrections coming from the marginal operator $J^A_RJ^A_L$. If the $\mathbb{Z}_3$ symmmetry 
is broken, but a parity symmetry is preserved, along the $\theta_1=-\theta_3=\theta$ line for example, we believe that the extra term should have the form $\tr g+ \tr  g ^\dagger $ 
since $g \to g^\dagger$ corresponds to the parity transformation. In this case the extra term in the  $\mbox{SU}(3)_1$ WZW model should have the form $\propto (\theta-2\pi/3) (\tr g +\tr g^\dagger)$. If we write the diagonal elements of $g$ as $e^{i \alpha_j}$ for $\alpha=1,2,3$ with $\sum_j \alpha_j \equiv 0 (\text{mod } 2\pi) $, the extra term takes the form
\begin{equation}
\begin{split}
V \propto  (\theta-2\pi/3) \sum_j \cos \alpha_j.
\end{split}
\end{equation}
For $\theta < 2\pi/3$, this term has a unique minimum with $\alpha_j=0$. But if $\theta>2\pi/3$, there are two minima, with $\alpha_j = 2\pi/3$ or $\alpha_j = -2\pi/3$ . As discussed in Sec.~\ref{subsec:ginfty}, the two cases correspond to spin chains with ...SSWSSW... and ...WWSWWS... bond patterns, with a unique and gapped ground state in the former, and gapped twofold degenerate ground states in the latter case. 
The situation is similar along the other two parity conserving lines ($-\theta_3= \pi -\theta_1/2$ and $ -\theta_3 = 2\pi -2\theta_1 $), which are connected to the $\theta_1=-\theta_3$ line by the $\mathbb{Z}_3$ transformation. Therefore along these lines the extra term in the WZW model should be $ \propto ( e^{\pm i2\pi/3} \tr g + e^{\mp i2\pi/3} \tr g^\dagger)$, which will also  have 1 or 2 minima depending on the sign.
For a generic point around $\theta_1=-\theta_3=2\pi/3$ the extra term should have the form $(\mu \tr g + \mu^*\tr g^\dagger)$, where $\mu$ is complex in general, and vanishes at the $\mathbb{Z}_3$ symmetric point.  
Based on the above considerations we believe the general form is  $\mu = \exp( i\theta_1) +\exp(i\theta_2)+\exp(i\theta_3)$, where in our discussion we fixed $\theta_2 = 0$. 


It is interesting to contrast these results with those of the $\mbox{CP}^2$ model with a topological term. If we ignore the non-topological term $\lambda$-term, the flag manifold $\sigma$-model we have studied can be seen as three copies of the $\mbox{CP}^2$ model coupled by the orthogonality
constraint. Now, it is well established that, as soon as $n>2$, the $\mbox{CP}^{n-1}$ model with a topological term is gapped for all values of the coupling constant
and of the topological angle \cite{SeibergPRL1984,WieseCPN1PRL2005,VicariPhysRep2009}. The model undergoes a phase transition with spontaneous breaking of charge conjugation symmetry at $\theta=\pi$, but it is first order, and the gap does not close.  So the coupling between the fields appears to be essential to produce the interferences that close the gap at $\theta=2\pi/3$.

\begin{figure}[h]
\begin{center}
\includegraphics[width=0.8\textwidth]{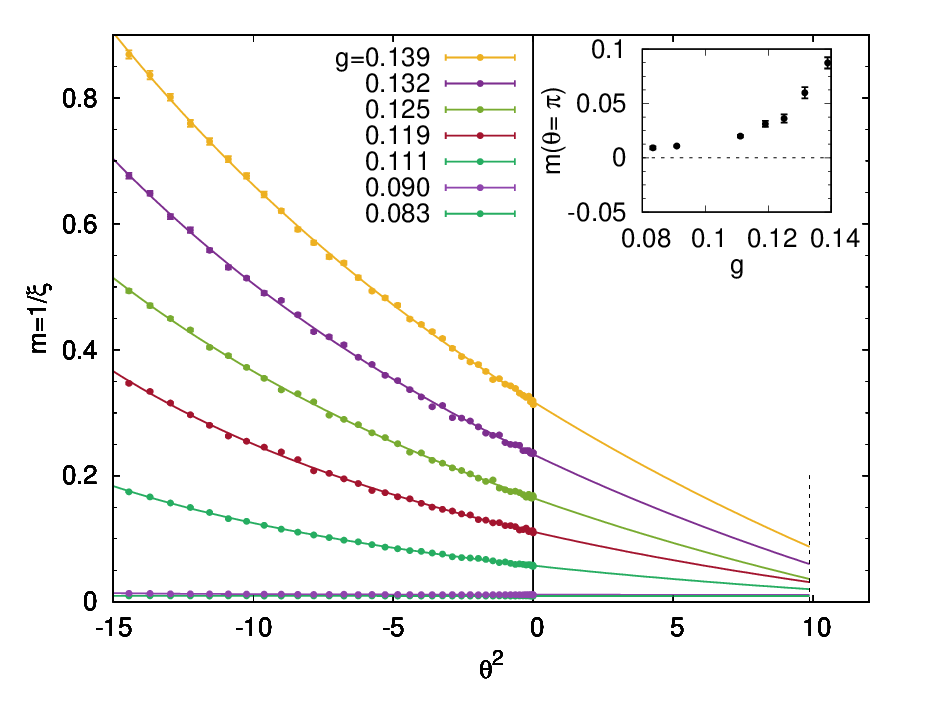}
\caption{Extrapolation of the inverse of the correlation length from Monte Carlo calculations of imaginary $\theta$ angles for the $\mbox{CP}^2$ model on a  $192 \times 192$ lattice. The results are consistent with the expectation, that the mass gap remains open for all values of $\theta$.  }
\label{fig:CP2}
\end{center}
\end{figure}

We also carried out shorter MC simulations (with $5\times10^4$ samples) for the $\mbox{CP}^2$ model, shown in Fig.~\ref{fig:CP2}. The correlation lengths are extracted the same way as for the $\mbox{SU}(3)$ case.  The extrapolation clearly shows that the gap remains open even for $\theta= \pi$.    
\section{Conclusions and open questions}
\label{sec:conclusion}

Let us first summarize the implications of our field theory results for the $\mbox{SU}(3)$ chains in the $p$-box symmetric representation. For a
translationally invariant system, the model can be defined by a single topological angle $\theta =2\pi p/3$. When $p=3m$, $\theta$ is a multiple of $2\pi$, so there is no topological
term in the action, and the model is expected to be gapped whatever the coupling constant. This prediction is the generalization of Haldane's
prediction of a gap for integer spin in the SU(2) case. When $p=3m\pm 1$, there is a nontrivial topological term in the action with 
topological angle $\theta =\pm 2\pi /3$. In that case, we have shown that there is a critical coupling constant $g_c$ below which the model
is gapless in the SU(3)$_1$ WZW universality class. Starting from the original SU(3) lattice model, the coupling constant is given by 
$g = (J_1+J_2)/p\sqrt{J_1J_2+ 2J_3 J_1+2 J_3J_2)}$, where the additional interactions $J_2$ and $J_3$ are effective couplings that  
have been included to account
for the effect of the zero point fluctuations, which select the three sublattice order. In the large $p$ limit, these additional exchange integrals $J_2,J_3\propto 1/p$,
and the coupling constant $g\propto 1/\sqrt{p}$ becomes very small. So, our field theory results predict that for $p=3m\pm 1$ and large $m$, the model 
of Eq.\ \ref{eq:H_nn} should be gapless. Since this is known to be true also for $p=1$ from Sutherland's Bethe ansatz solution, we conjecture that the same conclusion holds for all $p$ not multiple of 3.

The field theory also predicts that, if the topological angle $\theta =\pm 2\pi /3$, i.e.\ for $p=3m\pm 1$, there should be a phase transition
into a gapped phase upon increasing the coupling constant $g$ beyond a critical value $g_c$. According to the Lieb-Schulz-Mattis-Affleck theorem \cite{LSM1961,AffleckLieb1986}, this phase should be at least three-fold
degenerate. So we expect it to be spontaneously trimerized, with a gap turning on exponentially slowly. 
The form of the coupling constant $g = (J_1+J_2)/p\sqrt{J_1J_2+ 2J_3 J_1+2 J_3J_2)}$
suggests that this transition can be induced by increasing the next nearest neighbour coupling $J_2$. This prediction is consistent with the spontaneous
trimerization reported for the SU(3) chain with $p=1$ and next nearest neighbour interactions \cite{CorbozTsunetsugutrimerization}.

By contrast, the gapped phase that occurs for $g<g_c$ when $\theta$ is shifted from $\pm 2\pi /3$ should arise from breaking the 
translation symmetry by hand, with a period 3 exchange term. 
We expect that the coefficient of the periodic exchange coupling $\lambda_3$ will be proportional to $|\theta \mp 2\pi /3|$, and accordingly that 
the gap will scale as $\lambda_3^{3/4}$.

Let us now briefly review the issues that  deserve further investigation. The first one is about the additional couplings $J_2$ and $J_3$ that we have introduced to get rid of the zero modes. To introduce additional interactions to mimic the effect of zero point fluctuations is a standard approach in 2D and 3D frustrated magnets when the classical ground state is infinitely degenerate. In particular, if the system develops long-range magnetic order because of zero point fluctuations, an effect known as "order by disorder"\cite{Lacroixmagnetism2011}, introducing additional effective couplings is necessary to restore the appropriate structure of Goldstone modes. These couplings must come form higher order spin wave interactions, but even in the simpler context of SU(2) models, there is no known systematic way to calculate the $1/s$ expansion of these couplings. So how to actually calculate these couplings in the context of the present paper remains an open question. 

The other open issue concerns the determination of the critical exponents of the field theory at $\theta=2\pi/3$. For the $\mbox{CP}^1$ model, the critical exponents of the mass gap (main term and logarithmic correction) have been determined on the basis of the numerical results obtained for imaginary topological angles \cite{Azcoiti2003,AllesPapa2014}. This approach relies on a change of variable that is motivated by exact results obtained on the Ising model in imaginary magnetic field \cite{AzcoitiNuclPhys2011}. Whether a similar approach with an appropriate change of variable can be used in the present case to extract the exponents from the imaginary topological angle data remains to be seen. 

The numerical check of our predictions for $p>1$ in the context of the lattice SU(3) model is a real challenge for all types of simulations: Quantum Monte Carlo simulations
suffer from a severe minus sign problem, exact diagonalizations are limited to very small system sizes because the local Hilbert space grows very
fast, and DMRG simulations require to keep a huge number of states to reach convergence. As a consequence, the results obtained so far 
are not conclusive, and the nature of the ground state and of the low-energy spectrum is an open numerical issue as soon as the number of
boxes $p$ is larger than one. Let us briefly review the current status of the numerical investigation of these models to substantiate this conclusion.

The first numerical study of the SU(3) Heisenberg chain with $p=2$ and $3$ is a DMRG investigation by \citet{GreiterDMRG2009}. 
In this paper, the authors report on a calculation of the entanglement spectrum of a chain of 48 sites with periodic boundary conditions, keeping
5000 states for $p=2$ and 1650 states for $p=3$. For $p=2$, fitting the curve with the Calabrese-Cardy formula \cite{CalabreseCardy2009}, they found a central charge
equal to $2.48$, which they interpret as being  evidence that the model is in the $\mbox{SU}(3)_1$ universality class with central charge 2, the difference
being attributed to logarithmic corrections. For $p=3$, they found that the entanglement entropy saturates after a few sites, which was taken 
as an indication that the system is gapped. 

The second numerical study is an exact diagonalization investigation by \citet {NatafchainED2016} based on a new approach that allows one
to work directly in the irreducible representations of the global $\mbox{SU}(n)$ symmetry \cite{NatafEDPRL2014}. This has allowed one to investigate the properties of the SU(3) model 
with $p=2$ up to 15 sites and with $p=3$ up to 12 sites with periodic boundary conditions, and to extract the properties of the model using standard finite-size
analysis. Quite surprisingly, the results turn out to be consistent with a gapless spectrum in both cases, with a central charge $c=3.23$ for $p=2$, in
good agreement with the $\mbox{SU}(3)_2$ universality class (central charge 3.2), and with a central charge $c=4.09$ for $p=3$, in good agreement with the
$\mbox{SU}(3)_3$ universality class (central charge 4). These results have led the authors of Ref.\ \cite{NatafchainED2016} to suggest that there is a length scale in the problem beyond which 
the system might cross-over to another universality class or to a gapped behaviour, and below which the system looks $\mbox{SU}(3)_p$. 

At first sight, these conclusions might look in agreement with the conclusions of \citet{GreiterDMRG2009}, who worked on much larger systems 
(48 sites), but they are not. In particular, the saturation of the entanglement entropy of the $p=3$ case reported in Ref.~\cite{GreiterDMRG2009} points
to a correlation length of the order of 6 lattice sites, as for the spin-1 SU(2) chain, a result clearly incompatible with the ED results of 
Ref.~\cite{NatafchainED2016}, where no sign of gap (i.e.\ no curvature of the finite-size gap) could be detected up to 12 sites, in sharp contrast with the
spin-1 SU(2) chain.

According to recent DMRG results, the problem comes from the number of states kept in Ref.~\cite{GreiterDMRG2009}, which was too small to reach
convergence for the entanglement entropy. Using codes where the $\mbox{SU}(n)$ symmetry is fully implemented, \citet{WeichselbaumPC} 
and \citet{NatafPC} have been able to keep a much larger number of states. This has allowed them to reach convergence on very large systems with open boundary conditions, with conclusions that differ significantly from those of Ref.~\cite{GreiterDMRG2009}. For $p=2$, and up to at least 300 sites, the central charge is larger than 3, in agreement with ED and with the $\mbox{SU(3)}_2$ 
universality class \cite{WeichselbaumPC, NatafPC}. For $p=3$, the entanglement entropy does not saturate but is compatible with a central
charge larger than $4$ up to 120 sites, and larger systems are currently under investigation to see if the presence of a gap can be detected \cite{NatafPC}.

Finally, with the help of Monte Carlo calculations, we have provided strong evidence that the $\mbox{SU}(3)/[\mbox{U}(1)\times \mbox{U}(1)]$ nonlinear $\sigma$-model is massless for $\theta =\pm 2\pi /3$. 
This represents the first generalization of Haldane's argument for the $\mbox{SU}(2)/\mbox{U}(1)$ $\sigma$-model at $\theta =\pi$ and leaves open the intriguing possibility that a more general set of 
$\sigma$-models have massless phases for special values of their topological angles. [We are currently studying the generalization of our results to $\mbox{SU}(n)$ chains for general $n$.]
This may have implications for field theories with topological terms in higher dimensions 
[including QCD in (3+1) dimensions] and might be useful in finding exact exponents for the quantum Hall effect localization transition. 


\vspace{5mm}

At the time of publication, we were unaware of the following paper of Bykov, \cite{2012NuPhB.855..100B}, that previously derived the flag manifold sigma model, including topological and $\lambda$ terms, from the SU(3) chains considered here. We would also like to acknowledge the paper \cite{10.21468/SciPostPhys.6.2.017} of Ohmori et. al., that corrected our original calculations of the beta function for $\lambda$. Despite its relevance in the flag manifold sigma model, those authors argue that a flow to the SU(3)${}_1$ WZW model is still possible, so that our predicted phase diagram in Fig.~\ref{fig:SU23RGflow}\lcol{b} is still valid. The Monte Carlo calculations could be extended to include the $\lambda$-term as well by making simulations for both imaginary $\theta$ and $\lambda$, and then extrapolate in both variables. This, however, would require substantially more computational time, and we don't have any good indication of what form of extrapolation function would be appropriate either. Including the $\lambda$-term in the $g\to \infty$ limit
is currently under investigation. 

\section*{Acknowledgments}
We would like to thank Francisco Kim, Philippe Lecheminant, Nathan Seiberg, and Andrew Smerald for helpful discussions. We would also like to acknowledge the {\it Exotic states of matter with $\mbox{SU}(N)$ symmetry workshop} \cite{Kyotoworkshop}(YITP program YITP-T-16-03) for the opportunity to discuss and present our results. This research was supported in part by
  an NSERC  graduate scholarship (KW),  NSERC Discovery Grant 04033-2016 (IA),  CIFAR (IA), and by the Swiss National Science Foundation (ML, FM). 
 ML also acknowledges the computational time provided through the Hungarian OTKA Grant No. K106047. 

\appendix 

\section{Flavour wave theory}
\label{FWT}

The calculation of the harmonic fluctuations around an ordered state is most easily done using 
linear flavour wave theory, the extension of the usual SU(2) spin wave theory to $\mbox{SU}(n)$ models. It has been formulated in Refs.~\cite{N1984281} 
and \cite{papa1988} for the SU(3) case and in Ref.~\cite{JoshiZhang1999} for the SU(4) case. The notations used in this appendix are those of 
Ref.~\cite{bauer2012}, where the triangular and square lattices are treated.

To treat fluctuations around a three sublattice ordered state where the spins on the sites $l_\alpha$ belonging to sublattice $\subl_\alpha$ point in the direction $\alpha$,
we start from the $p \rightarrow \infty$ limit in which the bosons $\alpha$ have condensed at the sites of sublattice $\subl_\alpha$, and we do a $1/p$ expansion, by analogy with the spin wave theory that is a 1/s expansion for SU(2) systems. Starting from the ordered state we can use the following expansion for the spin operators $S^\beta_\gamma$ for sites $l_\alpha \in \subl_\alpha$  in the large--$p$ limit:
\begin{equation}
\begin{split}
  S^\alpha_\alpha(l_\alpha) &= p - \mu_\alpha(l_\alpha), \\
  S^\alpha_\beta(l_\alpha) &= b^{(\alpha)\dagger}_{\beta}(l_\alpha) \sqrt{p- \mu_\alpha(l_\alpha)}
     \simeq \sqrt{p} \ b^{(\alpha)\dagger}_{\beta}(l_\alpha) ,
   \\
  S^\beta_\alpha(l_\alpha) &= \sqrt{p -  \mu_\alpha(l_\alpha)} b^{(\alpha)}_{\beta}(l_\alpha)
   \simeq \sqrt{p} \ b^{(\alpha)}_{\beta}(l_\alpha), \\
  S^{\beta'}_\beta(l_\alpha) &= b^{(\alpha)\dagger}_{\beta}(l_\alpha) b^{(\alpha)}_{\beta'}(l_\alpha),
 \end{split}
\end{equation}
where we have introduced the shorthand notation
\begin{equation}
  \mu_\alpha(l_\alpha) = \sum_{\beta (\neq \alpha)} b^{(\alpha)\dagger}_{\beta}(l_\alpha) b^{(\alpha)}_{\beta}(l_\alpha) .
\end{equation}
The $b^{(\alpha)\dagger}_{\beta}(l_\alpha)$ operators with $\beta\neq \alpha$ now correspond to the Holstein--Primakoff bosons on sublattice $\subl_\alpha$, and the superscript $(\alpha)$ keeps track of the sublattice. 
Expanding in $1/p$ and keeping the quadratic terms only, the exchange
term between sites $l_\alpha \in \subl_\alpha$ and  $m_{\alpha'} \in \subl_{\alpha'}$ is given by
\begin{equation}
\begin{split}
 \sum_{\beta,\gamma} S_{\beta}^{\gamma}(l_\alpha) S_{\gamma}^{\beta}(m_{\alpha'}) = p
      \bigl[
         &b^{(\alpha)\dagger}_{\alpha'}(l_\alpha) b^{(\alpha)}_{\alpha'}(l_\alpha)
       + b^{(\alpha')\dagger}_{\alpha}(m_{\alpha'}) b^{(\alpha')}_{\alpha}(m_{\alpha'})  \\
 &+ b^{(\alpha)\dagger}_{\alpha'}(l_\alpha) b^{(\alpha')\dagger}_{\alpha}(m_{\alpha'})
       + b^{(\alpha)}_{\alpha'}(l_\alpha) b^{(\alpha')}_{\alpha}(m_{\alpha'})
      \bigr]
  \end{split}
\end{equation}
for $\alpha \neq \alpha'$, and by
\begin{equation}
\begin{split}
 \sum_{\beta,\gamma} S_{\beta}^{\gamma}(l_\alpha) S_{\gamma}^{\beta}(m_\alpha) = p \sum_{\beta(\neq\alpha)}
       \bigl[&-
         b^{(\alpha)\dagger}_{\beta}(l_\alpha) b^{(\alpha)}_{\beta}(l_\alpha)
       - b^{(\alpha)\dagger}_{\beta}(m_\alpha) b^{(\alpha)}_{\beta}(m_\alpha) \\
 &+ b^{(\alpha)\dagger}_{\beta}(l_\alpha) b^{(\alpha)}_{\beta}(m_\alpha)
   + b^{(\alpha)\dagger}_{\beta}(m_\alpha) b^{(\alpha)}_{\beta}(l_\alpha)    
      \bigr] 
\end{split}
\end{equation}
for $\alpha = \alpha'$. Note that when $\alpha \neq \alpha'$ the exchange term does not involve bosons with flavours different from the ordered ones $\alpha$ and $\alpha'$. 

Assuming a three sublattice ordered state, we further define the following Fourier transformation:
\begin{equation}
  b^{(\alpha)}_{\beta,k} = \sqrt{\frac{3}{\left|\subl \right|}} \sum_{l_\alpha \in \subl_\alpha} b^{(\alpha)}_\beta(l_\alpha)
  e^{i k l_\alpha}
\end{equation}
where the summation is over the $\left|\subl \right|/3$ sites of sublattice $\subl_\alpha$  ($\left|\subl \right|$ is the number of lattice sites).
The Hamiltonian involving bosons with subscripts and superscripts $\alpha$ and $\beta$ then reads
\begin{equation}
\begin{split}
  {\cal H }_{\alpha \beta} = p\sum_{k}
   \bigl[ 2(J_1+J_2+2(1-\cos (3ka))J_3)
          b^{(\beta)\dagger}_{\alpha,k} b^{(\beta)}_{\alpha,k}  + (J_1 \gamma_k^1+J_2 \gamma_k^2)  b^{(\alpha)\dagger}_{\beta,-k} b^{(\beta)\dagger}_{\alpha,k}
    +\text{H.c.}
          \bigr] ,
          \end{split}
\end{equation}
where the factors $\gamma_k^1$ and $\gamma_k^2$ are given by
\begin{equation}
 \gamma_k^1 = e^{ika},\ 
 \gamma_k^2 = e^{-2ika}.
\end{equation}

The full Hamiltonian is $\mathcal{H} = \mathcal{H}_{AB}+\mathcal{H}_{BC}+\mathcal{H}_{CA}$.
It can be diagonalized via a Bogoliubov transformation,
leading to
\begin{equation}
 \mathcal{H} = 
 p \sum_{k\in \textrm{RBZ}} \sum_{\alpha} \sum_{\beta\neq\alpha}
  \omega(k)
\tilde b^{(\alpha)\dagger}_{\beta,k} \tilde b^{(\alpha)}_{\beta,k}  + \text{const.}
   \;.
\end{equation}
The dispersion of the flavour waves is given by
\begin{equation}
\begin{split}
  \omega(k)
= p\sqrt{\Big(J_1+J_2+2\big[1-\cos (3ka)\big]J_3\Big)^2- \Big(J_1^2+J_2^2+2J_1J_2\cos(3ka)\Big)}
\end{split}
\end{equation}
There are 6 degenerate branches in the reduced Brillouin zone  $[-\pi/3,\pi/3]$, and 6 Goldstone modes. 

\section{Details of Lieb-Schultz-Mattis-Affleck theorem}
\label{LSMA}
To show that $U\ket{\psi}$ has low energy, consider the term in $\mathcal{H}$ involving the two neighbouring sites, $j$ and $j+1$:
\be \mathcal{H}_{j,j+1}\equiv J_1S^\alpha_\beta (j)S^\beta_\alpha (j+1).\ee
Since $[Q_j+Q_{j+1},\mathcal{H}_{j,j+1}]=0$, and the other $Q_i$'s commute with $\mathcal{H}_{j,j+1}$, 

\begin{equation}
\begin{split} 
U^\dagger \mathcal{H}_{j,j+1}U&=e^{-i(\pi / 3L)(Q_{j+1}-Q_{j})}\mathcal{H}_{j,j+1}e^{i(\pi / 3L)(Q_{j+1}-Q_{j})} \\
&\approx \mathcal{H}_{j,j+1}+i\frac{\pi}{3L}\big[\mathcal{H}_{j,j+1},Q_{j+1}-Q_{j}\big]+\mathcal{O}(1/L^2) \\
&=\mathcal{H}_{j,j+1} + i\frac{2\pi}{3L} \big[\mathcal{H}_{j,j+1}, \sum_j j Q_j\big] +\mathcal{O}(1/L^2)
\end{split}
\label{1/L}
\end{equation}
Summing over all terms in $\mathcal{H}$, 
\begin{equation}\begin{split} 
\bra{\psi} U^\dagger \mathcal{H}U\ket{\psi}=\bra{\psi}  \mathcal{H} \ket{\psi}  +i\frac{2\pi} {3L}\bra{\psi }[ \mathcal{H} ,\sum_{j=1}^LjQ_j]\ket{\psi}+\mathcal{O}(1/L).
\end{split}\end{equation}
The second term on the right hand side vanishes, since $\ket{\psi}$ is an eigenstate of $\mathcal{H}$. The remainder is seen to be $\mathcal{O}(1/L)$ because there are $\mathcal{O}(L)$ terms in $\mathcal{H}$ 
and each contributes a term $\propto 1/L^2$ as follows from Eq.\ \eqref{1/L}.
Now we wish to prove that $U\ket{\psi}$ is orthogonal to $\ket{\psi}$.  To do this, we can assume that $\ket{\psi}$ is translationally invariant, $T\ket{\psi }=\ket{\psi }$, where $T$ generates translations by 1 site. 
\be T^{-1}UT=\exp\big[ i\frac{2\pi}{3L}\sum_{j=1}^LjQ_{j+1}\big].\ee
Using periodic boundary conditions, $Q_{L+1}=Q_1$, we obtain:
\be T^{-1}UT=U\exp\big[ -i\frac{2\pi}{3L}\sum_{j=1}^LQ_{j}\big]\exp\big[ i\frac{2\pi}{3}Q_1\big].\ee
Next, we use the fact that $ (\sum_j Q_j) \ket{\psi}=0$, which follows from the $\mbox{SU}(3)$ invariance of the ground state,  so
\be \exp\big[-i\frac{2\pi}{ 3}\sum_{j=1}^LQ_j\big] \ket{\psi}=\ket{\psi }.\ee
Finally we use
\begin{equation}
\exp\big[{i(2\pi/3)Q_1}\big]= \exp\big[ i 2\pi p/3 \big] \exp \big[ -i 2\pi b^{\dagger}_3(1) b^{\phantom{\dagger}}_3(1)\big]  = e^{i2\pi p/3} \bf{I}.
\end{equation}

\section{Spin coherent state path integral of the SU(2) Heisenberg chain}
\label{app:su2}

Typically, when discussing Haldane's conjecture  a derivation of the $\mbox{O}(3)$ nonlinear $\sigma$-model from the Heisenberg model is presented (see for example Refs. \cite{fradkin2013,Cabra2004}). To make the connection with the calculations of Sec.~\ref{sec:FT} more explicit,  we provide here a derivation of the $\mbox{CP}^1$ nonlinear $\sigma$-model directly starting from the SU(2) nearest neighbour Heisenberg model. We also show the equivalence between the $\mbox{CP}^1$ and 
$\mbox{O}(3)$ formulations.

We start from the nearest neighbour antiferromagnetic Heisenberg model
\begin{equation}
\begin{split}
H=\sum_{i} J_1\vec{S}(i) \cdot \vec{S}(i+1)
\end{split}
\end{equation}
where  $\vec{S}_i = (S_i^x, S_i^y, S_i^z)$ are the conventional $\mbox{SU}(2)$ generators in the   $\vec{S}^2=s(s+1)$ representation. These are connected to the general form defined in Eq.\ \eqref{eq:Scommrel} as
\begin{equation}
\begin{split}
S^x = \frac{S^2_1+S^1_2}{2}, \quad S^y = \frac{S^2_1-S^1_2}{2i}, \quad S^z = \frac{S^1_1-S^2_2}{2}.
\end{split}
\end{equation}
In a Schwinger boson representation, $b^\dagger_1(b^\dagger_2)$ corresponds to creating an $\uparrow$$(\downarrow)$ spin, respectively.  The spin operators can be written as
\begin{equation}
\begin{split}
\vec{S}(i) =  \frac{1}{2}  b_\alpha^\dagger(i)\vec{\sigma}^{\alpha \beta} b_\beta^{\phantom{\dagger}}(i)\quad  \text{or} \quad S^\alpha_\beta (i)= b^\dagger_\beta(i) b_\alpha(i),
\end{split}
\end{equation}
where $\vec{\sigma}=(\sigma_x,\sigma_y, \sigma_z)$ is a vector of the Pauli matrices. In this representation, the Heisenberg interaction is given by
\begin{equation}
\begin{split}
\vec{S}(i) \cdot \vec{S}(i+1)=\frac{1}{2} S^\alpha_\beta(i) S^\beta_\alpha(i+1) +\text{const.}\;.
\end{split}
\end{equation}

To write down the imaginary time partition function, we use a spin coherent state path integral approach. In terms of Schwinger bosons, 
the $\mbox{SU}(2)$ spin coherent states can be written as
\begin{equation}
\begin{split}
\ket{\vec{\Phi}} = \frac{1}{\sqrt{(2s)!}}\big(\Phi_1^{\phantom{\dagger}} b_{1}^\dagger + \Phi_2^{\phantom{\dagger}} b_{2}^\dagger\big)^{2s} \ket{0}
\end{split}
\end{equation}
where  $\vec{\Phi}$ is a two component complex unit vector which is usually parametrized in terms of two angles as $\vec{\Phi} = (\cos(\theta/2) , \sin(\theta/2)e^{i \varphi})$, with $0\leq\theta \leq\pi$  and $0 \leq \varphi \leq 2\pi$  and with the  integral measure $d\Omega_{\vec \Phi} = \sin(\theta)/(4\pi) d\theta d\varphi$.  The partition function in the path integral language is formally the same as in Eq.\ \eqref{eq:SU3partitionfunction}
\begin{equation}
\begin{split}
 \Tr \big( e^ {- \beta H}\big)= \int \mathcal{D} [{\Phi}]  \exp \Bigg[ -\int \limits_0^\beta d\tau \Big[ &\bra{ \{\vec{\Phi}(\cdot,\tau) \} }
   H  \ket{ \{ \vec{\Phi}(\cdot,\tau) \}}  +  2s\sum_{j}\vec{\Phi}^*(j,\tau)  \cdot \partial_\tau \vec{\Phi}(j,\tau)  \Big] \Bigg],
\end{split}
\label{eq:su2partitionfunction}
\end{equation}
where $ \ket{ \{ \vec{\Phi}(\cdot,\tau) \}}$ represents a direct product of spin coherent states of each spin at time $\tau$. 
When calculating the expectation value of the Hamiltonian the spin operators can be replaced by classical three dimensional real vectors
\begin{equation}
\begin{split}
\bra{\vec \Phi} \vec{S}\ket{\vec \Phi} &= s \ \Big( \Phi^{1*} \Phi^{2}+\Phi^{2*} \Phi^1, -i\big(\Phi^{1*} \Phi^2 -\Phi^{2*} \Phi^{1}\big), \Phi^{1*} \Phi^{1} -\Phi^{2*} \Phi^2 \Big)\\
&= s \ \Big(\cos\varphi \sin\theta, \sin\varphi \sin\theta,  \cos\theta \Big)\\
&=s\ \big( n_1, n_2, n_3\big),
\end{split}
\label{eq:su2exp}
\end{equation}
where $\vec n$ has unit length and
\begin{equation}
\begin{split}
\vec{n}= {\Phi}^{\alpha*} \vec{\sigma}_{\alpha \beta} {\Phi}^{\beta}.
\end{split}
\label{eq:nphi}
\end{equation}
Similarly\begin{equation}
\begin{split}
\bra{\vec \Phi} S^\alpha_\beta \ket{\vec \Phi} = 2s\,\Phi^{\beta*} \Phi^{\alpha}.
\end{split} 
\end{equation}

The Heisenberg interaction in the path integral is given by
\begin{equation}
\begin{split}
\bra{\vec{\Phi}(i)}\vec{S}(i)\ket{\vec{\Phi}(i)}\cdot\bra{\vec{\Phi}(i+1)}\vec{S}(i+1)\ket{\vec{\Phi}(i+1)}= s^2 \vec{n}(i)\cdot \vec{n}(i+1) = \frac{4s^2}{2}\left| \vec{\Phi}(i)^*\cdot \vec{\Phi}(i+1)\right|^2 + \text{const.}\;.
\end{split}
\end{equation}
In the classical ground state, the $\vec{n}$ vectors on neighbouring sites should be antiparallel, or, equivalently, the $\vec{\Phi}$ vectors should be orthogonal. In the path integral, we consider fluctuations around the classical ground state manifold, which are conventionally  parametrized as
\begin{equation}
\begin{split}
\vec{n}(2j) &=  \sqrt{1-\frac{a^2}{s^2} |\vec{l}|^2}\,\vec{m}(j)  + \frac{a}{s} \vec{l}(j)\\
\vec{n}(2j+1) &= -\sqrt{1-\frac{a^2}{s^2} |\vec{l}|^2}\,\vec{m}(j)  +\frac{a}{s} \vec{l}(j),
\end{split}
\label{eq:ml}
\end{equation}
where the index  $j$ runs through the two site unit cells, the  vector $\vec{m}(j)$  (with $|\vec m_j|=1$) describes the local staggered magnetization, and the vector $\vec{l}(j)$, which is  perpendicular to $\vec{m}(j)$ 
describes the local uniform magnetization. Equivalently, the  complex vectors $\vec{\Phi}$ can be  parametrized as
\begin{equation}
\begin{split}
\left(\begin{array}{c} \vec{\Phi}(2j)^T\\ \vec{\Phi}(2j+1)^T\end{array}\right) =  \left( \begin{array}{cc} \sqrt{1-\frac{a^2}{s^2} |L_{12}(j)|^2}&\frac{a}{s}L(j)_{12} \\\frac{a}{s}L(j)_{12}^*& \sqrt{1-\frac{a^2}{s^2}
|L_{12}(j)|^2} \\ \end{array}\right)  U(j), 
\end{split}
\label{eq:UL}
\end{equation}
where, similarly to the $\mbox{SU}(3)$ case, the unitary $U(j)$ matrix describes the slow joint rotation, while the $L(j)$ matrix  parametrizes the non-orthogonality of spin states inside the unit cell. 
The form of the resulting spin matrices  $\mathcal{S}^{(\alpha\beta)}=\bra{{\vec \Phi} }\hat{S}^\alpha_\beta \ket{\vec \Phi} = 2s\,\Phi^{\beta*} \Phi^{\alpha}$ read as

\begin{equation}
\begin{split}
\mathbfcal{S}(2j) &=  2s U^\dagger \Lambda_1  U  + a U^\dagger  \left( \begin{array}{cc} 0&L_{12} \\L_{12}^*&0 \\ \end{array}\right) U + \frac{a^2}{s}U^\dagger  \left( \begin{array}{cc}-|L_{12}|^2&0\\ 0& 
|L_{12}|^2  \end{array}\right) U +\mathcal{O}(a^3/s^2)\\
\mathbfcal{S}(2j+1) &=  2s U^\dagger \Lambda_2  U  + a U^\dagger  \left( \begin{array}{cc} 0&L_{12} \\L_{12}^*&0 \\ \end{array}\right) U + \frac{a^2}{s}U^\dagger  \left( \begin{array}{cc}|L_{12}|^2&0\\ 0& 
-|L_{12}|^2  \end{array}\right) U +\mathcal{O}(a^3/s^2)
\end{split}
\label{eq:su2spinmx}
\end{equation}
where we omitted  the $(j)$ argument on the right hand side.

  The two formulations are equivalent up to $\mathcal{O}(a^2/s)$ (higher order terms  are neglected in the path integral anyway). 
From Eq.\ \eqref{eq:nphi}, by using the completeness property of the Pauli matrices (i.e.\ $\sigma_\mu^{\alpha\beta} \sigma_\mu^{\gamma\delta}=2 \delta_{\alpha\delta}\delta_{\beta\gamma}-\delta_{\alpha\beta}\delta_{\gamma\delta}$) and setting  $L_{12}(j)$ and $\vec{l}_j$ to 0 we have
\begin{equation}
\begin{split}
\vec{m}(j) \vec{\sigma} = \big(2U(j)^\dagger \Lambda_1 U(j) \big)^*-  \id= \big(U(j)^\dagger \sigma_z U(j)\big)^*
\end{split}
\label{eq:mUcompare}
\end{equation}
where 
\begin{equation}
\begin{split}
\Lambda_1=\left(\begin{array}{cc} 1&0\\ 0&0\end{array}\right),\quad  \Lambda_2=\left(\begin{array}{cc} 0&0\\ 0&1 \end{array}\right).
\end{split}
\end{equation}
The connection between $L(j)$ and $\vec{l}(j)$ is given by
\begin{equation}
\begin{split}
\vec{l}(j) \vec{\sigma} = 2\bigg(U(j)^\dagger \left( \begin{array}{cc} 0& L_{12}(j)\\L_{12}(j)^*&0\end{array}\right) U(j)\bigg)^*
\end{split}
\end{equation}
This implies in particular that $4|L_{12}(j)|^2= |\vec{l}(j)|^2$. Furthermore the orthogonality of $\vec m(j)$ and $\vec l(j)$ is equivalent to the orthogonality of the 0th and 1st order terms in Eq.\ \eqref{eq:su2spinmx}

\begin{equation}
\begin{split}
\tr \left[\Lambda_1\left( \begin{array}{cc} 0&L_{12} \\L_{12}^*&0 \\ \end{array}\right)\right] = l_\mu \sigma_\mu^{\alpha\beta} m_\nu \sigma_\nu^{\beta \alpha}=  2 \vec l \cdot\vec m
\end{split}
\end{equation}
Also the $\mathcal{O}(a^2/s)$ term in  Eq.\ \eqref{eq:su2spinmx} is  a normalization similarly to the $\sqrt{1-\frac{a^2}{s^2}}$  in Eq.\ \eqref{eq:ml} in the vector formalism, which guarantees that the spin matrices has the correct Casimir,
$\mathcal{S}^{\alpha \beta} \mathcal{S}^{\beta \alpha}=4s^2$.

In the following we use the matrix formulation of  Eq.\ \eqref{eq:UL} to write the action.
The terms in the partition function can be expanded up to  $\mathcal{O}(a^2/s^0)$ corrections  as
 \begin{equation}
\begin{split}
2s^2\left| \vec{\Phi}^*(2j)\cdot \vec{\Phi}(2j+1)\right|^2 =  {}&8a^2 |L_{12}|^2\\
2s^2\left| \vec{\Phi}^*(2j+1)\cdot\vec{\Phi}(2j+2)\right|^2 = {}&8a^2 |L_{12}|^2 + 8 s^2 a^2|(U \partial_x U^\dagger)_{12}|^2\\
&+8   a^2s  (\partial_xU U^\dagger)^{\phantom{*}}_{21}  L_{12}^{\phantom{*}}+8  a^2s (U\partial_x U^\dagger)^{\phantom{*}}_{12}  L_{12}^*\\
2s \Big(\vec{\Phi}^*(2j) \cdot \partial_\tau \vec \Phi(2j)+\vec{\Phi}^*(2j+1) \cdot \partial_\tau \vec \Phi(2j+1) \Big) =  {}&2s \,\tr(U^*\partial_{\tau} U^T) \\&+ 4 a \Big( L^{\phantom{*}}_{12} (\partial_\tau UU^\dagger)^{\phantom{*}}_{21}+ L^*_{12} (\partial_\tau UU^\dagger)^{\phantom{*}}_{12}\Big)
\end{split}
\end{equation}
On the right hand sides every quantity is taken at unit cell $j$, but we omitted the indices for compactness.
 Once again, $\tr (U^*\partial_\tau U^T)=0$ since  $U$ is unitary and its row vectors are orthogonal. After taking the continuum limit and replacing the sum with an integral ($2a \sum_j= \int dx$), the full action becomes
 \begin{equation}
\begin{split}
S[U,L] = a  \int& dx d\tau\Bigg[  4 s^2J_1 \tr \left[ \Lambda_1 U   \partial_x U ^\dagger \Lambda_2 \partial_xU U ^\dagger     \right]  +8 J_1 |L_{12}|^2\\ 
+{}& L_{12} \Big( 4 J_1 s (\partial_xUU^\dagger)_{21} + \frac{2}{a} (\partial_\tau UU^\dagger)_{21}\Big)
+L_{12}^* \Big( 4J_1 s (U\partial_xU^\dagger)_{12} + \frac{2}{a} (\partial_\tau UU^\dagger)_{12}\Big)\Bigg].
\end{split}
\end{equation}
At this point the $L_{12}$ fluctuations can be integrated out:
\begin{equation}
\begin{split}
\int dL_{12}d L_{12}^* \exp \Bigg[a \Big[ & -8J_1|L_{12}|^2 - L_{12} \Big( 4 J_1s (\partial_xUU^\dagger)_{21} + \frac{2}{a} (\partial_\tau UU^\dagger)_{21}\Big) 
\\&- L_{12}^* \Big( 4 J_1s(U\partial_xU^\dagger)_{12} + \frac{2}{a} (\partial_\tau UU^\dagger)_{12}\Big)\Big] \Bigg]\\
= \exp \Bigg[\frac{ a}{8J_1}\Big[& 16 s^2 J_1^2 (\partial_xUU^\dagger)_{21}(U\partial_xU^\dagger)_{12}+ \frac{4}{a^2} (\partial_\tau UU^\dagger)_{21} (\partial_\tau UU^\dagger)_{12}\\
&+\frac{8J_1s}{a} \Big( (\partial_xUU^\dagger)_{21} (\partial_\tau UU^\dagger)_{12} +( U\partial_xU^\dagger)_{12} (\partial_\tau UU^\dagger)_{21}\Big) \Big]\Bigg].
\end{split}
\label{eq:SU2gaussint}
\end{equation}
The resulting action only contains the $U$ matrices and reads 
\begin{equation}
\begin{split}
S[U] = \int dx d\tau\Bigg(&\frac{2}{g} \Biggl[ v\,\tr   \left[  \Lambda_{1} U \partial_x U^\dagger \Lambda_{2} \partial_x U  U^\dagger \right]  +\frac{1}{v} \tr \left[  \Lambda_{1} U  \partial_\tau U^\dagger \Lambda_{2} \partial_\tau U  U^\dagger \right]  \Biggr]\\
& +i\frac{\theta}{2\pi i} \varepsilon_{\mu\nu}\tr\left[ \Lambda_{1} U \partial_\mu U^\dagger \Lambda_{2} \partial_\nu U U^\dagger  \right]
 \Bigg),
\end{split}
\end{equation}
where $ 2/g = s$, $v=2 J_1 a$, $\theta= 2\pi s$ (we introduced $g$ this way to follow the convention).   The $\theta$-term is topological, with an integer topological charge
\begin{equation}
\begin{split}
Q_1 =\frac{1}{2\pi i} \int dxd\tau\,&\varepsilon_{\mu\nu}\tr\left[ \Lambda_{1} U \partial_\mu U^\dagger \Lambda_{2} \partial_\nu U U^\dagger  \right]\\
 =\frac{1}{2\pi i} \int dxd\tau\,&\Bigg(\varepsilon_{\mu\nu}\tr\left[ \Lambda_{1} U \partial_\mu U^\dagger \Lambda_{2} \partial_\nu U U^\dagger  \right]\\
&+\varepsilon_{\mu\nu}\tr\left[ \Lambda_{1} U \partial_\mu U^\dagger \Lambda_{1} \partial_\nu U U^\dagger  \right]\Bigg)\\
=\mathbin{\phantom{-}}\frac{1}{2\pi i} \int dxd\tau\,&\varepsilon_{\mu\nu}\tr\left[ \Lambda_{1} \partial_\mu U\partial_\nu U^\dagger  \right]\\
=-\frac{1}{2\pi i} \int dxd\tau\,&\varepsilon_{\mu\nu}\tr\left[ \Lambda_{2} \partial_\mu U\partial_\nu U^\dagger  \right]
=-Q_2.
\end{split}
\end{equation}
The second term in the second equation vanishes since the expression in the trace is symmetric under exchanging $\mu$ and $\nu$.
Comparing this result to the action of the $\mbox{SU}(3)$ case given in Eq.\ \eqref{eq:fullaction}, we see that there is no non-topological imaginary term. The term similar to $q_{12}$ defined in Eq.\ \eqref{eq:qterm} is equivalent to the topological charge in the SU(2) case. 

Writing $U$ in terms of two orthogonal row vectors
 \begin{equation}
\begin{split}
U= \left(\begin{array}{c}\vec{\phi}_1^T\\ \vec{\phi}_2^T\end{array}\right),
\end{split}
\end{equation}
leads to the action
\begin{equation}
\begin{split}
S = \int dx d\tau\Bigg(&\frac{2}{g} \left| \vec{\phi}_2^*\cdot \partial_\mu \vec{\phi}_1\right|^2 +i\frac{\theta}{2\pi i} \varepsilon_{\mu\nu} \partial_\mu \vec{\phi}_1\cdot \partial_\nu \vec{\phi}_1^* \Bigg)\\
= \int dx d\tau\Bigg(&\frac{2}{g}\bigg( \left| \partial_\mu \vec{\phi}_1\right|^2 - \left| \vec{\phi}_1^* \cdot\partial_\mu \vec{\phi}_1\right|^2\bigg)+i\frac{\theta}{2\pi i} \varepsilon_{\mu\nu} \partial_\mu \vec{\phi}_1 \cdot\partial_\nu \vec{\phi}_1^* \Bigg)\\
= \int dx d\tau\Bigg(&\frac{2}{g}\bigg( \left| \partial_\mu \vec{\phi}_1\right|^2 + \big( \vec{\phi}_1^*\cdot \partial_\mu \vec{\phi}_1\big)^2\bigg)+i\frac{\theta}{2\pi i} \varepsilon_{\mu\nu} \partial_\mu \vec{\phi}_1\cdot\partial_\nu \vec{\phi}_1^* \Bigg),\\
\end{split}
\label{eq:CP1}
\end{equation}
 This is just a $\mbox{CP}^1$ theory with a topological term \cite{AffleckLesHouches1988}. 
This theory is equivalent to the  $\mbox{O}(3)$ $\sigma$-model \cite{Cabra2004, fradkin2013}
\begin{equation}
\begin{split}
S = \int dx d\tau \Bigg( \frac{1}{2g} \big( \partial_\mu \vec{m}\big)^2 + i \frac{\theta}{8\pi} \varepsilon_{\mu\nu} \vec{m}\cdot\big( \partial_\mu \vec{m} \times \partial_\nu \vec{m}\big) \Bigg).
\end{split}
\label{eq:O3}
\end{equation}
Indeed, using identities of the Pauli matrices ($ \tr(\sigma_k \sigma_l)=2\delta_{kl}$ and $\tr(\sigma_{k} \sigma_l \sigma_m) = 2 i\, \varepsilon_{klm}$) and  Eq.\ \eqref{eq:mUcompare} we can write
\begin{equation}
\begin{split}
2 (\partial_\mu \vec{m})^2&=\tr\big(\partial_\mu(\vec{m}\cdot\vec{\sigma})\partial_\mu(\vec{m}\cdot\vec{\sigma})\big)\\
&=   4 \tr \big[\partial_\mu( U^\dagger\Lambda_1 U) \partial_\mu( U^\dagger\Lambda_1 U)\big]\\
&= 8 \tr \big[\partial_\mu U^\dagger \Lambda_1 U \partial_\mu U^\dagger \Lambda_1 U\big]+8 \tr \big[\partial_\mu U \partial_\mu U^\dagger \Lambda_1 \big]\\
&=8 (\vec{\phi}_1^{\phantom{*}} \cdot\partial_\mu \vec{\phi}_1^*)^2+8 |\partial_\mu \vec{\phi}_1^{\phantom{*}}|^2
\end{split}
\end{equation}
and
\begin{equation}
\begin{split}
2i\, \varepsilon_{\mu\nu}\vec{m}\cdot\big( \partial_\mu \vec{m} \times \partial_\nu \vec{m}\big)&=\varepsilon_{\mu\nu}\tr\big( (\vec{m}\cdot \vec{\sigma} ) \partial_\mu(\vec{m}\cdot\vec{ \sigma} )\partial_\nu(\vec{m}\cdot \vec{\sigma })\big)\\
&=4  \varepsilon_{\mu\nu} \tr \big[(2 U^\dagger\Lambda_1 U-I) \partial_\mu( U^\dagger\Lambda_1 U) \partial_\nu( U^\dagger\Lambda_1 U)\big]\\
&= 8 \varepsilon_{\mu\nu}\tr \big[\Lambda_1 \partial_\mu U \partial_\nu U^\dagger \big] +\varepsilon_{\mu \nu} \bigg (\text{terms symmetric in $\mu \nu$}\bigg)\\
&=8 \varepsilon_{\mu\nu} \partial_\mu \vec{\phi}_1^{\phantom{*}}\cdot \partial_\nu \vec{\phi}_1^*.\end{split}
\end{equation}

\begin{figure}[h]
\begin{center}
\includegraphics[width=0.8\textwidth]{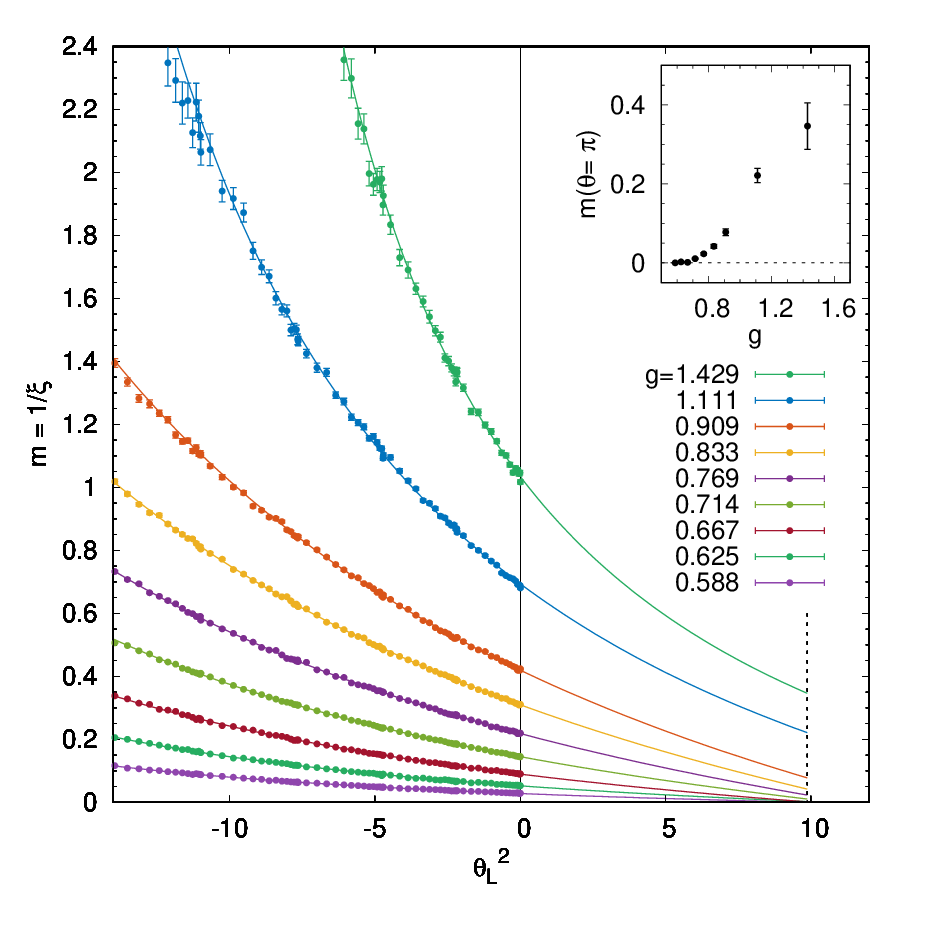}
\caption{Extrapolation of the inverse of the correlation length from Monte Carlo calculations of the $\mbox{CP}^1$ $\sigma$-model with imaginary $\theta$  for a system of $192 \times 192$ sites. The inset shows the extrapolated value of the mass gap at $\theta=\pi$.}
\label{fig:su2extrapolation}
\end{center}
\end{figure}

In the $\sigma$-model of Eqs.\ \eqref{eq:CP1} or \eqref{eq:O3}, the topological term is written $i \theta Q_1$. For translationally invariant spin models, $\theta=2\pi s$, which distinguishes integer and half integer spin models. For integer spins the topological term is trivial and the $\sigma$-model is gapped \cite{Cabra2004}, while for half integer spins $\theta\equiv \pi \ (\text{mod}\, 2\pi)$. As discussed in Sec.~\ref{sec:intro}, a phase transition is expected at $\theta=\pi$ for some critical coupling $g_c$. For $g<g_c$ the system is expected to be gapless, while for $g>g_c$ the system is gapped with a spontaneous breakdown of a $\mathbb{Z}_2$ symmetry. This transition corresponds to the gapless-dimerized transition in the $J_1 -J_2$, $s=1/2$ spin chain, \cite{Okamoto1992, EggertPRB1996}. Using the  extrapolation method of \citet{AllesPapa2008} this transition can be observed in the $\mbox{CP}^1$ $\sigma$-model as well. Fig.~\ref{fig:su2extrapolation} shows our Monte Carlo results for the mass gap at different couplings. The mass gap is extracted from the spin-spin correlation function,
\begin{equation}
\begin{split}
C(x) &= \frac{1}{L}\sum_{\tau}  \Bigg\langle \vec m(0,0) \cdot \vec m(x,\tau) \Bigg\rangle \\
&=2\frac{1}{L} \sum_{\tau}  \Bigg\langle  \left( U^\dagger(0,0) \Lambda_1 U(0,0) \right)^* \left( U^\dagger(x,\tau) \Lambda_1 U(x,\tau)\right)^*\Bigg\rangle \\
&=2\frac{1}{L}  \sum_{\tau} \Bigg\langle \left( \vec\phi_1^{\phantom{*}}(0,0) \cdot \vec\phi_1^* (x,\tau)  \right) \left(  \vec\phi_1^* (0,0) \cdot \vec\phi_1^{\phantom{*}}(x,\tau) \right) \Bigg\rangle.
\end{split}
\end{equation}
The equivalence of the above forms can be shown using Eq.~\eqref{eq:mUcompare} and the identity $\sigma^{\alpha\beta}_i \sigma_j ^{\beta\alpha}=2 \delta_{i,j}$. Notice that the correlation function used in the $\mbox{SU}(3)$ case (see Eq.~\eqref{eq:corrfct})  is a direct generalization of  this spin-spin correlation function.  The gap opening can be seen around $g_c \approx 0.7$. For $g< g_c$, the system is gapless at $\theta=\pi$, as reported in Ref.\ \cite{AllesPapa2008}, where calculations for $ 0.57<g < 0.67$ have been carried out, while for $g>g_c$ the system is gapped. The spontaneous breakdown of $\mathbb{Z}_2$ was explicitly shown in the $g\to\infty$ limit in  Refs.\ \cite{SeibergPRL1984, PlefkaSamuelPRD1997}. The Monte Carlo  results also confirm that the gap remains finite away for $\theta\neq\pi$ for all values of $g$.

\section{Supplementary calculations for the derivation of the $\mbox{SU}(3)/ [\mbox{U}(1)\times \mbox{U}(1)]$ $\sigma$-model.}
\label{sec:su3calc} 
We consider an $\mbox{SU}(3)$ spin chain with antiferromagnetic nearest and next nearest, and ferromagnetic third nearest Heisenberg interactions as defined in Eq.\ \eqref{eq:J1-J2model}
\begin{equation}
\begin{split}
\mathcal{H}= \sum_{i}  \left[J_1 S^\alpha_\beta(i) S^\beta_\alpha(i+1)+ J_2 S^\alpha_\beta(i) S^\beta_\alpha(i+2)- J_3 S^\alpha_\beta(i) S^\beta_\alpha(i+3)\right],
\end{split}
\end{equation}
We use a spin coherent state path integral approach \cite{KlauderPRD1979,GnutzmannKus1998, ShibataTakagi1999, MathurSen2001} to calculate the imaginary time partition function of the  model in Eq.\ \eqref{eq:J1-J2model}. In the bosonic picture,   spin coherent states in the $p$-box representation can be written as 
\begin{equation}
\begin{split}
\ket{ \vec{\Phi}}=\frac{1}{\sqrt{p!}} (\Phi^\mu b_\mu^\dagger )^p\ket{0}
\end{split}
\end{equation}
where $\vec{\Phi}= (\Phi^{(1)},\Phi^{(2)},\Phi^{(3)})$ is a three component complex unit vector (i.e.\ $|\vec{\Phi}|^2=1$), while $b_\mu^\dagger$ is the creation operator for a boson of flavour $\mu$ \cite{MathurSen2001, BroznanPRD1988}. These spin coherent states are normalized, and form a complete set over the fully symmetric $p$-box  representation:
\begin{equation}
\begin{split}
\braket{\vec{\Phi}}{\vec{\Phi}} = 1\\
 \frac{(p+2)(p+1)}{2}\int\limits d\Omega_{\vec \Phi} \ket{\vec{\Phi} } \bra{\vec \Phi} = \id.
\end{split}
\end{equation}
The overlap between two spin coherent states is given by
\begin{equation}
\begin{split}
\braket{\vec{\Psi}}{\vec{\Phi}}=( \Psi^{\mu*} \Phi^\mu)^p.
\end{split}
\end{equation}
Parametrizing $\vec \Phi$ as 
\begin{equation}
\begin{split}
\vec{\Phi}= (\sin\theta\cos\varphi \ e^{i\alpha_1}, \sin\theta\sin\varphi \ e^{i\alpha_2}, \cos\theta e^{i\alpha_3} )
\label{eq:csparam}
\end{split}
\end{equation}
with $0\leq\theta,\phi\leq \pi/2$, $0\leq \alpha_1,\alpha_2,\alpha_3 \leq 2\pi$, the  Haar-measure $d \Omega_{\vec{\Phi}}$ can be defined by
\begin{equation}
\begin{split}
\int\limits d\Omega_{\vec{\Phi}} = \int\limits \frac{1}{\pi^3} \sin^3 \theta \cos\theta \cos \varphi \sin\varphi\, d\theta\, d\varphi\, d\alpha_1\, d\alpha_2\, d\alpha_3 = 1.
\label{eq:intmeasure}
\end{split}
\end{equation}
Using the Lie-Trotter decomposition of the imaginary time partition function, one can insert a complete set of spin coherent states at each time slice for each spin
\begin{equation}
\begin{split}
\Tr \big( e^ {- \beta H}\big)= \int\limits \mathcal{D} [{\Phi}] \prod _{n=0}^{N-1}  \bra{ \{\vec{\Phi}(\cdot ,n+1) \} }   e^{-dt \mathcal{H}}  \ket{ \{ \vec{\Phi}(\cdot ,n) \}} \end{split}
\end{equation}
where $dt= \beta/N$ and
\begin{equation}
\begin{split}
 \ket{ \{ \vec{\Phi}(\cdot,n)\}} = \otimes_{j=1 }^L  \ket{\vec{\Phi}(j,n)}
\end{split}
\end{equation}
is the direct product of spin coherent states at each site of the chain at time step $n$ (note that  $\ket{ \{\vec{\Phi}(\cdot,0) \} } = \ket{ \{\vec{\Phi}(\cdot,N)\} }$), and  $\int \mathcal{D} [{\Phi}] = \int \prod \limits_{j=1 }^L \prod \limits_{n=1}^N d\Omega_{\vec{\Phi}(j,n)}$ is the integral over the spin coherent states at each site and each time step. 
Up to first order in $dt$, 
\begin{equation}
\begin{split}
 \bra{ \{\vec \Phi(\cdot,n+1) \} }&
   e^{-dt \mathcal{H}}  \ket{ \{ \vec{\Phi}(\cdot,n) \}} =\\
   &= \braket{ \{\vec{\Phi}(\cdot,n+1) \} }{ \{ \vec{\Phi}(\cdot,n) \}} - dt  \bra{ \{\vec{\Phi}(\cdot,n+1) \} }
   \mathcal{H}  \ket{ \{ \vec{\Phi}(\cdot,n) \}} +\mathcal{O}(dt^2) \\ 
   &=\prod_j \bigg(  \vec{\Phi}(j,n+1)^*  \vec{\Phi}(j,n) \bigg)^p {-dt\bra{ \{\vec{\Phi}(\cdot,n) \} }   \mathcal{H}  \ket{ \{ \vec{\Phi}(\cdot,n) \}} } +\mathcal{O}(dt^2)\\
   &=\prod_j \bigg( 1 + p\,\delta\vec{\Phi}(j,n)^*  \vec{\Phi}(j,n) \bigg) {-dt\bra{ \{\vec{\Phi}(\cdot,n) \} }   \mathcal{H}  \ket{ \{ \vec{\Phi}(\cdot,n) \}} } +\mathcal{O}(dt^2)\\
   &=\exp \Big[{-dt\bra{ \{\vec{\Phi}(\cdot,n) \} }   \mathcal{H}  \ket{ \{ \vec{\Phi}(\cdot,n) \}} }- p\sum_j  \vec{\Phi}^*(j,n) \cdot \delta \vec {\Phi}(j,n) \Big] +\mathcal{O}(dt^2),
\end{split}
\end{equation}
where $\delta \vec{\Phi}(j,n) =\vec{\Phi}(j,n+1) -\vec{\Phi}(j,n)$, and we used the fact that $\delta\vec\Phi^* \vec\Phi +\vec\Phi^*\delta \vec\Phi = \delta|\vec \Phi|^2=0$.
Thus, as  $N\to \infty$ the path integral will take the form
\begin{equation}
\begin{split}
 \Tr \big( e^ {- \beta \mathcal{H}}\big)= \int \mathcal{D} [{\Phi}]  \exp \Bigg[ -\int \limits_0^\beta d\tau \Big[ &\bra{ \{\vec{\Phi}(\cdot,\tau) \} }
   \mathcal{H}  \ket{ \{ \vec{\Phi}(\cdot,\tau) \}}  +  \sum_{j }p\vec{\Phi}^*(j,\tau)  \partial_\tau \vec{\Phi}(j,\tau)  \Big] \Bigg]
\end{split}
\label{eq:SU3partitionfunction}
\end{equation}
with  $\tau  \approx n dt$. The second part of the action is imaginary and describes the Berry phase for each spin coming from the overlap of spin coherent states of different time slices.

Upon calculating the expectation of the Hamiltonian  the spin operators are replaced by classical matrices as 
\begin{equation}
\begin{split}
\bra{\vec \Phi(j,\tau)} S^\alpha_\beta(j)\ket{\vec \Phi(j,\tau)} &= p{}\Phi^{\beta*}(j,\tau) \Phi^\alpha(j,\tau).
\end{split}
\end{equation} 
Using the parametrization of Eq.\ \eqref{eq:spinstates0} the spin matrices $\mathcal{S}_n(j)^{\beta \alpha}=\bra{\vec \Phi(3j+n)} S^\alpha_\beta\ket{\vec \Phi(3j+n)}$  take the form 
\begin{equation}
\begin{split} 
\mathbfcal{S}_1(j)={}& p \big(U^\dagger \Lambda_1U\big) +  a U^\dagger
\left( \begin{array}{ccc}
0& L_{12}^{\phantom{*}}&  L_{13}^{\phantom{*}} \\
 L_{12}^*& 0& 0 \\
 L_{13}^*& 0&0\\
 \end{array}\right) U\\
 &+  \frac{a^2}{p} U^\dagger
 \left( \begin{array}{ccc}
 - |L_{12}^{\phantom{*}}|^2-|L_{13}^{\phantom{*}}|^2&0& 0 \\
0&|L_{12}^{\phantom{*}}|^2& L_{12}^* L_{13}^{\phantom{*}} \\
 0& L_{12}^{\phantom{*}}L_{13}^*&|L_{13}^{\phantom{*}}|^2
 \end{array}\right) 
U +\mathcal{O}(a^3/p^2)\\
\\
\mathbfcal{S}_2(j)={}& p \big(U^\dagger \Lambda_2U\big) +  a U^\dagger
\left( \begin{array}{ccc}
0& L_{12}^{\phantom{*}}& 0 \\
 L_{12}^*& 0& L_{23}^{\phantom{*}} \\
 0& L_{23}^*&0\\
 \end{array}\right) U\\
 &+  \frac{a^2}{p} U^\dagger
 \left( \begin{array}{ccc}
 |L_{12}^{\phantom{*}}|^2 &0& L_{12}^{\phantom{*}}L_{23}^{\phantom{*}} \\
0& - |L_{12}^{\phantom{*}}|^2-|L_{23}^{\phantom{*}}|^2&0 \\
L_{12}^*L_{23}^*& 0&|L_{23}^{\phantom{*}}|^2
 \end{array}\right) 
U +\mathcal{O}(a^3/p^2) \\
\\
\mathbfcal{S}_3(j)={}& p \big(U^\dagger \Lambda_3 U\big) +  a U^\dagger
\left( \begin{array}{ccc}
0& 0&L_{13}^{\phantom{*} }\\
0& 0& L_{23}^{\phantom{*}} \\
 L_{13}^*& L_{23}^*&0\\
 \end{array}\right) U\\
 &+  \frac{a^2}{p} U^\dagger
 \left( \begin{array}{ccc}
|L_{13}^{\phantom{*}}|^2& L_{13}^{\phantom{*}}L_{23}^{*} &0\\
L_{13}^*L_{23}^{\phantom{*}} & |L_{23}^{\phantom{*}}|^2&0 \\
0& 0&- |L_{13}^{\phantom{*}}|^2-|L_{23}^{\phantom{*}}|^2
 \end{array}\right) 
U +\mathcal{O}(a^3/p^2)
\end{split}
\label{eq:spinmxs}
\end{equation}
where the  $\Lambda_n$ matrices are defined by
\begin{equation}
\begin{split}
\Lambda_1=\left(\begin{array}{ccc} 1&0&0\\ 0&0&0\\ 0&0&0 \end{array}\right),\quad  \Lambda_2=\left(\begin{array}{ccc} 0&0&0\\ 0&1&0\\ 0&0&0 \end{array}\right), \quad
\Lambda_3=\left(\begin{array}{ccc} 0&0&0\\ 0&0&0\\ 0&0&1 \end{array}\right).
\end{split}
\label{eq:Lambdamx}
\end{equation}
This form of the spin matrices can be compared to the  $\mbox{SU}(2)$ case discussed in \ref{app:su2}. The leading term $P_n(x,\tau)= U^\dagger(x,\tau)\Lambda_n U(x,\tau)$ corresponds to the staggered magnetization $\vec m$ in the $\mbox{SU}(2)$ case (see Eq. \eqref{eq:mUcompare}).  The first order term describes fluctuations orthogonal to the leading term. For example, 

\begin{equation}
\begin{split}
\tr \left[ \Lambda_1 \left( \begin{array}{ccc}
0& L_{12}^{\phantom{*}}&  L_{13}^{\phantom{*}} \\
 L_{12}^*& 0& 0 \\
 L_{13}^*& 0&0\\
 \end{array}\right) \right] =0,
\end{split}
\end{equation}
while the  $\mathcal{O}(a^2/p)$ terms can be thought of as a normalization which guarantees that the Casimirs of the spin matrices are correct: $\mathcal{S}^{\alpha \beta} \mathcal{S}^{\beta \alpha}=  p^2 +\mathcal{O}(a^4/p^2)$ and  $\mathcal{S}^{\alpha \beta} \mathcal{S}^{\beta \gamma} \mathcal{S}^{\gamma \alpha}=  p^3 +\mathcal{O}(a^4/p)$. Note that we use a non-traceless definition of the spin operators, therefore the Casimirs are different from those of the traceless convention.
The action of Eq.\ \eqref{eq:fullLUaction} can be rewritten in terms of the three fields inside a unit cell as
\begin{equation}
\begin{split}
S = \int_0^\beta d\tau \sum_j  &\Biggl[ p^2 \Bigl[J_1  \left|\vec  \Phi_1^*(j)\cdot \vec \Phi_2(j) \right|^2 +J_2  \left|\vec  \Phi_1^*(j) \cdot \vec \Phi_2(j-1) \right|^2 -J_3  \left|\vec  \Phi_1^*(j) \cdot \vec \Phi_1(j+1) \right|^2 \\
&+ J_1  \left|\vec  \Phi_2^*(j) \cdot \vec \Phi_3(j) \right|^2 +J_2  \left|\vec  \Phi_2^*(j) \cdot \vec \Phi_3(j-1) \right|^2 -J_3  \left|\vec  \Phi_2^*(j) \cdot \vec \Phi_2(j+1) \right|^2 \\
&+ J_1  \left|\vec  \Phi_3^*(j) \cdot \vec \Phi_1(j+1) \right|^2 +J_2  \left|\vec  \Phi_3^*(j) \cdot \vec \Phi_1(j) \right|^2-J_3  \left|\vec  \Phi_3^*(j) \cdot \vec \Phi_3(j+1) \right|^2\Bigr]
\\&+p \Big(\vec \Phi_1(j)^* \cdot \partial_\tau \vec \Phi_1(j) \Big)+\Big(\vec \Phi_2^*(j)\cdot \partial_\tau \vec \Phi_2(j) \Big)+\Big(\vec \Phi_3^*(j) \cdot \partial_\tau \vec \Phi_3(j) \Big) \Biggr]
\end{split}
\end{equation}
where the sum over $j$  runs through the three site unit cells.  In the continuum limit ($x \approx 3a j$) the sum can be replaced by an integral $ 3a\sum_j \to  \int dx$ and the fields/matrices on neighbouring unit cells can be expanded as $ U(j\pm1) = U(j) \pm 3a \partial_x U(j) +9a^2\partial_x^2 U(j)/2 +\mathcal{O}(a^3)$.   
Here we show an example of how each type of term reads after inserting the parametrization of Eq.\ \eqref{eq:spinstates0} into the action:
\begin{equation}
\openup 1\jot
\begin{split}
J_1p^2|\vec{\Phi}_1^*(j)\cdot \vec{\Phi}_2(j)|^2 ={}&  4 J_1 a^2 |L_{12}|^2 +\mathcal{O}(a^3)\\
J_2p^2|\vec{\Phi}_1^*(j)\cdot \vec{\Phi}_2(j-1)|^2 ={}&  4 J_2 a^2 |L_{12}|^2 + 9J_2 p^2 a^2|(U \partial_x U^\dagger)_{12}|^2\\
&-6  J_2a^2p (\partial_xU U^\dagger)_{21}  L_{12}- 6  J_2a^2p (U\partial_x U^\dagger)_{12}  L_{12}^* + \mathcal{O}(a^3)\\
-J_3p^2|\vec{\Phi}_1^*(j)\cdot \vec{\Phi}_1(j+1)|^2={}& -J_3p^2 -9J_3a^2p^2 |(U\partial_x U^\dagger)_{11}  |^2 \\
&-\frac{9}{2} J_3a^2 p^2\big[ (U\partial_x^2 U^\dagger)_{11} +(\partial_x^2 UU^\dagger)_{11} \big]+\mathcal{O}(a^3).
\end{split}
\label{eq:su3expansion}
\end{equation}
The other terms can be expanded similarly. 
Up to $\mathcal{O}(a^2/p)$ corrections, the  Berry phase term is given by
\begin{equation}
\begin{split}
\sum_{n=1}^3 p \Big(\Phi_n^*(j)  \cdot \partial_\tau \vec \Phi_n(j) \Big) 
&= p\, \tr(U^*\partial_{\tau} U^T) + 2a \sum_{n=1}^3  \Big( L^{\phantom{*}}_{n,n+1} (\partial_\tau UU^\dagger)^{\phantom{*}}_{n+1,n}+ L^*_{n,n+1} (\partial_\tau UU^\dagger)^{\phantom{*}}_{n,n+1}\Big).
\end{split}
\end{equation}
The first term on the right hand side is identically 0 since $U$ is a unitary matrix. Indeed, writing the rows of $U$ as $\vec{\phi}_{1,0}^{\phantom{*}}$, $\vec{\phi}_{2,0}^{\phantom{*}}$ and $\vec{\phi}_{3,0}^{\phantom{*}}=\vec{\phi}_{1,0} ^*\times \vec{\phi}_{2,0}^*$, then expanding the trace we get 
\begin{equation}
\begin{split}
\tr(U^*\partial_\tau U^T) = \vec{\phi}_{1,0}^* \cdot \partial_\tau \vec{\phi}_{1,0}^{\phantom{*}}+ \vec{\phi}_{1,0}^{\phantom{*}}\cdot \partial_\tau \vec{\phi}_{1,0}^*+\vec{\phi}_{2,0}^*\cdot \partial_\tau \vec{\phi}_{2,0}^{\phantom{*}}+\vec{\phi}_{2,0}^{\phantom{*}} \cdot \partial_\tau \vec{\phi}_{2,0}^*=\partial_\tau |\vec{\phi}_{1,0}^{\phantom{*}}|^2+\partial_\tau |\vec{\phi}_{2,0}^{\phantom{*}}|^2=0.
\end{split}
\label{eq:dUUzero}
\end{equation}

Using the following simple identities:
 \begin{equation}
\begin{split}
\id&=\Lambda_1+\Lambda_2+\Lambda_3\\ 
\partial_x( UU^\dagger)&=\partial_x U U^\dagger+ U\partial_x U^\dagger=0\\
 (U\partial_xU^\dagger)^*_{n,n+1} &= (\partial_x UU^\dagger)_{n+1,n} \\
\partial_x^2(UU^\dagger)&=\partial_x^2U U^\dagger+ U \partial_x^2 U^\dagger+2\partial_xU\partial_x U^\dagger=0
\end{split}
\end{equation} 
and rewriting the $J_3$ terms as
 \begin{equation}
\begin{split}
 \sum_{n=1}^3{}&{}\Big(  -|(U\partial_x U^\dagger)_{nn}|^2 - \frac{1}{2}(U\partial_x^2 U^\dagger)_{nn}- \frac{1}{2}(\partial_x^2 U U^\dagger)_{nn}\Big) \\
 &=  -\sum_{n=1}^3\tr \big[ \Lambda_n  U\partial_xU^\dagger \Lambda_n  \partial_xU U^\dagger\big] + \tr\big[\partial_xU\partial_xU^\dagger\big]\\
 &= -\sum_{n=1}^3\tr \big[ \Lambda_n  U\partial_xU^\dagger \Lambda_n  \partial_xU U^\dagger\big] +\sum_{n,m} \tr\big[\Lambda_n \partial_xU U^\dagger \Lambda_m U^\dagger \partial_xU^\dagger\big]\\
 &= \sum_{n=1}^3\Big(\tr \big[ \Lambda_n  U\partial_xU^\dagger \Lambda_{n+1}  \partial_xU U^\dagger\big] +\tr \big[ \Lambda_{n+1}  U\partial_xU^\dagger \Lambda_{n}  \partial_xU U^\dagger\big]\Big)\\
 &=2 \sum_{n=1}^3\tr \big[ \Lambda_n  U\partial_xU^\dagger \Lambda_{n+1}  \partial_xU U^\dagger\big] 
\end{split}
\end{equation}
 the action has the following form up to $\mathcal{O}(a^2)$ and  $\mathcal{O}(a/p)$ corrections
\begin{equation}
\begin{split}
S[U,L]
&= \frac{ a}{3}  \int dx d\tau\Bigg[  \\
&\mathbin{\phantom{+}}9 p^2(J_2+2J_3) \tr \big[ \Lambda_1 U   \partial_x U ^\dagger \Lambda_2 \partial_xU U ^\dagger     \big] + 9 p^2 (J_2+2 J_3) \tr \big[  \Lambda_2 U   \partial_x U ^\dagger \Lambda_3 \partial_xU  U ^\dagger  \big] \\&+ 9 p^2 (J_1+2J_3) \tr \big[  \Lambda_3 U   \partial_x U ^\dagger \Lambda_1 \partial_xU U ^\dagger   \big]  +4\big(J_1+J_2\big) \big( |L_{12}|^2+|L_{23}|^2+|L_{13}|^2\big)\\ 
&+L_{12} \Big( -6 J_2p (\partial_xUU^\dagger)_{21} + \frac{2}{a} (\partial_\tau UU^\dagger)_{21}\Big)
+L_{12}^* \Big( -6 J_2p (U\partial_xU^\dagger)_{12} + \frac{2}{a} (\partial_\tau UU^\dagger)_{12}\Big)\\
&+L_{23} \Big( -6 J_2 p(\partial_xUU^\dagger)_{32} + \frac{2}{a} (\partial_\tau UU^\dagger)_{32}\Big)
+L_{23}^* \Big( -6 J_2 p(U\partial_xU^\dagger)_{23} + \frac{2}{a} (\partial_\tau UU^\dagger)_{23}\Big)\\
&+L_{13}^* \Big( 6 J_1 p(\partial_x UU^\dagger)_{13} + \frac{2}{a} (\partial_\tau UU^\dagger)_{13}\Big)
+L_{13} \Big( 6 J_1p (U\partial_xU^\dagger)_{31} + \frac{2}{a} (\partial_\tau UU^\dagger)_{31}\Big) \Bigg].
\end{split}
\label{eq:S[UL]}
\end{equation}
 where we omitted the $(x, \tau)$ variables of the $U$ and $L$ matrices for the sake of compactness. 
 At this point the $L$ matrices can be integrated out in Eq.\ \eqref{eq:S[UL]} using the Gaussian identity
\begin{equation}
\begin{split}
\int dzdz^* \exp\big(- z^* \omega z + u^*z+vz^*\big) = \frac{\pi}{\omega} \exp\big(\frac{u^*v}{\omega}\big).
\end{split}
\end{equation}
For example,  integrating out $(L_{12}^{\phantom{\dagger}}, L_{12}^*)$ leads to
\begin{equation}
\begin{split}
\int dL_{12}d L_{12}^* \exp \Bigg[\frac{a}{3} \Big[ & -4(J_1+J_2)|L_{12}|^2 - L_{12} \Big( -6 J_2p (\partial_xUU^\dagger)_{21} + \frac{2}{a} (\partial_\tau UU^\dagger)_{21}\Big) 
\\&- L_{12}^* \Big( -6 J_2p (U\partial_xU^\dagger)_{12} + \frac{2}{a} (\partial_\tau UU^\dagger)_{12}\Big)\Big] \Bigg]\\
= \exp \Bigg[\frac{ a}{12(J_1+J_2)}\Big[& 36 p^2 J_2^2 (\partial_xUU^\dagger)_{21}(U\partial_xU^\dagger)_{12}+ \frac{4}{a^2} (\partial_\tau UU^\dagger)_{21} (\partial_\tau UU^\dagger)_{12}\\
&-\frac{12J_2p}{a} \Big( (\partial_xUU^\dagger)_{21} (\partial_\tau UU^\dagger)_{12} +( U\partial_xU^\dagger)_{12} (\partial_\tau UU^\dagger)_{21}\Big) \Big]\Bigg].
\end{split}
\label{eq:Sgaussint}
\end{equation}
Similar integrals can be done for the $L_{23}$ and $L_{13}$ terms as well. The maximum of the exponential occurs at  $aL_{12} /p  \propto \left( -6 aJ_2 (U\partial_xU^\dagger)_{12} + \frac{2}{p} (\partial_\tau UU^\dagger)_{12}\right)$ while the standard deviation for the $L_{12}$ variable is ${O}(1/\sqrt{a})$, or $\mathcal{O}( \sqrt{a}/p)$ for $aL_{12}/p$. So as we mentioned before large fluctuations in Eq.\ \eqref{eq:spinstates0} are indeed suppressed for large $p$ and $a \to 0$

After carrying out the integrals in the $L$ variables, we arrive at the action presented in Eq.\ \eqref{eq:fullaction}
\begin{equation*}
\begin{split}
S[U] = \int dx d\tau\Bigg( \sum_{n=1}^3  &\frac{1}{g} \Biggl[ v\,\tr   \left[  \Lambda_{n-1} U \partial_x U^\dagger \Lambda_{n} \partial_x U  U^\dagger \right]  +\frac{1}{v} \tr \left[  \Lambda_{n-1} U  \partial_\tau U^\dagger \Lambda_{n} \partial_\tau U  U^\dagger \right]  \Biggr]\\
& +i\frac{\theta}{2\pi i}\varepsilon_{\mu\nu}\tr\left[(\Lambda_1- \Lambda_3)\partial_\mu U\partial_\nu U^\dagger\right] +i\frac{\lambda}{2\pi i} \varepsilon_{\mu\nu}\sum_{n=1}^3 \tr\left[ \Lambda_{n-1} U \partial_\mu U^\dagger \Lambda_{n} \partial_\nu U U^\dagger  \right]
 \Bigg),
\end{split}
\end{equation*}
where the three topological charges  defined in Eq.\ \eqref{eq:Qdef}
\begin{equation*}
\begin{split}
Q_n =\frac{1}{2\pi i} \varepsilon_{\mu \nu} \int dxd\tau \tr\left[\Lambda_n\partial_\mu U\partial_\nu U^\dagger\right]
\end{split}
\end{equation*}
are not independent, namely $Q_1+Q_2+Q_3=0$. This can be also seen from
\begin{equation}
\begin{split}
Q_1+Q_2+Q_3 =\frac{1}{2\pi i} \varepsilon_{\mu \nu} \int dxd\tau \tr\left[\partial_\mu U\partial_\nu U^\dagger\right]= \frac{1}{2\pi i} \varepsilon_{\mu \nu} \int dxd\tau \partial_\nu \tr\left[\partial_\mu UU^\dagger\right],
\end{split}
\label{eq:sumQzero}
\end{equation}
where the $\tr \left[\partial_\mu UU^\dagger\right]$ is identically 0 as was shown in Eq.\ \eqref{eq:dUUzero}. Note that not only the sum of the topological charges is zero, but the sum of the charge densities also vanishes. 

The equivalence of the forms in Eq.\ \eqref{eq:fullaction} and Eq.\ \eqref{eq:fullactionphi} can be proven by rewriting the real part as
 \begin{equation}
\begin{split}
S_R={1\over g}\sum_{n=1}^3\tr   \left[  \Lambda_{n-1} U \partial_\mu U^\dagger \Lambda_{n} \partial_\mu U  U^\dagger \right] &= {1\over g}\sum_{n=1}^3\left|  \vec{\phi}_{n-1}^* \cdot \partial_\mu \vec{\phi}_{n}^{\phantom{*}} \right|^2 ={1\over 2g} \sum_{n=1}^3 \left(\left|  \vec{\phi}_{n-1}^* \cdot \partial_\mu \vec{\phi}_{n}^{\phantom{*}} \right|^2+ \left|  \vec{\phi}_{n+1}^* \cdot \partial_\mu \vec{\phi}_{n}^{\phantom{*}} \right|^2\right)\\
&= \frac{1}{2g}\sum_{n=1}^3 \left( \left| \partial_\mu \vec{\phi}_n^{\phantom{*}}\right|^2 -  \left| \vec{\phi}_n^*\cdot \partial_\mu \vec{\phi}_n^{\phantom{*}} \right|^2 \right),
\end{split}
\label{eq:cp2trans}
\end{equation}
where we used the fact that the three  fields $\vec{\phi}_1,\vec{\phi}_2, \vec{\phi}_3$ form an orthonormal basis on $\mathbb{C}^3$.
 
Another equivalent formulation of the action in Eqs.~\eqref{eq:fullaction} and ~\eqref{eq:fullactionphi} can be given  using the  $P_n(x,\tau) =U^\dagger(x,\tau) \Lambda_n U(x,\tau)$ matrices introduced below Eq. \eqref{eq:spinmxs} \cite{BergLuscher1981,PetcherLuscher1983}: 
\begin{equation}
\begin{split}
S[U] = &\int dx d\tau\Bigg( \sum_{n=1}^3  \frac{1}{4g} \tr\left[  \partial_\mu P_n \partial_\mu P_n  \right] \\
& +i\frac{\theta}{2\pi i}\varepsilon_{\mu\nu}\tr\left[P_1\partial_\mu P_1 \partial_\nu P_1- P_3\partial_\mu P_3 \partial_\nu P_3 \right]\\
& +i\frac{\lambda}{2\pi i} \varepsilon_{\mu\nu}\sum_{n=1}^3 \tr\left[ P_n\partial_\mu P_{n+1} \partial_\nu P_{n+1}   \right]
 \Bigg).
\end{split}
\label{eq:Paction}
\end{equation}

The equivalence between Eq.\ \eqref{eq:fullactionphi} (with $\lambda=0$)  and  Eq.\ \eqref{eq:cp2gaugeform} follows straightforwardly by integrating out the gauge fields. 
Only considering the real term
 \begin{equation}
S_R= \frac{1}{2g}\sum_{n=1}^3 \left|\left( \partial_\mu +iA_\mu^n\right) \vec{\phi}_n^{\phantom{*}}\right|^2
\label{eq:cp2gauge}
\end{equation}
this follows since
\begin{equation}
\begin{split}
 \left|\left( \partial_\mu +iA_\mu^n\right) \vec{\phi}_n^{\phantom{*}}\right|^2&=\left|\partial_\mu \vec \phi_n\right|^2+(A_\mu^n)^2-2iA_\mu^n\vec \phi_n^*\cdot \partial_\mu \vec \phi_n. \\
 &=\left|\partial_\mu \vec \phi_n\right|^2+\left( A_\mu^n-i\vec \phi_n^*\cdot \partial_\mu \vec \phi_n\right)^2-\left|\vec \phi_n^*\cdot \partial_\mu \vec \phi_n\right|^2.
 \end{split}
 \end{equation}
 Upon shifting $A_\mu^n$ by $i\vec \phi_n^*\cdot \partial_\mu \vec \phi_n$, the $A_\mu^n$ integral gives a constant factor. 
 
 Including the topological terms, the action becomes, as written in Eq.\ \eqref{eq:cp2gaugeform}
 \begin{equation*} S=\sum_{n=1}^3\left[  \frac{1}{2g} \left|\left( \partial_\mu +iA_\mu^n\right) \vec{\phi}_n^{\phantom{*}}\right|^2+{i\theta_n\over 2\pi}\epsilon_{\mu \nu}\partial_\mu A_\nu^n\right].
\end{equation*}
Once again, shifting $A_\mu^n$, as above, the imaginary term gives:
\be \epsilon_{\mu \nu}\partial_\mu A_\nu^n\to \epsilon_{\mu \nu}\partial_\mu A_\nu^n+i \epsilon_{\mu \nu}\partial_\mu \vec \phi^*_n\cdot \partial_\nu \vec \phi_n.\label{shift}\ee
The second term in Eq.\ \eqref{shift} gives the topological term $Q_n$, while integrating over the terms quadratic or linear in $A_\mu^n$  simply gives a constant, as they are decoupled from the $\vec \phi$ fields after the shift. 
 
The constraint
 \begin{equation} A_\mu^1(x)+A_\mu^2(x)+A_\mu^3(x)=0\ \  (\forall x,\mu )
 \end{equation}
 can be imposed since
 \be \sum_{n=1}^3\vec \phi_n^*\cdot \partial_\mu \vec \phi_n=0.
 \ee
 This follows because the terms in the real part of the action involving $A_\mu^1$ and $A_\mu^3$ become:
 \be 
 \begin{split}
 \left( A_\mu^1\right)^2&+ \left( A_\mu^3\right)^3+ \left( A_\mu^1+A_\mu^3\right)^2-2iA_\mu^1\vec \phi_1^*\cdot \partial_\mu \vec \phi_1
 -2iA_\mu^3\vec \phi_3^*\cdot \partial_\mu \vec \phi_3-2i\left( A_\mu^1+A_\mu^3\right)\left( \vec \phi_1^*\cdot \partial_\mu \vec \phi_1+ \vec \phi_3^*\cdot \partial_\mu \vec \phi_3\right)\\
 &=\left( A_\mu^1-i\vec \phi_1^*\cdot \partial_\mu \vec \phi_1\right)^3+\left( A_\mu^3-i\vec \phi_3^*\cdot \partial_\mu \vec \phi_3\right)^2
 +\left( A_\mu^1+A_\mu^3-i\vec \phi_1^*\cdot \partial_\mu \vec \phi_1-i\vec \phi_3^*\cdot \partial_\mu \vec \phi_3\right)^2\\
 &-\left|\vec \phi_1^*\cdot \partial_\mu \vec \phi_1\right|^2-\left|\vec \phi_3^*\cdot \partial_\mu \vec \phi_3\right|^2
 -\left|\vec \phi_1^*\cdot \partial_\mu \vec \phi_1+\vec \phi_3^*\cdot \partial_\mu \vec \phi_3\right|^2.
 \end{split}
 \ee
 Again, the integrals over $A_\mu^1$ and $A_\mu^3$ simply give constant factors after shifting them by $i\vec \phi_1^*\cdot \partial_\mu \vec \phi_1$ and 
 $i\vec \phi_3^*\cdot \partial_\mu \vec \phi_3$ respectively.  This continues to work when the topological terms are included.  Shifting the $A_\mu$ now gives the 
 real part of the action written in terms of the $\vec \phi_n$ together with $i(\theta_1-\theta_2)Q_1+i(\theta_3-\theta_2)Q_3$. 
We see  that the partition function is invariant  under shifting all three topological angles by the same constant.

\section{Additional calculations for Sec.~\ref{sec:symms}}
\subsection{Gauge invariance}
\label{gauge_invariance}

As mentioned in the main text,  the overall phases of the spin coherent states shouldn't change the form of the action. This manifests in the gauge invariance of the action in Eq.\ \eqref{eq:fullaction} under the  transformation $U'(x,\tau)= D(x,\tau) U(x,\tau)$, where 
\begin{equation}
\begin{split}
D (x,\tau)=\left( \begin{array}{ccc} e^{i\vartheta_1(x,\tau)}&0&0\\0&e^{i\vartheta_2(x,\tau)}&0\\0&0&e^{i\vartheta_3(x,\tau)}\end{array} \right)
\end{split}
\end{equation}
with $\vartheta_3(x,\tau)=-(\vartheta_1(x,\tau)+\vartheta_2(x,\tau))$.  In terms of the fields, this transformation corresponds to $\vec \phi'_n = e^{i \vartheta_n} \vec \phi_n$.
 
To check the invariance of the real part of Eq.\ \eqref{eq:fullactionphi} we use Eq.\ \eqref{eq:cp2trans} to rewrite it as
\begin{equation}
\begin{split}
 \left( \left| \partial_\mu \vec{\phi}_n^{\phantom{*}}\right|^2 -  \left| \vec{\phi}_n^* \cdot \partial_\mu \vec{\phi}_n^{\phantom{*}} \right|^2 \right)=  \left|  \vec{\phi}_{n+1}^* \cdot \partial_\mu \vec{\phi}_{n}^{\phantom{*}} \right|^2+  \left|  \vec{\phi}_{n-1}^* \cdot \partial_\mu \vec{\phi}_{n}^{\phantom{*}} \right|^2.
\end{split}
\end{equation} 
Now, the terms on the right hand side transform as
\begin{equation}
\begin{split}
\left|  \vec{\phi'}_{n\pm1}^* \cdot \partial_\mu \vec{\phi'}_{n}^{\phantom{*}} \right|^2 = \left|  \vec{\phi}_{n\pm1}^*e^{-i \vartheta_{n\pm1}}  \cdot \left( \partial_\mu   \vec{\phi}_{n}^{\phantom{*}} e^{i \vartheta_{n}}+ \vec{\phi}_n^{\phantom{*}}\partial_\mu e^{i \vartheta_{n}} \right)   \right|^2.
 \end{split}
\end{equation}
Since $\vec{\phi}_n$ and $\vec{\phi}_{n\pm1}$ are orthogonal the second term on the right hand side gives 0. So these terms are invariant under gauge transformation. The $\lambda$-term can be similarly shown to be invariant since the phase factors cancel out. Finally, the topological charges  transform as
\begin{equation}
\begin{split}
\varepsilon_{\mu \nu} \left(\partial_\mu \vec{\phi'}_n^{\phantom{*}}\cdot \partial_\nu \vec{\phi'}^*_n\right) = &\mathbin{\phantom{+}} \varepsilon_{\mu \nu} \left(\partial_\mu \vec{\phi}_n^{\phantom{*}}\cdot \partial_\nu \vec{\phi}^*_n\right) + \varepsilon_{\mu \nu} \left(\partial_\mu (i \vartheta_n )  \partial_\nu (-i \vartheta_n) \right)\\
&+\varepsilon_{\mu \nu} \left(\partial_\mu \vec{\phi}_n^{\phantom{*}}\cdot \vec{\phi}^*_n \partial_\nu (-i \vartheta_n)\right) +  \varepsilon_{\mu \nu} \left(\partial_\mu (i \vartheta_n) \vec{\phi}_n^{\phantom{*}}\cdot \partial_\nu \vec{\phi}^*_n\right).
\end{split}
\end{equation}
 The second term on the right hand side is 0 since $(\partial_\mu \vartheta_n \partial_\nu \vartheta_n)$ is symmetric under exchanging $\mu$ and $\nu$. Also, noticing that ${\partial_\mu \vec{\phi}_n^{\phantom{*}}\cdot \vec{\phi}^*_n + \vec{\phi}_n^{\phantom{*}}\cdot \partial_\mu\vec{\phi}^*_n =\partial_\mu (\vec{\phi}_n^{\phantom{*}}\cdot \vec{\phi}^*_n )=0}$, one can easily show that the third and fourth terms cancel each other, thus proving that the topological term is also invariant under the gauge transformation.
 
\subsection{Breaking lattice symmetries of the spin model and the general form of the action}
 \label{symmbreak}
If the spin model is not invariant under the translation and the mirror symmetries, the $\mathbb{Z}_3$ and parity symmetries of the $\sigma$-model will be broken. This can be achieved, for example, by setting the strength of nearest neighbour interactions to be different between different sublattices, i.e.\ $J_1^{(1,2)} $, $J_1^{(2,3)}$ and $J_1^{(3,1)}$ between sublattices 1 and 2, 2 and 3, and 3 and 1, respectively. In this case the field theory can be derived as in Sec.~\ref{sec:FT}, and the action becomes
\begin{equation}
\begin{split}
S[U] = \int dx d\tau\Bigg( \sum_{n=1}^3  &\frac{1}{g_{n-1,n}} \Biggl[ v_{n-1,n}\,\tr   \left[  \Lambda_{n-1} U \partial_x U^\dagger \Lambda_{n} \partial_x U  U^\dagger \right]  +\frac{1}{v_{n-1,n}} \tr \left[  \Lambda_{n-1} U  \partial_\tau U^\dagger \Lambda_{n} \partial_\tau U  U^\dagger \right]  \Biggr]\\
& +i\sum_{n=1}^3\frac{\theta_n}{2\pi i}\varepsilon_{\mu\nu}\tr\left[\Lambda_n \partial_\mu U\partial_\nu U^\dagger\right] +i\frac{\lambda}{2\pi i} \varepsilon_{\mu\nu}\sum_{n=1}^3 \tr\left[ \Lambda_{n-1} U \partial_\mu U^\dagger \Lambda_{n} \partial_\nu U U^\dagger  \right]
 \Bigg),
\end{split}
\label{eq:fullactiongenform}
\end{equation}
where the coupling constants and velocities are given by
\begin{equation}
\begin{split}
{1 \over g_{n-1,n}} &= p{\sqrt{J_1^{(n-1,n)}J_2^{\phantom{()}}+ 2J_3^{\phantom{()}} J_1^{(n-1,n)}+2 J_3^{\phantom{()}}J_2^{\phantom{()}}} \over(J_1^{(n-1,n)}+J_2^{\phantom{()}})},\\
v_{n-1,n}&=3 a p \sqrt{J_1^{(n-1,n)}J_2^{\phantom{()}} + 2J_3^{\phantom{()}} J_1^{(n-1,n)}+2J_3^{\phantom{()}}J_2^{\phantom{()}}}.
\end{split}
\label{eq:gform}
\end{equation}

Still using the convention $\theta_2=0$, the topological angles  become
\begin{equation}
\begin{split}
\theta_1&=\mathbin{\phantom{-}}  p\frac{2\pi}{3} \left( \frac{  2J_2^{\phantom{()}} }{J_1^{(1,2)}+J_2^{\phantom{()}} }     +     \frac{ J_1^{(3,1)}}{J_1^{(3,1)}+J_2^{\phantom{()}} } - \frac{  J_2}{J_1^{(2,3)}+J_2^{\phantom{()}} } \right)\\
\theta_3&=  -p\frac{2\pi}{3} \left( \frac{  2J_2^{\phantom{()}} }{J_1^{(2,3)}+J_2^{\phantom{()}} }     +     \frac{ J_1^{(3,1)}}{J_1^{(3,1)}+J_2^{\phantom{()}} } - \frac{  J_2}{J_1^{(1,2)}+J_2^{\phantom{()}} } \right)\\
\end{split}
\end{equation}
while the $\lambda$ coefficient takes the value
\begin{equation}
\begin{split}
\lambda= p \frac{2\pi}{3} \left( \frac{  J_2^{\phantom{()}} }{J_1^{(1,2)}+J_2^{\phantom{()}} }     +    \frac{  J_2}{J_1^{(2,3)}+J_2^{\phantom{()}} } -  \frac{ J_1^{(3,1)}}{J_1^{(3,1)}+J_2^{\phantom{()}} }  \right).
\end{split}
\end{equation}
The action written in terms of the $\phi$ fields still has the form of three copies of $\mbox{CP}^2$, as in Eq.\ \eqref{eq:fullactionphi}, but this time the three theories have
different parameters: 
\begin{equation}
\begin{split}
S =\int dx d\tau\Bigg[& \sum_{n=1}^3\Bigg(  \frac{v_n}{2g_{n}} \,   \left( \left| \partial_x^{\phantom{*}} \vec{\phi}_n^{\phantom{*}}\right|^2 -  \left| \vec{\phi}_n^* \cdot\partial_x^{\phantom{*}} \vec{\phi}_n^{\phantom{*}} \right|^2 \right)+\frac{1}{2 v_n g_{n}} \,   \left( \left| \partial_\tau^{\phantom{*}} \vec{\phi}_n^{\phantom{*}}\right|^2 -  \left| \vec{\phi}_n^* \cdot\partial_\tau^{\phantom{*}} \vec{\phi}_n^{\phantom{*}} \right|^2 \right)  \Bigg)\\
&+i\sum_{n=1}^3 \theta_n Q_n  + i \lambda (q_{12}+q_{23}+q_{31})\Bigg].
\label{eq:fullactionphigenform_app}
\end{split}
\end{equation}
The coupling constants and velocities in this language are related to those in Eq.\ \eqref{eq:fullactiongenform} by the following equations:
\begin{equation}
\begin{split}
\frac{v_n}{g_n} &=  \frac{v_{n-1,n}}{g_{n-1,n}}+\frac{v_{n,n+1}}{g_{n,n+1}}-  \frac{v_{n+1,n-1}}{g_{n+1,n-1}}\\
\frac{1}{v_n g_n} &=  \frac{1}{v_{n-1,n} g_{n-1,n}}+\frac{1}{v_{n,n+1} g_{n,n+1}}-  \frac{1}{v_{n+1,n-1} g_{n+1,n-1}}.\\
\end{split}
\label{eq:gtransform}
\end{equation}
Note that, since there are multiple velocities in the field theory, one can no longer set all three of them to 1 by rescaling the space and time directions. 
In particular, this simple calculation shows that if all symmetries are broken, the topological angles $\theta_1$ and $\theta_3$ can take arbitrary values.

In the special case discussed in Sec.~\ref{subsec:symmbreak}  the translational symmetry is broken, but one of the mirror symmetries is conserved.  As shown in Fig.~\ref{fig:Tbreak}, the nearest neighbour couplings take the values $J_1^{(1,2)}=J_1^{(2,3)}=J_1$ and $J_1^{(3,1)}=J_1(1+\delta) $.
In this case the  coupling constants  and  velocities are given by

  \begin{equation}
\begin{split}
\frac{1}{g_{1,2}}=\frac{1}{g_{2,3}} &=p \frac{\sqrt{J_1J_2+ 2J_3 J_1+2 J_3J_2}}{(J_1+J_2)},\\
\frac{1}{g_{3,1}}&=p \frac{\sqrt{J_1J_2(1+\delta)+ 2J_3 J_1(1+\delta)+2 J_3J_2}}{(J_1(1+\delta)+J_2)},\\
v_{1,2}=v_{2,3}&=3 a p \sqrt{J_1J_2 + 2J_3 J_1+2J_3J_2)},\\
v_{3,1}&=3 a p \sqrt{J_1J_2 (1+\delta)+ 2J_3 J_1(1+\delta)+2J_3J_2}.
\end{split}
\end{equation}
Based on Eq.\ \eqref{eq:gtransform}, the parameters of the three $\mbox{CP}^2$ theories become

  \begin{equation}
\begin{split}
\frac{1}{g_{1}}=\frac{1}{g_{3}} &=\frac{1}{g_{3,1}},\\
v_1=v_3&=v_{1,3},
\end{split}
\end{equation}
while $g_2$ and $v_2$ will be different:
\begin{equation}
\begin{split}
\frac{v_2}{g_2}&=\frac{v_{1,2}}{g_{1,2}}+\frac{v_{2,3}}{g_{2,3}}-\frac{v_{3,1}}{g_{3,1}},\\
\frac{1}{v_2g_2}&=\frac{1}{v_{1,2}g_{1,2}}+\frac{1}{v_{2,3}g_{2,3}}-\frac{1}{v_{3,1}g_{3,1}}.
\end{split}
\end{equation}
Since  $g_1=g_3$ (and $v_1=v_3$), the  $R_{13}$ parity symmetry is conserved (corresponding to the remaining mirror symmetry), but  the $\mathbb{Z}_3$ and the other two parity symmetries  are explicitly broken since $g_2$ takes a different value. 


In that case, the topological term is given by
\begin{equation}
\begin{split}
 ip \frac{2\pi}{3} \left(1 +\frac{J_2}{J_1+J_2} -\frac{J_2}{J_1 (1+\delta)+J_2} \right)(Q_1-Q_3).
\end{split}
\end{equation}
So $\theta_1 =-\theta_3$. As a consequence, $R_{13}$ is still a symmetry. However, the topological angle deviates from $p 2\pi/3$. If $\delta<0$, the bond is weakened and $\theta_1 =-\theta_3 < p2\pi/3$, while if $\delta>0$, the bond is strengthened and $\theta_1 =-\theta_3 > p2\pi/3$.   Since $\theta \neq p 2\pi/3$,  the topological term is no longer invariant under $\mathbb{Z}_3$, or  under $R_{12}$ and $R_{23}$. 

If instead of $R_{13}$, $R_{12}$ or $R_{23}$ was conserved, the topological term would become ${i \theta \big(Q_2- Q_1\big) }$ $\equiv {i \theta \big(-2Q_1- Q_3\big) } $ or ${i \theta \big(Q_3- Q_2\big)}\equiv {i \theta \big(Q_1+ 2Q_3\big)} $, respectively. 
Note that if two parity symmetries are conserved, then the $\mathbb{Z}_3$ and the third parity are conserved as well. This is  easy to understand in terms of the symmetries of the spin model: if two different mirror symmetries are conserved, then the translation symmetry is also necessarily present\footnote{As we mentioned before, spin models are always invariant under a three site translation. If on top of that two inequivalent mirror symmetries are present (i.e.\ not connected by the three site translation), invariance under two site translation and thus under one site translation follows as well.}. And vice versa, if $\mathbb{Z}_3$ symmetry is present, it necessarily means that the action is invariant under all three parities. In terms of the spin model, the translation symmetry implies the mirror symmetries as well. This is due to the form of the Heisenberg interaction, which satisfy $S^\alpha_\beta(i) S^\beta_\alpha(j)$ = $S^\alpha_\beta(j) S^\beta_\alpha(i)$. 

The coefficient of the $\lambda$-term is also modified and becomes
\begin{equation}
\begin{split}
p \frac{2\pi}{3}\left(\frac{2J_2}{J_1+J_2} -\frac{J_1(1+\delta)}{J_1(1+\delta)+J_2}  \right)\left(q_{12} +q_{23}+q_{31}\right),
\end{split}
\end{equation}
but since it was non-universal to begin with, this change is of no particular importance. 

 \subsection{General form of the action}
 \label{app:genform}
 Assuming $\mbox{SU}(3)$, gauge, and time reversal invariance of the action, one can show that there are no other possible terms with two derivatives apart from  those appearing in Eqs.\ \eqref{eq:fullactiongenform}, or \eqref{eq:fullactionphigenform_app}. A general term can be written $\phi_m^{\alpha (*)} \phi_n^{\beta (*)} \partial_\mu \phi^{\gamma (*)}_o \partial_\nu \phi_p^{\delta (*)}$, where  $^{(*)}$ means that it is either complex conjugated or not. $m,n,o,p$ index the three different fields, while the $\alpha, \beta, \gamma, \delta$ indices run through the components of the fields.   Such a  term would transform under the gauge transformation as
 \begin{equation}
\begin{split}
\phi_n^{\alpha (*)} \phi_m^{\beta (*)} \partial_\mu \phi^{\gamma (*)}_o \partial_\nu \phi_p^{\delta (*)} \to
&\mathbin{\phantom{+}}\phi_m^{\alpha (*)} \phi_n^{\beta (*)} \partial_\mu \phi^{\gamma (*)}_o \partial_\nu \phi_p^{\delta (*)} e^{ \pm i \vartheta_m} e^{\pm i\vartheta_n}e^{\pm i\vartheta_o} e^{\pm i \vartheta_p}  \\
& +\phi_m^{\alpha (*)} \phi_n^{\beta (*)} \phi^{\gamma (*)}_o \partial_\nu \phi_p^{\delta (*)} e^{ \pm i \vartheta_m} e^{\pm i\vartheta_n} (\pm i \partial_\mu \vartheta_o) e^{\pm i\vartheta_o} e^{\pm i \vartheta_p}  \\
& +\phi_m^{\alpha (*)} \phi_n^{\beta (*)} \partial_\mu \phi^{\gamma (*)}_o  \phi_p^{\delta (*)} e^{ \pm i \vartheta_m} e^{\pm i\vartheta_n}e^{\pm i\vartheta_o} (\pm i \partial_\nu \vartheta_p) e^{\pm i \vartheta_p}  \\
& +\phi_m^{\alpha (*)} \phi_n^{\beta (*)}  \phi^{\gamma (*)}_o \phi_p^{\delta (*)} e^{ \pm i \vartheta_m}  e^{\pm i\vartheta_n} (\pm i \partial_\mu \vartheta_o)e^{\pm i\vartheta_o} (\pm i \partial_\nu \vartheta_p) e^{\pm i \vartheta_p}.\end{split}
\label{eq:gaugerule}
\end{equation}
 For this term to be invariant there must be exactly two complex conjugates, and the field index of each complex conjugated field should be the same as one of the non conjugated ones, otherwise the $e^{\pm i \vartheta}$ phases wouldn't cancel in the first term on the right hand side.  Furthermore to conserve $\mbox{SU}(3)$, the fields should form scalar products.  With
 these restrictions, the possible terms are:
 \begin{equation}
\begin{split}
&A_{n,m}^{\mu,\nu} \big( \partial_\mu \vec \phi_n^*  \cdot \partial_\nu\vec \phi_n \big)\big( \vec \phi_m^* \cdot \vec \phi_m \big)= A_{n}^{\mu,\nu}\big( \partial_\mu \vec \phi_n^*  \cdot \partial_\nu\vec \phi_n \big),\\
&B_{n,m}^{\mu, \nu} \big( \partial_\mu \vec \phi_n^*  \cdot \vec \phi_n \big)\big( \partial_\nu \vec \phi_m^* \cdot \vec \phi_m \big),\\
&C_{n,m}^{\mu, \nu} \big( \partial_\mu \vec \phi_n^*  \cdot \partial_\nu\vec \phi_m \big)\big( \vec \phi_m^* \cdot \vec \phi_n \big)= \delta_{m,n}C_{n,n}^{\mu, \nu}\big( \partial_\mu \vec \phi_n^*  \cdot \partial_\nu\vec \phi_n \big),  \\
&D_{n,m}^{\mu, \nu} \big( \partial_\mu \vec \phi_n^*  \cdot \vec \phi_m \big)\big( \partial_\nu \vec \phi_m^* \cdot \vec \phi_n \big).\\
\end{split}
\end{equation}
Note that the  $C_{n,n}^{\mu, \nu}$ term gives the same term as the $A_{n}^{\mu,\nu}$ . The $D$-term is  gauge invariant if $m\neq n$, while the $m=n$ case is already considered in the $B$-term. The other terms transform under the gauge transformation as
 \begin{equation}
\begin{split}
A_{n}^{\mu,\nu}\big( \partial_\mu \vec \phi_n^*  \cdot \partial_\nu\vec \phi_n \big)  &+ A_{n}^{\mu,\nu}\big(i \partial_\nu \vartheta_n \partial_\mu \vec \phi_n^*  \cdot \vec \phi_n  +  i \partial_\mu \vartheta_n   \partial_\nu\vec \phi_n^*  \cdot \vec \phi_n + \partial_\mu \vartheta_n \partial_\nu \vartheta_n  \big), \\ 
B_{n,m}^{\mu, \nu} \big( \partial_\mu \vec \phi_n^*  \cdot \vec \phi_n \big)\big( \partial_\nu \vec \phi_m^* \cdot \vec \phi_m \big) &+ B_{n,m}^{\mu, \nu} \big(-i\partial_\mu \vartheta_n  \partial_\nu \vec \phi_m^* \cdot \vec \phi_m -i \partial_\nu\vartheta_m  \partial_\mu \vec \phi_n^*  \cdot \vec \phi_n  - \partial_\mu \vartheta_n \partial_\nu \vartheta_m\big).
\\
\end{split}
\end{equation}
The $A$-term is gauge invariant in itself if  $A_{n}^{\mu,\nu}$ is antisymmetric in $\mu, \nu$. 
If $m\neq n$, the $B$-term can't be made gauge invariant, and there is no other possible term with which it would give a gauge invariant combination either. Note, however, that in the case  $m=n$, setting $A_{n}^{\mu,\nu} = B_{n,n}^{\mu,\nu}$ leads to a  gauge invariant combination.  So the most general form of an $\mbox{SU}(3)$ and gauge invariant term is given by
\begin{equation}
\begin{split}
&\epsilon_{\mu \nu} A_n \big( \partial_\mu \vec \phi_n^*  \cdot \partial_\nu\vec \phi_n \big) + B_{n}^{\mu, \nu}\big[  \big( \partial_\mu \vec \phi_n^*  \cdot \partial_\nu\vec \phi_n \big) +\big( \partial_\mu \vec \phi_n^*  \cdot \vec \phi_n \big)\big( \partial_\nu \vec \phi_n^* \cdot \vec \phi_n \big) \big]  \\&+ D_{n,m}^{\mu, \nu}  (1-\delta_{m,n}) \big( \partial_\mu \vec \phi_n^*  \cdot \vec \phi_m \big)\big( \partial_\nu \vec \phi_m^* \cdot \vec \phi_n \big).
\end{split}
\end{equation}
Under time reversal symmetry (for real time) this would transform as
\begin{equation}
\begin{split}
-\epsilon_{\mu \nu} ( A_n)^* \big( \partial_\mu \vec \phi_n^*  \cdot \partial_\nu\vec \phi_n \big) &+ (-1)^{\varepsilon_{\mu,\nu} }( B_{n}^{\mu, \nu})^*\big[  \big( \partial_\mu \vec \phi_n^*  \cdot \partial_\nu\vec \phi_n \big) +\big( \partial_\mu \vec \phi_n^*  \cdot \vec \phi_n \big)\big( \partial_\nu \vec \phi_n^* \cdot \vec \phi_n \big) \big]  \\&+ (-1)^{\varepsilon_{\mu,\nu} } (D_{n,m}^{\mu, \nu})^*  (1-\delta_{m,n}) \big( \partial_\mu \vec \phi_n^*  \cdot \vec \phi_m \big)\big( \partial_\nu \vec \phi_m^* \cdot \vec \phi_n \big).
\end{split}
\end{equation}
The factor $( -1)$  only appears if there is exactly one time derivative term. Based on these, for real time, we have $A_n= -A_n^*$, $(B_n^{\mu,\nu})^*=(-1)^{\varepsilon_{\mu,\nu} } (B_n^{\mu,\nu})$, and $(D_{m,n}^{\mu,\nu})^*=(-1)^{\varepsilon_{\mu,\nu} } D_{m,n}^{\mu,\nu}$.  So $A_n$ should be imaginary for real time, hence real for imaginary time. Similarly, $D^{\mu,\mu}$ and $B^{\mu,\mu}$ are real, while $D^{\mu,\nu}$ and $B^{\mu,\nu}$ for $\mu\neq \nu$ are imaginary for real time (therefore all the elements of $D^{\mu,\nu}_{m,n}$ and $B^{\mu,\nu}_n$ are real in imaginary time).
 Furthermore, every term in the action should be real for real time $t =- i\tau$. This constrains  $D^{x,t}_{m,n}=-D^{t,x}_{m,n}$, as they are  pure imaginary and they are coupled to terms which are complex conjugates of each other. Similarly 
 $B^{x,t}_{n}=-B^{t,x}_{n}$ has to be fullfilled, but  these terms drop out since they are coupled to terms which are symmetric under  $x \leftrightarrow t$.   Hence finally we arrive at a general form 
\begin{equation}
\begin{split}
& D^{\mu,\mu}_{m,n} (1-\delta_{m,n}) \big( \partial_\mu \vec \phi_n^*  \cdot \vec \phi_m \big)\big( \partial_\mu \vec \phi_m^* \cdot \vec \phi_n \big) + B^{\mu,\mu}_{n}   \big[\big( \partial_\mu \vec \phi_n^*  \cdot \partial_\nu\vec \phi_n \big) +\big( \partial_\mu \vec \phi_n^*  \cdot \vec \phi_n \big)\big( \partial_\nu \vec \phi_n^* \cdot \vec \phi_n \big) \big] \\
&+\epsilon_{\mu \nu} A_n \big( \partial_\mu \vec \phi_n^*  \cdot \partial_\nu\vec \phi_n \big)  +  \varepsilon_{m,n} D^{x,\tau}_{m,n}  \big( \partial_\mu \vec \phi_n^*  \cdot \vec \phi_m \big)\big( \partial_\nu \vec \phi_m^* \cdot \vec \phi_n \big).
\end{split}
\end{equation}
Note however that some of these terms are redundant. The $D^{\mu, \mu}_{m,n}$- and the $B^{\mu,\mu}_{n}$-terms both express the  real part of the action. The $D^{\mu, \mu}_{m,n}$-terms correspond to the formulation in Eq.\ \eqref{eq:fullactiongenform}, while the $B^{\mu,\mu}_n$-terms actually give the form of the real part in Eq.\ \eqref{eq:fullactionphigenform_app}. The transformation  between the $D^{\mu,\mu}_{m,n}$ and $B_n^{\mu,\nu}$ parameters is similar to that in Eq.\ \eqref{eq:gtransform}. 
The $A_n$-terms give the three topological charges (up to a factor of $2\pi i$). As we mentioned before, since the three charges sum up to zero only two independent  parameters remain. The $D^{x,\tau}_{m,n}$-terms give the $q_{m,n}$ terms as defined in Eq.\ \eqref{eq:qterm}. 
 In general, the $D^{x,\tau}_{1,2},D^{x,\tau}_{2,3},D^{x,\tau}_{3,1}$ can have different values. Using the relation between the $q_{mn}$ quantities and the topological charges (Eq.\ \eqref{eq:qQrel}) the $D^{x,\tau}_{m,n}$-terms give 
  \begin{equation} 
\begin{split}
2\pi i \left(D^{x,\tau}_{1,2}q_{12} +D^{x,\tau}_{2,3} q_{23}+D^{x,\tau}_{3,1} q_{31}\right) = &2\pi i\frac{D^{x,\tau}_{1,2}+D^{x,\tau}_{2,3}+D^{x,\tau}_{3,1}}{3} (q_{12}+q_{23}+q_{31})  \\&+2\pi i \frac{2 D^{x,\tau}_{1,2}-D^{x,\tau}_{2,3} -D^{x,\tau}_{3,1}}{3}Q_1 +2\pi i \frac{ D^{x,\tau}_{1,2}-2D^{x,\tau}_{2,3}+D^{x,\tau}_{3,1}}{3}Q_3
\end{split}
\end{equation}
Thus, the $D^{x\tau}_{m,n}$-terms actually give both the unusual imaginary $\lambda$-term and the topological terms. Therefore the $A_n$-terms are also redundant.  Reviewing Eqs.\ \eqref{eq:fullactiongenform} and \eqref{eq:fullactionphigenform_app}, we find that those already have the most general form compatible with $\mbox{SU}(3)$, gauge and time reversal invariance. 

\section{Details of renormalization group calculation}

Throughout this appendix, lowercase Roman letters index the off-diagonal Gell-Mann Matrices (GM), lowercase Greek letters index the diagonal GM, uppercase Roman letters index all eight GM, and repeated indices are summed over.

\subsection{Definitions and identities of $\mbox{SU}(3)$ structure constants}

We label the Gell-Mann matrices $T_A$ of $\mbox{SU}(3)$, according to
\be \label{gm}
\begin{split}
T_1 &= \bp 0 & 1 & 0 \\
1 & 0 & 0 \\
0 & 0 & 0 \\
\ep \hspace{5mm} 
T_2 = \bp 0 & -i & 0 \\
i & 0 & 0 \\
0 & 0 & 0 \\
\ep \hspace{5mm} 
T_3 = \bp 1 & 0 & 0 \\
0 & -1 & 0 \\
0 & 0 & 0 \\
\ep 
\\
T_4 &= \bp 0 & 0 & 1 \\
0& 0 & 0 \\
1 & 0 & 0 \\
\ep\hspace{5mm} 
T_5 = \bp 0 & 0 & -i \\
0 & 0 & 0 \\
i & 0 & 0 \\
\ep \hspace{5mm} \\
T_6 &= \bp 0 & 0& 0 \\
0 & 0 & 1 \\
0 & 1 & 0 \\
\ep \hspace{5mm} 
T_7 = \bp 0 & 0 & 0 \\
0 & 0 & -i \\
0 & i & 0 \\
\ep \hspace{5mm} 
T_8 = \frac{1}{\sqrt{3}} \bp 1 & 0 & 0 \\
0& 1& 0 \\
0 & 0 & -2 \\
\ep
\end{split}
\end{equation}
These matrices satisfy the $\mathfrak{su}(3)$ algebra
\be \label{commute}
	[T_A, T_B] = 2i f_{ABC} T_C,
\ee
where the $f_{ABC}$ structure constants are fully antisymmetric. In what follows we prove various identities involving $f_{ABC}$. By construction, the Gell-Mann matrices satisfy
\be \label{norm}
\tr T_A T_B = 2\delta_{AB}.
\ee
Using the completeness of $\mbox{SU}(3)$ generators, 
\be \label{gencomplete} (T_A)^i_j(T_A)^k_l=2\delta^i_l\delta^k_j-\frac{2}{3}\delta^i_j\delta^k_l,\ee
we prove
\be \label{complete}
f_{ABC}f_{ABD} = 3\delta_{CD}.
\ee
\emph{Proof:} From \eqref{commute} and \eqref{norm} 
\be	\tr \big([T_A,T_B][T_A,T_C]\big)= \tr\big(2T_AT_BT_AT_C-T_A^2T_CT_B-T_A^2T_BT_C\big)=-8f_{ABD}f_{ACD},
\ee
then using \eqref{gencomplete} on the middle term:
\begin{equation}
\begin{split}
	\tr\big(2T_AT_BT_AT_C-T_A^2T_CT_B-T_A^2T_BT_C\big)
	=2 (T_A)^i_j (T_B)^j_k (T_A)^k_l (T_C)^l_i - (T_A)^i_j(T_A)^j_k (T_CT_B + T_BT_C)^k_i\\
	=- \frac{4}{3}\tr T_BT_C
	- 6 \tr (T_CT_B + T_BT_C)
	+\frac{2}{3} \tr (T_CT_B +T_BT_C) = -12\tr T_BT_C = -24\delta_{BC},
\end{split}
\end{equation}
where in the last step, we used \eqref{norm}. This completes the proof.

Now we prove two partial completeness results:
\be \label{partial}
	f_{c\gamma a} f_{c\gamma b} = \delta_{ab}.
\ee
\emph{Proof:} We write 
\be \label{prepartial}
	f_{c\gamma a}f_{c\gamma b} = f_{c3a}f_{c3b} + f_{c8a}f_{c8b},
\ee
with the structure constants  $f_{123}  =1, f_{345}=f_{376} = \frac{1}{2}, f_{458}=f_{678} = \frac{\sqrt{3}}{2}$. The first term in (\ref{prepartial}) equals
\be \label{t1}
	 f_{c3a}f_{c3b}= \delta_{a2}\delta_{b2} + \delta_{a1}\delta_{b1} +\frac{1}{4}\delta_{a5}\delta_{b5} + \frac{1}{4}\delta_{a4}\delta_{b4}
	+ \frac{1}{4} \delta_{a6}\delta_{b6} + \frac{1}{4}\delta_{a7}\delta_{b7}
\ee
and the second term equals:
\be \label{t2}
	f_{c8a}f_{c8b}=\frac{3}{4}\left( \delta_{a5}\delta_{b5} + \delta_{a4}\delta_{b4} +\delta_{a6}\delta_{b6}+\delta_{a7}\delta_{b7}\right).
\ee
Adding \eqref{t1}  and \eqref{t2} completes the proof.

The second partial completeness result is 
\be \label{cases}
	 f_{abC}f_{abD} = \begin{cases} \delta_{CD} & \text{ if $C=c$ is off diagonal},\\
	3\delta_{CD} & \text{ if $C =\gamma$ is diagonal}. \\
\end{cases}
\ee
\emph{Proof:} Expand \eqref{complete} as	
\be
	3\delta_{CD} = f_{ABC}f_{ABD} = f_{abC}f_{abD} + 2f_{a\beta C} f_{a\beta D}.
\ee
If $C=\gamma$ is a diagonal index, the second term vanishes, and the second case of \eqref{cases} follows.  If $C$ is off-diagonal, the second term gives $2\delta_{CD}$ according to (\ref{partial}), which proves the first case of \eqref{cases}.

Next, the Jacobi Identity 
\be
	[T_A, [T_B,T_C]] + [T_B,[T_C,T_A]] + [T_C,[T_A,T_B]] = 0
\ee
gives us an identity for the structure constants,
\be
	 f_{ADE} f_{BCD}  + f_{BDE}f_{CAD} + f_{CDE}f_{ABD}=0. 
\ee

Finally, using this Jacobi identity, we prove 
\be \label{cases2}
	f_{a\gamma d}f_{b\gamma e} f_{Cde} = \begin{cases} \frac{1}{2}f_{abC} & \text{ if $C=c$ is off-diagonal}, \\
	f_{abC} & \text{ if $C=\rho$ is diagonal}. \\
	\end{cases}
\ee
\emph{Proof:} A special case of the Jacobi identity is:
\be
f_{Cde}f_{b\gamma e}+f_{C\gamma e}f_{dbe} = f_{Cbe}f_{d\gamma e}.
\ee
Now we multiply this Jacoby identity by $f_{d\gamma a}$ and sum over $d$, $\gamma$,
\be \label{above}
f_{d\gamma a}f_{Cde}f_{b\gamma e}+f_{d\gamma a} f_{C\gamma e}f_{dbe} = f_{d \gamma a}f_{Cbe}f_{d\gamma e} = f_{Cba},
\ee
where we used (\ref{partial}) on the RHS. Now, if $C$ is diagonal, the second term on the LHS vanishes, and we prove case 2. If $C$ is off-diagonal, the two terms on the LHS of (\ref{above}) are equal, which can be checked, term by term. This proves case 1.

\subsection{Polyakov's renormalization of O(N) nonlinear $\sigma$-model} \label{nlsmrg}
We review Polyakov's calculation of the beta function for the $\mbox{O}(N)$ nonlinear $\sigma$-model \cite{Polyakov:1975rr},
\be
	\fL = \frac{1}{2g} |\partial_\mu \vn|^2.
\ee 
The idea is to construct a `slow' unit vector $\vn_s$ out of $\vn$'s momentum modes below $b\Lambda$, where $\Lambda$ is a reduced cutoff, and $b\lesssim1 $. The remaining modes of $\vn$ can then be written in terms of an orthonormal basis $\{\ve_a\}$, orthogonal to $\vn_s$:
\be \label{exp}
	\vn = \vn_s(1-\phi^2)^{1/2} + \sum_{a=1}^{N-1}\phi_a \ve_a,
\ee
where $\phi^2= \sum_{a=1}^{N-1}\phi_a\phi_a$. The fields $\phi_a$ consist entirely of `fast' modes, with momentum lying in the Wilson shell $[b\Lambda,\Lambda)$. Integration over the shell is then equivalent to integrating out the fields $\phi_a$. Inserting this expansion (\ref{exp}) into $\fL$ gives (to quadratic order in $\phi_a$)
\be
	2g\fL 
	= (\partial_\mu \vn_s)^2(1-\phi^2) + (\partial_\mu \phi_a)^2 + \phi_a\phi_b\partial_\mu \ve_a\cdot \partial_\mu \ve_b +2\phi_a \partial_\mu \phi_b \partial_\mu \ve_a \cdot \ve_b
	- 2\phi_a\partial^2_\mu \vn_s\cdot   \ve_a\;.
\ee
A naive argument would claim that the term linear in $\phi_a$ can be neglected, since $n_s,e_a$ contain slow modes only. However, their product will generically have modes lying in the Wilson shell; a better argument is presented in (\ref{linear}). Dropping linear terms, we are left to evaluate the Gaussian integral
\be \label{nlsmgaussian}
	\int \fD[ \phi ]\exp \Bigg[-\frac{1}{2g}\int d^2 x \left[ (\partial_\mu \vn_s)^2(1-\phi^2) + (\partial_\mu \phi_a)^2 + \phi_a\phi_b\partial_\mu \ve_a\cdot \partial_\mu \ve_b +2\phi_a \partial_\mu \phi_b \partial_\mu \ve_a\cdot \ve_b 
\right]\Bigg].
\ee
Using
\be
	\int \fD [\phi] = \int \prod_{b\Lambda < k <\Lambda} \fD \phi(k)\fD \phi(-k)
\ee
the effective Lagrangian is 
\be \label{eq:nlsmlog}
	2g\fL_{\text{eff}} =(\partial_\mu \vn_s)^2 + g \,\tr \left[ \log  \big(\delta_{ab}  -G_{ab}\delta_{ab}(\partial_\mu\vn_s)^2+G_{ab}\partial_\mu \ve_a\cdot \partial_\mu \ve_b +2\partial_\mu G_{ab} \partial_\mu \ve_a\cdot \ve_b\big) \right],
\ee
where $G_{ab}(x)$ is the Green's function of the fields $\phi_a$. Since terms involving more than two derivatives of slow fields are irrelevant, we can expand the trace-logarithm:

\begin{equation}
\label{eq:averages}
\begin{split}
	&\tr\log\left[ \delta_{ab} - G_{ab}\delta_{ab}(\partial_\mu \vn_s)^2 + G_{ab}\partial_\mu \ve_a \cdot \partial_\mu \ve_b + 2\partial_\mu G_{ab} \partial_\mu \ve_a \cdot \ve_b\right]\\
&= -\int d^2 x \,G_{ab}(0)\delta_{ab} (\partial_\mu \vn_s)^2 + \int d^2 x \,G_{ab}(0) (\partial_\mu \ve_a\cdot\partial_\mu \ve_b)
	+ 2 \int d^2 x\, \partial_\mu G_{ab}(0) \partial_\mu \ve_a\cdot \ve_b \\
		&\mathbin{\phantom{=}}-\frac{1}{2} 4\int d^2 xd^2y \,\partial_\mu G(x-y)[\partial_\mu \ve_a\cdot\ve_b](x),\ \partial_\nu G(y-x) [\partial_\nu \ve_b\cdot\ve_a](y)  + \text{irrelevant},  
\end{split}
\end{equation}
where we  defined $G(x)$ by $G_{ab}(x) = \delta_{ab}G(x)$.) The third term vanishes since
\be \label{eq:rotation}
	\partial_\mu G(0) = -i\int \frac{d^2 k}{(2\pi)^2}  \frac{k_\mu}{k^2} = 0.
\ee
Rewriting $G(0) = \int_{b\Lambda < k < \Lambda} \frac{d^2k}{(2\pi)^2} \frac{\delta_{ab}}{k^2}$, the  expansion in \eqref{eq:averages} reduces to
\begin{equation}
\label{eq:dkdq}
\begin{split}
	={}& \int_{b\Lambda < k < \Lambda} \frac{d^2k}{(2\pi)^2} \frac{1}{k^2} \int d^2 x \Big(-(N-1) (\partial_\mu \vn_s)^2 + (\partial_\mu \ve_a\cdot\partial_\mu \ve_a)\Big)\\
	&-2 \int \frac{d^2k d^2q}{(2\pi)^4} \frac{k_\mu }{k^2} \frac{(q+k)_\nu}{(q+k)^2} [\partial_\mu \ve_a\cdot\ve_b](q)[ \partial_\nu \ve_b\cdot\ve_a](-q) + \text{irrelevant,} 
\end{split}
\end{equation}
where we Fourier transformed the last term, and both $\vk$ and $\vk+\vq$ lie in the Wilson shell $[b\Lambda ,\Lambda)$. Now we expand $\int_{b\Lambda < |\vq+\vk|<\Lambda} \frac{(q+k)_\nu}{(q+k)^2}$ in powers of $q$, and keep the zeroth order term only, since terms with more powers of $q$ will correspond to irrelevant operators:
\be \label{eq:zerothorder}
	 \int \frac{d^2k d^2q}{(2\pi)^4} \frac{k_\mu }{k^2} \frac{(q+k)_\nu}{(q+k)^2}
	 = \int \frac{d^2k d^2q}{(2\pi)^4} \frac{k_\mu k_\nu }{k^4}
	 + \text{irrelevant} 
	 =\frac{1}{2} \int \frac{d^2k d^2q}{(2\pi)^4} \frac{1 }{k^2}
	 + \text{irrelevant} .
\ee
Inserting this expansion into \eqref{eq:dkdq}, and integrating over $k$, the effective Lagrangian becomes
\be
	2g\fL_{\text{eff}} = (\partial_\mu n_s)^2 +	\frac{g\log b}{\pi} \left[(N-1)(\partial_\mu \vn_s)^2 - \partial_\mu \ve_a\cdot \partial_\mu \ve_a  +  (\partial_\mu \ve_a\cdot \ve_b)^2 \right].
\ee
Finally, we insert a complete set of states to obtain the identity
\be
(\partial_\mu \ve_a)^2 = \partial_\mu e_a^i \partial_\mu e_a^j\left[ e_b^i e_b^j + n_s^in_s^j\right]
= (\partial_\mu \ve_a \cdot \ve_b)^2 + (\partial_\mu \ve_a\cdot \vn_s)^2
= (\partial_\mu \ve_a \cdot\ve_b)^2 + (\partial_\mu \vn_s)^2
\ee
since $\vn_s \cdot \ve_a = 0$. Thus,
\be
	\fL_{\text{eff}} = \frac{1}{2g} (\partial_\mu \vn_s)^2 \left[ 1 + g\frac{(N-2)}{2\pi }\log b \right].
\ee
From this, we conclude
\be
	\beta(g) = \frac{dg}{d\log b} =  -\frac{(N-2)}{2\pi} g^2.
\ee

\subsection{Polyakov's renormalization in $\mbox{SU}(2)$ language}  \label{polyakov}
In \ref{nlsmrg}, the RG calculation relied on an expansion of the field $\vn$ in terms of `slow' and `fast' components:
\be 
	\vn = \vn_s(1-\phi^2)^{1/2} + \sum_{a=1}^2 \phi_a \ve_a.
\ee
We would like to generalize this to matrix field theories, by writing
\be \label{exp2}
	U = U_f U_s
\ee
for $U,U_f,U_s \in \mbox{SU}(n)$ , where $U_f$ contains the fast modes of $U$, and $U_s$ the slow modes. Rewriting the O(3) nonlinear $\sigma$-model in terms of $\mbox{SU}(2)$ matrices,
\be
	\vn \cdot \vsig = U^\dag \sig_z U \hspace{5mm} U \in \text{SU(2)},
\ee
we prove that (\ref{exp2}) is equivalent to (\ref{exp}) in the SU(2) case. We expand the fast matrix as
\be
	U_f =\id + i\theta_a \sig_a - \theta_a\theta_b \sig_a \sig_b + \fO(\theta^3)
\ee
and only keep terms up to quadratic order in $\theta$ (higher order terms will correspond to diagrams beyond one loop).  Now
\begin{equation}
\begin{split}
U^\dag \sig_z U &=  U_s^\dag U_f \sig_z U_fU_s
	= U_s^\dag \left(\id - i\theta_a T_a - \frac{1}{2}\theta_a\theta_b T_a T_b \right) \sig_z \left(\id + i\theta_a T_a - \frac{1}{2}\theta_a\theta_b T_a T_b \right) U_s\\
&=U_s^\dag \sig_z U_s  -\theta_a\theta_b U_s^\dag \sig_z T_a T_b U_s + i\theta_a U_s^\dag [\sig_z, T_a] U_s
	-\frac{1}{2}\theta_a\theta_b U_s^\dag  \{T_a T_b ,\sig_z\} U_s \\
	&= U_S^\dag \sig_z U_s(1-2\theta^2) +-2\theta_x U_s^\dag  \sig_y U_s + 2\theta_y U_s^\dag \sig_x U_s.
\end{split}
\end{equation}
Defining 
\be
	 \ve_1 =  \tr \frac{1}{2} \vec{\sig} U_s^\dag \sig_y U_s,
\hspace{10mm}
\ve_2 = \frac{1}{2}\tr \vec{\sig} U_s^\dag \sig_x U_s,
\hspace{10mm} 
(\phi_1,\phi_2) = (2\theta_y,-2\theta_x),
\ee
we find 
\be
	\vsig \cdot \vn = \vsig\cdot \vn_s (1-\frac{1}{2}\phi^2) + 
	\phi_1 U_s^\dag \sig_y U_s 
	+ \phi_2 U_s^\dag \sig_x U_s.
\ee
To read off the components of $n$, we use
\be
	n^a = \frac{1}{2} \tr \sig^a \vsig\cdot \vn= \frac{1}{2} \tr \sig^a U^\dag \sig_z U .
\ee
So applying $\frac{1}{2} \tr \sig$ to the above expression, we find 
\be
	\vn = \vn_s (1- \frac{1}{2} \phi^2) +\phi_1 \frac{1}{2} \tr \vec{\sig} U_s^\dag \sig_y U_s
	+ \phi_2 \frac{1}{2}\tr \vec{\sig} U_s^\dag \sig_x U_s
	=\vn_s(1-\phi^2)^{1/2} + \sum_{a=1}^2 \phi_a \ve_a + \fO(\phi^3).
\ee
Finally, we check that we've found an orthonormal basis:
\begin{equation}
\begin{split}
	\ve_1\cdot \ve_2 &= \frac{1}{2} \tr (\ve_1\cdot \vsig) (\ve_2\cdot \vsig) = \frac{1}{2} \tr U_s^\dag \sig_y U_s  U_s^\dag \sig_x U_s = 0,\\
	\vn_s \cdot \ve_a &= \frac{1}{2} \tr (\sig\cdot \vn_s \vsig\cdot \ve_a) = \frac{1}{2} \tr  U_s^\dag \sig_z U_s U_s^\dag \sig_a U_s= 0.
\end{split}
\end{equation}
Therefore, our expansion (\ref{exp2}) is equivalent to Polyakov's expansion (\ref{exp}).

\subsection{Rewriting the Lagrangian} \label{equivalent}
We rewrite $\Lambda_j$ in terms of $T_3,T_8$ and $\id$, and expand the two terms of the imaginary time Lagrangian. Our results are (\ref{ian}) and (\ref{eq:ian2}).

\subsubsection*{Real part:}
We start with
\be 
{\cal L} =-\sum_{j=1}^3\hbox{tr}\partial_\mu UU^\dagger \Lambda_j\partial_\mu UU^\dagger \Lambda_{j+1}.\ee
This can be rewritten as
\be {\cal L}=-(1/2)\sum_{j=1}^3\hbox{tr}\partial_\mu (U^\dagger \Lambda_jU)\partial_\mu (U^\dagger \Lambda_{j+1}U)
\label{L}\ee
since $\tr \big(\partial_\mu U\partial_\mu U^\dagger \Lambda_j\Lambda_{j+1}\big)=0$. Note that
\begin{equation}
\begin{split}
6\Lambda_1&=\sqrt{3}T_8+2\id+3T_3, \\
6\Lambda_2&= \sqrt{3}T_8+2\id-3T_3, \\
3\Lambda_3&= \id-\sqrt{3}T_8,
\end{split} \label{Lamexp}
\end{equation}
where $I$ is the identity matrix. Substituting this into Eq.\ (\ref{L}), we may drop the $\id$ terms 
since $\partial_\mu (U^\dagger U)=0$. Thus
\begin{equation}
\begin{split}
36{\cal L}={}&-(1/2)\partial_\mu [U^\dagger (\sqrt{3}T_8+3T_3)U]\partial_\mu [U^\dagger (\sqrt{3}T_8-3T_3)U] \\
&-\partial_\mu [U^\dagger (\sqrt{3}T_8-3T_3)U]\partial_\mu [U^\dagger (-\sqrt{3}T_8)U] \\
&-\partial_\mu [U^\dagger (-\sqrt{3}T_8)U]\partial_\mu [U^\dagger (\sqrt{3}T_8+3T_3)U].
\end{split}
\end{equation}Collecting terms,
\be \label{ian} {\cal L}=(1/8)\hbox{tr}\partial_\mu [U^\dagger T_\gamma U)]\partial_\mu [U^\dagger T_\gamma U)].\ee
This result doesn't depend on how we choose  the diagonal Gell-Mann matrices. Note that  
the diagonal matrix elements of the two diagonal Gell-Mann matrices together with $\sqrt{2/3} \id$
form a complete orthogonal set of real vectors with norm $\sqrt{2}$. Thus:
\be \sum_{\gamma}(T_\gamma )^i_i(T_\gamma )^j_j=2\delta_{ij}-\frac{2}{3}.\ee
Thus we may also write:
\be {\cal L}=(1/8)\sum_{i,j,k}\partial_\mu [(U^\dagger )^i_{\ j}U^j_{\ k}]\partial_\mu [(U^\dagger )^k_{\ j}U^j_{\ i}].\ee
The $2/3$ term can be dropped because it gives a term containing $\partial_\mu (U^\dagger U)$.
 The same result is obtained with any basis of diagonal Gell-Mann matrices which obey the same completeness condition 
and the same normalization.

\subsubsection*{Imaginary $\lambda$-term:} We start with 
\be \label{firstproof}
	 -\epsilon_{\mu\nu} \sum_{j=1}^3 \tr \partial_\mu U U^\dag \Lambda_j \partial_\nu U U^\dag \Lambda_{j+1} 
\ee
and use (\ref{Lamexp}). Defining $N_\mu := \partial_\mu U U^\dag$, we have

\begin{equation}
\begin{split}
	36\tr N_\mu \Lambda_1 N_\nu \Lambda_2 ={}& \tr N_\mu ( \sqrt{3}T_8 + 2\id + 3T_3)N_\nu (\sqrt{3}T_8 + 2\id - 3T_3)\\
	={}&2\sqrt{3} \tr N_\mu N_\nu T_8 
	+ 3\sqrt{3} \tr N_\mu T_3N_\nu T_8 
	+2\sqrt{3} \tr N_\mu T_8 N_\nu \\
	&+6 \tr N_\mu T_3N_\nu 
	-3\sqrt{3} \tr N_\mu T_8 N_\nu T_3
	-6 \tr N_\mu N_\nu T_3\\
	={}& -12 \tr N_\mu N_\nu T_3
	-6\sqrt{3} \tr N_\mu T_8 N_\nu T_3
\end{split}
\end{equation}
	
and
\begin{equation}
\begin{split}
	36 \tr N_\mu \Lambda_2 N_\nu \Lambda_3 ={}& 2\tr N_\mu (\sqrt{3}T_8 + 2\id - 3T_3) N_\nu (\id - \sqrt{3}T_8 )\\
	= {}& 2\sqrt{3} \tr N_\mu T_8 N_\nu
	 -6\tr N_\mu T_3 N_\nu \\
	&-4\sqrt{3} \tr N_\mu N_\nu T_8
	+6\sqrt{3} \tr N_\mu T_3 N_\nu T_8\\
	={}&  
	 6 \tr N_\mu N_\nu T_3 
	-6\sqrt{3} \tr N_\mu N_\nu T_8
	-6\sqrt{3} \tr N_\mu T_8 N_\nu T_3
\end{split}
\end{equation}
and
\begin{equation}
\begin{split}
	36 \tr N_\mu \Lambda_3 N_\nu \Lambda_1 ={}& 2\tr N_\mu(\id - \sqrt{3}T_8) N_\nu (\sqrt{3}T_8 + 2\id + 3T_3)\\
	={}& 6\sqrt{3} \tr N_\mu  N_\nu T_8
	+6 \tr N_\mu  N_\nu T_3
	-6\sqrt{3} \tr N_\mu T_8 N_\nu T_3.
\end{split}
\end{equation}
Taking the sum of all three, we find
\be \label{eq:ian2}
	 36\epsilon_{\mu\nu} \sum_{j=1}^3 \tr \partial_\mu U U^\dag \Lambda_j \partial_\nu U U^\dag \Lambda_{j+1} 
	 = -18\sqrt{3} \tr \partial_\mu U U^\dag T_8 \partial_\mu U U^\dag T_3.
\ee

\subsection{Discussion of linear terms} \label{linear}
Both terms in the Lagrangian in \eqref{step2} have pieces linear in the fast fields $\theta_a$. After integration by parts, they are of the form
\be
	\theta_a F_a(\vx)
\ee
where $F_a(\vx)$ is a function of the slow matrices $U_s$, and involves two derivatives. Naively, we may argue that $\int d^2 x \theta_a F_a(\vx)=0$, since $\theta_a$ only contains fast modes, while $F_a$ is made of slow functions. However, since $F_a$ contains \emph{products} of slow modes, it will generically have some fast modes; thus a different argument is required to justify neglecting these terms.  

\emph{Argument 1:} Upon integrating out the fast fields, the contribution to $\fL_{\text{eff}}$ is 
\be \label{deltaL}
	\delta\fL = \frac{1}{2}\int d^2\vx d^2 \vy F_a(\vx)G_{ab}(\vx-\vy)F_b(\vy)
\ee
where $G_{ab}(\vx)$ is the Green's function of the $\theta$ fields. In $\vk$-space, this is 
\be
	\delta \fL(\vk) 
	=\frac{1}{2} \int_{b\Lambda <|\vk| < \Lambda} \frac{ d^2\vk }{(2\pi)^2}\frac{1}{k^2}F_a(\vk)F_a(-\vk )
\ee
because the $\theta$ fields only have momentum modes in the Wilson shell. Since $F(\vk)$ involves no more than four slow fields, $k$ is restricted to $|k| < 4\tilde\Lambda$, where $\tilde\Lambda := b\Lambda$ and $\Lambda$ is a reduced cutoff.  Following \citet{Polyakov:1975rr}, we will take the limit $\tilde\Lambda \to 0$, and argue that $|\delta \fL(\vk) | \to 0$.  Indeed, since $F_a(\vk)$ contains two derivatives of slow modes, 
\be
	|F(\vk) | < \tilde \Lambda^2 |\tilde F(\vk)| 
\ee
for some operator $\tilde F(\vk)$ involving up to four slow fields, without derivatives. Therefore,
\be
	|\delta \fL| <\frac{3}{2} \tilde\Lambda^3 \max \limits_{\tilde \Lambda < |k| < 4\tilde \Lambda} |\tilde F(\vk)||\tilde F(-\vk)|.
\ee
Since $\tilde F(\vk)$ just involves products of slow fields, this maximum should be bounded by some $\tilde\Lambda$ independent constant. $\tilde\Lambda \to 0$, say $\text{max}_{0< |k| < \Lambda}|F(\vk)||\tilde F(-\vk)|$. Therefore $|\delta \fL|$ vanishes as $\tilde\Lambda^3$. This is to be compared with the marginal kinetic term, $(\partial_\mu \theta_a)^2$, which only vanishes as $\tilde\Lambda^2$. This agrees with our naive intuition, that since $F_a(\vx)$ involves two derivatives, $\delta \fL$ should consist of irrelevant operators; however, we are more careful here, since the momenta of $G_{ab}(\vx-\vy)$ in (\ref{deltaL}) is restricted to the Wilson shell.

\emph{Argument 2:} 

In the perturbative Lagrangian in \eqref{step1}, the leading term that arises from $\theta_a(x)F_a(x)$ will correspond to a Feynman diagram with a single internal line. Since the leading $g$-dependent interaction is a four-point vertex, the simplest $\fO(\lambda)$ diagram arising from $\theta_a(x)F_a(x)$ will serve to renormalize a five-point or six-point interaction in \eqref{step1}. Such diagrams can be excluded from a first-order perturbative calculation of $\beta(g)$ and $\beta(\lambda)$, which consider the four-point and three-point interactions, respectively.

\subsection{Calculation of $\beta$ functions} \label{beta}
We begin by studying the two terms of \eqref{step2} separately. 
\subsubsection*{First term of (\ref{step2}):} Start by inserting $U = U_f U_s$ into the first term of (\ref{step2}), and expanding $U_f = \id+ i\theta_a T_a - \frac{1}{2}\theta_a\theta_b T_a T_b +\fO(\theta^3)$. Then 
\be 
	U_f^\dag T_\gamma U_f =
	T_\gamma  - i\theta_a[ T_a, T_\gamma]
	+ \theta_a \theta_b T_a T_\gamma T_b 
	-\frac{1}{2} \theta_a\theta_b\{T_aT_b, T_\gamma\}.
\ee
The quadratic term can be simplified by noting that since $\theta_a\theta_b$ is symmetric in $a$ and $b$, 
\be
	\theta_a\theta_b \left( T_a T_\gamma T_b - \frac{1}{2} \{ T_a T_b, T_\gamma \}\right)
	=\frac{1}{4} \theta_a\theta_b \left( [T_a, [T_\gamma, T_b]] + [T_b, [T_\gamma,T_a]]	 \right).
\ee
Now using (\ref{commute}), and defining 
\be
	h_{ab\gamma D} = \left(  f_{acD} f_{\gamma bc} +  f_{bcD} f_{\gamma ac}\right) \hspace{10mm} M_A := U_s^\dag T_A U_s
\ee
we obtain 
\be \label{key}
	U^\dag T_\gamma U = M_\gamma  +2f_{a\gamma b}\theta_a M_b
	 - \theta_a\theta_b h_{ab\gamma D}M_D.
\ee
Derivatives and traces are now taken, resulting in 

\begin{equation}
\openup 1\jot
\begin{split}
\tr (\partial_\mu[ U^\dag T_\gamma U])^2
	={}& \tr (\partial_\mu M_\gamma)^2 
	+4 f_{a\gamma c}f_{b \gamma d} \partial_\mu \theta_a \partial_\mu \theta_b \tr M_c M_d
	+4 f_{a\gamma c}f_{b\gamma d} \theta_a \theta_b \tr \partial_\mu M_c \partial_\mu M_d\\
	&+ 4f_{a\gamma b} \partial_\mu \theta_a \tr \partial_\mu M_\gamma  M_b
	+4 f_{a\gamma b} \theta_a \tr \partial_\mu M_\gamma \partial_\mu M_b
	+8 f_{a\gamma c}f_{b\gamma d} \partial_\mu \theta_a \theta_b \tr M_c \partial_\mu M_d\\
	&-4\partial_\mu \theta_a \theta_b h_{ab\gamma D} \tr \partial_\mu M_\gamma M_D
	-2 \theta_a\theta_b  h_{ab\gamma D} \tr \partial_\mu M_\gamma \partial_\mu M_D
	+ \fO(\theta^3).
\end{split}
\end{equation}
We can simplify some terms. Using \eqref{norm} 
\be
	\tr M_c M_d = \tr U_s^\dag T_c U_s U_s^\dag T_d U_s = \tr T_c T_d = 2\delta_{cd}.
\ee
Additionally,
\begin{equation}
\label{nice}
\begin{split}
\tr M_A \partial_\mu M_B &= \tr  U_s \partial_\mu U_s^\dag T_B T_A
	+ \tr  \partial_\mu U_s U_s^\dag T_A  T_B\\
	&= \tr \partial_\mu U_s U_s^\dag [T_A,T_B]\\
	&= 2i f_{ABC}\tr \partial_\mu U_s U_s^\dag T_C
\end{split}
\end{equation}
and 
\be
	h_{ab\gamma D} f_{D\gamma C}
	= -f_{DaE} f_{b \gamma E} f_{D\gamma C} -   f_{DbE} f_{a\gamma E}f_{D\gamma C}
	=  0
\ee
imply that the term proportional to $\partial_\mu \theta_a\theta_b  h_{ab\gamma D}$ vanishes.  The result is
\begin{equation}
\openup 1\jot
\begin{split}
(\partial_\mu[ U^\dag T_\gamma U])^2
	= {}&\tr (\partial_\mu M_\gamma)^2 
	+8 (\partial_\mu \theta_a)^2
	+4 f_{a\gamma c}f_{b\gamma d} \theta_a \theta_b \tr \partial_\mu M_c \partial_\mu M_d\\
		&+8i   f_{abe}  \partial_\mu \theta_a \theta_b\tr \partial_\mu U_s U_s^\dag T_e
	+16i f_{ab\rho}  \partial_\mu \theta_a \theta_b\tr \partial_\mu U_s U_s^\dag T_\rho\\
	&-2 \theta_a\theta_b  h_{ab\gamma D} \tr \partial_\mu M_\gamma \partial_\mu M_D
\end{split}
\end{equation}
plus linear terms in $\theta_a$ and higher order corrections. Based on \ref{linear}, these linear terms can be ignored.  Therefore, to quadratic order in $\theta_a$, the first term of (\ref{step2}) can be written as   
\begin{equation}
 \label{long1}
\begin{split}
	\frac{1}{8g} \tr \partial_\mu [U^\dag T_\gamma U] \partial_\mu [U^\dag T_\gamma U] =  \frac{1}{8g} \tr (\partial_\mu M_\gamma)^2
	+\frac{1}{8g} \Bigg[ 8 (\partial_\mu \theta_a)^2
	+4 f_{a\gamma C}f_{b\gamma D} \theta_a \theta_b \tr \partial_\mu M_C \partial_\mu M_D\\
	+8i f_{abe}  \partial_\mu \theta_a \theta_b\tr \partial_\mu U_s U_s^\dag T_e
	+16i f_{ab\rho}  \partial_\mu \theta_a \theta_b\tr \partial_\mu U_s U_s^\dag T_\rho
	-2 \theta_a\theta_b  h_{ab\gamma D} \tr \partial_\mu M_\gamma \partial_\mu M_D\Bigg].
\end{split}
\end{equation}

\subsubsection*{Second term of (\ref{step2}):}
We now want to perform the same expansion of $U_f$ to simplify the second term of (\ref{step2}). Define
\be
	\fL_q:=  \epsilon_{\mu\nu} \tr \partial_\mu U U^\dag T_8  \partial_\nu U U^\dag T_3,
	\hspace{10mm} N_\mu := \partial_\mu U_s U_s^\dag.
\ee
Then we have
 \begin{equation}
\begin{split}
\fL_q
 ={}&\mathbin{\phantom{+}}	\epsilon_{\mu\nu}\tr \partial_\mu U_f U_f^\dag T_8 \partial_\nu U_f U_f^\dag T_3
 +\epsilon_{\mu\nu}\tr \partial_\mu U_f U_f^\dag T_8 U_f  N_\nu U_f^\dag T_3 \\
&+ \epsilon_{\mu\nu}\tr N_\mu U_f^\dag T_8 \partial_\nu U_f  U_f^\dag T_3 U_f
+ \epsilon_{\mu\nu}\tr  N_\mu U_f^\dag T_8 U_f N_\nu  U_f^\dag T_3 U_f.
\end{split}
\end{equation}
 Using (\ref{key}), this is 
\begin{equation}
 \label{eight}
\begin{split}
\fL_q
 ={}&	\epsilon_{\mu\nu}\tr \partial_\mu U_f U_f^\dag T_8 \partial_\nu U_f U_f^\dag T_3
 +\epsilon_{\mu\nu}\tr \partial_\mu U_f \left( T_8  +2 f_{a8 b} \theta_a T_b
 - \theta_a\theta_b h_{ab8 D} T_D\right)  N_\nu U_f^\dag T_3 \\
&+ \epsilon_{\mu\nu}\tr N_\mu U_f^\dag T_8 \partial_\nu U_f \left( T_3 +2 f_{a3 b} \theta_a T_b
- \theta_a\theta_b h_{ab3 D} T_D\right)\\
&+ \epsilon_{\mu\nu}\tr  N_\mu \left( T_8  +2 f_{a8 b} \theta_a T_b
- \theta_a\theta_b h_{ab8 D} T_D\right) N_\nu \left( T_3 +2 f_{a3 b} \theta_a T_b
- \theta_a\theta_b h_{ab3 D} T_D\right).
\end{split}
\end{equation}
 Expanding the remaining $U_f$, and dropping a term proportional to$\partial_\mu\theta_a\partial_\nu\theta_b \epsilon_{\mu\nu}$ which vanishes after Fourier transforming, we find 
\begin{equation}
\label{long2}
\begin{split}
	\lambda \fL_q = \epsilon_{\mu\nu} \lambda \frac{\sqrt{3}}{2}\Bigg[{}& \tr N_\mu T_8 N_\nu T_3 +\partial_\mu \theta_a \theta_b \tr N_\nu T_b \left[  T_3 T_a T_8 -  T_8 T_a T_3\right]\\
	&+ \frac{1}{2}\partial_\mu(\theta_a\theta_b) \tr N_\nu [T_8T_aT_bT_3 - T_3T_aT_bT_8]\\
	&+ \theta_a\theta_b \big(4f_{b3d}f_{a8c}\tr N_\mu T_c N_\nu T_d + \tr N_\nu T_D N_\mu\left[ h_{ab8D} T_3 - h_{ab3D}T_8\right]\big)\\
	&+ 2i \partial_\mu\theta_a\theta_b \tr T_a T_cN_\nu \left[ f_{b8c} T_3   - f_{b3c}T_8\right]
	\Bigg].
\end{split}
\end{equation}

\subsubsection*{Integration:} Using (\ref{long1}) and (\ref{long2}) in (\ref{step2}), we will evaluate the following Gaussian integral:
\be
	\int \fD [\phi] e^{-\int d ^2x(\fL -\fL_0)} =  N e^{-\frac{1}{2}\tr \log \int \fO}
\ee
where $\fO$ is a formal expression involving the Green's functions of $\theta_a$. Following the steps in \ref{nlsmrg}, we expand the $\tr \log$ to second order, since higher-order terms will involve more than two derivatives acting on the $U_s$, and correspond to irrelevant operators. The slow functions appearing in the linear term of this expansion can be replaced by their averages, as in (\ref{eq:averages}). Meanwhile, the quadratic terms of the $\tr\log$ can be approximated using (\ref{eq:zerothorder}). Notice that terms proportional to $\partial_\mu\theta_a\theta_b$ do not contribute to linear order, according to (\ref{eq:rotation}). The result is
\begin{equation}
\begin{split}
	\tr \log &\fO= \\ \phantom{=}&	 \int d^2x\Bigg( \int \frac{d^2k}{(2\pi)^2}\Bigg[ \frac{1}{2k^2} f_{a\gamma C}f_{a\gamma D} \tr \partial_\mu M_C \partial_\mu M_D
	- \frac{1}{4k^2} h_{aa\gamma D} \tr \partial_\mu M_\gamma \partial_\mu M_D\Bigg]
\\
	&- \frac{1}{2} \int \frac{d^2k}{(2\pi)^2} \frac{k_\mu k_\nu  }{k^4}\tr \left( f_{abE}\tr N_\mu T_E + f_{ab\rho} \tr N_\mu T_\rho\right)^2
\\
	&+\sqrt{3} \int \frac{d^2k}{(2\pi)^2}\Big[ \frac{2}{gk^2} \lambda f_{a3d}f_{a8c} \epsilon_{\mu\nu} \tr N_\mu T_c N_\nu T_d
	+ \frac{g}{2k^2} \lambda \epsilon_{\mu\nu} \tr N_\nu T_D N_\mu [ h_{aa8d} T_3 - h_{aa3D}T_8]\Big]
\\
	 &+ \sqrt{3}\lambda g  \int \frac{d^2k}{(2\pi)^2} \frac{k_\mu k_\rho \epsilon_{\rho \nu} }{k^4} \Big[  f_{abE}  \tr N_\mu T_E
	+ f_{ab\gamma }  \tr N_\mu T_\gamma \Big]
\\
	&\times \Big[ + \frac{i}{2} \left[ \tr  T_b T_8 N_\nu  T_a T_3 
 - \tr N_\nu  T_a T_8 T_b T_3\right]
 	- \tr    N_\nu \left[ f_{a8c} T_3T_b  T_c-f_{a3c} T_8 T_b T_c\right]
\\
	&+ \frac{i}{4} \tr N_\nu \left[T_8\{ T_a,T_b\}T_3
	- T_3 \{T_a,T_b\}T_8\right]
	\Big]\Bigg)
	+\text{higher order corrections}.
\end{split}
\end{equation}
Here the integration over $k$ is restricted to the Wilson shell. The \emph{first line} can be simplified using $h_{aa\gamma D} = 6\delta_{\gamma D}$, and  (\ref{partial}). The \emph{second line} can be simplified using
\be
	\tr \left( f_{abE}\tr N_\mu T_E + f_{ab\rho} \tr N_\mu  T_\rho\right)^2
	= -12( \tr N_\mu T_\gamma)^2
	- (\tr N_\mu T_a)^2.
\ee
The \emph{third line} can be simplified too. Since $f_{a8c}f_{a3d}$ vanishes unless $c=d$, the first term is proportional to $\epsilon_{\mu\nu} \tr N_\mu T_c N_\nu T_c = 0$. Using $h_{aa\gamma D} = 6\delta_{\gamma D}$,  the whole third line is
\be
 6\sqrt{3} g\lambda \epsilon_{\mu\nu}   \int \frac{d^2k}{(2\pi)^2}  \frac{1}{k^2}   \tr N_\nu T_8 N_\mu   T_3.
\ee
The \emph{last line} can be dropped, since it is symmetric in $a,b$ and multiplies terms proportional to $f_{ab\gamma}$. Finally,  we use the momentum-shell integrals
\be
	\int_{b\Lambda < k <\Lambda} \frac{d^2k}{(2\pi)^2} \frac{1}{k^2} = \frac{-1}{2\pi} \log b
\hspace{10mm}
	\int_{b\Lambda < k <\Lambda} d^2k \frac{k_\mu k_\nu}{k^4} g(k^2) = \int d^2k \frac{1}{2k^2 }g(k^2)
\ee
for a general function $g(k^2)$.
At last, we arrive at
\begin{equation} \label{almost}
\begin{split}
	 \tr \log \fO 
	\approx{}& \int d^2 x\Bigg[ -\frac{\log b}{4\pi} \Big[  \tr (\partial_\mu M_c )^2
	- 3\tr( \partial_\mu M_\gamma )^2\Big]
	- \frac{\log b}{8\pi} \tr \left(12( \tr N_\mu T_A)^2
	-11 (\tr N_\mu T_a)^2 \right)\\
& - 
3\sqrt{3}(\log b) g\lambda \epsilon_{\mu\nu}   \tr N_\nu T_8 N_\mu   T_3
-	i\sqrt{3}\frac{\epsilon_{\mu\nu} \lambda g \log b }{8\pi }   \Big[  f_{abE}  \tr N_\mu T_E
	+ f_{ab\gamma }  \tr N_\mu T_\gamma \Big]
\\
&	\times \Big[  \left[ \tr   N_\nu  T_a T_3  T_b T_8
 - \tr N_\nu  T_a T_8 T_b T_3\right]
 	+ 2i \tr    N_\nu \left[ f_{a8c} T_3T_b  T_c-f_{a3c} T_8 T_b T_c\right] 
	\Big]\Bigg].
\end{split}
\end{equation}
To proceed, we use a variety of identities, proven in \ref{ids}. The identities (\ref{identity1}) and (\ref{identity2}) let us rewrite the first line as
\be \label{gresult}
	\frac{5 \log b}{16 \pi} \tr (\partial_\mu M_\gamma)^2.
\ee
Now consider the last term:
\begin{equation}
\begin{split}
	A:=	- {}&i\sqrt{3} \frac{\epsilon_{\mu\nu} \lambda g \log b }{8\pi }   \Bigg[  f_{abE}  \tr N_\mu T_E
	+ f_{ab\gamma }  \tr N_\mu T_\gamma \Bigg]\\
	&\times \Bigg[  \left[ \tr   N_\nu  T_a T_3  T_b T_8
 - \tr N_\nu  T_a T_8 T_b T_3\right]
 	+ 2i \tr    N_\nu \left[ f_{a8c} T_3T_b  T_c-f_{a3c} T_8 T_b T_c\right] 
	\Bigg].
\end{split}
\end{equation}
The first term of the second line is simplified using (\ref{lid1}). Then antisymmetry in $a$ and $b$ imply
\begin{equation}
\begin{split}
	A ={}& \sqrt{3} \frac{\epsilon_{\mu\nu}   \lambda g \log b }{8\pi }   \Bigg[  f_{abE}  \tr N_\mu T_E
	+ f_{ab\gamma }  \tr N_\mu T_\gamma \Bigg]\\
	&\times \Bigg[ f_{3bc} \tr  N_\nu [ T_a T_c T_8  - T_8 T_aT_c]
		+ f_{8bc} \tr    N_\nu [T_3T_a  T_c - T_a T_c T_3]\\
	&+ f_{3ac} \tr    N_\nu[ T_8 T_b T_c - T_bT_c T_8]
	+f_{8ac} \tr  N_\nu [T_bT_c T_3 - T_3 T_bT_c]
	\Bigg].
\end{split}
\end{equation}
Now we use (\ref{lid2}) to simplify this to:
\be
	A= 2\sqrt{3} i \frac{\epsilon_{\mu\nu}  \lambda g \log b }{8\pi }   \Bigg[  f_{abE}  \tr N_\mu T_E
	+ f_{ab\gamma }  \tr N_\mu T_\gamma \Bigg]
	\times  \tr N_\nu T_d T_c \Bigg[ 
	f_{3bc}f_{a8d}
		+ f_{8bc} f_{3ad} 
	+ f_{3ac}f_{8bd} 
		+f_{8ac}f_{b3d} 
	\Bigg].
\ee
Recognizing the antisymmetry, this is 
\be
	A = \sqrt{3} 4i \frac{\epsilon_{\mu\nu}  \lambda g \log b }{4\pi }   \tr N_\nu T_d T_c   \Bigg[  f_{abE}  \tr N_\mu T_E
	+ f_{ab\gamma }  \tr N_\mu T_\gamma \Bigg]\Big(
		 f_{8bc} f_{3ad} 
	+ f_{3ac}f_{8bd} \Big).
\ee
Now we use (\ref{lid3}). The result is 
\be
	A = - \sqrt{3} i \frac{\epsilon_{\mu\nu}  \sqrt{3} \lambda g \log b }{2\pi }   \Big[ \tr N_\mu T_1 \tr N_\nu T_2
	- \tr N_\mu T_4 \tr N_\nu T_5 
	+ \tr  N_\mu T_6 \tr N_\nu T_7\Big].
\ee
Now we need to prove that the operator appear here is proportional to $\fL_0^q$. This is (\ref{lid4}):
 \be 
 		\epsilon_{\mu\nu} \Big[ \tr N_\mu T_1 \tr N_\nu T_2
	- \tr N_\mu T_4 \tr N_\nu T_5 
	+ \tr  N_\mu T_6 \tr N_\nu T_7\Big]  =-i\sqrt{3} \tr N_\mu T_8 N_\nu T_3.
\ee
Therefore,
\be \label{aresult}
	A =- \lambda \frac{\sqrt{3}}{2} \frac{3g\log b}{4\pi} \epsilon_{\mu\nu} \tr N_\mu T_8 N_\nu T_3.
\ee

Now we return to the Lagrangian, which requires dividing (\ref{almost}) by 2. Using (\ref{aresult}) and (\ref{gresult}), this is
\be
	\fL = \frac{1}{8g}\left( 1 +g \frac{5 \log b}{16\pi}\right) \tr \partial_\mu M_\gamma \partial_\mu M_\gamma + 
	\frac{\sqrt{3}}{2}\lambda \epsilon_{\mu\nu} \tr N_\mu T_8 N_\nu T_3 \Big[ 1 +2\frac{3g}{2\pi} \log b -2 \frac{ g \log b}{4\pi} 3\Big].
\ee
This allows for the identification of 
\be
	\lambda_{\text{eff}} =  \lambda\left(1 + \frac{3g}{2\pi} \log b\right)
	\hspace{10mm}
	g_{\text{eff}} = g\left(1 + \frac{5g \log b}{4\pi}\right)^{-1}
\ee
from which we can read off the $\beta$ functions.

\subsection{Additional identities} \label{ids}

\subsubsection*{Identity 1}
\be \label{identity1}
	 \sum_A \tr (\partial_\mu M_A)^2  = -6\sum_A  (\tr N_\mu T_A)^2.
\ee
\emph{Proof:} Using (\ref{gencomplete}),
\be
	(M_B)^i_j(M_B)^k_l = (U_s^\dag)^i_n (T_B)^n_m (U_s)^m_j(U_s^\dag)^k_p (T_B)^p_q (U_s)^q_l
\ee
\[
	=  (U_s^\dag)^i_n (U_s)^q_l (U_s)^m_j(U_s^\dag)^k_p\left[ 2\delta^n_q\delta^p_m - \frac{2}{3}\delta^n_m\delta^p_q\right]=2 \delta^i_l \delta^k_j
	-\frac{2}{3}\delta^i_j\delta^k_l.
\]
Also, using (\ref{nice}):
\be
	 \tr M_B \partial_\mu M_A = 2i f_{BAC} \tr \partial_\mu U_s U_s^\dag T_C.
\ee
On the one hand:
\be
		\sum_{A,B} (\tr M_B \partial_\mu M_A)^2 = \sum_{A,B} (M_B)^i_j (\partial_\mu M_A)^j_i (M_B)^k_l (\partial_\mu M_A)^l_k
\ee
\be
	=\sum_{A} (\partial_\mu M_A)^j_i (\partial_\mu M_A)^l_k \left[ 2 \delta^i_l \delta^k_j - \frac{2}{3} \delta^i_j\delta^k_l\right]
	=2 \sum_A \tr (\partial_\mu M_A)^2 - \frac{2}{3} (\tr \partial_\mu M_A)^2 = 2\sum_A \tr (\partial_\mu M_A)^2 
\ee
since $\tr \partial_\mu M_a = \partial_\mu \tr T_a = 0$. On the other hand
\be
	\sum_{A,B} (\tr M_B \partial_\mu M_A)^2 = 
	-4\sum_{A,B,C,D} f_{BAC} f_{BAD} ( \tr N_\mu T_C )  (\tr N_\mu T_D)
	= -12 \sum_C (\tr \partial_\mu U_s U_s^\dag T_C)^2.
\ee
This proves (\ref{identity1}).

\subsubsection*{Identity 2}
\be \label{identity2}
	-2 \sum_c (\tr N_\mu T_c)^2 = \sum_\gamma \tr (\partial_\mu M_\gamma)^2 .
\ee
\emph{Proof:} Follow the proof of (\ref{identity1}), but reduce the summation over $A$ to being a diagonal summation. On one hand,
\begin{equation}
\begin{split}
	\sum_{\gamma, B} (\tr M_B \partial_\mu M_\gamma)^2 &= \sum_{\gamma,B} (M_B)^i_j (\partial_\mu M_\gamma)^j_i (M_B)^k_l (\partial_\mu M_\gamma)^l_k
\\
	&=\sum_{\gamma} (\partial_\mu M_\gamma)^j_i (\partial_\mu M_\gamma)^l_k \left[ 2 \delta^i_l \delta^k_j - \frac{2}{3} \delta^i_j\delta^k_l\right]\\
	&=2 \sum_A \tr (\partial_\mu M_\gamma)^2 - \frac{2}{3} (\tr \partial_\mu M_\gamma)^2 = 2\sum_\gamma \tr (\partial_\mu M_\gamma)^2 .
\end{split}
\end{equation}
On the other hand
\be
	\sum_{\gamma,B} (\tr M_B \partial_\mu M_\gamma)^2 = 
	-4\sum_{\gamma,B,C,D} f_{B\gamma c} f_{B\gamma d} ( \tr N_\mu T_c )  (\tr N_\mu T_d)
	= -4 \sum_c (\tr \partial_\mu U_s U_s^\dag T_c)^2
\ee
where we used (\ref{partial}). This proves (\ref{identity2}).

\subsubsection*{Identity 3}

\be \label{lid1}
	\tr N_\nu T_a \left[ T_3T_b T_8 - T_8T_bT_3\right] = 2i\tr N_\nu T_a \left( f_{3bc} T_c T_8 - f_{8bc}T_cT_3\right).
\ee
\emph{Proof:} 
This follows from:
\be
T_3 T_a T_8 =[T_3,T_a]T_8 + T_a T_3 T_8 = 2if_{3ac}T_cT_8 + \frac{1}{\sqrt{3}}T_aT_3.
\ee

\subsubsection*{Identity 4}
\begin{equation}
 \label{lid2}
\begin{split}
	 &\mathbin{\phantom{+}}f_{3bc} \tr  N_\nu [ T_a T_c T_8  - T_8 T_aT_c]
		+ f_{8bc} \tr    N_\nu [T_3T_a  T_c - T_a T_c T_3]
\\
	&+ f_{3ac} \tr    N_\nu[ T_8 T_b T_c - T_bT_c T_8]
	+f_{8ac} \tr  N_\nu [T_bT_c T_3 - T_3 T_bT_c]
\\
	&= 2i\tr N_\nu T_d T_c \left[ f_{3bc}f_{a8d} + f_{8bc}f_{3ad} + f_{3ac} f_{8bd} + f_{8ac}f_{b3d}\right].
\end{split}
\end{equation}

\emph{Proof:}
Rewrite the LHS as
\be
 f_{3bc} \tr  N_\nu [ T_a T_c, T_8  ]
		+ f_{8bc} \tr    N_\nu [T_3, T_aT_c]
	+ f_{3ac} \tr    N_\nu[ T_8, T_bT_c]
	+f_{8ac} \tr  N_\nu [T_bT_c ,T_3]
\ee
which can be rewritten as 
\begin{equation*}
\begin{split}
& f_{3bc} \tr  N_\nu\left(T_a [T_c,T_8] + [T_a,T_8]T_c\right)
		+ f_{8bc} \tr    N_\nu\left( [T_3,T_a]T_c + T_a[T_3,T_c]\right)\\
	&+ f_{3ac} \tr    N_\nu\left( [T_8,T_b]T_c + T_b[T_8,T_c]\right)
		+f_{8ac} \tr  N_\nu\left( T_b[T_c,T_3] + [T_b,T_3]T_c\right).
\end{split}
\end{equation*}
Replacing commutators with structure constants gives 
\begin{equation*}
\begin{split}
	= 2i 
 \Big[& f_{3bc} \tr  N_\nu\left(T_a f_{c8d}T_d + f_{a8d}T_dT_c\right)
		+ f_{8bc} \tr    N_\nu\left( f_{3ad}T_dT_c + T_af_{3cd}T_d\right)\\
	&+ f_{3ac} \tr    N_\nu\left( f_{8bd}T_dT_c + T_bf_{8cd}T_d \right)
		+f_{8ac} \tr  N_\nu\left( T_bf_{c3d}T_d + f_{b3d}T_dT_c\right)
	\Big]
\end{split}
\end{equation*}
which can be reorganized into
\begin{equation*}
\begin{split}
	=2i     \Big[& f_{3bc}f_{c8d} \tr  N_\nu T_a T_d 
	+f_{3bc}f_{a8d} \tr  N_\nu T_dT_c
		+ f_{8bc} f_{3ad} \tr    N_\nu T_dT_c 
		+ f_{8bc}f_{3cd} \tr    N_\nu  T_aT_d\\
	&+ f_{3ac}f_{8bd} \tr    N_\nu  T_dT_c 
	+ f_{3ac} f_{8cd}\tr    N_\nu  T_bT_d 
		+f_{8ac}f_{c3d} \tr  N_\nu T_bT_d 
		+f_{8ac}f_{b3d} \tr  N_\nu  T_dT_c
	\Big].
\end{split}
\end{equation*}
Now, terms of the form $f_{3bc}f_{c8d}$ vanish unless $b=d$. This means the first and fourth terms cancel. Likewise, the sixth and seventh terms cancel. What remains is
 \begin{equation*}
	=2i     \Big[ 
	f_{3bc}f_{a8d} \tr  N_\nu T_dT_c
		+ f_{8bc} f_{3ad} \tr    N_\nu T_dT_c 
	+ f_{3ac}f_{8bd} \tr    N_\nu  T_dT_c 
		+f_{8ac}f_{b3d} \tr  N_\nu  T_dT_c
	\Big].
\end{equation*}

\subsubsection*{Identity 5}

\begin{equation}
 \label{lid3}
\begin{split}
	&\epsilon_{\mu\nu}	\tr N_\nu T_d T_c   \Big[  f_{abE}  \tr N_\mu T_E
	+ f_{ab\gamma }  \tr N_\mu T_\gamma \Big]\Big(
		 f_{8bc} f_{3ad} 
	+ f_{3ac}f_{8bd} \Big)\\
	&=-\epsilon_{\mu\nu}\sqrt{3} \Big[ \tr N_\mu T_1 \tr N_\nu T_2
	- \tr N_\mu T_4 \tr N_\nu T_5 
	+ \tr  N_\mu T_6 \tr N_\nu T_7\Big]
\end{split}
\end{equation}
\emph{Proof:} We check explicitly that the term
\begin{equation}
	f_{abE}\Big(
		 f_{8bc} f_{3ad} 
	+ f_{3ac}f_{8bd} \Big)
\end{equation}
vanishes unless $E$ is diagonal. In fact, it is either zero or $\pm \frac{\sqrt{3}}{4}$. We find the LHS of (\ref{lid3}) equals
\begin{equation}
\begin{split}
	=\frac{\sqrt{3}}{4}\epsilon_{\mu\nu} \Big[ &\tr N_\mu T_1  \tr N_\nu(\{T_4, T_7\} -  \{T_5,T_6\})
	+  \tr N_\mu T_2  \tr N_\nu( \{T_4,T_6\}  + \{T_5,T_7\})
\\	&+  \tr N_\mu T_4 \tr N_\nu ( \{T_1,T_7\} + \{T_2,T_6\})
	+ \tr N_\mu T_5 \tr N_\nu (\{T_2,T_7\} - \{T_1,T_6\})
\\
	&+\tr N_\mu T_6 \tr N_\nu ( \{T_2,T_4\} - \{T_1,T_5\})
	+\tr N_\mu T_7 \tr N_\nu (\{T_1,T_4\} + \{T_5,T_2\})\Big].
\end{split}
\end{equation}
Using
\begin{equation}
\begin{split}
	\{T_5, T_6\} &= -\{T_4,T_7\} =  T_2, \hspace{10mm}
	\{T_4, T_6\} = \{T_5,T_7\}=  T_1, \\
	\{T_1,T_7\} &=\{T_2,T_6\} = T_5, \hspace{12.2mm} 
	\{T_1,T_6 \} = -\{T_2,T_7\} = T_4, \\
	\{T_1,T_5\} &=- \{T_2,T_4\} = T_7, \hspace{10mm}
	\{T_1,T_4 \} = \{T_5,T_2\}= T_6, 
\end{split}
\end{equation}
the  LHS of \eqref{lid3} becomes
\begin{equation}
\begin{split}
	=\frac{\sqrt{3}}{2}\epsilon_{\mu\nu} \Bigg[&- \tr N_\mu T_1  \tr N_\nu T_2	+  \tr N_\mu T_2  \tr N_\nu T_1
	+ \tr N_\mu T_4 \tr N_\nu T_5
	- \tr N_\mu T_5 \tr N_\nu T_4
\\
	&-\tr N_\mu T_6 \tr N_\nu T_7
	+\tr N_\mu T_7 \tr N_\nu T_6\Bigg].
\end{split}
\end{equation}
The $\epsilon_{\mu\nu}$ tensor allows us to combine these terms, proving (\ref{lid3}).
 
\subsubsection*{Identity 6}
 
 \be \label{lid4}
 		\epsilon_{\mu\nu} \Big[ \tr N_\mu T_1 \tr N_\nu T_2
	- \tr N_\mu T_4 \tr N_\nu T_5 
	+ \tr  N_\mu T_6 \tr N_\nu T_7\Big]  =-i\sqrt{3} \tr N_\mu T_8 N_\nu T_3.
\ee

\emph{Proof:} First note that
\begin{equation}
\begin{split}
\tr N_\mu T_1 &= [N_\mu]_{ij}[T_1]_{ji} = [N_\mu]_{21} + [N_\mu]_{12},\\
	\tr N_\nu T_2 &= [N_\nu]_{ij}[T_2]_{ji} = -i[N_\nu]_{21} + i[N_\nu]_{12},
\end{split}
\end{equation}
so

\begin{equation}
\begin{split}
\epsilon_{\mu\nu}  \tr N_\mu T_1 \tr N_\nu T_2
	&= i\epsilon_{\mu\nu}\left[ [N_\mu]_{21} + [N_\mu]_{12}\right]\left[ -[N_\nu]_{21} + [N_\nu]_{12}\right]
\\
	&=i\epsilon_{\mu\nu}\left( [N_\mu]_{21}[N_\nu]_{12} - [N_\mu]_{21}[N_\nu]_{21} + [N_\mu]_{12}[N_\nu]_{12} - [N_\mu]_{12}[N_\nu]_{21}\right)
\\	&=2i\epsilon_{\mu\nu} [N_\mu]_{21}[N_\nu]_{12}.
\end{split}
\end{equation}	
Similar results hold for $T_4,T_5,T_6,T_7$. Therefore the LHS of (\ref{lid4}) is 
\be
	2i \epsilon_{\mu\nu}\Big[  [N_\mu]_{21}[N_\nu]_{12}
	- [N_\mu]_{31}[N_\nu]_{13}
	+ [N_\mu]_{32}[N_\nu]_{23}\Big]
=
	2i\epsilon_{\mu\nu} \sum_{j=1}^3 \tr N_\mu \Lambda_j N_\nu \Lambda_{j+1}.
\ee
Now, the proof of (\ref{ian}) with $U$ replaced by $U_s$, shows that this equals
\be
		-i\sqrt{3} \epsilon_{\mu\nu} \tr N_\mu T_8 N_\nu T_3.
\ee

\section{Factorization of SU(3) matrices}
\label{sec:factorization}

Since the Lagrangian (\ref{eq:fullaction}) is invariant under (\ref{eq:gauge}), the factorization
\be \label{factorization}
	U = e^{i\theta_\gamma T_\gamma } e^{i\theta_a T_a} \hspace{5mm} U \in \mbox{SU}(3)
\ee
would allow for $\fL$ to be written purely in terms of the $\theta_a$. Though we believe (\ref{factorization}) is true in general, our calculations only require a factorization to hold to cubic order in $\theta_a$, and this is what we prove here. We have

\begin{equation}
\begin{split}
 e^{i\theta_AT_A}-e^{i\theta_\alpha T_\alpha}e^{i\theta_aT_a}&=-(1/2)\theta_A\theta_BT_AT_B+(1/2)\theta_\alpha \theta_\beta T_\alpha T_\beta +(1/2)\theta_a\theta_bT_aT_b
+\theta_\alpha \theta_aT_\alpha T_a +O(\theta^3) \\
&=(1/2)\theta_a\theta_\alpha [T_\alpha ,T_a]+O(\theta^3)=if_{\alpha ab}\theta_a\theta_\alpha T_b+O(\theta^3).\label{e}
\end{split}
\end{equation}

To correct this $O(\theta^2)$ error, add a correction to $\theta_a$ in $e^{i\theta_aT_a}$ of $O(\theta^2)$called $\delta \theta_a$. Then:
\be e^{i\theta_AT_A}-e^{i\theta_\alpha T_\alpha}e^{i(\theta_a+\delta\theta_a)T_a}=if_{\alpha ab}\theta_a\theta_\alpha T_b-i\delta \theta_bT_b +O(\theta^3).\label{dt}\ee
By choosing
\be \delta \theta_b=f_{\alpha ab}\theta_a\theta_\alpha\ee
we make the identity true to second order.

Now, let's make it true to 3rd order. Let's calculate the $O(\theta^3)$ term in Eq.\ (\ref{dt}). The 3rd order term in Eq.\ (\ref{e}) is:
\begin{equation}
\begin{split}
&{i\over 6}[-\theta_A\theta_B\theta_CT_AT_BT_C+\theta_\alpha\theta_\beta\theta_\gamma T_\alpha T_\beta T_\gamma +\theta_a\theta_b\theta_cT_aT_bT_c
+3\theta_\alpha\theta_\beta \theta_aT_\alpha T_\beta T_a+3\theta_\alpha \theta_a\theta_bT_\alpha T_aT_b] \\
& ={i\over 6}[\theta_\alpha \theta_\beta \theta_a(2T_\alpha T_\beta T_a-T_\alpha T_aT_\beta -T_aT_\alpha T_\beta )+\theta_\alpha \theta_a\theta_b
(2T_\alpha T_aT_b-T_aT_\alpha T_b-T_aT_bT_\alpha )]\\
&={i\over 6}\{\theta_\alpha \theta_\beta \theta_a([2T_\alpha [T_\beta ,T_a]+[T_\alpha ,T_a]T_\beta )+\theta_\alpha \theta_a\theta_b(2[T_\alpha T_a]T_b+T_a[T_\alpha ,T_b ])\} .
\end{split}
\end{equation}
This can be written:
\be {i\over 3}[\theta_\alpha \theta_\beta \theta_a(2if_{\beta a c}T_\alpha T_c+if_{\alpha ac}T_cT_\beta )+\theta_\alpha \theta_a\theta_b
(2if_{\alpha ac}T_cT_b+if_{\alpha bc}T_aT_c)]
\ee
The $\delta \theta_a\theta_B$ cross term in  Eq.\ (\ref{dt}) is:
\be \theta_\alpha \delta \theta_aT_\alpha T_a+(1/2)\theta_a\delta \theta_b\{T_a,T_b\}=\theta_\alpha f_{\beta ca}\theta_\beta \theta_cT_\alpha T_a
+(1/2)\theta_af_{\beta cb}\theta_\beta \theta_c\{T_a,T_b\}
\ee
The sum of of $O(\theta^3)$ terms in Eq.\ (2) is:
\be (1/3)\theta_\alpha\theta_\beta \theta_af_{\beta ac}[T_\alpha ,T_c]+(1/6)\theta_\alpha \theta_a\theta_bf_{\alpha ac}[T_b,T_c]
=[(2/3)f_{\beta ac}f_{\alpha cD}\theta_\alpha \theta_\beta \theta_a+(1/6)f_{\alpha ac}f_{bcD}\theta_\alpha \theta_a\theta_b]T_D.
\ee
(In the first term, $D$ can be restricted to $d$, off-diagonal terms.) The important thing about this result is that all terms are proportional to $T_D$. Therefore, we can add a cubic correction to Eq.\ (\ref{dt}) to made the factorization work:
\be e^{i\theta_AT_A}=e^{i\phi_\alpha T_\alpha}e^{i\phi_aT_a}\ee
where 
\begin{equation}
\begin{split}
\phi_\alpha &=\theta_\alpha +(1/6)f_{\beta ac}f_{bc\alpha}\theta_\beta \theta_a\theta_b+O(\theta^4)
 \\
\phi_a&=\theta_a+f_{\alpha ba}\theta_b\theta_\alpha +[(2/3)f_{\beta dc}f_{\alpha ca}\theta_\alpha \theta_\beta \theta_d+(1/6)f_{\alpha dc}f_{bca}\theta_\alpha \theta_d\theta_b]+O(\theta^4)
\end{split}
\end{equation}

\bibliography{su3qft}

\begin{thebibliography}{120}%
\makeatletter
\providecommand \@ifxundefined [1]{%
 \@ifx{#1\undefined}
}%
\providecommand \@ifnum [1]{%
 \ifnum #1\expandafter \@firstoftwo
 \else \expandafter \@secondoftwo
 \fi
}%
\providecommand \@ifx [1]{%
 \ifx #1\expandafter \@firstoftwo
 \else \expandafter \@secondoftwo
 \fi
}%
\providecommand \natexlab [1]{#1}%
\providecommand \enquote  [1]{``#1''}%
\providecommand \bibnamefont  [1]{#1}%
\providecommand \bibfnamefont [1]{#1}%
\providecommand \citenamefont [1]{#1}%
\providecommand \href@noop [0]{\@secondoftwo}%
\providecommand \href [0]{\begingroup \@sanitize@url \@href}%
\providecommand \@href[1]{\@@startlink{#1}\@@href}%
\providecommand \@@href[1]{\endgroup#1\@@endlink}%
\providecommand \@sanitize@url [0]{\catcode `\\12\catcode `\$12\catcode
  `\&12\catcode `\#12\catcode `\^12\catcode `\_12\catcode `\%12\relax}%
\providecommand \@@startlink[1]{}%
\providecommand \@@endlink[0]{}%
\providecommand \url  [0]{\begingroup\@sanitize@url \@url }%
\providecommand \@url [1]{\endgroup\@href {#1}{\urlprefix }}%
\providecommand \urlprefix  [0]{URL }%
\providecommand \Eprint [0]{\href }%
\providecommand \doibase [0]{http://dx.doi.org/}%
\providecommand \selectlanguage [0]{\@gobble}%
\providecommand \bibinfo  [0]{\@secondoftwo}%
\providecommand \bibfield  [0]{\@secondoftwo}%
\providecommand \translation [1]{[#1]}%
\providecommand \BibitemOpen [0]{}%
\providecommand \bibitemStop [0]{}%
\providecommand \bibitemNoStop [0]{.\EOS\space}%
\providecommand \EOS [0]{\spacefactor3000\relax}%
\providecommand \BibitemShut  [1]{\csname bibitem#1\endcsname}%
\let\auto@bib@innerbib\@empty
\bibitem [{\citenamefont {Haldane}(1983{\natexlab{a}})}]{HaldanePRL1983}%
  \BibitemOpen
  \bibfield  {author} {\bibinfo {author} {\bibfnamefont {F.~D.~M.}\
  \bibnamefont {Haldane}},\ }\href {\doibase 10.1103/PhysRevLett.50.1153}
  {\bibfield  {journal} {\bibinfo  {journal} {Phys. Rev. Lett.}\ }\textbf
  {\bibinfo {volume} {50}},\ \bibinfo {pages} {1153} (\bibinfo {year}
  {1983}{\natexlab{a}})}\BibitemShut {NoStop}%
\bibitem [{\citenamefont {Haldane}(1983{\natexlab{b}})}]{HaldanePLA1983}%
  \BibitemOpen
  \bibfield  {author} {\bibinfo {author} {\bibfnamefont {F.}~\bibnamefont
  {Haldane}},\ }\href {\doibase
  https://dx.doi.org/10.1016/0375-9601(83)90631-X} {\bibfield  {journal}
  {\bibinfo  {journal} {Physics Letters A}\ }\textbf {\bibinfo {volume} {93}},\
  \bibinfo {pages} {464 } (\bibinfo {year} {1983}{\natexlab{b}})}\BibitemShut
  {NoStop}%
\bibitem [{\citenamefont {Mermin}\ and\ \citenamefont
  {Wagner}(1966)}]{MerminWagner1966}%
  \BibitemOpen
  \bibfield  {author} {\bibinfo {author} {\bibfnamefont {N.~D.}\ \bibnamefont
  {Mermin}}\ and\ \bibinfo {author} {\bibfnamefont {H.}~\bibnamefont
  {Wagner}},\ }\href {\doibase 10.1103/PhysRevLett.17.1133} {\bibfield
  {journal} {\bibinfo  {journal} {Phys. Rev. Lett.}\ }\textbf {\bibinfo
  {volume} {17}},\ \bibinfo {pages} {1133} (\bibinfo {year}
  {1966})}\BibitemShut {NoStop}%
\bibitem [{\citenamefont {Coleman}(1973)}]{Coleman1973}%
  \BibitemOpen
  \bibfield  {author} {\bibinfo {author} {\bibfnamefont {S.}~\bibnamefont
  {Coleman}},\ }\href {\doibase 10.1007/BF01646487} {\bibfield  {journal}
  {\bibinfo  {journal} {Communications in Mathematical Physics}\ }\textbf
  {\bibinfo {volume} {31}},\ \bibinfo {pages} {259} (\bibinfo {year}
  {1973})}\BibitemShut {NoStop}%
\bibitem [{\citenamefont {Elitzur}(1983)}]{Elitzur1983}%
  \BibitemOpen
  \bibfield  {author} {\bibinfo {author} {\bibfnamefont {S.}~\bibnamefont
  {Elitzur}},\ }\href {\doibase http://dx.doi.org/10.1016/0550-3213(83)90682-X}
  {\bibfield  {journal} {\bibinfo  {journal} {Nuclear Physics B}\ }\textbf
  {\bibinfo {volume} {212}},\ \bibinfo {pages} {501 } (\bibinfo {year}
  {1983})}\BibitemShut {NoStop}%
\bibitem [{\citenamefont {Zamolodchikov}\ and\ \citenamefont
  {Zamolodchikov}(1979)}]{Zamolodchikov1979}%
  \BibitemOpen
  \bibfield  {author} {\bibinfo {author} {\bibfnamefont {A.~B.}\ \bibnamefont
  {Zamolodchikov}}\ and\ \bibinfo {author} {\bibfnamefont {A.~B.}\ \bibnamefont
  {Zamolodchikov}},\ }\href {\doibase
  http://dx.doi.org/10.1016/0003-4916(79)90391-9} {\bibfield  {journal}
  {\bibinfo  {journal} {Annals of Physics}\ }\textbf {\bibinfo {volume}
  {120}},\ \bibinfo {pages} {253 } (\bibinfo {year} {1979})}\BibitemShut
  {NoStop}%
\bibitem [{\citenamefont {Bhanot}\ \emph
  {et~al.}(1984{\natexlab{a}})\citenamefont {Bhanot}, \citenamefont
  {Rabinovici}, \citenamefont {Seiberg},\ and\ \citenamefont
  {Woit}}]{BhanotNuclPhys1984}%
  \BibitemOpen
  \bibfield  {author} {\bibinfo {author} {\bibfnamefont {G.}~\bibnamefont
  {Bhanot}}, \bibinfo {author} {\bibfnamefont {E.}~\bibnamefont {Rabinovici}},
  \bibinfo {author} {\bibfnamefont {N.}~\bibnamefont {Seiberg}}, \ and\
  \bibinfo {author} {\bibfnamefont {P.}~\bibnamefont {Woit}},\ }\href {\doibase
  http://dx.doi.org/10.1016/0550-3213(84)90214-1} {\bibfield  {journal}
  {\bibinfo  {journal} {Nuclear Physics B}\ }\textbf {\bibinfo {volume}
  {230}},\ \bibinfo {pages} {291 } (\bibinfo {year}
  {1984}{\natexlab{a}})}\BibitemShut {NoStop}%
\bibitem [{\citenamefont {Bhanot}\ \emph
  {et~al.}(1984{\natexlab{b}})\citenamefont {Bhanot}, \citenamefont {Dashen},
  \citenamefont {Seiberg},\ and\ \citenamefont {Levine}}]{BhanotPRL1984}%
  \BibitemOpen
  \bibfield  {author} {\bibinfo {author} {\bibfnamefont {G.}~\bibnamefont
  {Bhanot}}, \bibinfo {author} {\bibfnamefont {R.}~\bibnamefont {Dashen}},
  \bibinfo {author} {\bibfnamefont {N.}~\bibnamefont {Seiberg}}, \ and\
  \bibinfo {author} {\bibfnamefont {H.}~\bibnamefont {Levine}},\ }\href
  {\doibase 10.1103/PhysRevLett.53.519} {\bibfield  {journal} {\bibinfo
  {journal} {Phys. Rev. Lett.}\ }\textbf {\bibinfo {volume} {53}},\ \bibinfo
  {pages} {519} (\bibinfo {year} {1984}{\natexlab{b}})}\BibitemShut {NoStop}%
\bibitem [{\citenamefont {Renard}\ \emph {et~al.}(2003)\citenamefont {Renard},
  \citenamefont {Regnault},\ and\ \citenamefont
  {Verdaguer}}]{RenardRegnault2003review}%
  \BibitemOpen
  \bibfield  {author} {\bibinfo {author} {\bibfnamefont {J.-P.}\ \bibnamefont
  {Renard}}, \bibinfo {author} {\bibfnamefont {L.-P.}\ \bibnamefont
  {Regnault}}, \ and\ \bibinfo {author} {\bibfnamefont {M.}~\bibnamefont
  {Verdaguer}},\ }\enquote {\bibinfo {title} {Haldane quantum spin chains},}\
  in\ \href {\doibase 10.1002/9783527620548.ch2} {\emph {\bibinfo {booktitle}
  {Magnetism: Molecules to Materials}}}\ (\bibinfo  {publisher} {Wiley-VCH
  Verlag GmbH \& Co. KGaA},\ \bibinfo {year} {2003})\ pp.\ \bibinfo {pages}
  {49--93},\ \bibinfo {note} {and references therein}\BibitemShut {NoStop}%
\bibitem [{\citenamefont {Botet}\ \emph {et~al.}(1983)\citenamefont {Botet},
  \citenamefont {Jullien},\ and\ \citenamefont {Kolb}}]{BotetPRB1983}%
  \BibitemOpen
  \bibfield  {author} {\bibinfo {author} {\bibfnamefont {R.}~\bibnamefont
  {Botet}}, \bibinfo {author} {\bibfnamefont {R.}~\bibnamefont {Jullien}}, \
  and\ \bibinfo {author} {\bibfnamefont {M.}~\bibnamefont {Kolb}},\ }\href
  {\doibase 10.1103/PhysRevB.28.3914} {\bibfield  {journal} {\bibinfo
  {journal} {Phys. Rev. B}\ }\textbf {\bibinfo {volume} {28}},\ \bibinfo
  {pages} {3914} (\bibinfo {year} {1983})}\BibitemShut {NoStop}%
\bibitem [{\citenamefont {Nightingale}\ and\ \citenamefont
  {Bl\"ote}(1986)}]{NightingalePRB1986}%
  \BibitemOpen
  \bibfield  {author} {\bibinfo {author} {\bibfnamefont {M.~P.}\ \bibnamefont
  {Nightingale}}\ and\ \bibinfo {author} {\bibfnamefont {H.~W.~J.}\
  \bibnamefont {Bl\"ote}},\ }\href {\doibase 10.1103/PhysRevB.33.659}
  {\bibfield  {journal} {\bibinfo  {journal} {Phys. Rev. B}\ }\textbf {\bibinfo
  {volume} {33}},\ \bibinfo {pages} {659} (\bibinfo {year} {1986})}\BibitemShut
  {NoStop}%
\bibitem [{\citenamefont {Kennedy}(1990)}]{KennedyJPhys1990}%
  \BibitemOpen
  \bibfield  {author} {\bibinfo {author} {\bibfnamefont {T.}~\bibnamefont
  {Kennedy}},\ }\href {http://stacks.iop.org/0953-8984/2/i=26/a=010} {\bibfield
   {journal} {\bibinfo  {journal} {Journal of Physics: Condensed Matter}\
  }\textbf {\bibinfo {volume} {2}},\ \bibinfo {pages} {5737} (\bibinfo {year}
  {1990})}\BibitemShut {NoStop}%
\bibitem [{\citenamefont {White}\ and\ \citenamefont
  {Huse}(1993)}]{WhitePRB1993}%
  \BibitemOpen
  \bibfield  {author} {\bibinfo {author} {\bibfnamefont {S.~R.}\ \bibnamefont
  {White}}\ and\ \bibinfo {author} {\bibfnamefont {D.~A.}\ \bibnamefont
  {Huse}},\ }\href {\doibase 10.1103/PhysRevB.48.3844} {\bibfield  {journal}
  {\bibinfo  {journal} {Phys. Rev. B}\ }\textbf {\bibinfo {volume} {48}},\
  \bibinfo {pages} {3844} (\bibinfo {year} {1993})}\BibitemShut {NoStop}%
\bibitem [{\citenamefont {Schollw\"ock}\ \emph {et~al.}(1996)\citenamefont
  {Schollw\"ock}, \citenamefont {Golinelli},\ and\ \citenamefont
  {Jolic\oe{}ur}}]{Schollwockspin2PRB1996}%
  \BibitemOpen
  \bibfield  {author} {\bibinfo {author} {\bibfnamefont {U.}~\bibnamefont
  {Schollw\"ock}}, \bibinfo {author} {\bibfnamefont {O.}~\bibnamefont
  {Golinelli}}, \ and\ \bibinfo {author} {\bibfnamefont {T.}~\bibnamefont
  {Jolic\oe{}ur}},\ }\href {\doibase 10.1103/PhysRevB.54.4038} {\bibfield
  {journal} {\bibinfo  {journal} {Phys. Rev. B}\ }\textbf {\bibinfo {volume}
  {54}},\ \bibinfo {pages} {4038} (\bibinfo {year} {1996})}\BibitemShut
  {NoStop}%
\bibitem [{\citenamefont {Todo}\ and\ \citenamefont
  {Kato}(2001)}]{todoPRL2001}%
  \BibitemOpen
  \bibfield  {author} {\bibinfo {author} {\bibfnamefont {S.}~\bibnamefont
  {Todo}}\ and\ \bibinfo {author} {\bibfnamefont {K.}~\bibnamefont {Kato}},\
  }\href {\doibase 10.1103/PhysRevLett.87.047203} {\bibfield  {journal}
  {\bibinfo  {journal} {Phys. Rev. Lett.}\ }\textbf {\bibinfo {volume} {87}},\
  \bibinfo {pages} {047203} (\bibinfo {year} {2001})}\BibitemShut {NoStop}%
\bibitem [{\citenamefont {Bietenholz}\ \emph {et~al.}(1995)\citenamefont
  {Bietenholz}, \citenamefont {Pochinsky},\ and\ \citenamefont
  {Wiese}}]{WieseMC1995}%
  \BibitemOpen
  \bibfield  {author} {\bibinfo {author} {\bibfnamefont {W.}~\bibnamefont
  {Bietenholz}}, \bibinfo {author} {\bibfnamefont {A.}~\bibnamefont
  {Pochinsky}}, \ and\ \bibinfo {author} {\bibfnamefont {U.~J.}\ \bibnamefont
  {Wiese}},\ }\href {\doibase 10.1103/PhysRevLett.75.4524} {\bibfield
  {journal} {\bibinfo  {journal} {Phys. Rev. Lett.}\ }\textbf {\bibinfo
  {volume} {75}},\ \bibinfo {pages} {4524} (\bibinfo {year}
  {1995})}\BibitemShut {NoStop}%
\bibitem [{\citenamefont {Azcoiti}\ \emph {et~al.}(2003)\citenamefont
  {Azcoiti}, \citenamefont {Carlo}, \citenamefont {Galante},\ and\
  \citenamefont {Laliena}}]{Azcoiti2003}%
  \BibitemOpen
  \bibfield  {author} {\bibinfo {author} {\bibfnamefont {V.}~\bibnamefont
  {Azcoiti}}, \bibinfo {author} {\bibfnamefont {G.~D.}\ \bibnamefont {Carlo}},
  \bibinfo {author} {\bibfnamefont {A.}~\bibnamefont {Galante}}, \ and\
  \bibinfo {author} {\bibfnamefont {V.}~\bibnamefont {Laliena}},\ }\href
  {\doibase http://dx.doi.org/10.1016/S0370-2693(03)00601-4} {\bibfield
  {journal} {\bibinfo  {journal} {Physics Letters B}\ }\textbf {\bibinfo
  {volume} {563}},\ \bibinfo {pages} {117 } (\bibinfo {year}
  {2003})}\BibitemShut {NoStop}%
\bibitem [{\citenamefont {All\'es}\ and\ \citenamefont
  {Papa}(2008)}]{AllesPapa2008}%
  \BibitemOpen
  \bibfield  {author} {\bibinfo {author} {\bibfnamefont {B.}~\bibnamefont
  {All\'es}}\ and\ \bibinfo {author} {\bibfnamefont {A.}~\bibnamefont {Papa}},\
  }\href {\doibase 10.1103/PhysRevD.77.056008} {\bibfield  {journal} {\bibinfo
  {journal} {Phys. Rev. D}\ }\textbf {\bibinfo {volume} {77}},\ \bibinfo
  {pages} {056008} (\bibinfo {year} {2008})}\BibitemShut {NoStop}%
\bibitem [{\citenamefont {Azcoiti}\ \emph {et~al.}(2012)\citenamefont
  {Azcoiti}, \citenamefont {Di~Carlo}, \citenamefont {Follana},\ and\
  \citenamefont {Giordano}}]{Azcoiti2012}%
  \BibitemOpen
  \bibfield  {author} {\bibinfo {author} {\bibfnamefont {V.}~\bibnamefont
  {Azcoiti}}, \bibinfo {author} {\bibfnamefont {G.}~\bibnamefont {Di~Carlo}},
  \bibinfo {author} {\bibfnamefont {E.}~\bibnamefont {Follana}}, \ and\
  \bibinfo {author} {\bibfnamefont {M.}~\bibnamefont {Giordano}},\ }\href
  {\doibase 10.1103/PhysRevD.86.096009} {\bibfield  {journal} {\bibinfo
  {journal} {Phys. Rev. D}\ }\textbf {\bibinfo {volume} {86}},\ \bibinfo
  {pages} {096009} (\bibinfo {year} {2012})}\BibitemShut {NoStop}%
\bibitem [{\citenamefont {de~Forcrand}\ \emph {et~al.}(2012)\citenamefont
  {de~Forcrand}, \citenamefont {Pepe},\ and\ \citenamefont
  {Wiese}}]{WiesePRD2012}%
  \BibitemOpen
  \bibfield  {author} {\bibinfo {author} {\bibfnamefont {P.}~\bibnamefont
  {de~Forcrand}}, \bibinfo {author} {\bibfnamefont {M.}~\bibnamefont {Pepe}}, \
  and\ \bibinfo {author} {\bibfnamefont {U.~J.}\ \bibnamefont {Wiese}},\ }\href
  {\doibase 10.1103/PhysRevD.86.075006} {\bibfield  {journal} {\bibinfo
  {journal} {Phys. Rev. D}\ }\textbf {\bibinfo {volume} {86}},\ \bibinfo
  {pages} {075006} (\bibinfo {year} {2012})}\BibitemShut {NoStop}%
\bibitem [{\citenamefont {All\'es}\ \emph {et~al.}(2014)\citenamefont
  {All\'es}, \citenamefont {Giordano},\ and\ \citenamefont
  {Papa}}]{AllesPapa2014}%
  \BibitemOpen
  \bibfield  {author} {\bibinfo {author} {\bibfnamefont {B.}~\bibnamefont
  {All\'es}}, \bibinfo {author} {\bibfnamefont {M.}~\bibnamefont {Giordano}}, \
  and\ \bibinfo {author} {\bibfnamefont {A.}~\bibnamefont {Papa}},\ }\href
  {\doibase 10.1103/PhysRevB.90.184421} {\bibfield  {journal} {\bibinfo
  {journal} {Phys. Rev. B}\ }\textbf {\bibinfo {volume} {90}},\ \bibinfo
  {pages} {184421} (\bibinfo {year} {2014})}\BibitemShut {NoStop}%
\bibitem [{\citenamefont {Zamolodchikov}\ and\ \citenamefont
  {Zamolodchikov}(1992)}]{Zamolodchikov1992}%
  \BibitemOpen
  \bibfield  {author} {\bibinfo {author} {\bibfnamefont {A.}~\bibnamefont
  {Zamolodchikov}}\ and\ \bibinfo {author} {\bibfnamefont {A.}~\bibnamefont
  {Zamolodchikov}},\ }\href {\doibase
  http://dx.doi.org/10.1016/0550-3213(92)90136-Y} {\bibfield  {journal}
  {\bibinfo  {journal} {Nuclear Physics B}\ }\textbf {\bibinfo {volume}
  {379}},\ \bibinfo {pages} {602 } (\bibinfo {year} {1992})}\BibitemShut
  {NoStop}%
\bibitem [{\citenamefont {Wu}\ \emph {et~al.}(2003)\citenamefont {Wu},
  \citenamefont {Hu},\ and\ \citenamefont {Zhang}}]{WuPRL2003}%
  \BibitemOpen
  \bibfield  {author} {\bibinfo {author} {\bibfnamefont {C.}~\bibnamefont
  {Wu}}, \bibinfo {author} {\bibfnamefont {J.-p.}\ \bibnamefont {Hu}}, \ and\
  \bibinfo {author} {\bibfnamefont {S.-c.}\ \bibnamefont {Zhang}},\ }\href
  {\doibase 10.1103/PhysRevLett.91.186402} {\bibfield  {journal} {\bibinfo
  {journal} {Phys. Rev. Lett.}\ }\textbf {\bibinfo {volume} {91}},\ \bibinfo
  {pages} {186402} (\bibinfo {year} {2003})}\BibitemShut {NoStop}%
\bibitem [{\citenamefont {Honerkamp}\ and\ \citenamefont
  {Hofstetter}(2004)}]{HonerkampHofstetter2004}%
  \BibitemOpen
  \bibfield  {author} {\bibinfo {author} {\bibfnamefont {C.}~\bibnamefont
  {Honerkamp}}\ and\ \bibinfo {author} {\bibfnamefont {W.}~\bibnamefont
  {Hofstetter}},\ }\href {\doibase 10.1103/PhysRevLett.92.170403} {\bibfield
  {journal} {\bibinfo  {journal} {Phys. Rev. Lett.}\ }\textbf {\bibinfo
  {volume} {92}},\ \bibinfo {pages} {170403} (\bibinfo {year}
  {2004})}\BibitemShut {NoStop}%
\bibitem [{\citenamefont {Cazalilla}\ \emph {et~al.}(2009)\citenamefont
  {Cazalilla}, \citenamefont {Ho},\ and\ \citenamefont {Ueda}}]{Cazalilla2009}%
  \BibitemOpen
  \bibfield  {author} {\bibinfo {author} {\bibfnamefont {M.~A.}\ \bibnamefont
  {Cazalilla}}, \bibinfo {author} {\bibfnamefont {A.~F.}\ \bibnamefont {Ho}}, \
  and\ \bibinfo {author} {\bibfnamefont {M.}~\bibnamefont {Ueda}},\ }\href
  {http://stacks.iop.org/1367-2630/11/i=10/a=103033} {\bibfield  {journal}
  {\bibinfo  {journal} {New Journal of Physics}\ }\textbf {\bibinfo {volume}
  {11}},\ \bibinfo {pages} {103033} (\bibinfo {year} {2009})}\BibitemShut
  {NoStop}%
\bibitem [{\citenamefont {Gorshkov}\ \emph {et~al.}(2010)\citenamefont
  {Gorshkov}, \citenamefont {Hermele}, \citenamefont {Gurarie}, \citenamefont
  {Xu}, \citenamefont {Julienne}, \citenamefont {Ye}, \citenamefont {Zoller},
  \citenamefont {Demler}, \citenamefont {Lukin},\ and\ \citenamefont
  {Rey}}]{gorshkov2010}%
  \BibitemOpen
  \bibfield  {author} {\bibinfo {author} {\bibfnamefont {A.~V.}\ \bibnamefont
  {Gorshkov}}, \bibinfo {author} {\bibfnamefont {M.}~\bibnamefont {Hermele}},
  \bibinfo {author} {\bibfnamefont {V.}~\bibnamefont {Gurarie}}, \bibinfo
  {author} {\bibfnamefont {C.}~\bibnamefont {Xu}}, \bibinfo {author}
  {\bibfnamefont {P.~S.}\ \bibnamefont {Julienne}}, \bibinfo {author}
  {\bibfnamefont {J.}~\bibnamefont {Ye}}, \bibinfo {author} {\bibfnamefont
  {P.}~\bibnamefont {Zoller}}, \bibinfo {author} {\bibfnamefont
  {E.}~\bibnamefont {Demler}}, \bibinfo {author} {\bibfnamefont {M.~D.}\
  \bibnamefont {Lukin}}, \ and\ \bibinfo {author} {\bibfnamefont {A.~M.}\
  \bibnamefont {Rey}},\ }\href {\doibase 10.1038/nphys1535} {\bibfield
  {journal} {\bibinfo  {journal} {Nat Phys}\ }\textbf {\bibinfo {volume} {6}},\
  \bibinfo {pages} {289} (\bibinfo {year} {2010})}\BibitemShut {NoStop}%
\bibitem [{\citenamefont {Bieri}\ \emph {et~al.}(2012)\citenamefont {Bieri},
  \citenamefont {Serbyn}, \citenamefont {Senthil},\ and\ \citenamefont
  {Lee}}]{BieriSU32012}%
  \BibitemOpen
  \bibfield  {author} {\bibinfo {author} {\bibfnamefont {S.}~\bibnamefont
  {Bieri}}, \bibinfo {author} {\bibfnamefont {M.}~\bibnamefont {Serbyn}},
  \bibinfo {author} {\bibfnamefont {T.}~\bibnamefont {Senthil}}, \ and\
  \bibinfo {author} {\bibfnamefont {P.~A.}\ \bibnamefont {Lee}},\ }\href
  {\doibase 10.1103/PhysRevB.86.224409} {\bibfield  {journal} {\bibinfo
  {journal} {Phys. Rev. B}\ }\textbf {\bibinfo {volume} {86}},\ \bibinfo
  {pages} {224409} (\bibinfo {year} {2012})}\BibitemShut {NoStop}%
\bibitem [{\citenamefont {Scazza}\ \emph {et~al.}(2014)\citenamefont {Scazza},
  \citenamefont {Hofrichter}, \citenamefont {H\"{o}fer}, \citenamefont {{De
  Groot}}, \citenamefont {Bloch},\ and\ \citenamefont
  {F\"{o}lling}}]{Scazza2014}%
  \BibitemOpen
  \bibfield  {author} {\bibinfo {author} {\bibfnamefont {F.}~\bibnamefont
  {Scazza}}, \bibinfo {author} {\bibfnamefont {C.}~\bibnamefont {Hofrichter}},
  \bibinfo {author} {\bibfnamefont {M.}~\bibnamefont {H\"{o}fer}}, \bibinfo
  {author} {\bibfnamefont {P.~C.}\ \bibnamefont {{De Groot}}}, \bibinfo
  {author} {\bibfnamefont {I.}~\bibnamefont {Bloch}}, \ and\ \bibinfo {author}
  {\bibfnamefont {S.}~\bibnamefont {F\"{o}lling}},\ }\href {\doibase
  10.1038/nphys3061} {\bibfield  {journal} {\bibinfo  {journal} {Nature
  Physics}\ } (\bibinfo {year} {2014}),\ 10.1038/nphys3061}\BibitemShut
  {NoStop}%
\bibitem [{\citenamefont {Taie}\ \emph {et~al.}(2012)\citenamefont {Taie},
  \citenamefont {Yamazaki}, \citenamefont {Sugawa},\ and\ \citenamefont
  {Takahashi}}]{takahashi2012}%
  \BibitemOpen
  \bibfield  {author} {\bibinfo {author} {\bibfnamefont {S.}~\bibnamefont
  {Taie}}, \bibinfo {author} {\bibfnamefont {R.}~\bibnamefont {Yamazaki}},
  \bibinfo {author} {\bibfnamefont {S.}~\bibnamefont {Sugawa}}, \ and\ \bibinfo
  {author} {\bibfnamefont {Y.}~\bibnamefont {Takahashi}},\ }\href {\doibase
  10.1038/nphys2430} {\bibfield  {journal} {\bibinfo  {journal} {Nat Phys}\
  }\textbf {\bibinfo {volume} {8}},\ \bibinfo {pages} {825} (\bibinfo {year}
  {2012})}\BibitemShut {NoStop}%
\bibitem [{\citenamefont {Pagano}\ \emph {et~al.}(2014)\citenamefont {Pagano},
  \citenamefont {Mancini}, \citenamefont {Cappellini}, \citenamefont
  {Lombardi}, \citenamefont {Sch\"{a}fer}, \citenamefont {Hu}, \citenamefont
  {Liu}, \citenamefont {Catani}, \citenamefont {Sias}, \citenamefont
  {Inguscio},\ and\ \citenamefont {Fallani}}]{Pagano2014}%
  \BibitemOpen
  \bibfield  {author} {\bibinfo {author} {\bibfnamefont {G.}~\bibnamefont
  {Pagano}}, \bibinfo {author} {\bibfnamefont {M.}~\bibnamefont {Mancini}},
  \bibinfo {author} {\bibfnamefont {G.}~\bibnamefont {Cappellini}}, \bibinfo
  {author} {\bibfnamefont {P.}~\bibnamefont {Lombardi}}, \bibinfo {author}
  {\bibfnamefont {F.}~\bibnamefont {Sch\"{a}fer}}, \bibinfo {author}
  {\bibfnamefont {H.}~\bibnamefont {Hu}}, \bibinfo {author} {\bibfnamefont
  {X.-J.}\ \bibnamefont {Liu}}, \bibinfo {author} {\bibfnamefont
  {J.}~\bibnamefont {Catani}}, \bibinfo {author} {\bibfnamefont
  {C.}~\bibnamefont {Sias}}, \bibinfo {author} {\bibfnamefont {M.}~\bibnamefont
  {Inguscio}}, \ and\ \bibinfo {author} {\bibfnamefont {L.}~\bibnamefont
  {Fallani}},\ }\href {\doibase 10.1038/nphys2878} {\bibfield  {journal}
  {\bibinfo  {journal} {Nature Physics}\ }\textbf {\bibinfo {volume} {10}},\
  \bibinfo {pages} {198} (\bibinfo {year} {2014})}\BibitemShut {NoStop}%
\bibitem [{\citenamefont {Zhang}\ \emph {et~al.}(2014)\citenamefont {Zhang},
  \citenamefont {Bishof}, \citenamefont {Bromley}, \citenamefont {Kraus},
  \citenamefont {Safronova}, \citenamefont {Zoller}, \citenamefont {Rey},\ and\
  \citenamefont {Ye}}]{ZhangScience2014}%
  \BibitemOpen
  \bibfield  {author} {\bibinfo {author} {\bibfnamefont {X.}~\bibnamefont
  {Zhang}}, \bibinfo {author} {\bibfnamefont {M.}~\bibnamefont {Bishof}},
  \bibinfo {author} {\bibfnamefont {S.~L.}\ \bibnamefont {Bromley}}, \bibinfo
  {author} {\bibfnamefont {C.~V.}\ \bibnamefont {Kraus}}, \bibinfo {author}
  {\bibfnamefont {M.~S.}\ \bibnamefont {Safronova}}, \bibinfo {author}
  {\bibfnamefont {P.}~\bibnamefont {Zoller}}, \bibinfo {author} {\bibfnamefont
  {A.~M.}\ \bibnamefont {Rey}}, \ and\ \bibinfo {author} {\bibfnamefont
  {J.}~\bibnamefont {Ye}},\ }\href {\doibase 10.1126/science.1254978}
  {\bibfield  {journal} {\bibinfo  {journal} {Science}\ }\textbf {\bibinfo
  {volume} {345}},\ \bibinfo {pages} {1467} (\bibinfo {year} {2014})},\ \Eprint
  {http://arxiv.org/abs/http://www.sciencemag.org/content/345/6203/1467.full.pdf}
  {http://www.sciencemag.org/content/345/6203/1467.full.pdf} \BibitemShut
  {NoStop}%
\bibitem [{\citenamefont {Cazalilla}\ and\ \citenamefont
  {Rey}(2014)}]{CazalillaReyreview2014}%
  \BibitemOpen
  \bibfield  {author} {\bibinfo {author} {\bibfnamefont {M.~A.}\ \bibnamefont
  {Cazalilla}}\ and\ \bibinfo {author} {\bibfnamefont {A.~M.}\ \bibnamefont
  {Rey}},\ }\href {http://stacks.iop.org/0034-4885/77/i=12/a=124401} {\bibfield
   {journal} {\bibinfo  {journal} {Reports on Progress in Physics}\ }\textbf
  {\bibinfo {volume} {77}},\ \bibinfo {pages} {124401} (\bibinfo {year}
  {2014})}\BibitemShut {NoStop}%
\bibitem [{\citenamefont {Capponi}\ \emph {et~al.}(2016)\citenamefont
  {Capponi}, \citenamefont {Lecheminant},\ and\ \citenamefont
  {Totsuka}}]{Capponi_SUNreview_AnnPhys2016}%
  \BibitemOpen
  \bibfield  {author} {\bibinfo {author} {\bibfnamefont {S.}~\bibnamefont
  {Capponi}}, \bibinfo {author} {\bibfnamefont {P.}~\bibnamefont
  {Lecheminant}}, \ and\ \bibinfo {author} {\bibfnamefont {K.}~\bibnamefont
  {Totsuka}},\ }\href {\doibase http://dx.doi.org/10.1016/j.aop.2016.01.011}
  {\bibfield  {journal} {\bibinfo  {journal} {Annals of Physics}\ }\textbf
  {\bibinfo {volume} {367}},\ \bibinfo {pages} {50 } (\bibinfo {year}
  {2016})}\BibitemShut {NoStop}%
\bibitem [{\citenamefont {Levine}\ \emph {et~al.}(1983)\citenamefont {Levine},
  \citenamefont {Libby},\ and\ \citenamefont {Pruisken}}]{LevinePRL1983}%
  \BibitemOpen
  \bibfield  {author} {\bibinfo {author} {\bibfnamefont {H.}~\bibnamefont
  {Levine}}, \bibinfo {author} {\bibfnamefont {S.~B.}\ \bibnamefont {Libby}}, \
  and\ \bibinfo {author} {\bibfnamefont {A.~M.~M.}\ \bibnamefont {Pruisken}},\
  }\href {\doibase 10.1103/PhysRevLett.51.1915} {\bibfield  {journal} {\bibinfo
   {journal} {Phys. Rev. Lett.}\ }\textbf {\bibinfo {volume} {51}},\ \bibinfo
  {pages} {1915} (\bibinfo {year} {1983})}\BibitemShut {NoStop}%
\bibitem [{\citenamefont {Affleck}(1986)}]{Affleck1986}%
  \BibitemOpen
  \bibfield  {author} {\bibinfo {author} {\bibfnamefont {I.}~\bibnamefont
  {Affleck}},\ }\href {\doibase http://dx.doi.org/10.1016/0550-3213(86)90167-7}
  {\bibfield  {journal} {\bibinfo  {journal} {Nuclear Physics B}\ }\textbf
  {\bibinfo {volume} {265}},\ \bibinfo {pages} {409} (\bibinfo {year}
  {1986})}\BibitemShut {NoStop}%
\bibitem [{\citenamefont {Evers}\ and\ \citenamefont
  {Mirlin}(2008)}]{FerdinandMirlinRevModPhys2008}%
  \BibitemOpen
  \bibfield  {author} {\bibinfo {author} {\bibfnamefont {F.}~\bibnamefont
  {Evers}}\ and\ \bibinfo {author} {\bibfnamefont {A.~D.}\ \bibnamefont
  {Mirlin}},\ }\href {\doibase 10.1103/RevModPhys.80.1355} {\bibfield
  {journal} {\bibinfo  {journal} {Rev. Mod. Phys.}\ }\textbf {\bibinfo {volume}
  {80}},\ \bibinfo {pages} {1355} (\bibinfo {year} {2008})}\BibitemShut
  {NoStop}%
\bibitem [{\citenamefont {Chalker}\ and\ \citenamefont
  {Coddington}(1988)}]{ChalkerCoddington1988}%
  \BibitemOpen
  \bibfield  {author} {\bibinfo {author} {\bibfnamefont {J.~T.}\ \bibnamefont
  {Chalker}}\ and\ \bibinfo {author} {\bibfnamefont {P.~D.}\ \bibnamefont
  {Coddington}},\ }\href {http://stacks.iop.org/0022-3719/21/i=14/a=008}
  {\bibfield  {journal} {\bibinfo  {journal} {Journal of Physics C: Solid State
  Physics}\ }\textbf {\bibinfo {volume} {21}},\ \bibinfo {pages} {2665}
  (\bibinfo {year} {1988})}\BibitemShut {NoStop}%
\bibitem [{\citenamefont {Huckestein}\ and\ \citenamefont
  {Kramer}(1990)}]{HuckesteinKramerPRL1990}%
  \BibitemOpen
  \bibfield  {author} {\bibinfo {author} {\bibfnamefont {B.}~\bibnamefont
  {Huckestein}}\ and\ \bibinfo {author} {\bibfnamefont {B.}~\bibnamefont
  {Kramer}},\ }\href {\doibase 10.1103/PhysRevLett.64.1437} {\bibfield
  {journal} {\bibinfo  {journal} {Phys. Rev. Lett.}\ }\textbf {\bibinfo
  {volume} {64}},\ \bibinfo {pages} {1437} (\bibinfo {year}
  {1990})}\BibitemShut {NoStop}%
\bibitem [{\citenamefont {Lee}\ \emph {et~al.}(1993)\citenamefont {Lee},
  \citenamefont {Wang},\ and\ \citenamefont
  {Kivelson}}]{LeeWangKivelsonPRL1993}%
  \BibitemOpen
  \bibfield  {author} {\bibinfo {author} {\bibfnamefont {D.-H.}\ \bibnamefont
  {Lee}}, \bibinfo {author} {\bibfnamefont {Z.}~\bibnamefont {Wang}}, \ and\
  \bibinfo {author} {\bibfnamefont {S.}~\bibnamefont {Kivelson}},\ }\href
  {\doibase 10.1103/PhysRevLett.70.4130} {\bibfield  {journal} {\bibinfo
  {journal} {Phys. Rev. Lett.}\ }\textbf {\bibinfo {volume} {70}},\ \bibinfo
  {pages} {4130} (\bibinfo {year} {1993})}\BibitemShut {NoStop}%
\bibitem [{\citenamefont {Zirnbauer}(1994)}]{ZimbauerAnnPhys1994}%
  \BibitemOpen
  \bibfield  {author} {\bibinfo {author} {\bibfnamefont {M.~R.}\ \bibnamefont
  {Zirnbauer}},\ }\href {\doibase 10.1002/andp.19945060702} {\bibfield
  {journal} {\bibinfo  {journal} {Annalen der Physik}\ }\textbf {\bibinfo
  {volume} {506}},\ \bibinfo {pages} {513} (\bibinfo {year}
  {1994})}\BibitemShut {NoStop}%
\bibitem [{\citenamefont {Zirnbauer}(1997)}]{ZimbauerJMathPhys1997}%
  \BibitemOpen
  \bibfield  {author} {\bibinfo {author} {\bibfnamefont {M.~R.}\ \bibnamefont
  {Zirnbauer}},\ }\bibfield  {booktitle} {\emph {\bibinfo {booktitle} {Journal
  of Mathematical Physics}},\ }\href {\doibase 10.1063/1.531921} {\bibfield
  {journal} {\bibinfo  {journal} {Journal of Mathematical Physics}\ }\textbf
  {\bibinfo {volume} {38}},\ \bibinfo {pages} {2007} (\bibinfo {year}
  {1997})}\BibitemShut {NoStop}%
\bibitem [{\citenamefont {Slevin}\ and\ \citenamefont
  {Ohtsuki}(2009)}]{SlevinOhtsukiPRB2009}%
  \BibitemOpen
  \bibfield  {author} {\bibinfo {author} {\bibfnamefont {K.}~\bibnamefont
  {Slevin}}\ and\ \bibinfo {author} {\bibfnamefont {T.}~\bibnamefont
  {Ohtsuki}},\ }\href {\doibase 10.1103/PhysRevB.80.041304} {\bibfield
  {journal} {\bibinfo  {journal} {Phys. Rev. B}\ }\textbf {\bibinfo {volume}
  {80}},\ \bibinfo {pages} {041304} (\bibinfo {year} {2009})}\BibitemShut
  {NoStop}%
\bibitem [{\citenamefont {Amado}\ \emph {et~al.}(2011)\citenamefont {Amado},
  \citenamefont {Malyshev}, \citenamefont {Sedrakyan},\ and\ \citenamefont
  {Dom\'{\i}nguez-Adame}}]{AmadoPRL2011}%
  \BibitemOpen
  \bibfield  {author} {\bibinfo {author} {\bibfnamefont {M.}~\bibnamefont
  {Amado}}, \bibinfo {author} {\bibfnamefont {A.~V.}\ \bibnamefont {Malyshev}},
  \bibinfo {author} {\bibfnamefont {A.}~\bibnamefont {Sedrakyan}}, \ and\
  \bibinfo {author} {\bibfnamefont {F.}~\bibnamefont {Dom\'{\i}nguez-Adame}},\
  }\href {\doibase 10.1103/PhysRevLett.107.066402} {\bibfield  {journal}
  {\bibinfo  {journal} {Phys. Rev. Lett.}\ }\textbf {\bibinfo {volume} {107}},\
  \bibinfo {pages} {066402} (\bibinfo {year} {2011})}\BibitemShut {NoStop}%
\bibitem [{\citenamefont {Greiter}\ \emph {et~al.}(2007)\citenamefont
  {Greiter}, \citenamefont {Rachel},\ and\ \citenamefont
  {Schuricht}}]{GreiterRachelSchuricht2007}%
  \BibitemOpen
  \bibfield  {author} {\bibinfo {author} {\bibfnamefont {M.}~\bibnamefont
  {Greiter}}, \bibinfo {author} {\bibfnamefont {S.}~\bibnamefont {Rachel}}, \
  and\ \bibinfo {author} {\bibfnamefont {D.}~\bibnamefont {Schuricht}},\ }\href
  {\doibase 10.1103/PhysRevB.75.060401} {\bibfield  {journal} {\bibinfo
  {journal} {Phys. Rev. B}\ }\textbf {\bibinfo {volume} {75}},\ \bibinfo
  {pages} {060401} (\bibinfo {year} {2007})}\BibitemShut {NoStop}%
\bibitem [{\citenamefont {Greiter}\ and\ \citenamefont
  {Rachel}(2007)}]{GreiterRachel2007}%
  \BibitemOpen
  \bibfield  {author} {\bibinfo {author} {\bibfnamefont {M.}~\bibnamefont
  {Greiter}}\ and\ \bibinfo {author} {\bibfnamefont {S.}~\bibnamefont
  {Rachel}},\ }\href {\doibase 10.1103/PhysRevB.75.184441} {\bibfield
  {journal} {\bibinfo  {journal} {Phys. Rev. B}\ }\textbf {\bibinfo {volume}
  {75}},\ \bibinfo {pages} {184441} (\bibinfo {year} {2007})}\BibitemShut
  {NoStop}%
\bibitem [{\citenamefont {Polyakov}(1975)}]{Polyakov:1975rr}%
  \BibitemOpen
  \bibfield  {author} {\bibinfo {author} {\bibfnamefont {A.~M.}\ \bibnamefont
  {Polyakov}},\ }\href {\doibase 10.1016/0370-2693(75)90161-6} {\bibfield
  {journal} {\bibinfo  {journal} {Phys. Lett.}\ }\textbf {\bibinfo {volume}
  {B59}},\ \bibinfo {pages} {79} (\bibinfo {year} {1975})}\BibitemShut
  {NoStop}%
\bibitem [{\citenamefont
  {Affleck}(1988{\natexlab{a}})}]{AffleckLesHouches1988}%
  \BibitemOpen
  \bibfield  {author} {\bibinfo {author} {\bibfnamefont {I.}~\bibnamefont
  {Affleck}},\ }in\ \href@noop {} {\emph {\bibinfo {booktitle} {Les Houches
  Summer School in Theoretical Physics: Fields, Strings, Critical Phenomena Les
  Houches, France, June 28-August 5, 1988}}}\ (\bibinfo {year} {1988})\ pp.\
  \bibinfo {pages} {0563--640}\BibitemShut {NoStop}%
\bibitem [{\citenamefont {Shankar}\ and\ \citenamefont
  {Read}(1990)}]{ShankarReadNuclPhysB1990}%
  \BibitemOpen
  \bibfield  {author} {\bibinfo {author} {\bibfnamefont {R.}~\bibnamefont
  {Shankar}}\ and\ \bibinfo {author} {\bibfnamefont {N.}~\bibnamefont {Read}},\
  }\href {\doibase http://dx.doi.org/10.1016/0550-3213(90)90437-I} {\bibfield
  {journal} {\bibinfo  {journal} {Nuclear Physics B}\ }\textbf {\bibinfo
  {volume} {336}},\ \bibinfo {pages} {457 } (\bibinfo {year}
  {1990})}\BibitemShut {NoStop}%
\bibitem [{\citenamefont {Affleck}\ and\ \citenamefont
  {Haldane}(1987)}]{AffleckHaldane1987}%
  \BibitemOpen
  \bibfield  {author} {\bibinfo {author} {\bibfnamefont {I.}~\bibnamefont
  {Affleck}}\ and\ \bibinfo {author} {\bibfnamefont {F.~D.~M.}\ \bibnamefont
  {Haldane}},\ }\href {\doibase 10.1103/PhysRevB.36.5291} {\bibfield  {journal}
  {\bibinfo  {journal} {Phys. Rev. B}\ }\textbf {\bibinfo {volume} {36}},\
  \bibinfo {pages} {5291} (\bibinfo {year} {1987})}\BibitemShut {NoStop}%
\bibitem [{\citenamefont {Majumdar}\ and\ \citenamefont
  {Ghosh}(1969)}]{MajumdarGhosh1969}%
  \BibitemOpen
  \bibfield  {author} {\bibinfo {author} {\bibfnamefont {C.~K.}\ \bibnamefont
  {Majumdar}}\ and\ \bibinfo {author} {\bibfnamefont {D.~K.}\ \bibnamefont
  {Ghosh}},\ }\href {\doibase 10.1063/1.1664978} {\bibfield  {journal}
  {\bibinfo  {journal} {J. Math. Phys.}\ }\textbf {\bibinfo {volume} {10}},\
  \bibinfo {pages} {1388} (\bibinfo {year} {1969})}\BibitemShut {NoStop}%
\bibitem [{\citenamefont {Haldane}(1982)}]{Haldane_dimerization_PRB1982}%
  \BibitemOpen
  \bibfield  {author} {\bibinfo {author} {\bibfnamefont {F.~D.~M.}\
  \bibnamefont {Haldane}},\ }\href {\doibase 10.1103/PhysRevB.25.4925}
  {\bibfield  {journal} {\bibinfo  {journal} {Phys. Rev. B}\ }\textbf {\bibinfo
  {volume} {25}},\ \bibinfo {pages} {4925} (\bibinfo {year}
  {1982})}\BibitemShut {NoStop}%
\bibitem [{\citenamefont {Okamoto}\ and\ \citenamefont
  {Nomura}(1992)}]{Okamoto1992}%
  \BibitemOpen
  \bibfield  {author} {\bibinfo {author} {\bibfnamefont {K.}~\bibnamefont
  {Okamoto}}\ and\ \bibinfo {author} {\bibfnamefont {K.}~\bibnamefont
  {Nomura}},\ }\href {\doibase https://doi.org/10.1016/0375-9601(92)90823-5}
  {\bibfield  {journal} {\bibinfo  {journal} {Physics Letters A}\ }\textbf
  {\bibinfo {volume} {169}},\ \bibinfo {pages} {433 } (\bibinfo {year}
  {1992})}\BibitemShut {NoStop}%
\bibitem [{\citenamefont {Eggert}(1996)}]{EggertPRB1996}%
  \BibitemOpen
  \bibfield  {author} {\bibinfo {author} {\bibfnamefont {S.}~\bibnamefont
  {Eggert}},\ }\href {\doibase 10.1103/PhysRevB.54.R9612} {\bibfield  {journal}
  {\bibinfo  {journal} {Phys. Rev. B}\ }\textbf {\bibinfo {volume} {54}},\
  \bibinfo {pages} {R9612} (\bibinfo {year} {1996})}\BibitemShut {NoStop}%
\bibitem [{\citenamefont {Hulth\'en}(1938)}]{Hulthen1938}%
  \BibitemOpen
  \bibfield  {author} {\bibinfo {author} {\bibfnamefont {L.}~\bibnamefont
  {Hulth\'en}},\ }\href@noop {} {\bibfield  {journal} {\bibinfo  {journal}
  {Ark. Mat. Astron. Fysik A}\ }\textbf {\bibinfo {volume} {26}},\ \bibinfo
  {pages} {106 p.} (\bibinfo {year} {1938})}\BibitemShut {NoStop}%
\bibitem [{\citenamefont {Lieb}\ \emph {et~al.}(1961)\citenamefont {Lieb},
  \citenamefont {Schultz},\ and\ \citenamefont {Mattis}}]{LSM1961}%
  \BibitemOpen
  \bibfield  {author} {\bibinfo {author} {\bibfnamefont {E.}~\bibnamefont
  {Lieb}}, \bibinfo {author} {\bibfnamefont {T.}~\bibnamefont {Schultz}}, \
  and\ \bibinfo {author} {\bibfnamefont {D.}~\bibnamefont {Mattis}},\ }\href
  {\doibase 10.1016/0003-4916(61)90115-4} {\bibfield  {journal} {\bibinfo
  {journal} {Annals of Physics}\ }\textbf {\bibinfo {volume} {16}},\ \bibinfo
  {pages} {407 } (\bibinfo {year} {1961})}\BibitemShut {NoStop}%
\bibitem [{\citenamefont {Affleck}\ and\ \citenamefont
  {Lieb}(1986)}]{AffleckLieb1986}%
  \BibitemOpen
  \bibfield  {author} {\bibinfo {author} {\bibfnamefont {I.}~\bibnamefont
  {Affleck}}\ and\ \bibinfo {author} {\bibfnamefont {E.~H.}\ \bibnamefont
  {Lieb}},\ }\href {\doibase 10.1007/BF00400304} {\bibfield  {journal}
  {\bibinfo  {journal} {Letters in Mathematical Physics}\ }\textbf {\bibinfo
  {volume} {12}},\ \bibinfo {pages} {57} (\bibinfo {year} {1986})}\BibitemShut
  {NoStop}%
\bibitem [{\citenamefont {Affleck}\ \emph {et~al.}(1988)\citenamefont
  {Affleck}, \citenamefont {Kennedy}, \citenamefont {Lieb},\ and\ \citenamefont
  {Tasaki}}]{AKLT1988}%
  \BibitemOpen
  \bibfield  {author} {\bibinfo {author} {\bibfnamefont {I.}~\bibnamefont
  {Affleck}}, \bibinfo {author} {\bibfnamefont {T.}~\bibnamefont {Kennedy}},
  \bibinfo {author} {\bibfnamefont {E.}~\bibnamefont {Lieb}}, \ and\ \bibinfo
  {author} {\bibfnamefont {H.}~\bibnamefont {Tasaki}},\ }\href {\doibase
  10.1007/BF01218021} {\bibfield  {journal} {\bibinfo  {journal}
  {Communications in Mathematical Physics}\ }\textbf {\bibinfo {volume}
  {115}},\ \bibinfo {pages} {477} (\bibinfo {year} {1988})}\BibitemShut
  {NoStop}%
\bibitem [{\citenamefont {Sutherland}(1975)}]{Sutherland1975}%
  \BibitemOpen
  \bibfield  {author} {\bibinfo {author} {\bibfnamefont {B.}~\bibnamefont
  {Sutherland}},\ }\href {\doibase 10.1103/PhysRevB.12.3795} {\bibfield
  {journal} {\bibinfo  {journal} {Phys. Rev. B}\ }\textbf {\bibinfo {volume}
  {12}},\ \bibinfo {pages} {3795} (\bibinfo {year} {1975})}\BibitemShut
  {NoStop}%
\bibitem [{\citenamefont {Tsvelik}\ and\ \citenamefont
  {Wiegmann}(1983)}]{TsvelikWiegmann1983}%
  \BibitemOpen
  \bibfield  {author} {\bibinfo {author} {\bibfnamefont {A.}~\bibnamefont
  {Tsvelik}}\ and\ \bibinfo {author} {\bibfnamefont {P.}~\bibnamefont
  {Wiegmann}},\ }\href {\doibase 10.1080/00018738300101581} {\bibfield
  {journal} {\bibinfo  {journal} {Advances in Physics}\ }\textbf {\bibinfo
  {volume} {32}},\ \bibinfo {pages} {453} (\bibinfo {year} {1983})}\BibitemShut
  {NoStop}%
\bibitem [{\citenamefont {Andrei}\ \emph {et~al.}(1983)\citenamefont {Andrei},
  \citenamefont {Furuya},\ and\ \citenamefont
  {Lowenstein}}]{AndreiFuruyaLowensteinRMP1983}%
  \BibitemOpen
  \bibfield  {author} {\bibinfo {author} {\bibfnamefont {N.}~\bibnamefont
  {Andrei}}, \bibinfo {author} {\bibfnamefont {K.}~\bibnamefont {Furuya}}, \
  and\ \bibinfo {author} {\bibfnamefont {J.~H.}\ \bibnamefont {Lowenstein}},\
  }\href {\doibase 10.1103/RevModPhys.55.331} {\bibfield  {journal} {\bibinfo
  {journal} {Rev. Mod. Phys.}\ }\textbf {\bibinfo {volume} {55}},\ \bibinfo
  {pages} {331} (\bibinfo {year} {1983})}\BibitemShut {NoStop}%
\bibitem [{\citenamefont {Affleck}(1988{\natexlab{b}})}]{AffleckSUn1988}%
  \BibitemOpen
  \bibfield  {author} {\bibinfo {author} {\bibfnamefont {I.}~\bibnamefont
  {Affleck}},\ }\href {\doibase
  https://dx.doi.org/10.1016/0550-3213(88)90117-4} {\bibfield  {journal}
  {\bibinfo  {journal} {Nuclear Physics B}\ }\textbf {\bibinfo {volume}
  {305}},\ \bibinfo {pages} {582 } (\bibinfo {year}
  {1988}{\natexlab{b}})}\BibitemShut {NoStop}%
\bibitem [{\citenamefont {Katsura}\ \emph {et~al.}(2008)\citenamefont
  {Katsura}, \citenamefont {Hirano},\ and\ \citenamefont
  {Korepin}}]{Katsura2008}%
  \BibitemOpen
  \bibfield  {author} {\bibinfo {author} {\bibfnamefont {H.}~\bibnamefont
  {Katsura}}, \bibinfo {author} {\bibfnamefont {T.}~\bibnamefont {Hirano}}, \
  and\ \bibinfo {author} {\bibfnamefont {V.~E.}\ \bibnamefont {Korepin}},\
  }\href {http://stacks.iop.org/1751-8121/41/i=13/a=135304} {\bibfield
  {journal} {\bibinfo  {journal} {Journal of Physics A: Mathematical and
  Theoretical}\ }\textbf {\bibinfo {volume} {41}},\ \bibinfo {pages} {135304}
  (\bibinfo {year} {2008})}\BibitemShut {NoStop}%
\bibitem [{\citenamefont {Duivenvoorden}\ and\ \citenamefont
  {Quella}(2013)}]{QuellaPRB2013}%
  \BibitemOpen
  \bibfield  {author} {\bibinfo {author} {\bibfnamefont {K.}~\bibnamefont
  {Duivenvoorden}}\ and\ \bibinfo {author} {\bibfnamefont {T.}~\bibnamefont
  {Quella}},\ }\href {\doibase 10.1103/PhysRevB.87.125145} {\bibfield
  {journal} {\bibinfo  {journal} {Phys. Rev. B}\ }\textbf {\bibinfo {volume}
  {87}},\ \bibinfo {pages} {125145} (\bibinfo {year} {2013})}\BibitemShut
  {NoStop}%
\bibitem [{\citenamefont {Nonne}\ \emph {et~al.}(2013)\citenamefont {Nonne},
  \citenamefont {Moliner}, \citenamefont {Capponi}, \citenamefont
  {Lecheminant},\ and\ \citenamefont {Totsuka}}]{LecheminantEPL2013}%
  \BibitemOpen
  \bibfield  {author} {\bibinfo {author} {\bibfnamefont {H.}~\bibnamefont
  {Nonne}}, \bibinfo {author} {\bibfnamefont {M.}~\bibnamefont {Moliner}},
  \bibinfo {author} {\bibfnamefont {S.}~\bibnamefont {Capponi}}, \bibinfo
  {author} {\bibfnamefont {P.}~\bibnamefont {Lecheminant}}, \ and\ \bibinfo
  {author} {\bibfnamefont {K.}~\bibnamefont {Totsuka}},\ }\href
  {http://stacks.iop.org/0295-5075/102/i=3/a=37008} {\bibfield  {journal}
  {\bibinfo  {journal} {EPL (Europhysics Letters)}\ }\textbf {\bibinfo {volume}
  {102}},\ \bibinfo {pages} {37008} (\bibinfo {year} {2013})}\BibitemShut
  {NoStop}%
\bibitem [{\citenamefont {Morimoto}\ \emph {et~al.}(2014)\citenamefont
  {Morimoto}, \citenamefont {Ueda}, \citenamefont {Momoi},\ and\ \citenamefont
  {Furusaki}}]{MorimotoFurusakiPRB2014}%
  \BibitemOpen
  \bibfield  {author} {\bibinfo {author} {\bibfnamefont {T.}~\bibnamefont
  {Morimoto}}, \bibinfo {author} {\bibfnamefont {H.}~\bibnamefont {Ueda}},
  \bibinfo {author} {\bibfnamefont {T.}~\bibnamefont {Momoi}}, \ and\ \bibinfo
  {author} {\bibfnamefont {A.}~\bibnamefont {Furusaki}},\ }\href {\doibase
  10.1103/PhysRevB.90.235111} {\bibfield  {journal} {\bibinfo  {journal} {Phys.
  Rev. B}\ }\textbf {\bibinfo {volume} {90}},\ \bibinfo {pages} {235111}
  (\bibinfo {year} {2014})}\BibitemShut {NoStop}%
\bibitem [{\citenamefont {Nataf}\ and\ \citenamefont
  {Mila}(2016)}]{NatafchainED2016}%
  \BibitemOpen
  \bibfield  {author} {\bibinfo {author} {\bibfnamefont {P.}~\bibnamefont
  {Nataf}}\ and\ \bibinfo {author} {\bibfnamefont {F.}~\bibnamefont {Mila}},\
  }\href {\doibase 10.1103/PhysRevB.93.155134} {\bibfield  {journal} {\bibinfo
  {journal} {Phys. Rev. B}\ }\textbf {\bibinfo {volume} {93}},\ \bibinfo
  {pages} {155134} (\bibinfo {year} {2016})}\BibitemShut {NoStop}%
\bibitem [{\citenamefont {Weichselbaum}\ \emph {et~al.}()\citenamefont
  {Weichselbaum} \emph {et~al.}}]{WeichselbaumPC}%
  \BibitemOpen
  \bibfield  {author} {\bibinfo {author} {\bibfnamefont {A.}~\bibnamefont
  {Weichselbaum}} \emph {et~al.},\ }\href@noop {} {}\bibinfo {note}
  {Unpublished}\BibitemShut {NoStop}%
\bibitem [{\citenamefont {White}(1992)}]{WhiteDMRG1992}%
  \BibitemOpen
  \bibfield  {author} {\bibinfo {author} {\bibfnamefont {S.~R.}\ \bibnamefont
  {White}},\ }\href {\doibase 10.1103/PhysRevLett.69.2863} {\bibfield
  {journal} {\bibinfo  {journal} {Phys. Rev. Lett.}\ }\textbf {\bibinfo
  {volume} {69}},\ \bibinfo {pages} {2863} (\bibinfo {year}
  {1992})}\BibitemShut {NoStop}%
\bibitem [{\citenamefont {Corboz}\ \emph {et~al.}(2012)\citenamefont {Corboz},
  \citenamefont {Lajk\'o}, \citenamefont {L\"auchli}, \citenamefont {Penc},\
  and\ \citenamefont {Mila}}]{CorbozPRX2012}%
  \BibitemOpen
  \bibfield  {author} {\bibinfo {author} {\bibfnamefont {P.}~\bibnamefont
  {Corboz}}, \bibinfo {author} {\bibfnamefont {M.}~\bibnamefont {Lajk\'o}},
  \bibinfo {author} {\bibfnamefont {A.~M.}\ \bibnamefont {L\"auchli}}, \bibinfo
  {author} {\bibfnamefont {K.}~\bibnamefont {Penc}}, \ and\ \bibinfo {author}
  {\bibfnamefont {F.}~\bibnamefont {Mila}},\ }\href {\doibase
  10.1103/PhysRevX.2.041013} {\bibfield  {journal} {\bibinfo  {journal} {Phys.
  Rev. X}\ }\textbf {\bibinfo {volume} {2}},\ \bibinfo {pages} {041013}
  (\bibinfo {year} {2012})}\BibitemShut {NoStop}%
\bibitem [{\citenamefont {Papanicolaou}(1984)}]{N1984281}%
  \BibitemOpen
  \bibfield  {author} {\bibinfo {author} {\bibfnamefont {N.}~\bibnamefont
  {Papanicolaou}},\ }\href {\doibase 10.1016/0550-3213(84)90268-2} {\bibfield
  {journal} {\bibinfo  {journal} {Nuclear Physics B}\ }\textbf {\bibinfo
  {volume} {240}},\ \bibinfo {pages} {281 } (\bibinfo {year}
  {1984})}\BibitemShut {NoStop}%
\bibitem [{\citenamefont {Papanicolaou}(1988)}]{papa1988}%
  \BibitemOpen
  \bibfield  {author} {\bibinfo {author} {\bibfnamefont {N.}~\bibnamefont
  {Papanicolaou}},\ }\href {\doibase 10.1016/0550-3213(88)90073-9} {\bibfield
  {journal} {\bibinfo  {journal} {Nuclear Physics B}\ }\textbf {\bibinfo
  {volume} {305}},\ \bibinfo {pages} {367 } (\bibinfo {year}
  {1988})}\BibitemShut {NoStop}%
\bibitem [{\citenamefont {Joshi}\ \emph {et~al.}(1999)\citenamefont {Joshi},
  \citenamefont {Ma}, \citenamefont {Mila}, \citenamefont {Shi},\ and\
  \citenamefont {Zhang}}]{JoshiZhang1999}%
  \BibitemOpen
  \bibfield  {author} {\bibinfo {author} {\bibfnamefont {A.}~\bibnamefont
  {Joshi}}, \bibinfo {author} {\bibfnamefont {M.}~\bibnamefont {Ma}}, \bibinfo
  {author} {\bibfnamefont {F.}~\bibnamefont {Mila}}, \bibinfo {author}
  {\bibfnamefont {D.~N.}\ \bibnamefont {Shi}}, \ and\ \bibinfo {author}
  {\bibfnamefont {F.~C.}\ \bibnamefont {Zhang}},\ }\href {\doibase
  10.1103/PhysRevB.60.6584} {\bibfield  {journal} {\bibinfo  {journal} {Phys.
  Rev. B}\ }\textbf {\bibinfo {volume} {60}},\ \bibinfo {pages} {6584}
  (\bibinfo {year} {1999})}\BibitemShut {NoStop}%
\bibitem [{\citenamefont {Andrei}\ and\ \citenamefont
  {Johannesson}(1984)}]{AndreiJohannessonPLA1984}%
  \BibitemOpen
  \bibfield  {author} {\bibinfo {author} {\bibfnamefont {N.}~\bibnamefont
  {Andrei}}\ and\ \bibinfo {author} {\bibfnamefont {H.}~\bibnamefont
  {Johannesson}},\ }\href {\doibase
  http://dx.doi.org/10.1016/0375-9601(84)90819-3} {\bibfield  {journal}
  {\bibinfo  {journal} {Physics Letters A}\ }\textbf {\bibinfo {volume}
  {104}},\ \bibinfo {pages} {370 } (\bibinfo {year} {1984})}\BibitemShut
  {NoStop}%
\bibitem [{\citenamefont
  {Johannesson}(1986{\natexlab{a}})}]{JohannessonPLA1986}%
  \BibitemOpen
  \bibfield  {author} {\bibinfo {author} {\bibfnamefont {H.}~\bibnamefont
  {Johannesson}},\ }\href {\doibase
  http://dx.doi.org/10.1016/0375-9601(86)90300-2} {\bibfield  {journal}
  {\bibinfo  {journal} {Physics Letters A}\ }\textbf {\bibinfo {volume}
  {116}},\ \bibinfo {pages} {133 } (\bibinfo {year}
  {1986}{\natexlab{a}})}\BibitemShut {NoStop}%
\bibitem [{\citenamefont
  {Johannesson}(1986{\natexlab{b}})}]{JohannessonNuclPhysB1986}%
  \BibitemOpen
  \bibfield  {author} {\bibinfo {author} {\bibfnamefont {H.}~\bibnamefont
  {Johannesson}},\ }\href {\doibase
  http://dx.doi.org/10.1016/0550-3213(86)90554-7} {\bibfield  {journal}
  {\bibinfo  {journal} {Nuclear Physics B}\ }\textbf {\bibinfo {volume}
  {270}},\ \bibinfo {pages} {235 } (\bibinfo {year}
  {1986}{\natexlab{b}})}\BibitemShut {NoStop}%
\bibitem [{\citenamefont {Alcaraz}\ and\ \citenamefont
  {Martins}(1989)}]{AlcarazMartins_JPhysA1989}%
  \BibitemOpen
  \bibfield  {author} {\bibinfo {author} {\bibfnamefont {F.~C.}\ \bibnamefont
  {Alcaraz}}\ and\ \bibinfo {author} {\bibfnamefont {M.~J.}\ \bibnamefont
  {Martins}},\ }\href {http://stacks.iop.org/0305-4470/22/i=18/a=002}
  {\bibfield  {journal} {\bibinfo  {journal} {Journal of Physics A:
  Mathematical and General}\ }\textbf {\bibinfo {volume} {22}},\ \bibinfo
  {pages} {L865} (\bibinfo {year} {1989})}\BibitemShut {NoStop}%
\bibitem [{\citenamefont {F{\"u}hringer}\ \emph {et~al.}(2008)\citenamefont
  {F{\"u}hringer}, \citenamefont {Rachel}, \citenamefont {Thomale},
  \citenamefont {Greiter},\ and\ \citenamefont
  {Schmitteckert}}]{FuhringerDMRG_AnnPhys2008}%
  \BibitemOpen
  \bibfield  {author} {\bibinfo {author} {\bibfnamefont {M.}~\bibnamefont
  {F{\"u}hringer}}, \bibinfo {author} {\bibfnamefont {S.}~\bibnamefont
  {Rachel}}, \bibinfo {author} {\bibfnamefont {R.}~\bibnamefont {Thomale}},
  \bibinfo {author} {\bibfnamefont {M.}~\bibnamefont {Greiter}}, \ and\
  \bibinfo {author} {\bibfnamefont {P.}~\bibnamefont {Schmitteckert}},\ }\href
  {\doibase 10.1002/andp.200810326} {\bibfield  {journal} {\bibinfo  {journal}
  {Annalen der Physik}\ }\textbf {\bibinfo {volume} {17}},\ \bibinfo {pages}
  {922} (\bibinfo {year} {2008})}\BibitemShut {NoStop}%
\bibitem [{\citenamefont {Zamolodchikov}(1986)}]{Zamolodchikov1986}%
  \BibitemOpen
  \bibfield  {author} {\bibinfo {author} {\bibfnamefont {A.~B.}\ \bibnamefont
  {Zamolodchikov}},\ }\href@noop {} {\bibfield  {journal} {\bibinfo  {journal}
  {JETP Lett.}\ }\textbf {\bibinfo {volume} {43}},\ \bibinfo {pages} {730}
  (\bibinfo {year} {1986})},\ \bibinfo {note} {[Pisma Zh. Eksp. Teor.
  Fiz.43,565(1986)]}\BibitemShut {NoStop}%
\bibitem [{\citenamefont {Klauder}(1979)}]{KlauderPRD1979}%
  \BibitemOpen
  \bibfield  {author} {\bibinfo {author} {\bibfnamefont {J.~R.}\ \bibnamefont
  {Klauder}},\ }\href {\doibase 10.1103/PhysRevD.19.2349} {\bibfield  {journal}
  {\bibinfo  {journal} {Phys. Rev. D}\ }\textbf {\bibinfo {volume} {19}},\
  \bibinfo {pages} {2349} (\bibinfo {year} {1979})}\BibitemShut {NoStop}%
\bibitem [{\citenamefont {Gnutzmann}\ and\ \citenamefont
  {Kus}(1998)}]{GnutzmannKus1998}%
  \BibitemOpen
  \bibfield  {author} {\bibinfo {author} {\bibfnamefont {S.}~\bibnamefont
  {Gnutzmann}}\ and\ \bibinfo {author} {\bibfnamefont {M.}~\bibnamefont
  {Kus}},\ }\href {http://stacks.iop.org/0305-4470/31/i=49/a=011} {\bibfield
  {journal} {\bibinfo  {journal} {Journal of Physics A: Mathematical and
  General}\ }\textbf {\bibinfo {volume} {31}},\ \bibinfo {pages} {9871}
  (\bibinfo {year} {1998})}\BibitemShut {NoStop}%
\bibitem [{\citenamefont {Shibata}\ and\ \citenamefont
  {Takagi}(1999)}]{ShibataTakagi1999}%
  \BibitemOpen
  \bibfield  {author} {\bibinfo {author} {\bibfnamefont {J.}~\bibnamefont
  {Shibata}}\ and\ \bibinfo {author} {\bibfnamefont {S.}~\bibnamefont
  {Takagi}},\ }\href {\doibase 10.1142/S0217979299000096} {\bibfield  {journal}
  {\bibinfo  {journal} {International Journal of Modern Physics B}\ }\textbf
  {\bibinfo {volume} {13}},\ \bibinfo {pages} {107} (\bibinfo {year} {1999})},\
  \Eprint
  {http://arxiv.org/abs/http://www.worldscientific.com/doi/pdf/10.1142/S0217979299000096}
  {http://www.worldscientific.com/doi/pdf/10.1142/S0217979299000096}
  \BibitemShut {NoStop}%
\bibitem [{\citenamefont {Mathur}\ and\ \citenamefont
  {Sen}(2001)}]{MathurSen2001}%
  \BibitemOpen
  \bibfield  {author} {\bibinfo {author} {\bibfnamefont {M.}~\bibnamefont
  {Mathur}}\ and\ \bibinfo {author} {\bibfnamefont {D.}~\bibnamefont {Sen}},\
  }\href {\doibase 10.1063/1.1385563} {\bibfield  {journal} {\bibinfo
  {journal} {Journal of Mathematical Physics}\ }\textbf {\bibinfo {volume}
  {42}},\ \bibinfo {pages} {4181} (\bibinfo {year} {2001})},\ \Eprint
  {http://arxiv.org/abs/http://dx.doi.org/10.1063/1.1385563}
  {http://dx.doi.org/10.1063/1.1385563} \BibitemShut {NoStop}%
\bibitem [{\citenamefont {Ueda}\ \emph {et~al.}(2016)\citenamefont {Ueda},
  \citenamefont {Akagi},\ and\ \citenamefont {Shannon}}]{UedaShannon2016}%
  \BibitemOpen
  \bibfield  {author} {\bibinfo {author} {\bibfnamefont {H.~T.}\ \bibnamefont
  {Ueda}}, \bibinfo {author} {\bibfnamefont {Y.}~\bibnamefont {Akagi}}, \ and\
  \bibinfo {author} {\bibfnamefont {N.}~\bibnamefont {Shannon}},\ }\href
  {\doibase 10.1103/PhysRevA.93.021606} {\bibfield  {journal} {\bibinfo
  {journal} {Phys. Rev. A}\ }\textbf {\bibinfo {volume} {93}},\ \bibinfo
  {pages} {021606} (\bibinfo {year} {2016})}\BibitemShut {NoStop}%
\bibitem [{\citenamefont {Smerald}\ and\ \citenamefont
  {Shannon}(2013)}]{SmeraldPRB2013}%
  \BibitemOpen
  \bibfield  {author} {\bibinfo {author} {\bibfnamefont {A.}~\bibnamefont
  {Smerald}}\ and\ \bibinfo {author} {\bibfnamefont {N.}~\bibnamefont
  {Shannon}},\ }\href {\doibase 10.1103/PhysRevB.88.184430} {\bibfield
  {journal} {\bibinfo  {journal} {Phys. Rev. B}\ }\textbf {\bibinfo {volume}
  {88}},\ \bibinfo {pages} {184430} (\bibinfo {year} {2013})}\BibitemShut
  {NoStop}%
\bibitem [{\citenamefont {Andrew}(2013)}]{Smeraldthesis2013}%
  \BibitemOpen
  \bibfield  {author} {\bibinfo {author} {\bibfnamefont {S.}~\bibnamefont
  {Andrew}},\ }\href@noop {} {\emph {\bibinfo {title} {Theory of the Nuclear
  Magnetic 1/T1 Relaxation Rate in Conventional and Unconventional Magnets}}},\
  Springer Theses\ (\bibinfo  {publisher} {Springer International Publishing},\
  \bibinfo {year} {2013})\BibitemShut {NoStop}%
\bibitem [{\citenamefont {{Pimenov}}\ and\ \citenamefont
  {{Punk}}(2017)}]{PimenovPunk2017}%
  \BibitemOpen
  \bibfield  {author} {\bibinfo {author} {\bibfnamefont {D.}~\bibnamefont
  {{Pimenov}}}\ and\ \bibinfo {author} {\bibfnamefont {M.}~\bibnamefont
  {{Punk}}},\ }\href@noop {} {\bibfield  {journal} {\bibinfo  {journal} {ArXiv
  e-prints}\ } (\bibinfo {year} {2017})},\ \Eprint
  {http://arxiv.org/abs/1703.01308} {arXiv:1703.01308 [cond-mat.str-el]}
  \BibitemShut {NoStop}%
\bibitem [{\citenamefont {D'Adda}\ \emph {et~al.}(1978)\citenamefont {D'Adda},
  \citenamefont {L{\"u}scher},\ and\ \citenamefont
  {Vecchia}}]{DaddaLuscher1978}%
  \BibitemOpen
  \bibfield  {author} {\bibinfo {author} {\bibfnamefont {A.}~\bibnamefont
  {D'Adda}}, \bibinfo {author} {\bibfnamefont {M.}~\bibnamefont {L{\"u}scher}},
  \ and\ \bibinfo {author} {\bibfnamefont {P.~D.}\ \bibnamefont {Vecchia}},\
  }\href {\doibase http://dx.doi.org/10.1016/0550-3213(78)90432-7} {\bibfield
  {journal} {\bibinfo  {journal} {Nuclear Physics B}\ }\textbf {\bibinfo
  {volume} {146}},\ \bibinfo {pages} {63 } (\bibinfo {year}
  {1978})}\BibitemShut {NoStop}%
\bibitem [{\citenamefont {Bykov}(2013)}]{Bykov2013}%
  \BibitemOpen
  \bibfield  {author} {\bibinfo {author} {\bibfnamefont {D.}~\bibnamefont
  {Bykov}},\ }\href {\doibase 10.1007/s00220-013-1702-5} {\bibfield  {journal}
  {\bibinfo  {journal} {Communications in Mathematical Physics}\ }\textbf
  {\bibinfo {volume} {322}},\ \bibinfo {pages} {807} (\bibinfo {year}
  {2013})}\BibitemShut {NoStop}%
\bibitem [{\citenamefont {Murnaghan}(1962)}]{murnaghan1962}%
  \BibitemOpen
  \bibfield  {author} {\bibinfo {author} {\bibfnamefont {F.~D.}\ \bibnamefont
  {Murnaghan}},\ }\href@noop {} {\emph {\bibinfo {title} {The unitary and
  rotation groups}}},\ Vol.~\bibinfo {volume} {3}\ (\bibinfo  {publisher}
  {Spartan books},\ \bibinfo {year} {1962})\BibitemShut {NoStop}%
\bibitem [{\citenamefont {Wilcox}(1967)}]{Wilcox1967}%
  \BibitemOpen
  \bibfield  {author} {\bibinfo {author} {\bibfnamefont {R.~M.}\ \bibnamefont
  {Wilcox}},\ }\href {\doibase 10.1063/1.1705306} {\bibfield  {journal}
  {\bibinfo  {journal} {Journal of Mathematical Physics}\ }\textbf {\bibinfo
  {volume} {8}},\ \bibinfo {pages} {962} (\bibinfo {year} {1967})},\ \Eprint
  {http://arxiv.org/abs/http://dx.doi.org/10.1063/1.1705306}
  {http://dx.doi.org/10.1063/1.1705306} \BibitemShut {NoStop}%
\bibitem [{\citenamefont {Puri}(2001)}]{Puri2001}%
  \BibitemOpen
  \bibfield  {author} {\bibinfo {author} {\bibfnamefont {R.~R.}\ \bibnamefont
  {Puri}},\ }\href@noop {} {\emph {\bibinfo {title} {Mathematical Methods of
  Quantum Physics}}},\ \bibinfo {series} {Springer Series in Optical Sciences},
  Vol.~\bibinfo {volume} {79}\ (\bibinfo  {publisher} {Springer Berlin
  Heidelberg},\ \bibinfo {year} {2001})\ \bibinfo {note} {(see section
  2.4)}\BibitemShut {NoStop}%
\bibitem [{\citenamefont {Seiberg}(1984)}]{SeibergPRL1984}%
  \BibitemOpen
  \bibfield  {author} {\bibinfo {author} {\bibfnamefont {N.}~\bibnamefont
  {Seiberg}},\ }\href {\doibase 10.1103/PhysRevLett.53.637} {\bibfield
  {journal} {\bibinfo  {journal} {Phys. Rev. Lett.}\ }\textbf {\bibinfo
  {volume} {53}},\ \bibinfo {pages} {637} (\bibinfo {year} {1984})}\BibitemShut
  {NoStop}%
\bibitem [{\citenamefont {Plefka}\ and\ \citenamefont
  {Samuel}(1997)}]{PlefkaSamuelPRD1997}%
  \BibitemOpen
  \bibfield  {author} {\bibinfo {author} {\bibfnamefont {J.~C.}\ \bibnamefont
  {Plefka}}\ and\ \bibinfo {author} {\bibfnamefont {S.}~\bibnamefont
  {Samuel}},\ }\href {\doibase 10.1103/PhysRevD.55.3966} {\bibfield  {journal}
  {\bibinfo  {journal} {Phys. Rev. D}\ }\textbf {\bibinfo {volume} {55}},\
  \bibinfo {pages} {3966} (\bibinfo {year} {1997})}\BibitemShut {NoStop}%
\bibitem [{\citenamefont {Stone}(1979)}]{Stone1981}%
  \BibitemOpen
  \bibfield  {author} {\bibinfo {author} {\bibfnamefont {M.}~\bibnamefont
  {Stone}},\ }\href {\doibase http://dx.doi.org/10.1016/0550-3213(79)90081-6}
  {\bibfield  {journal} {\bibinfo  {journal} {Nuclear Physics B}\ }\textbf
  {\bibinfo {volume} {152}},\ \bibinfo {pages} {97 } (\bibinfo {year}
  {1979})}\BibitemShut {NoStop}%
\bibitem [{\citenamefont {Vecchia}\ \emph {et~al.}(1981)\citenamefont
  {Vecchia}, \citenamefont {Holtkamp}, \citenamefont {Musto}, \citenamefont
  {Nicodemi},\ and\ \citenamefont {Pettorino}}]{DiVecchia1981}%
  \BibitemOpen
  \bibfield  {author} {\bibinfo {author} {\bibfnamefont {P.~D.}\ \bibnamefont
  {Vecchia}}, \bibinfo {author} {\bibfnamefont {A.}~\bibnamefont {Holtkamp}},
  \bibinfo {author} {\bibfnamefont {R.}~\bibnamefont {Musto}}, \bibinfo
  {author} {\bibfnamefont {F.}~\bibnamefont {Nicodemi}}, \ and\ \bibinfo
  {author} {\bibfnamefont {R.}~\bibnamefont {Pettorino}},\ }\href {\doibase
  http://dx.doi.org/10.1016/0550-3213(81)90047-X} {\bibfield  {journal}
  {\bibinfo  {journal} {Nuclear Physics B}\ }\textbf {\bibinfo {volume}
  {190}},\ \bibinfo {pages} {719 } (\bibinfo {year} {1981})}\BibitemShut
  {NoStop}%
\bibitem [{\citenamefont {Rabinovici}\ and\ \citenamefont
  {Samuel}(1981)}]{RabinoviciSamuel1981}%
  \BibitemOpen
  \bibfield  {author} {\bibinfo {author} {\bibfnamefont {E.}~\bibnamefont
  {Rabinovici}}\ and\ \bibinfo {author} {\bibfnamefont {S.}~\bibnamefont
  {Samuel}},\ }\href {\doibase http://dx.doi.org/10.1016/0370-2693(81)90054-X}
  {\bibfield  {journal} {\bibinfo  {journal} {Physics Letters B}\ }\textbf
  {\bibinfo {volume} {101}},\ \bibinfo {pages} {323} (\bibinfo {year}
  {1981})}\BibitemShut {NoStop}%
\bibitem [{\citenamefont {Corboz}\ \emph {et~al.}(2007)\citenamefont {Corboz},
  \citenamefont {L\"auchli}, \citenamefont {Totsuka},\ and\ \citenamefont
  {Tsunetsugu}}]{CorbozTsunetsugutrimerization}%
  \BibitemOpen
  \bibfield  {author} {\bibinfo {author} {\bibfnamefont {P.}~\bibnamefont
  {Corboz}}, \bibinfo {author} {\bibfnamefont {A.~M.}\ \bibnamefont
  {L\"auchli}}, \bibinfo {author} {\bibfnamefont {K.}~\bibnamefont {Totsuka}},
  \ and\ \bibinfo {author} {\bibfnamefont {H.}~\bibnamefont {Tsunetsugu}},\
  }\href {\doibase 10.1103/PhysRevB.76.220404} {\bibfield  {journal} {\bibinfo
  {journal} {Phys. Rev. B}\ }\textbf {\bibinfo {volume} {76}},\ \bibinfo
  {pages} {220404} (\bibinfo {year} {2007})}\BibitemShut {NoStop}%
\bibitem [{\citenamefont {Affleck}(1990)}]{Afflecknnbar1990}%
  \BibitemOpen
  \bibfield  {author} {\bibinfo {author} {\bibfnamefont {I.}~\bibnamefont
  {Affleck}},\ }\href {http://stacks.iop.org/0953-8984/2/i=2/a=016} {\bibfield
  {journal} {\bibinfo  {journal} {Journal of Physics: Condensed Matter}\
  }\textbf {\bibinfo {volume} {2}},\ \bibinfo {pages} {405} (\bibinfo {year}
  {1990})}\BibitemShut {NoStop}%
\bibitem [{\citenamefont {S\o{}rensen}\ and\ \citenamefont
  {Young}(1990)}]{SorensenPRB1990}%
  \BibitemOpen
  \bibfield  {author} {\bibinfo {author} {\bibfnamefont {E.~S.}\ \bibnamefont
  {S\o{}rensen}}\ and\ \bibinfo {author} {\bibfnamefont {A.~P.}\ \bibnamefont
  {Young}},\ }\href {\doibase 10.1103/PhysRevB.42.754} {\bibfield  {journal}
  {\bibinfo  {journal} {Phys. Rev. B}\ }\textbf {\bibinfo {volume} {42}},\
  \bibinfo {pages} {754} (\bibinfo {year} {1990})}\BibitemShut {NoStop}%
\bibitem [{\citenamefont {Imachi}\ \emph {et~al.}(2006)\citenamefont {Imachi},
  \citenamefont {Kambayashi}, \citenamefont {Shinno},\ and\ \citenamefont
  {Yoneyama}}]{ImachiYoneyamaPTP2006}%
  \BibitemOpen
  \bibfield  {author} {\bibinfo {author} {\bibfnamefont {M.}~\bibnamefont
  {Imachi}}, \bibinfo {author} {\bibfnamefont {H.}~\bibnamefont {Kambayashi}},
  \bibinfo {author} {\bibfnamefont {Y.}~\bibnamefont {Shinno}}, \ and\ \bibinfo
  {author} {\bibfnamefont {H.}~\bibnamefont {Yoneyama}},\ }\href {\doibase
  10.1143/PTP.116.181} {\bibfield  {journal} {\bibinfo  {journal} {Progress of
  Theoretical Physics}\ }\textbf {\bibinfo {volume} {116}},\ \bibinfo {pages}
  {181} (\bibinfo {year} {2006})}\BibitemShut {NoStop}%
\bibitem [{\citenamefont {Berg}\ and\ \citenamefont
  {L{\"u}scher}(1981)}]{BergLuscher1981}%
  \BibitemOpen
  \bibfield  {author} {\bibinfo {author} {\bibfnamefont {B.}~\bibnamefont
  {Berg}}\ and\ \bibinfo {author} {\bibfnamefont {M.}~\bibnamefont
  {L{\"u}scher}},\ }\href {\doibase
  http://dx.doi.org/10.1016/0550-3213(81)90568-X} {\bibfield  {journal}
  {\bibinfo  {journal} {Nuclear Physics B}\ }\textbf {\bibinfo {volume}
  {190}},\ \bibinfo {pages} {412 } (\bibinfo {year} {1981})}\BibitemShut
  {NoStop}%
\bibitem [{\citenamefont {Metropolis}\ \emph {et~al.}(1953)\citenamefont
  {Metropolis}, \citenamefont {Rosenbluth}, \citenamefont {Rosenbluth},
  \citenamefont {Teller},\ and\ \citenamefont {Teller}}]{Metropolis1953}%
  \BibitemOpen
  \bibfield  {author} {\bibinfo {author} {\bibfnamefont {N.}~\bibnamefont
  {Metropolis}}, \bibinfo {author} {\bibfnamefont {A.~W.}\ \bibnamefont
  {Rosenbluth}}, \bibinfo {author} {\bibfnamefont {M.~N.}\ \bibnamefont
  {Rosenbluth}}, \bibinfo {author} {\bibfnamefont {A.~H.}\ \bibnamefont
  {Teller}}, \ and\ \bibinfo {author} {\bibfnamefont {E.}~\bibnamefont
  {Teller}},\ }\href {\doibase 10.1063/1.1699114} {\bibfield  {journal}
  {\bibinfo  {journal} {The Journal of Chemical Physics}\ }\textbf {\bibinfo
  {volume} {21}},\ \bibinfo {pages} {1087} (\bibinfo {year} {1953})},\ \Eprint
  {http://arxiv.org/abs/http://dx.doi.org/10.1063/1.1699114}
  {http://dx.doi.org/10.1063/1.1699114} \BibitemShut {NoStop}%
\bibitem [{\citenamefont {Bronzan}(1988)}]{BroznanPRD1988}%
  \BibitemOpen
  \bibfield  {author} {\bibinfo {author} {\bibfnamefont {J.~B.}\ \bibnamefont
  {Bronzan}},\ }\href {\doibase 10.1103/PhysRevD.38.1994} {\bibfield  {journal}
  {\bibinfo  {journal} {Phys. Rev. D}\ }\textbf {\bibinfo {volume} {38}},\
  \bibinfo {pages} {1994} (\bibinfo {year} {1988})}\BibitemShut {NoStop}%
\bibitem [{\citenamefont {Hasenbusch}\ and\ \citenamefont
  {Meyer}(1992)}]{Hasenbuschmultigrid}%
  \BibitemOpen
  \bibfield  {author} {\bibinfo {author} {\bibfnamefont {M.}~\bibnamefont
  {Hasenbusch}}\ and\ \bibinfo {author} {\bibfnamefont {S.}~\bibnamefont
  {Meyer}},\ }\href {\doibase 10.1103/PhysRevLett.68.435} {\bibfield  {journal}
  {\bibinfo  {journal} {Phys. Rev. Lett.}\ }\textbf {\bibinfo {volume} {68}},\
  \bibinfo {pages} {435} (\bibinfo {year} {1992})}\BibitemShut {NoStop}%
\bibitem [{\citenamefont {Ambegaokar}\ and\ \citenamefont
  {Troyer}(2010)}]{Troyerbinning2010}%
  \BibitemOpen
  \bibfield  {author} {\bibinfo {author} {\bibfnamefont {V.}~\bibnamefont
  {Ambegaokar}}\ and\ \bibinfo {author} {\bibfnamefont {M.}~\bibnamefont
  {Troyer}},\ }\href {\doibase 10.1119/1.3247985} {\bibfield  {journal}
  {\bibinfo  {journal} {American Journal of Physics}\ }\textbf {\bibinfo
  {volume} {78}},\ \bibinfo {pages} {150} (\bibinfo {year} {2010})},\ \Eprint
  {http://arxiv.org/abs/http://dx.doi.org/10.1119/1.3247985}
  {http://dx.doi.org/10.1119/1.3247985} \BibitemShut {NoStop}%
\bibitem [{\citenamefont {Beard}\ \emph {et~al.}(2005)\citenamefont {Beard},
  \citenamefont {Pepe}, \citenamefont {Riederer},\ and\ \citenamefont
  {Wiese}}]{WieseCPN1PRL2005}%
  \BibitemOpen
  \bibfield  {author} {\bibinfo {author} {\bibfnamefont {B.~B.}\ \bibnamefont
  {Beard}}, \bibinfo {author} {\bibfnamefont {M.}~\bibnamefont {Pepe}},
  \bibinfo {author} {\bibfnamefont {S.}~\bibnamefont {Riederer}}, \ and\
  \bibinfo {author} {\bibfnamefont {U.-J.}\ \bibnamefont {Wiese}},\ }\href
  {\doibase 10.1103/PhysRevLett.94.010603} {\bibfield  {journal} {\bibinfo
  {journal} {Phys. Rev. Lett.}\ }\textbf {\bibinfo {volume} {94}},\ \bibinfo
  {pages} {010603} (\bibinfo {year} {2005})}\BibitemShut {NoStop}%
\bibitem [{\citenamefont {Vicari}\ and\ \citenamefont
  {Panagopoulos}(2009)}]{VicariPhysRep2009}%
  \BibitemOpen
  \bibfield  {author} {\bibinfo {author} {\bibfnamefont {E.}~\bibnamefont
  {Vicari}}\ and\ \bibinfo {author} {\bibfnamefont {H.}~\bibnamefont
  {Panagopoulos}},\ }\href {\doibase
  http://doi.org/10.1016/j.physrep.2008.10.001} {\bibfield  {journal} {\bibinfo
   {journal} {Physics Reports}\ }\textbf {\bibinfo {volume} {470}},\ \bibinfo
  {pages} {93} (\bibinfo {year} {2009})}\BibitemShut {NoStop}%
\bibitem [{\citenamefont {Lacroix}\ \emph {et~al.}(2011)\citenamefont
  {Lacroix}, \citenamefont {Mendels},\ and\ \citenamefont
  {Mila}}]{Lacroixmagnetism2011}%
  \BibitemOpen
  \bibinfo {editor} {\bibfnamefont {C.}~\bibnamefont {Lacroix}}, \bibinfo
  {editor} {\bibfnamefont {P.}~\bibnamefont {Mendels}}, \ and\ \bibinfo
  {editor} {\bibfnamefont {F.}~\bibnamefont {Mila}},\ eds.,\ \href {\doibase
  10.1007/978-3-642-10589-0} {\emph {\bibinfo {title} {Introduction to
  Frustrated Magnetism: Materials, Experiments, Theory}}},\ Vol.\ \bibinfo
  {volume} {164}\ (\bibinfo  {publisher} {Springer Berlin Heidelberg},\
  \bibinfo {year} {2011})\BibitemShut {NoStop}%
\bibitem [{\citenamefont {Azcoiti}\ \emph {et~al.}(2011)\citenamefont
  {Azcoiti}, \citenamefont {Follana},\ and\ \citenamefont
  {Vaquero}}]{AzcoitiNuclPhys2011}%
  \BibitemOpen
  \bibfield  {author} {\bibinfo {author} {\bibfnamefont {V.}~\bibnamefont
  {Azcoiti}}, \bibinfo {author} {\bibfnamefont {E.}~\bibnamefont {Follana}}, \
  and\ \bibinfo {author} {\bibfnamefont {A.}~\bibnamefont {Vaquero}},\ }\href
  {\doibase http://dx.doi.org/10.1016/j.nuclphysb.2011.05.023} {\bibfield
  {journal} {\bibinfo  {journal} {Nuclear Physics B}\ }\textbf {\bibinfo
  {volume} {851}},\ \bibinfo {pages} {420 } (\bibinfo {year}
  {2011})}\BibitemShut {NoStop}%
\bibitem [{\citenamefont {Rachel}\ \emph {et~al.}(2009)\citenamefont {Rachel},
  \citenamefont {Thomale}, \citenamefont {F\"uhringer}, \citenamefont
  {Schmitteckert},\ and\ \citenamefont {Greiter}}]{GreiterDMRG2009}%
  \BibitemOpen
  \bibfield  {author} {\bibinfo {author} {\bibfnamefont {S.}~\bibnamefont
  {Rachel}}, \bibinfo {author} {\bibfnamefont {R.}~\bibnamefont {Thomale}},
  \bibinfo {author} {\bibfnamefont {M.}~\bibnamefont {F\"uhringer}}, \bibinfo
  {author} {\bibfnamefont {P.}~\bibnamefont {Schmitteckert}}, \ and\ \bibinfo
  {author} {\bibfnamefont {M.}~\bibnamefont {Greiter}},\ }\href {\doibase
  10.1103/PhysRevB.80.180420} {\bibfield  {journal} {\bibinfo  {journal} {Phys.
  Rev. B}\ }\textbf {\bibinfo {volume} {80}},\ \bibinfo {pages} {180420}
  (\bibinfo {year} {2009})}\BibitemShut {NoStop}%
\bibitem [{\citenamefont {Calabrese}\ and\ \citenamefont
  {Cardy}(2009)}]{CalabreseCardy2009}%
  \BibitemOpen
  \bibfield  {author} {\bibinfo {author} {\bibfnamefont {P.}~\bibnamefont
  {Calabrese}}\ and\ \bibinfo {author} {\bibfnamefont {J.}~\bibnamefont
  {Cardy}},\ }\href {http://stacks.iop.org/1751-8121/42/i=50/a=504005}
  {\bibfield  {journal} {\bibinfo  {journal} {Journal of Physics A:
  Mathematical and Theoretical}\ }\textbf {\bibinfo {volume} {42}},\ \bibinfo
  {pages} {504005} (\bibinfo {year} {2009})}\BibitemShut {NoStop}%
\bibitem [{\citenamefont {Nataf}\ and\ \citenamefont
  {Mila}(2014)}]{NatafEDPRL2014}%
  \BibitemOpen
  \bibfield  {author} {\bibinfo {author} {\bibfnamefont {P.}~\bibnamefont
  {Nataf}}\ and\ \bibinfo {author} {\bibfnamefont {F.}~\bibnamefont {Mila}},\
  }\href {\doibase 10.1103/PhysRevLett.113.127204} {\bibfield  {journal}
  {\bibinfo  {journal} {Phys. Rev. Lett.}\ }\textbf {\bibinfo {volume} {113}},\
  \bibinfo {pages} {127204} (\bibinfo {year} {2014})}\BibitemShut {NoStop}%
\bibitem [{\citenamefont {Nataf}\ \emph {et~al.}()\citenamefont {Nataf},
  \citenamefont {Mila} \emph {et~al.}}]{NatafPC}%
  \BibitemOpen
  \bibfield  {author} {\bibinfo {author} {\bibfnamefont {P.}~\bibnamefont
  {Nataf}}, \bibinfo {author} {\bibfnamefont {F.}~\bibnamefont {Mila}},  \emph
  {et~al.},\ }\href@noop {} {}\bibinfo {note} {Private
  communication}\BibitemShut {NoStop}%
\bibitem [{\citenamefont {{Bykov}}(2012)}]{2012NuPhB.855..100B}%
  \BibitemOpen
  \bibfield  {author} {\bibinfo {author} {\bibfnamefont {D.}~\bibnamefont
  {{Bykov}}},\ }\href {\doibase 10.1016/j.nuclphysb.2011.10.005} {\bibfield
  {journal} {\bibinfo  {journal} {Nuclear Physics B}\ }\textbf {\bibinfo
  {volume} {855}},\ \bibinfo {pages} {100} (\bibinfo {year} {2012})},\ \Eprint
  {http://arxiv.org/abs/1104.1419} {arXiv:1104.1419 [hep-th]} \BibitemShut
  {NoStop}%
\bibitem [{\citenamefont {Ohmori}\ \emph {et~al.}(2019)\citenamefont {Ohmori},
  \citenamefont {Seiberg},\ and\ \citenamefont
  {Shao}}]{10.21468/SciPostPhys.6.2.017}%
  \BibitemOpen
  \bibfield  {author} {\bibinfo {author} {\bibfnamefont {K.}~\bibnamefont
  {Ohmori}}, \bibinfo {author} {\bibfnamefont {N.}~\bibnamefont {Seiberg}}, \
  and\ \bibinfo {author} {\bibfnamefont {S.-H.}\ \bibnamefont {Shao}},\ }\href
  {\doibase 10.21468/SciPostPhys.6.2.017} {\bibfield  {journal} {\bibinfo
  {journal} {SciPost Phys.}\ }\textbf {\bibinfo {volume} {6}},\ \bibinfo
  {pages} {17} (\bibinfo {year} {2019})}\BibitemShut {NoStop}%
\bibitem [{Kyo(2016)}]{Kyotoworkshop}%
  \BibitemOpen
  \href@noop {} {\enquote {\bibinfo {title} {Exotic states of matter with
  $\mbox{SU}(n)$ symmetry},}\ }\bibinfo {howpublished}
  {\url{http://www2.yukawa.kyoto-u.ac.jp/~sun2016/index.html}} (\bibinfo {year}
  {2016})\BibitemShut {NoStop}%
\bibitem [{\citenamefont {Bauer}\ \emph {et~al.}(2012)\citenamefont {Bauer},
  \citenamefont {Corboz}, \citenamefont {L\"{a}uchli}, \citenamefont {Messio},
  \citenamefont {Penc}, \citenamefont {Troyer},\ and\ \citenamefont
  {Mila}}]{bauer2012}%
  \BibitemOpen
  \bibfield  {author} {\bibinfo {author} {\bibfnamefont {B.}~\bibnamefont
  {Bauer}}, \bibinfo {author} {\bibfnamefont {P.}~\bibnamefont {Corboz}},
  \bibinfo {author} {\bibfnamefont {A.~M.}\ \bibnamefont {L\"{a}uchli}},
  \bibinfo {author} {\bibfnamefont {L.}~\bibnamefont {Messio}}, \bibinfo
  {author} {\bibfnamefont {K.}~\bibnamefont {Penc}}, \bibinfo {author}
  {\bibfnamefont {M.}~\bibnamefont {Troyer}}, \ and\ \bibinfo {author}
  {\bibfnamefont {F.}~\bibnamefont {Mila}},\ }\href {\doibase
  10.1103/PhysRevB.85.125116} {\bibfield  {journal} {\bibinfo  {journal}
  {Physical Review B}\ }\textbf {\bibinfo {volume} {85}},\ \bibinfo {pages}
  {125116} (\bibinfo {year} {2012})}\BibitemShut {NoStop}%
\bibitem [{\citenamefont {Fradkin}(2013)}]{fradkin2013}%
  \BibitemOpen
  \bibfield  {author} {\bibinfo {author} {\bibfnamefont {E.}~\bibnamefont
  {Fradkin}},\ }\href {https://books.google.ch/books?id=x7\_6MX4ye\_wC} {\emph
  {\bibinfo {title} {Field Theories of Condensed Matter Physics}}},\ Field
  Theories of Condensed Matter Physics\ (\bibinfo  {publisher} {Cambridge
  University Press},\ \bibinfo {year} {2013})\BibitemShut {NoStop}%
\bibitem [{\citenamefont {Cabra}\ and\ \citenamefont
  {Pujol}(2004)}]{Cabra2004}%
  \BibitemOpen
  \bibfield  {author} {\bibinfo {author} {\bibfnamefont {D.~C.}\ \bibnamefont
  {Cabra}}\ and\ \bibinfo {author} {\bibfnamefont {P.}~\bibnamefont {Pujol}},\
  }\enquote {\bibinfo {title} {Field-theoretical methods in quantum
  magnetism},}\ in\ \href {\doibase 10.1007/BFb0119596} {\emph {\bibinfo
  {booktitle} {Quantum Magnetism}}},\ \bibinfo {editor} {edited by\ \bibinfo
  {editor} {\bibfnamefont {U.}~\bibnamefont {Schollw{\"o}ck}}, \bibinfo
  {editor} {\bibfnamefont {J.}~\bibnamefont {Richter}}, \bibinfo {editor}
  {\bibfnamefont {D.~J.~J.}\ \bibnamefont {Farnell}}, \ and\ \bibinfo {editor}
  {\bibfnamefont {R.~F.}\ \bibnamefont {Bishop}}}\ (\bibinfo  {publisher}
  {Springer Berlin Heidelberg},\ \bibinfo {address} {Berlin, Heidelberg},\
  \bibinfo {year} {2004})\ pp.\ \bibinfo {pages} {253--305}\BibitemShut
  {NoStop}%
\bibitem [{\citenamefont {Petcher}\ and\ \citenamefont
  {L{\"u}scher}(1983)}]{PetcherLuscher1983}%
  \BibitemOpen
  \bibfield  {author} {\bibinfo {author} {\bibfnamefont {D.}~\bibnamefont
  {Petcher}}\ and\ \bibinfo {author} {\bibfnamefont {M.}~\bibnamefont
  {L{\"u}scher}},\ }\href {\doibase
  http://dx.doi.org/10.1016/0550-3213(83)90012-3} {\bibfield  {journal}
  {\bibinfo  {journal} {Nuclear Physics B}\ }\textbf {\bibinfo {volume}
  {225}},\ \bibinfo {pages} {53 } (\bibinfo {year} {1983})}\BibitemShut
  {NoStop}%
\end{thebibliography}%
\bibliographystyle{apsrev4-1}

\end{document}